\pdfoutput=1

\documentclass[11pt,twoside,a4paper,cmspaper,final,collab]{cms-tdr}
\def\svnVersion{177391}\def\cmsCernNo{2012-173}\def\cmsCernDate{\today}\def\cmsMessage{Submitted to the Journal of Instrumentation}

\begin{document}\cmsNoteHeader{MUO-10-004}

\hyphenation{had-ron-i-za-tion}
\hyphenation{cal-or-i-me-ter}
\hyphenation{de-vices}
\RCS$Revision: 130610 $
\RCS$HeadURL: svn+ssh://svn.cern.ch/reps/tdr2/papers/MUO-10-004/trunk/MUO-10-004.tex $
\RCS$Id: MUO-10-004.tex 130610 2012-06-18 20:44:51Z alverson $
\def\Zo{\ensuremath{\mathrm {Z}}}
\def\W{\ensuremath{\mathrm{W}}}%
\def\mm{\mu^+\mu^-}%
\def\ppZ{\pp\to\Zo + X}%
\def\ppbarW{\ppbar\to\W + X}%
\def\ppbarZ{\ppbar\to\Zo + X}%
\def\ppW{\pp\to\W + X}%
\def\ppZGmm{\pp\to\Zo(\gamma^*) + X \to \mm + X}%
\def\ppZmm{\pp\to\Zo + X \to \mm + X}%
\def\ppWmn{\pp\to\W + X \to \mu\nu + X}%
\def\ppWtn{\pp\to\W + X \to \tau\nu + X}%
\def\Zll{\Zo\to \ell\ell}%
\def\Zmm{\Zo\to\mm}%
\def\Umm{\Upsilon \to\mm}%
\def\Wln{\W\to\ell\nu}%
\def\Wmn{\W\to\mu\nu}%
\def\Wpmn{\W^+\to\mu^+\nu_\mu}%
\def\Wmmn{\W^-\to\mu^-\bar{\nu}_\mu}%
\newcommand{\jpsi}{\ensuremath{\mathrm{J}/\!\psi}}
\def\jpsimm{\jpsi\to\mm}%
\newcommand{\IRelComb} {I^{\text{rel}}_{\text{comb}}}%
\newcommand{\ITRK}     {I_{\text{trk}}}%
\newcommand{\IECAL}    {I_{\text{ECAL}}}%
\newcommand{\IHCAL}    {I_{\text{HCAL}}}%
\newcommand{\IRelTrk} {I^{\text{rel}}_{\text{trk}}}%
\newcommand{\IPF} {I^{\text{rel}}_{\textrm{PF}}}%
\newcommand{\fix}[1]{{\bf <<< #1 !!!} }
\cmsNoteHeader{MUO-10-004} % This is over-written in the CMS environment: useful as preprint no. for export versions
\title{Performance of CMS muon reconstruction in pp collision events at $\sqrt{s} = 7$\TeV}% Force line breaks with \\

\date{\today}

\abstract{
The performance of muon reconstruction, identification, and triggering in CMS
has been studied using 40\pbinv of data collected in pp collisions
at $\sqrt{s}$ = 7\TeV at the LHC in 2010.
A few benchmark sets of selection criteria
covering a wide range of physics analysis needs have been examined.
For all considered selections, the efficiency to reconstruct and identify
a muon with a transverse momentum $\pt$ larger than a few \GeVc is above 95\%
over the whole region of pseudorapidity covered by the CMS muon system,
$|\eta|<2.4$,
while the probability to misidentify a hadron as a muon is well below 1\%.
The efficiency to trigger on single muons
with $\pt$ above a few \GeVc is higher than 90\% over the full $\eta$ range, and typically substantially better.
The overall momentum scale is measured to a precision of 0.2\% with muons from Z decays.
The transverse momentum resolution varies from 1\% to 6\% depending on
pseudorapidity for muons with $\pt$ below $100\GeVc$ and, using
cosmic rays, it is shown to be better than 10\% in the
central region up to $\pt = 1\TeVc$.
Observed distributions of all quantities are
well reproduced by the Monte Carlo simulation.
 }

\hypersetup{%
pdfauthor={CMS Collaboration},%
pdftitle={Performance of CMS muon reconstruction in pp collision events at sqrt(s) = 7 TeV},%
pdfsubject={CMS},%
pdfkeywords={CMS, muon, physics, reconstruction, trigger}}

\maketitle %maketitle comes after all the front information has been supplied

\section{Introduction}

The primary aim of the Compact Muon Solenoid (CMS) Collaboration is to
discover physics underlying electro-weak symmetry breaking with the
favoured mechanism being the Higgs mechanism. Many diverse experimental
signatures from other potential new physics should also be detectable.
In order to cleanly detect these signatures the identification and
precise energy measurement of muons, electrons, photons and jets over a
large energy range and at high luminosities is essential.

In this paper we report on the performance of
muon reconstruction, identification, and triggering evaluated
using the data collected by the CMS detector
at the Large Hadron Collider (LHC) at CERN during 2010.
During that period the CMS experiment recorded a sample of events produced
in proton--proton collisions at a centre-of-mass energy of $\sqrt{s} = 7\TeV$
with an integrated luminosity of 40~pb$^{-1}$.
Muon reconstruction in CMS has been previously studied in great
detail using muons from cosmic rays~\cite{CRAFT08,CMS_CFT_09_014}. 
The first studies using 60~nb$^{-1}$ of 2010 proton--proton collision
data were reported in Ref.~\cite{MUO-10-002}.

\begin{figure}[hbtp]
  \begin{center}
    \includegraphics[width=0.7\textwidth]{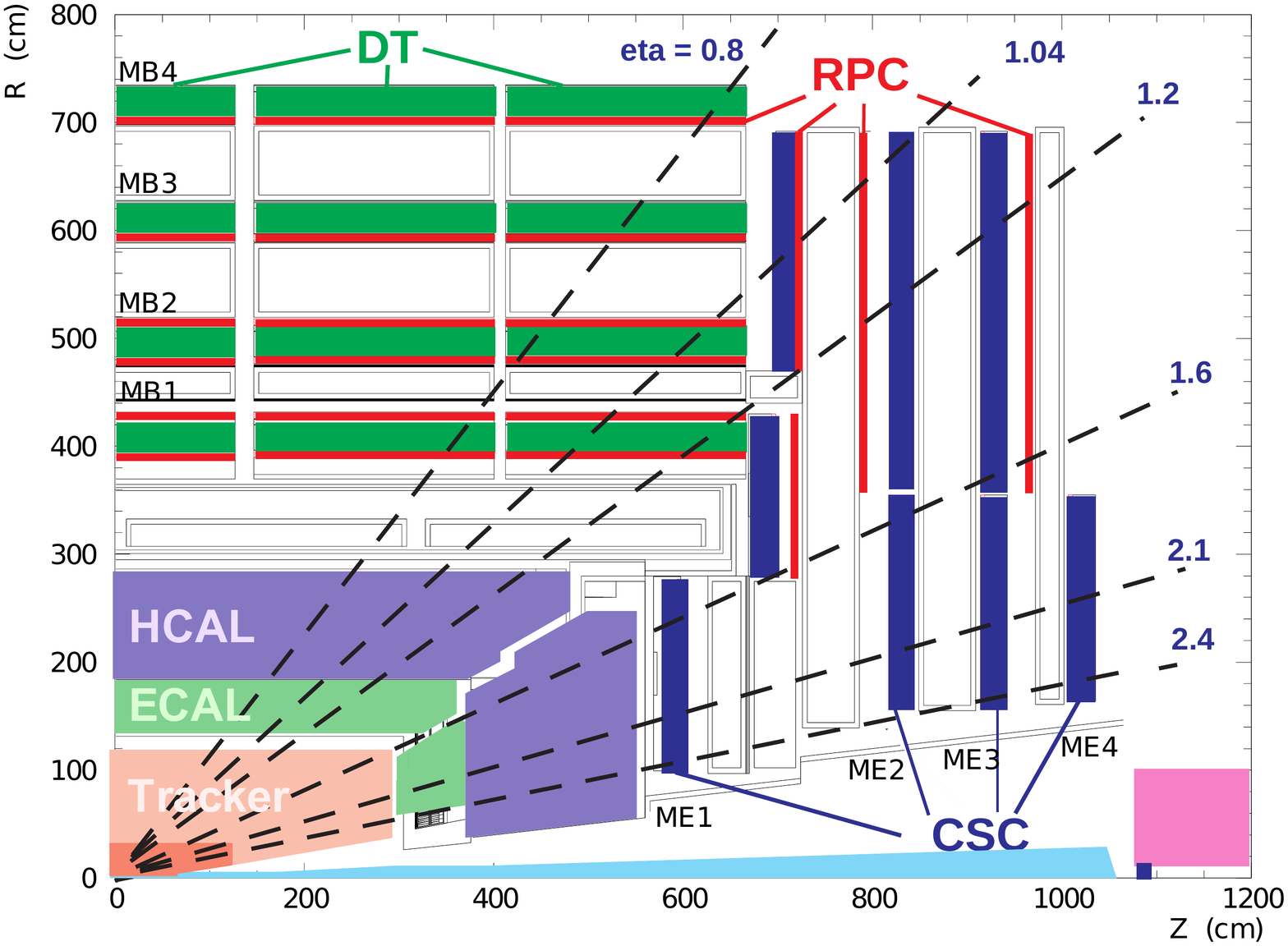}
    \caption{Longitudinal layout of one quadrant of the CMS detector.  The four DT stations
     in the barrel (MB1--MB4, green), the four CSC stations in the endcap (ME1--ME4, blue),
     and the RPC stations (red) are shown.}
    \label{fig:Display}
  \end{center}
\end{figure}

\begin{figure}[hbt!]
  \begin{center}
    \vspace*{-0.3cm}
    \hspace*{-0.7cm}\includegraphics[width=0.56\textwidth]{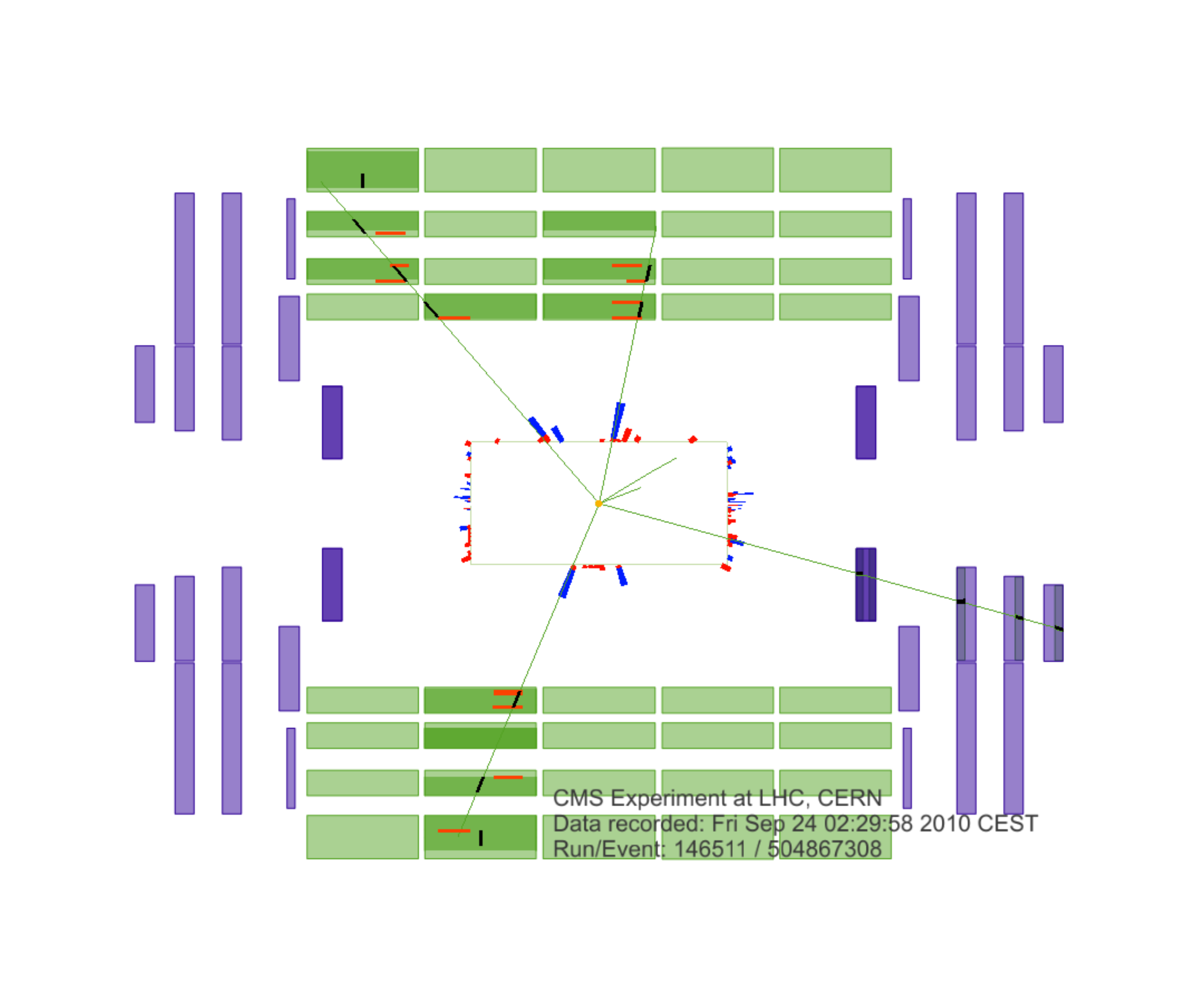}
    \put(-130,3){(a)}
    \hspace*{-1.0cm}\includegraphics[width=0.54\textwidth]{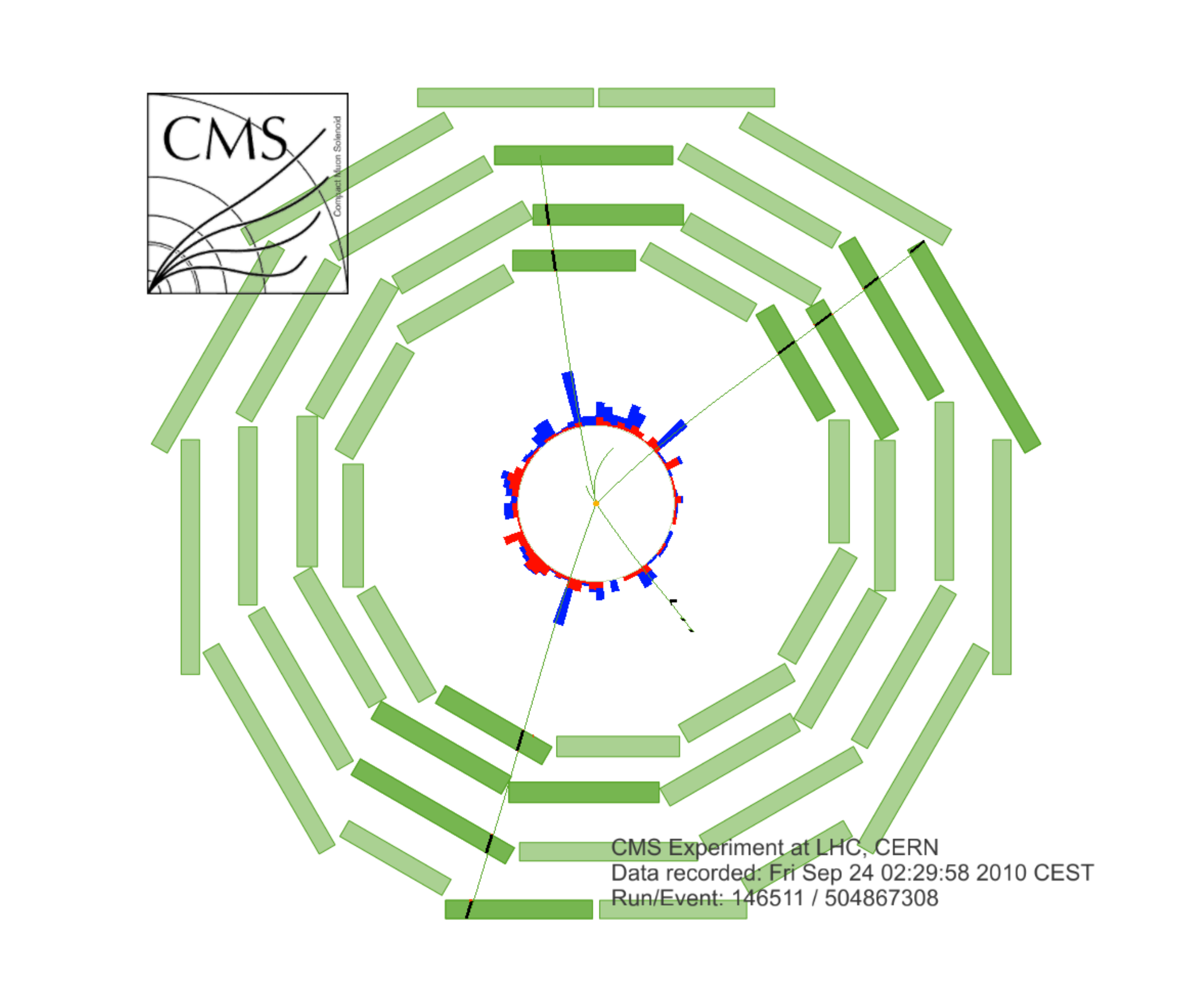}
    \put(-130,2){(b)}
    \caption{(a) The longitudinal ($r$-$z$) and b) the transverse
     ($r$-$\phi$) views of a collision event in which four muons were
     reconstructed.  The green (thin) curves in the inner cylinder represent
     tracks of charged particles reconstructed in the inner tracker with
     transverse momentum $\pt > 1\GeVc$; those extending to the muon
     system represent tracks of muons reconstructed using hits in both
     inner tracker and the muon system.  Three muons were identified
     by the DTs and RPCs, the fourth one by the CSCs.  Short black
     stubs in the muon system show fitted muon-track segments; as the $z$
     position is not measured in the outer barrel station, the
     segments in it are drawn at the $z$ centre of the wheel, with
     their directions perpendicular to the chamber.  Short red (light)
     horizontal lines in the $r$-$z$ view indicate positions of RPC
     hits; energy depositions in the ECAL and HCAL are shown as red (light)
     and blue (dark) bars, respectively.}
    \label{fig:EvDisplay}
  \end{center}
\end{figure}

A detailed description of the CMS detector can be found in Ref.~\cite{cms}.
A schematic view of the detector is shown in Fig.~\ref{fig:Display}.
Muon reconstruction is performed using the all-silicon inner tracker
at the centre of the detector immersed in a 3.8~T solenoidal magnetic
field, and with up to four stations of gas-ionization
muon detectors %~\cite{muonTDR}
installed outside the solenoid and sandwiched
between the layers of the steel return yoke.
The inner tracker is composed of a pixel detector and a silicon strip tracker,
and measures charged-particle trajectories in the pseudorapidity range
$|\eta|<2.5$\footnote{A
right-handed coordinate system is used in CMS, with the origin at the
nominal collision point, the $x$ axis pointing to the centre of the
LHC ring, the $y$ axis pointing up (perpendicular to the LHC plane),
and the $z$ axis along the anticlockwise-beam direction. The
pseudorapidity $\eta$ is defined as $\eta = -\ln \tan (\theta/2)$,
where $\cos \theta = p_z/p$.  The radius $r$ is the distance from the
$z$ axis; the azimuthal angle $\phi$ is the angle relative to the positive
$x$ axis measured in the $x$-$y$ plane.}.
The muon system covers the pseudorapidity region $|\eta|<2.4$
and performs three main tasks: triggering on muons, identifying muons, and
improving the momentum measurement and charge determination of high-$\pt$
muons.  Drift tube (DT) chambers and cathode strip
chambers (CSC) detect muons in the $\eta$ regions of $|\eta|<1.2$ and
$0.9 <|\eta|< 2.4$, respectively, and are complemented by a system of resistive
plate chambers (RPC) covering the range of $|\eta|<1.6$.
The use of these different technologies defines three regions in the detector, referred to as barrel ($|\eta|<0.9$), overlap ($0.9 <|\eta|< 1.2$), and endcap ($1.2 < |\eta| < 2.4$).
Muon energy deposits in the electromagnetic calorimeter (ECAL), hadronic
calorimeter (HCAL), and outer hadronic calorimeter (HO) are also used for
muon identification purposes.  An event in which four muons were reconstructed
involving all main CMS subdetectors is shown in Fig.~\ref{fig:EvDisplay}.

For the measurement of muons the single most important aspect is the
choice of magnetic field configuration. The requirement of a good
momentum resolution, specified to be $\sigma(\pt)/\pt$ $\sim$1\% at 100\GeVc and $\sim$10\% at
1\TeVc, without making stringent demands on spatial resolution and the
alignment of muon chambers leads to the choice of a high magnetic field.
CMS chose a high-field solenoid. The magnetic flux generated by the
central field is large enough to saturate a sufficient amount of steel
in the return yoke to allow the installation of four muon stations. This
provides a good level of redundancy in the measurement of muons. The
favourable length to radius ratio allows efficient muon measurement up
to pseudorapidity of $|\eta|<2.4$. The strong magnetic field also enables an
efficient first-level trigger with an acceptable rate.

The CMS experiment uses a two-level trigger system. The \Lone
trigger, composed of custom hardware processors, selects events
of interest using information from the calorimeters and muon
detectors and reduces the read-out rate from the 20~MHz bunch-crossing 
frequency to a maximum of 100~kHz~\cite{L1TDR}.  The high-level trigger
(HLT) is software-based and further decreases the recorded event rate
to around 300~Hz by using the full event
information, including that from the inner tracker~\cite{HLTTDR}.

Several types of triggers implemented for the 2010 data taking have been used for the present studies.
These are discussed in Section~\ref{sec:samples}, together with the resulting data and simulated samples.
Muon reconstruction and identification algorithms are described in
Section~\ref{sec:reco}. The measured distributions of various kinematic variables of selected muons are compared with simulation in Section~\ref{sec:kinematics}.
Section~\ref{sec:muonideff} presents muon reconstruction and identification
efficiencies for exclusive samples of prompt muons, kaons, pions, and
protons. Section~\ref{sec:momentumscale} summarizes the results on
muon momentum scale and resolution for different muon momentum ranges.
Backgrounds from cosmic rays and beam-halo muons are discussed in
Section~\ref{sec:cosmics}.  Section~\ref{sec:isolation} describes the
performance of different isolation algorithms. Muon trigger
performance is discussed in Section~\ref{sec:trigger}.
Section~\ref{sec:conclusions} gives a summary of our conclusions.

\section{Data and Monte Carlo Samples}
\label{sec:samples}

The data samples used for the muon performance studies reported in this paper
were collected with the following types of triggers:
\begin{itemize}
\item{{\em The zero-bias trigger}, defined by the coincidence of signals
in two dedicated beam position monitors (Beam Position and Timing for LHC
eXperiments, BPTX) in the same bunch crossing.
Its rate was kept constant at around 20~Hz throughout the year by adjusting
the prescale factor to compensate for the rising instantaneous luminosity.
Events collected by this trigger do not suffer from any muon-detection bias at trigger level and are used to define an inclusive muon sample for the study of muon kinematic distributions as discussed in Section~\ref{sec:kinematics}.}
\item{{\em Single-muon triggers.}  Muon candidates are reconstructed at the trigger level using information from the muon detectors and the inner tracker. Events containing a muon candidate with online-reconstructed transverse momentum $\pt$ greater than a predefined threshold (luminosity-dependent, 15\GeVc or lower in the year 2010) are recorded. All muon triggers used in 2010 data taking covered the full muon detector acceptance corresponding to $|\eta|<2.4$.
These triggers were the main source of intermediate-$\pt$ and high-$\pt$ muons during 2010 data taking, efficiently selecting, \eg, muonic decays of $\W$ and $\Z$ bosons.
Furthermore, to collect cosmic-ray data during breaks in LHC
operation, a trigger requiring at least two loosely matched segments
in the bottom half of the barrel muon system or a single segment in
the endcap muon system was implemented. The triggers from the top half
were disabled to avoid the need for special synchronization.}
\item{{\em Muon-plus-track triggers.} To improve the efficiency of collecting $\jpsi$ events, a specialized high-level trigger was implemented.
This trigger selected events in which the muon track can be paired with an inner-tracker track of opposite charge yielding an invariant mass close to that of the $\jpsi$.
To sample the efficiencies in the whole $\pt$ region evenly, multiple instances of the trigger were deployed with different thresholds on the transverse momentum of the inner-tracker track.
In addition, another set of specific $\jpsi$ triggers was implemented, using only the muon system for the reconstruction of one of the two muons.
The muon-plus-track triggers were used to measure identification and trigger efficiencies for low-$\pt$ muons, as described in Sections~\ref{sec:muonideff} and \ref{sec:trigger}}.
\item{{\em Jet and missing transverse energy ($\MET$) triggers.} Using calorimeter information, jets and missing transverse energy are reconstructed online. Triggers with different thresholds on jet transverse energy and $\MET$ were implemented.
These events were used to select a sample of muons that was unbiased by the requirements of the muon trigger.}
\end{itemize}

\begin{figure}[htbp]
\begin{center}
\includegraphics[width=0.8\textwidth]{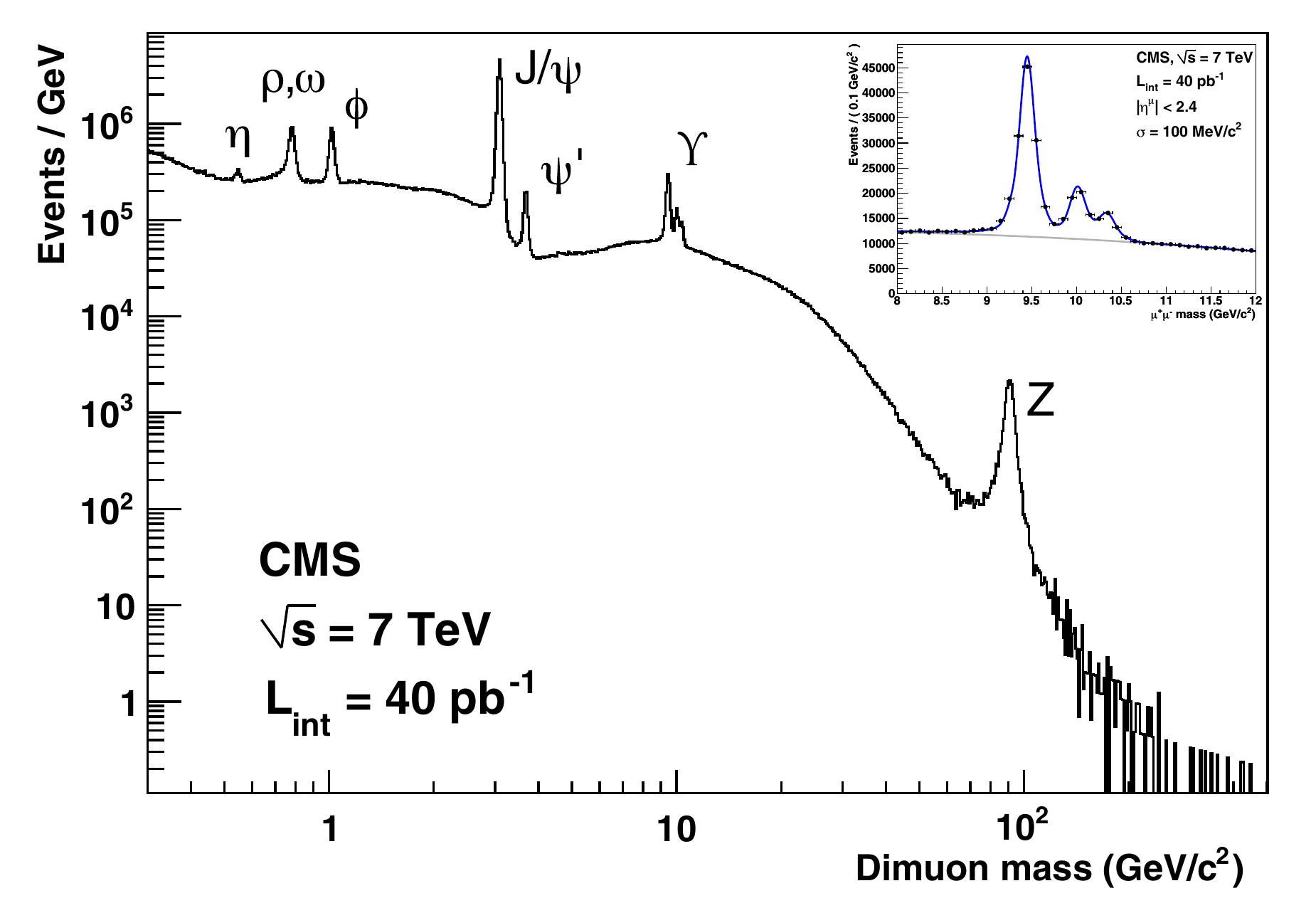}
  \caption{Invariant mass spectrum of dimuons in events collected with the
    loose double-muon trigger in 2010.  The inset is a zoom of the
    8--12\GeVcc region, showing the three $\Upsilon(nS)$ peaks clearly
    resolved owing to a good mass resolution, about 100\MeVcc in the entire
    pseudorapidity range and 70\MeVcc when both muons are within the range $|\eta| < 1$.}
  \label{fig:dimuon_mass_spectrum}
\end{center}
\end{figure}

In addition, a loose {\em double-muon trigger} requiring two or more
muon candidates reconstructed online and not applying any additional
selection criteria was implemented, taking advantage of the relatively
low luminosity during 2010 data taking.  This trigger selected
dimuons in the invariant mass region spanning more than three
orders of magnitude, from a few hundred \MeVcc to a few hundred
\GeVcc, as shown in Fig.~\ref{fig:dimuon_mass_spectrum}.  The events
collected with this trigger were used in both the detector
commissioning and physics studies.

All collision data samples studied in this paper were filtered by requiring at least one well-reconstructed primary vertex
to reduce the contamination from non-collision backgrounds. Techniques to further suppress the non-collision backgrounds according to
the needs of physics analysis are discussed in Section~\ref{sec:cosmics}.

To compare the results obtained in data to predictions, a number of simulated samples were produced using Monte Carlo (MC) techniques.
All MC samples were produced with the $\textsc{cteq6l}$~\cite{cteq6l} set of parton distribution functions and different event generators were used depending on the process considered.
Samples of $\ttbar$ and QCD multijet events were generated using \PYTHIA 6~\cite{PYTHIA} with the Z2 tune~\cite{QCD-10-010}, as well as
inclusive muon-enriched samples, in which only events containing at least one muon with transverse momentum greater than a given threshold were selected at generation level.
Samples of prompt $\jpsi$ mesons as well as $\jpsi$ particles originating from the decays of b hadrons were generated with \PYTHIA interfaced to $\textsc{evtgen}$~\cite{evtgen}.
Inclusive $\W$ and $\Z$ samples and non-resonant Drell--Yan events were produced using the $\textsc{powheg}$~\cite{powheg} event generator, interfaced with \PYTHIA for the simulation of parton showering and hadronization processes.
For $\W$+jets and $\Z$+jets samples with a given number of jets, the $\textsc{madgraph}$~\cite{Madgraph} event generator was used, combined with \PYTHIA for showering and hadronization.

Generated events were processed through a full
\GEANTfour-based~\cite{Agostinelli:2002hh, Allison:2006ve} detector
simulation, trigger emulation, and event reconstruction chain.  A
realistic misalignment scenario based on the knowledge of positions
of different elements of the inner-tracker and muon systems was used to describe the detector
geometry.  The positions of the tracker modules were evaluated by
applying the track-based alignment procedure to a sample of 2.2
million cosmic-ray muons and 3.3 million minimum-bias events collected
in 2010.  The residual uncertainties in the positions of individual
tracker modules were measured to be smaller than 6\micron in the pixel
detector and less than 10\micron in the silicon strip tracker.  The
procedure used to emulate the remaining misalignment effects in the
simulation closely followed that employed to align the tracker using
data, and used as input the module displacements determined from cosmic-ray data available at the
time of the MC sample production.  As
a result, the simulated geometry of the tracker included coherent
displacements and rotations of tracker modules that could
bias the track reconstruction without affecting the local alignment
precision.  These coherent movements were estimated not to exceed
200\micron.  The alignment precision for muon chambers was estimated
by comparison with photogrammetry to be about 500\micron for DT
chambers and between 300 and 600\micron (depending on the ring) for
CSCs, in the $r$-$\phi$ plane.  The misalignment scenario for the
muon chambers used in the simulation was consistent
with this precision.

Unless stated otherwise, additional proton--proton interactions in the
same bunch crossing (pile-up) were not simulated because of their relatively
small rate in 2010 (2.7 interactions on average). Such pile-up is expected to have
a negligible effect on the results presented here.

\section{Muon Reconstruction and Identification}
\label{sec:reco}

In the standard CMS reconstruction for pp
collisions~\cite{CMS_CFT_09_014,PTDR1}, tracks are first  
reconstructed independently in the inner tracker ({\em tracker
track}) and in the muon system ({\em standalone-muon track}). Based on
these objects, two reconstruction approaches are used:

\begin{itemize}
  \item {\em Global Muon reconstruction (outside-in).}  For each
    standalone-muon track, a matching tra\-cker track is found by comparing
    parameters of the two tracks propagated onto a common surface. A
    {\em global-muon track} is fitted combining hits from the tracker track
    and standalone-muon track, using the Kalman-filter
    technique~\cite{Fruhwirth:1987fm}.
    At large transverse momenta, $\pt \gtrsim 200\GeVc$,
    the global-muon fit can improve the momentum resolution compared
    to the tracker-only fit~\cite{CMS_CFT_09_014,PTDR1}. 
  \item {\em Tracker Muon reconstruction (inside-out).} In this approach, all
    tracker tracks with $\pt >$ 0.5\GeVc and total momentum
    $p >$ 2.5\GeVc are
    considered as possible muon candidates and are extra\-po\-lated to the muon system taking into
    account the magnetic field, the average expected energy losses, and multiple Coulomb
    scattering in the detector material. If at least one muon segment (\ie, a short track stub made of DT or CSC hits)
    matches the extrapolated track,
    the corresponding tracker track qualifies as a Tracker Muon.
    Track-to-segment matching is performed in a local (chamber) coordinate
    system, where local $x$ is the best-measured coordinate (in the
    $r$-$\phi$ plane) and local $y$ is the coordinate orthogonal to it.
    The extrapolated track and the segment are considered to be matched if the distance
    between them in local $x$ is less than 3~cm or if the value of the pull
    for local $x$ is less than 4, where the pull is defined as the difference
    between the position of the matched segment and the position
    of the extrapolated track, divided by their combined
    uncertainties~\cite{CMS_CFT_09_014}.
\end{itemize}

Tracker Muon reconstruction is more efficient than the Global Muon
reconstruction at low momenta, $p \lesssim 5\GeVc$, because it requires
only a single muon segment in the muon system, whereas Global Muon
reconstruction is designed to have high efficiency for muons
penetrating through more than one muon station and typically requires
segments in at least two muon stations.

Owing to the high efficiency of the tracker-track reconstruction~\cite{TRK-10-002} and the very
high efficiency of reconstructing segments in the muon system, about 99\%
of muons produced in pp collisions within the geometrical acceptance of 
the muon system and having sufficiently high momentum
are reconstructed either as a Global Muon or a Tracker Muon, and very
often as both.  Candidates found both by the Global Muon and the Tracker
Muon approaches that share the same tracker track are merged into
a single candidate.  Muons reconstructed only as standalone-muon tracks
have worse momentum resolution and higher admixture of
cosmic-ray muons than the Global and Tracker Muons and are
usually not used in physics analyses.

The combination of different algorithms provides robust and
efficient muon reconstruction. Physics analyses can set
the desired balance between identification efficiency and purity by
applying a selection based on various muon identification variables.
In this paper we study the performance of three basic muon
identification algorithms:
\begin{itemize}
  \item {\em Soft Muon selection.} This selection requires the
    candidate to be a Tracker Muon, with the additional requirement
    that a muon segment is matched in both $x$ and $y$ coordinates with
    the extrapolated tracker track,
    such that the pull for local $x$ and $y$ is less than 3.
    Segments that form a better match %in position
    with a different tracker track are not considered.  These
    additional requirements are optimized for low $\pt$
    ($<10\GeVc$) muons.  This selection is used in quarkonia and
    B-physics analyses in CMS~\cite{BPH-10-002}.
  \item {\em Tight Muon selection.} For this selection, the candidate
    must be reconstructed outside-in as a Global Muon with the
    $\chi^2/d.o.f.$ of the
    global-muon track fit less than 10 and at least one muon chamber hit
    included in the global-muon track fit.  In addition, its corresponding
    tracker track is required to be matched to muon segments in at
    least two muon stations
    (this implies that the muon is also reconstructed inside-out as a Tracker Muon),
    use more than 10 inner-tracker hits
    (including at least one pixel hit), and have a transverse impact
    parameter $|d_{\rm xy}| < 2$~mm with respect to the primary vertex.
    With this selection, the rate of muons from decays in flight is
    significantly reduced (see Section~\ref{sec:kinematics}), at the
    price of a few percent loss in efficiency for prompt muons such as
    those from $\W$ and $\Z$ decays (see Section~\ref{sec:muonideff}).  The
    Tight Muon selection is used in many physics analyses in CMS, in
    particular in the measurements of inclusive $\W$ and $\Z$ cross
    sections~\cite{EWK-10-002, EWK-10-005}.
  \item {\em Particle-Flow Muon selection.}
    The CMS particle-flow event reconstruction algorithm~\cite{PFT-09-001} combines information from all CMS subdetectors to identify and reconstruct individual particles like electrons, hadrons or muons.  For muons, the particle-flow approach applies particular selection criteria to the muon candidates reconstructed with the Global and Tracker Muon algorithms described above. Depending on the environment of the muon (for example, whether it is isolated or not) the selection criteria are adjusted making use of information from other subdetectors (for example, the energy deposition in the calorimeters). In general, the selection is optimized in order to identify muons within jets with high efficiency, while maintaining a low rate for the misidentification of charged hadrons as muons. The details of the particle-flow muon selection are described in Ref.~\cite{PFT-10-003}.
\end{itemize}

The default algorithm for muon momentum assignment in CMS is called the ``sigma switch''.
This algorithm chooses from the momentum estimates given by
the tracker-only fit and by
the global fit. 
The global fit is chosen when both fits 
yield muon $\pt$ above 200\GeVc and 
give the charge-to-momentum ratios $q/p$ that
agree to within $2\sigma_{q/p}$ of the tracker-only fit;
in all other cases the tracker-only fit is taken.

In addition, CMS has developed specialized algorithms for high-$\pt$ muon 
reconstruction and momentum assignment.
As the muon passes through
the steel of the magnet return yoke, multiple scattering and radiative
processes can alter the muon trajectory.  While the former is not so
important for high-momentum muons, the latter can result in large energy
losses and can also produce electromagnetic showers giving rise to additional
hits in the muon chambers.  As a consequence, the estimate of the muon
momentum at the production vertex 
can be significantly different from its true value.
Therefore, several different strategies for including information
from the muon system have been developed and studied using cosmic
rays~\cite{CMS_CFT_09_014}:

\begin{itemize}
 \item {\em Tracker-Plus-First-Muon-Station (TPFMS) fit.} This algorithm 
  refits the global-muon track ignoring hits in all muon stations
  except the innermost one containing hits, for reduced sensitivity
  to possible showering deeper in the muon system.
\item {\em The Picky fit.}  This algorithm again starts with the hit list of
  the global-muon track, but, in chambers appearing to have hits from showers
  (determined by the hit occupancy of the chamber), retains only the hits
  that, based on a $\chi^2$ comparison, are compatible with the
  extrapolated trajectory.
\end{itemize}

To further improve the resolution at high $\pt$, mainly by 
reducing the tails of the momentum resolution distribution,
combinations of the above can be used.
In particular, the {\em Tune P} algorithm chooses, on a muon-by-muon basis,
between the tracker-only, TPFMS, and Picky fits. The algorithm starts with the Picky fit, then
switches to the tracker-only fit if the goodness of fit of the latter is
significantly better. Then it compares the goodness of fit of the
chosen track with that of TPFMS; TPFMS is chosen if it is found to
be better. 
For high-$\pt$ muons, TPFMS and Picky algorithms are selected by Tune P in most of the cases, in approximately equal amounts, while the tracker-only fit is selected only in a few percent of events.
For most analyses of the 2010 LHC data involving high-$\pt$ muons,
Tune P was used for the determination of the muon momentum.

\section{General Comparisons between Data and Simulation}
\label{sec:kinematics}

In this section we present data-to-simulation comparisons
for two samples of muons:
1) a fully inclusive sample of low-$\pt$ muons collected
with the zero-bias trigger,
and 2) an inclusive sample of intermediate- and high-$\pt$ muons collected with the single-muon trigger requiring a minimum transverse momentum of 15\GeVc.

Events collected with the zero-bias trigger were required to contain
at least one reconstructed primary vertex within 24~cm of the geometric
centre of the detector along the beamline and within a transverse distance
from the beam axis of less than 2 cm.
The efficiency of this requirement for simulated pp collisions having
at least one reconstructed muon was found to be 99\%.
About 14 million minimum-bias events were thus selected from a total
data sample of events corresponding to 0.47 nb$^{-1}$ of integrated luminosity;
the contamination from cosmic-ray muons in the sample was estimated to
be negligible.
The corresponding MC sample consists of about 36 million minimum-bias
events generated using \PYTHIA. %, surviving the PV requirement.

The sample of events collected with the single-muon trigger with the $\pt$
threshold of 15\GeVc consists of about 20 million events corresponding
to an integrated luminosity of 31~pb$^{-1}$.
Monte Carlo samples used for the comparison correspond to about
10 times larger integrated luminosity and
include the simulation of QCD processes, quarkonia production,
electroweak processes such as $\W$ and $\Z$ boson production, non-resonant
Drell--Yan processes, and top-pair production.  The total cross sections
for $\W$ and $\Z$ production were rescaled to match the
next-to-next-to-leading-order (NNLO) calculations; the cross sections
for b-hadron and $\ttbar$ production were rescaled to the
next-to-leading-order (NLO) calculations (see Section~\ref{sec:kincomp}).

In both cases the simulation was normalized according to the integrated
luminosity of the data sample.
The uncertainty in the absolute value of luminosity was estimated to be 4\%~\cite{Burkhardt:1347440}.

\subsection{Classification of muon sources in simulation}
In the range of $\pt \lesssim 30\GeVc$, the most abundant source of muons
is semileptonic decays of heavy-flavour hadrons.
This contribution is accompanied by a high rate of muon candidates arising from light-flavour hadron decays and
hadron showers not fully contained in the calorimeters.
The relative weights of these background contributions are quite sensitive to the details of the muon selection.
Muons from decays of $\W$ and $\Z$ bosons dominate the $\pt$ spectrum
in the region $\pt \gtrsim 30\GeVc$.

In the simulation, for each reconstructed muon, the hits in the muon system
can be associated unambiguously
with the simulated particle that produced them. This allows the
classification of reconstructed muons into the following categories:
\begin{itemize}
  \item {\em Prompt muons.} Here the majority of muon chamber hits associated with the reconstructed muon candidate
           were produced by a muon, arising either from decays of $\W$, $\Z$, and promptly produced quarkonia states,
           or other sources such as Drell--Yan processes or top quark production.
           These individual sources are shown separately where appropriate.
  \item {\em Muons from heavy flavour.} Here the majority of muon chamber hits of the muon
          candidate were again produced by a muon, but
          the muon's parent particle was a beauty or charmed hadron, or a $\tau$ lepton.
          This class of events has been split according to the heaviest flavour generated in the event.
          Hence, {\em beauty} includes muons from direct b-hadron decays, from cascade b $\rightarrow$ c hadron decays, as well as
          cascade decays of $\tau$ leptons from b hadrons.
  \item {\em Muons from light flavour.} In this category, the majority of muon chamber hits of
    the muon candidate were produced by a muon
    arising from a decay in flight of light hadrons ($\pi$ and K)
    or, less frequently,
    from the decay of particles produced in nuclear interactions in the detector material.
    This category includes hadrons whose tracks reconstructed in the tracker were mistakenly matched
    to the muon chamber hits.
   \item {\em Hadron punch-through.} Here the majority of muon chamber hits
    of the misidentified muon candidate were
    produced by a particle that was not a muon. "Punch-through"
    (\ie, hadron shower remnants penetrating through the calorimeters and
    reaching the muon system) is the most common source of these candidates,
    although "sail-through" (\ie, particles not undergoing nuclear interactions upstream of the muon system) is present as well.
  \item {\em Duplicate.} If one simulated particle gives rise
    to more than one reconstructed muon candidate, that with the
    largest number of matched hits is assigned to one
    of the above categories, and any others are labeled as ``duplicate''.
    Duplicate candidates can arise either from failures of the pattern recognition of the reconstruction software, or from patterns that mimic multiple candidates.
\end{itemize}

\subsection{Kinematic distributions of muons} \label{sec:kincomp}

From the 2010 data sample of zero-bias events with a well-established primary vertex, we obtain 318\,713 muon candidates passing the
Soft Muon selection and 24\,334 passing the Tight Muon selection.
The overall ratio of the number of muon candidates in data to the prediction of the \PYTHIA MC generator
normalized to the same integrated luminosity is 1.05 for Soft Muons and 1.01 for Tight Muons.

\begin{figure}[p]
  \centering
  \includegraphics[width=0.48\textwidth]{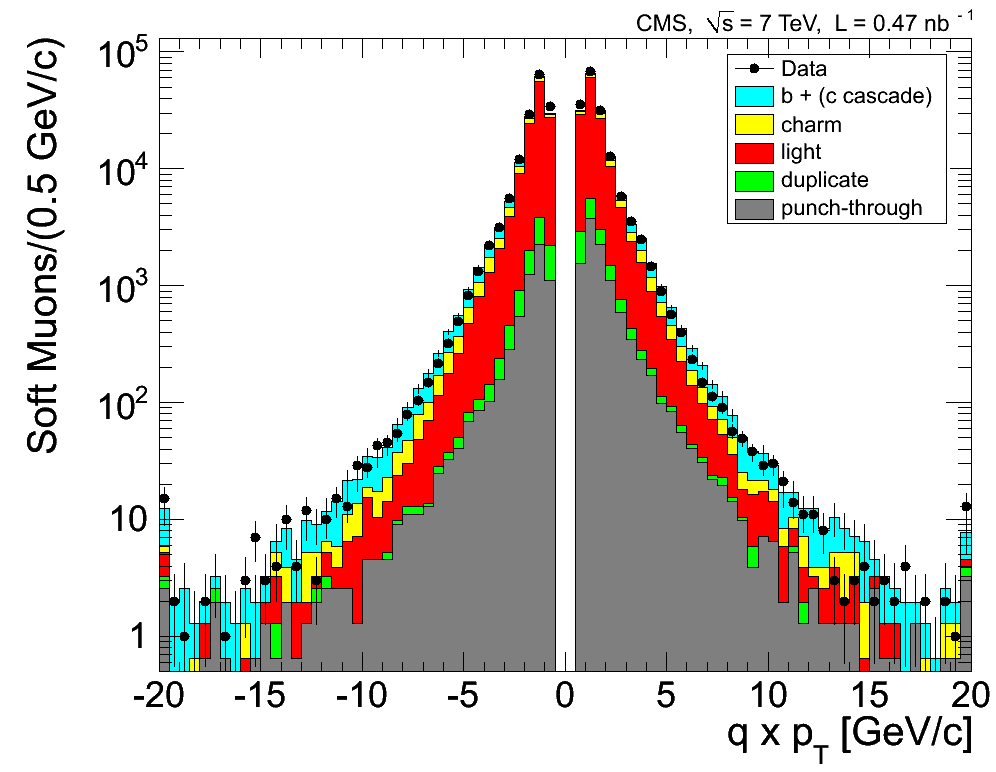}
  \includegraphics[width=0.48\textwidth]{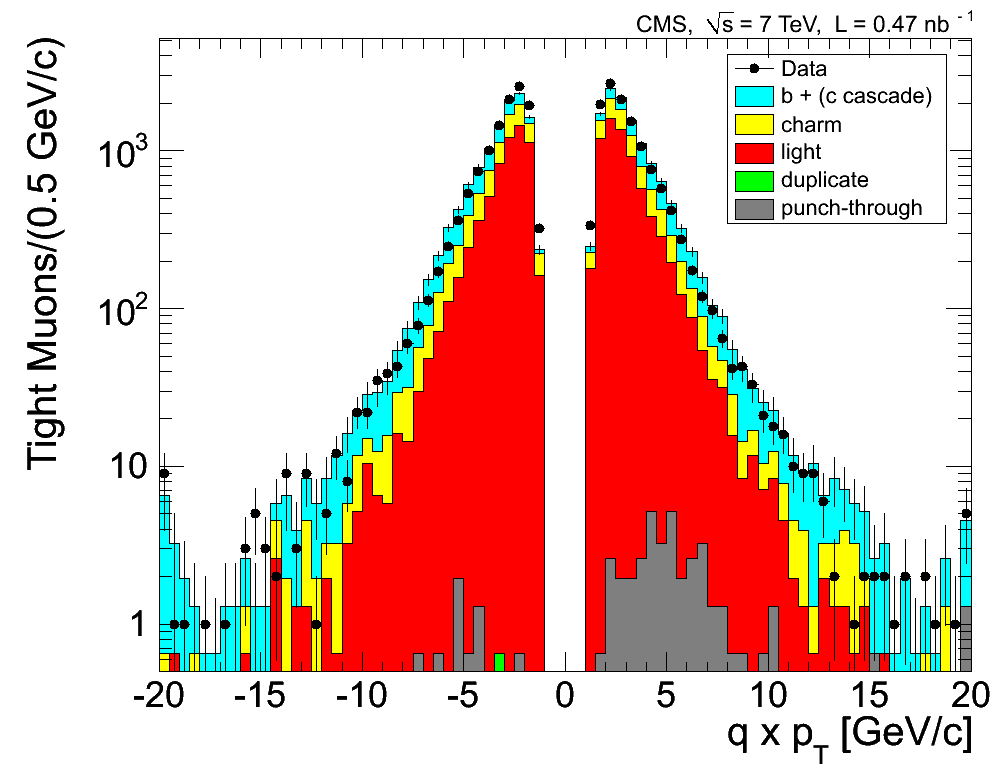}\\ \vspace{0.5cm}
  \includegraphics[width=0.48\textwidth]{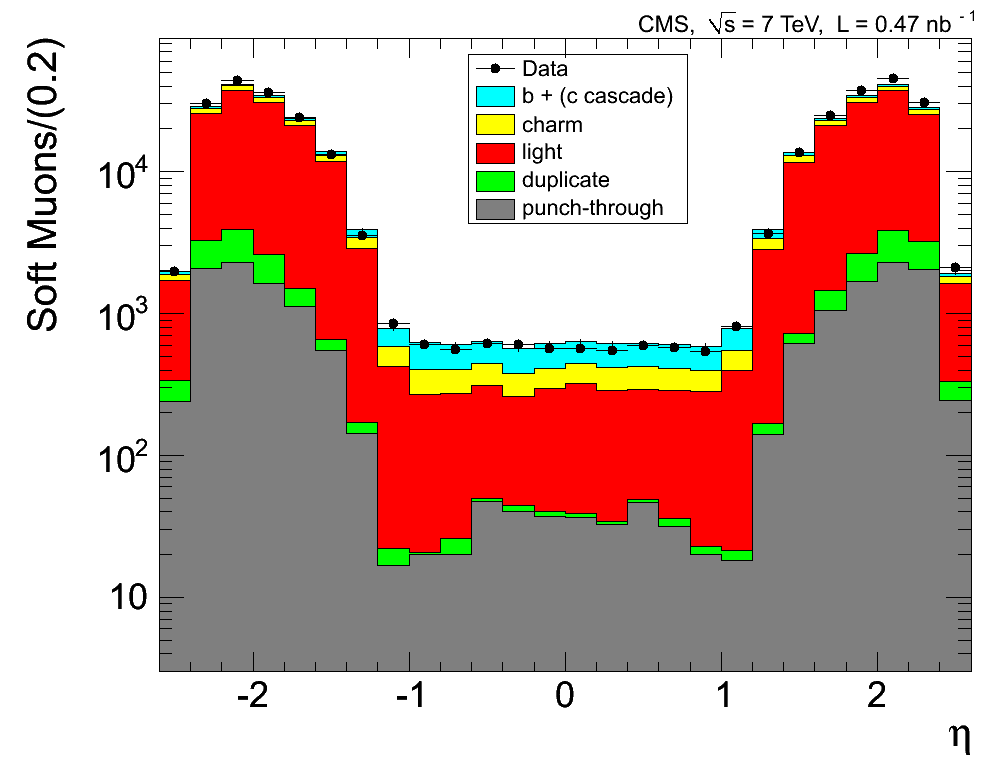}
  \includegraphics[width=0.48\textwidth]{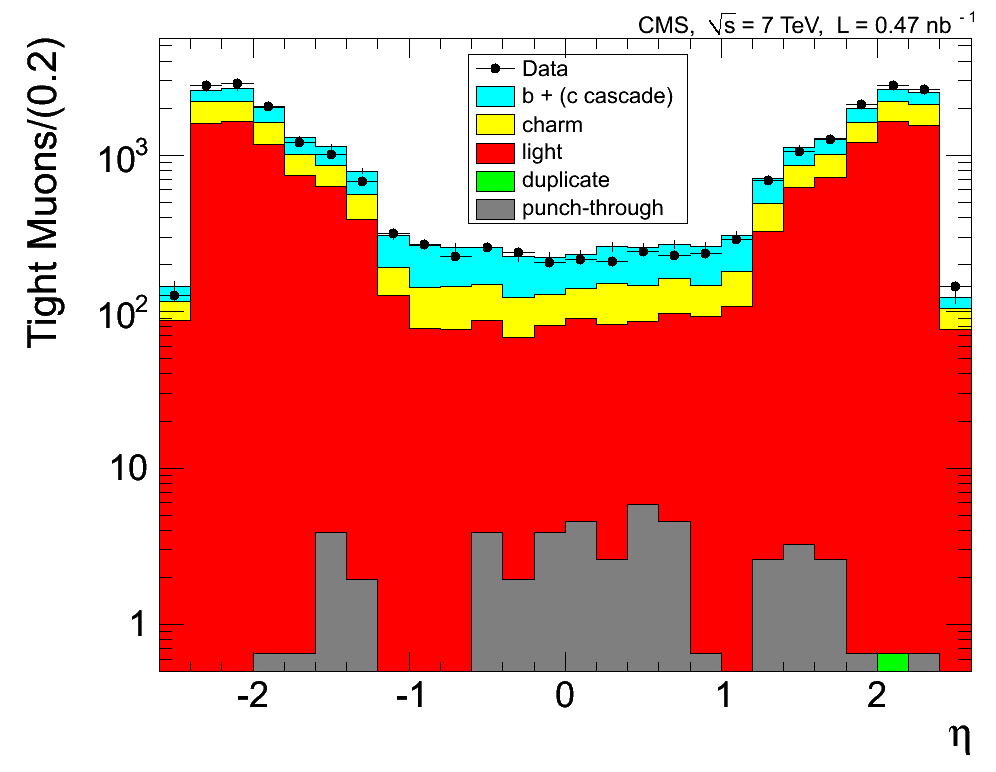}\\ \vspace{0.5cm}
  \includegraphics[width=0.48\textwidth]{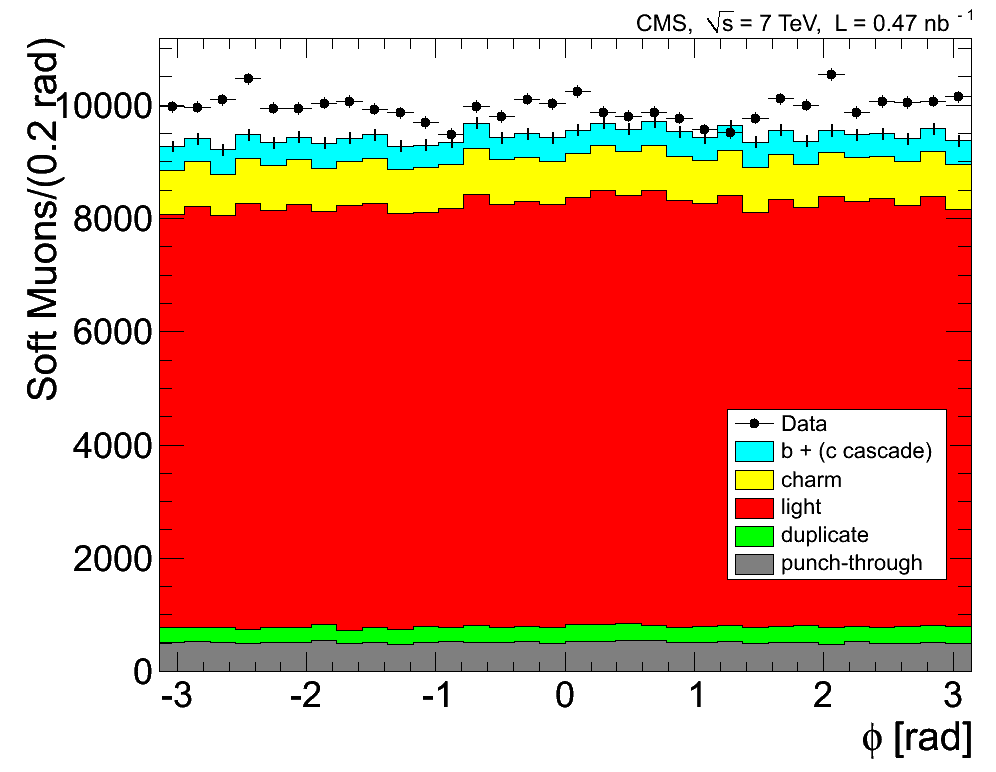}
  \includegraphics[width=0.48\textwidth]{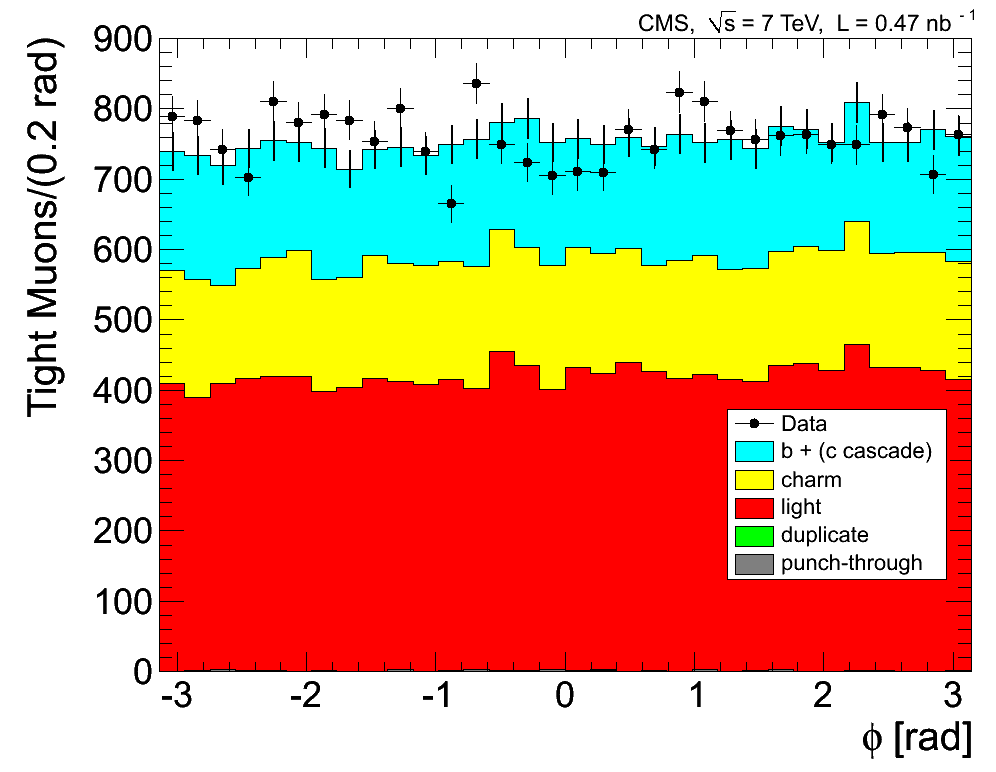}
  \caption{Distributions of kinematic variables for a sample of muons
    selected by the zero-bias trigger, for data (points) and for simulation
    subdivided into the different categories of contributing muons
    (histograms).  The kinematic variables are the muon transverse
    momentum multiplied by the charge (top),
             pseudorapidity (middle), and azimuthal angle (bottom).
             For each variable, the left plot shows the distribution for Soft Muons, and the right plot that for Tight Muons.
             The first (last) bin in the $q \times \pt$ distributions
             includes the underflow (the overflow).
             The error bars indicate the statistical uncertainty, for both data and
	     MC samples.
  }
  \label{fig:Kinematics_ZeroBias}
\end{figure}

The distributions of the muon transverse momentum $\pt$ multiplied
by its charge $q$, pseudorapidity $\eta$, and azimuthal angle
$\phi$ for Soft and Tight Muons in zero-bias events are shown in
Fig.~\ref{fig:Kinematics_ZeroBias}.
The pseudorapidity distribution is peaked in the forward region because there the
minimum $\pt$ required to reach the muon stations is lower than in the barrel: in
the endcaps the threshold in $\pt$ is about 0.5\GeVc, while in the
barrel it is about 3--4\GeVc.
Overall, there is good agreement between data and simulation both in the
number of events and in the shapes of the distributions.
Some discrepancies result from imperfect simulation of local detector conditions,
affecting for example the muon identification efficiency at low $\pt$, as shown in Section~\ref{sec:muonideff}.
Furthermore, the leading-order QCD predictions by \PYTHIA have large uncertainties.

Table~\ref{tab:composition} lists the sources of muons according
to simulation. % for reconstructed muons in zero-bias events.
The majority of reconstructed muon candidates originate
from decays in flight of pions and kaons (``light flavour'').
This is particularly evident for Soft Muons, while Tight Muons have larger heavy-flavour components.
For both selections the contribution of muons from heavy-flavour decays increases with $\pt$.
The Tight Muon selection reduces the hadron punch-through contribution to 0.2\% while it is
about 5\% in Soft Muons. %, according to the Monte Carlo simulation.
The measurements of muon misidentification probabilities presented in Section~\ref{sec:ExclusiveRates}
confirm that the simulation correctly estimates the probability for light hadrons to
be misidentified as muons.

\begin{table}[hbtp]
\begin{center}
\topcaption{Composition by source of the low-$\pt$ muon candidates reconstructed
in zero-bias events, according to simulation for the Soft
and Tight Muon selections.}
\label{tab:composition}
\begin{tabular}{|l|c|c|}
\hline
Muon source & Soft Muons [\%] & Tight Muons [\%]\\
\hline \hline
beauty                 &  4.4  & 22.2  \\ \hline
charm                  &  8.3  & 21.9  \\ \hline
light flavour          & 79.0  & 55.7  \\ \hline
hadron punch-through   &  5.4  &  0.2  \\ \hline
duplicate              &  2.9  & $<$0.01 \\ \hline
prompt                 & $\lesssim$0.1& $\lesssim$0.1 \\
\hline
\end{tabular}
\end{center}
\end{table}

Among all single-muon triggers used in 2010, the trigger with
a $\pt$ threshold of 15\GeVc was the lowest-threshold unprescaled
trigger during a period when most of the data, corresponding to an
integrated luminosity of about 31~pb$^{-1}$, were collected.
The kinematic distributions of muons collected with this trigger have been compared to the Monte Carlo expectations
after applying a selection on the reconstructed muon $\pt$ of 20\GeVc, for which the trigger efficiency has reached the plateau.
The Tight Muon selection applied in this kinematic range has a high efficiency for prompt muons,
removing most of the background from light-hadron decays and hadron punch-through.
After the Tight Muon selection, 824\,007 muon candidates remain.

For comparison with these events, the beauty production cross section given by \PYTHIA
has been rescaled to the NLO QCD
predictions~\cite{MC@NLO_main, MC@NLO_HeavyQuarks},
which were shown to describe recent CMS measurements well~\cite{BPH-10-004, BPH-10-005}.
Without this rescaling, an excess in the predicted beauty component is observed in the inclusive $\pt$ distribution
for muon $\pt$ lower than about 40\GeVc, and also in the muon impact-parameter distribution
for transverse distances characteristic of b-hadron decays.
No corrections accounting for differences between the measured and expected muon trigger and identification efficiencies have been applied.
Such corrections could lead to effects of up to 5\%, dependent on pseudorapidity,
as discussed in Sections~\ref{sec:muonideff} and \ref{sec:trigger}.
The overall ratio of the muon yield in data to the Monte Carlo predictions normalized to the same integrated luminosity is found to be 1.02.

\begin{figure}[hbt!]
  \centering
  \includegraphics[width=0.48\textwidth]{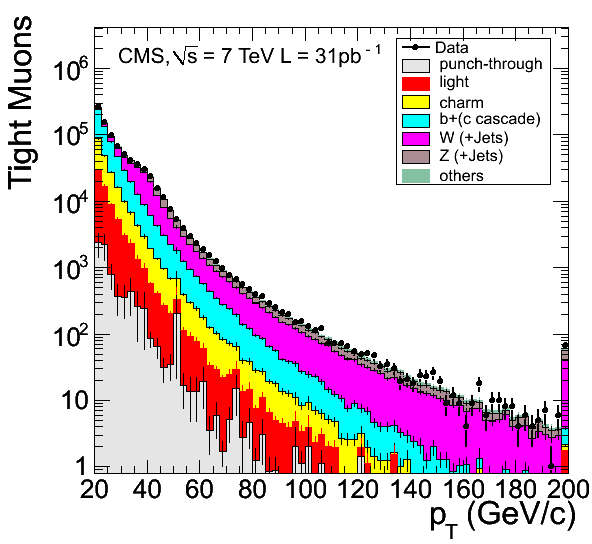}
  \includegraphics[width=0.48\textwidth]{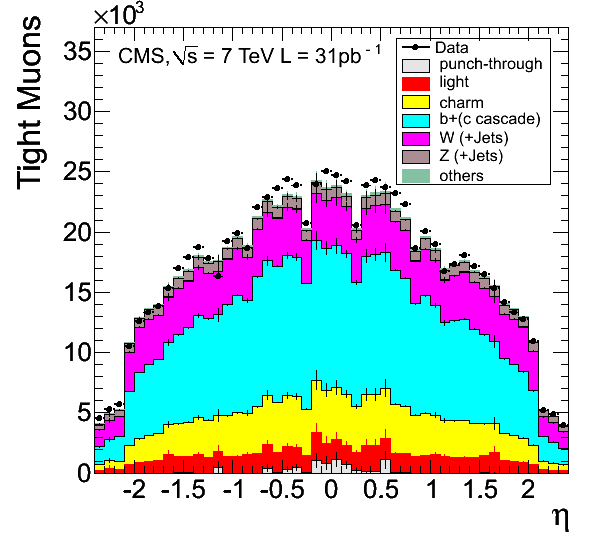}\\
  \includegraphics[width=0.48\textwidth]{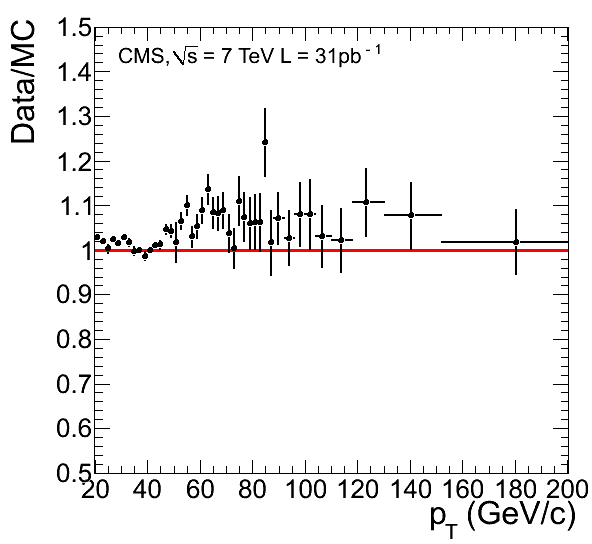}
  \includegraphics[width=0.48\textwidth]{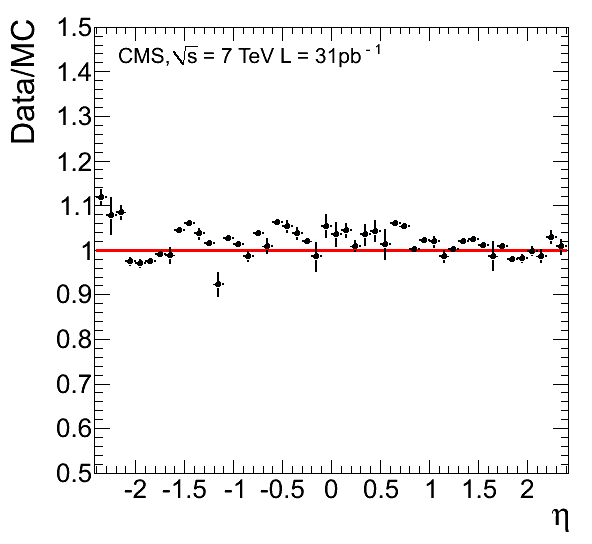}
  \caption{Distributions of transverse momentum (top left) and pseudorapidity (top right) for Tight Muons with $\pt > 20\GeVc$,
    comparing data (points with error bars) to Monte Carlo simulation broken down into its different components.
    The last bin in the $\pt$ distribution includes the overflow.
    Dips in the $\eta$ distribution are due to inefficiencies related
    to the muon detector geometry.
    The corresponding ratios of data and MC distributions are shown in the bottom row.
    The error bars include statistical uncertainties only.
  }
  \label{fig:kinematics_HighPt}
\end{figure}

The muon transverse momentum and pseudorapidity distributions for
$\pt > 20\GeVc$ are compared to the expectations
from the Monte Carlo simulation in Fig.~\ref{fig:kinematics_HighPt}.
The estimated composition of the sample obtained from the simulation-based studies is shown in Table~\ref{tab:compositionHighPt}.
Muons from light-hadron decays are predicted to contribute less than 10\%,
while the hadron punch-through is suppressed to about 1\%.
The beauty contribution dominates up to muon transverse momentum of about 30\GeVc,
where the $\W$ contribution starts to prevail, leading to a shoulder in the falling $\pt$ spectrum.
The inclusive muon yield agrees with the expectations within a few percent up to a transverse momentum of 50\GeVc.
At higher momenta the leading processes are $\W$ and $\Z$ production, occasionally associated with hard jets.
In this $\pt$ region, the data agree with the predictions within 10\%.
This has been verified to be fully consistent with theoretical uncertainties related to missing higher-order QCD contributions,
by comparing the $\textsc{madgraph}$ generator used to simulate $\W$ and $\Z$ with other Monte Carlo programs for the $\W$($\Z$)+jets processes.
In conclusion, given the known experimental and theoretical uncertainties,
the agreement between the data and simulation is satisfactory over the
entire momentum range of $\pt$ $\lesssim$ 200\GeVc.

\begin{table}[hbt!]
\begin{center}
\topcaption{Composition by source of Tight Muons with $\pt > 20$\GeVc according to simulation.}
\label{tab:compositionHighPt}
\begin{tabular}{|l|c|}
\hline
Muon source       & Tight Muons with $\pt > 20\GeVc$ [\%]\\
\hline \hline
 $\W$ (+ jets)        & 20.8  \\ \hline %powheg  21.1  \\
 $\Z$/Drell--Yan (+ jets) &  4.7  \\ \hline %powheg   4.9  \\
 top                  &  0.1  \\ \hline
 quarkonia            &  0.7  \\ \hline
 beauty               & 47.6  \\ \hline %powheg  47.3  \\
 charm                & 17.4  \\ \hline %powheg  17.3  \\
 light flavour        &  7.8  \\ \hline
 hadron punch-through &  0.9  \\ \hline
 duplicate            & $<$0.01 \\
\hline
\end{tabular}
\end{center}
\end{table}

\subsection{Muon identification variables}
\label{sec:muonidvar}

The basic selections discussed in Section~\ref{sec:reco} can be further refined for specific 
purposes using additional information available for each reconstructed muon.
  
\begin{figure}[hbt!]
  \begin{center}
    \subfigure[]{\includegraphics[width=0.45\textwidth]{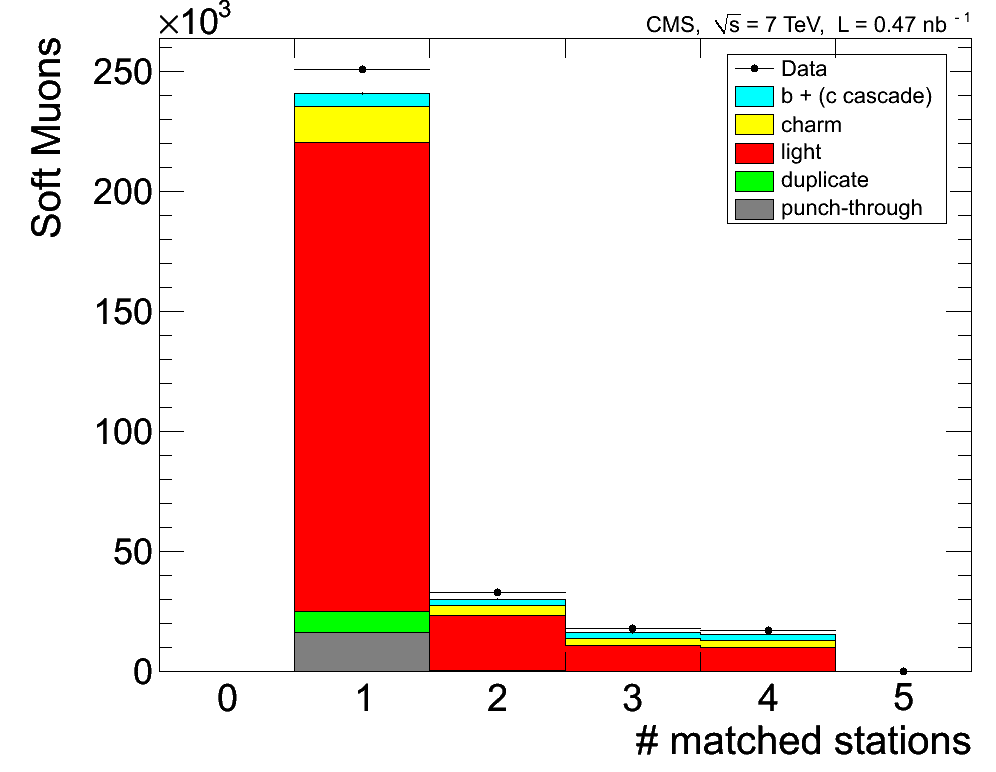}}
    \subfigure[]{\includegraphics[width=0.45\textwidth]{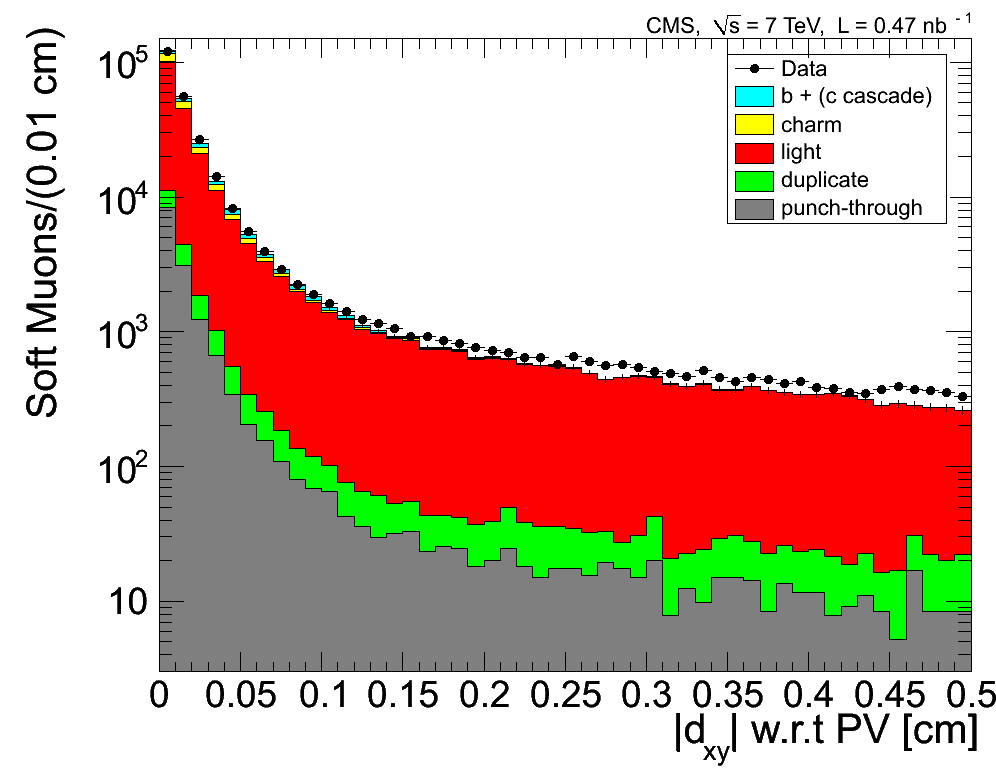}}
    \caption{Comparison of data and simulation for distributions of Soft Muons in zero-bias events: 
(a) number of muon stations with matched segments; 
(b) transverse impact parameter $d_{\rm xy}$ of the muon with respect to the primary vertex (PV).
The MC distributions are normalized to the integrated luminosity of the data sample.
The error bars indicate the statistical uncertainty.
}
    \label{fig:SoftID}
  \end{center}
\end{figure}

The Tracker Muon segment matching
is a powerful tool to reject hadron punch-through.
The number of muon stations with matched segments (see Section~\ref{sec:reco})
is shown in Fig.~\ref{fig:SoftID}(a) for Soft Muons in zero-bias events.
The contamination from hadron punch-through is evident in the expectation for one matched station. 
The probability for a punch-through to be identified as a Soft Muon is drastically reduced by requiring matched
segments in at least two stations. The contribution from low-momentum muons from light-quark decays is also suppressed by this requirement.
Figure~\ref{fig:SoftID}(b) shows the distribution of the transverse impact parameter, 
where the long tail is dominated by pion and kaon decays in flight.
Requirements on both variables are used in the Tight Muon selection designed
to select prompt muons such as those from $\W$ and $\Z$ decays (see
Section~\ref{sec:reco}).

\begin{figure}[hbt!]
  \begin{center}
    \subfigure[]{\includegraphics[width=0.45\textwidth]{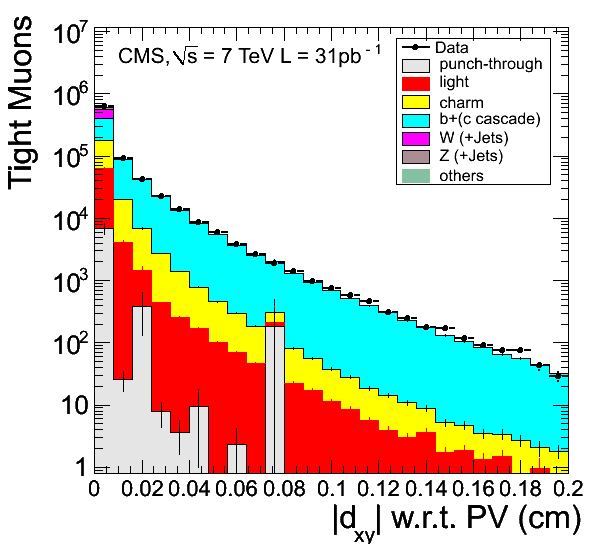}}
    \subfigure[]{\includegraphics[width=0.45\textwidth]{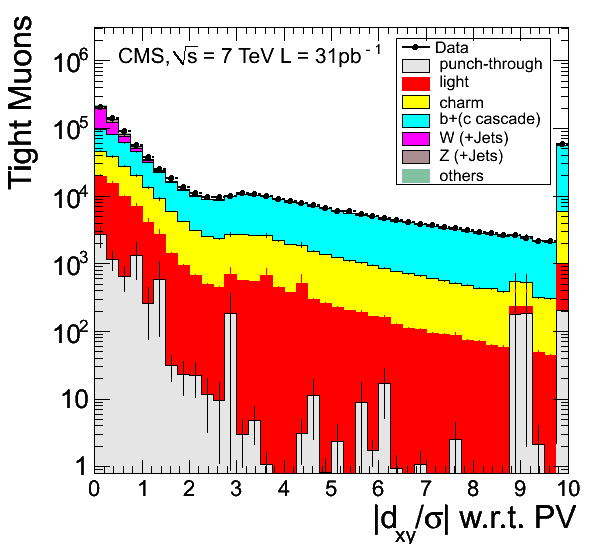}}\\
    \subfigure[]{\includegraphics[width=0.45\textwidth]{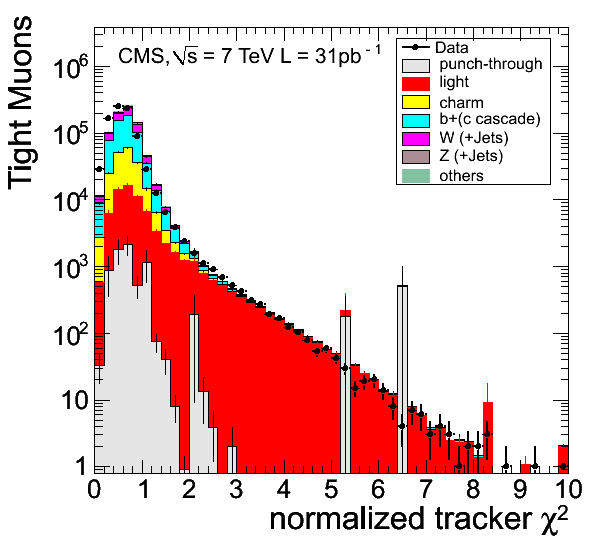}}
    \subfigure[]{\includegraphics[width=0.45\textwidth]{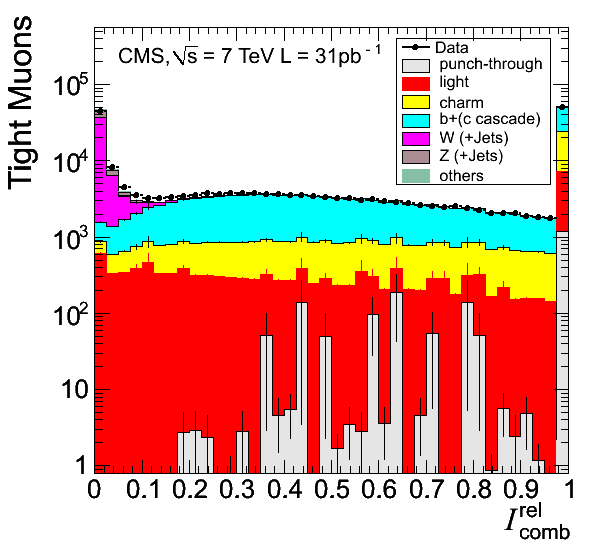}}
    \caption{Identification variables for Tight Muons with $\pt > 20 \GeVc$: 
      (a) transverse impact parameter $d_{\rm xy}$ with respect to the primary vertex (PV);
      (b) significance of the transverse impact parameter;
      (c) $\chi^2/d.o.f.$ of the fit of the track in the inner tracker;
      (d) relative combined isolation (tracker+calorimeters), with a cone size $\Delta R=0.3$,
      for events with a single reconstructed PV.  The MC distributions are
      normalized to the integrated luminosity of the data sample in (a), (b),
      and (c), and to the number of events in the data sample in (d).
      The last bin in (b) and in (d) includes the overflow.
      Error bars indicate statistical uncertainties.
    }
    \label{fig:ID1}
  \end{center}
\end{figure}

A few other examples of identification variables are shown in Fig.~\ref{fig:ID1} for Tight Muons with $\pt > 20\GeVc$ 
collected with the single-muon trigger.
 The transverse impact parameter and its significance in Figs.~\ref{fig:ID1}(a) and (b) are useful to select either prompt muons or, by inverting the requirement, muons from heavy-flavour decays.
The $\chi^2/d.o.f.$ of the tracker-track fit is also a good discriminant to suppress muons from decays in flight, 
as can be seen from the composition of the tail of the distribution in Fig.~\ref{fig:ID1}(c).
The muon isolation is a simple quantity to select prompt muons with high purity.
The scalar sum of the transverse momenta of tracks in the inner tracker
and the transverse energies in calorimeter cells (both in the ECAL and HCAL)
within a cone of radius $\Delta R =\sqrt{(\Delta \varphi)^2+(\Delta \eta)^2}
= 0.3$ centred on the direction vector of the muon candidate is calculated,
excluding the contribution from the candidate itself.  The relative
combined isolation $\IRelComb$ (further discussed in
Section~\ref{sec:isolation}) is defined as the ratio of this scalar sum to the transverse momentum of the muon candidate.
Figure~\ref{fig:ID1}(d) shows the distribution of the $\IRelComb$ variable
in events with a single reconstructed primary vertex, compared to the
simulation with no pile-up effects.
Isolated muons promptly produced in decays of $\W$ and $\Z$ bosons dominate the region $\IRelComb < 0.1$.
Overall, the agreement between data and Monte Carlo predictions for the muon identification variables is good both for zero-bias events 
and for events recorded with the single-muon trigger.

The accuracy of the propagation of the tracker tracks to the muon system and the performance of the
track-to-segment match have been further studied using Tracker Muons with $\pt > 20\GeVc$, with 
tight selection requirements on the tracker variables only, to avoid possible biases.
To further purify the muon sample, an isolation requirement $\IRelComb<$ 0.1
has been applied.

\begin{figure}[thb!]
\centering
\includegraphics[width=0.45\textwidth]{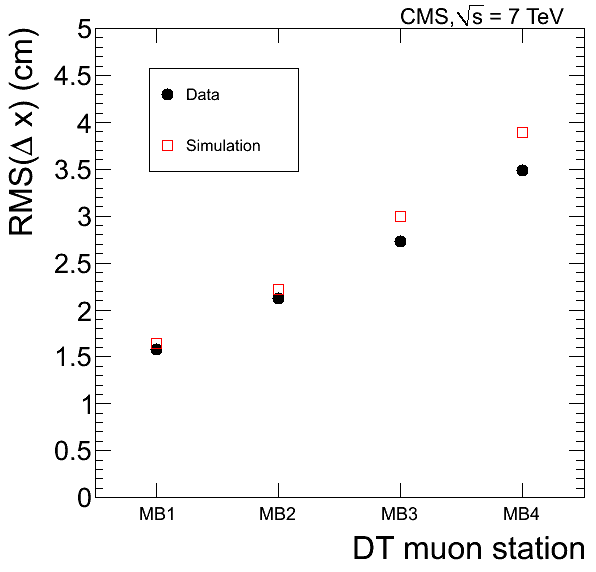}
\includegraphics[width=0.45\textwidth]{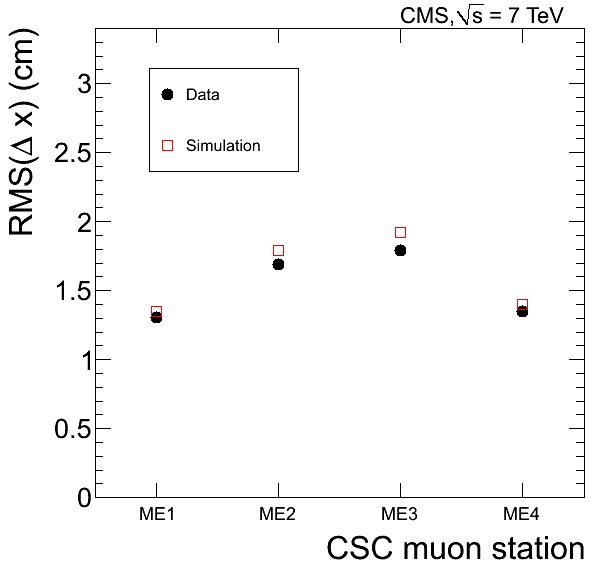}
\caption{RMS width of residuals of the local $x$ position, given by the position of the muon segment
  with respect to the extrapolated tracker track, as a function of the muon station, for
  DT chambers in the barrel region (left) and CSCs in the endcap regions (right).
  Data are compared with MC expectations.
}
\label{fig:Xresiduals}
\end{figure}

The distribution of distance in local $x$ between the position of the extrapolated tracker track and the position of the muon segment
has been compared between data and simulation, for successful track-to-segment matches (distance less than 3 cm or pull less than 4, see Section~\ref{sec:reco}).
The RMS width of residuals is shown in Fig.~\ref{fig:Xresiduals} as a function of the muon-station number,
for the DT and CSC systems. 
As expected, the width of the distributions increases with the amount of material upstream of the muon station 
and with the distance over which the track is extrapolated, from the innermost to the outermost muon stations
(from MB1 to MB4 and from ME1 to ME4 in the DT and the CSC systems, respectively). 
The general trend is well reproduced by the Monte Carlo simulation, 
although the increase of the width is a bit larger in the simulation.
As the outer ring of ME4 is only partially instrumented with chambers
(see Fig.~\ref{fig:Display}),
the residual for ME4 is not directly comparable with the measurements in the
other stations because muons traversing the installed ME4 chambers have
a higher average momentum.

\begin{figure}[htb]
\centering
\subfigure[]{\includegraphics[width=0.45\textwidth]{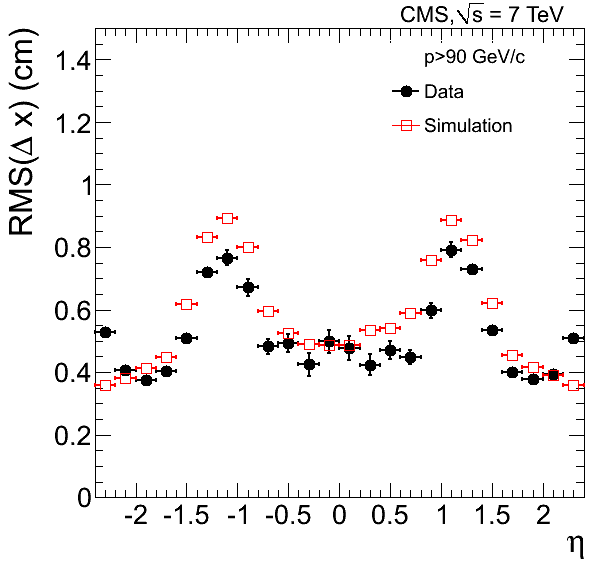}}
\subfigure[]{\includegraphics[width=0.45\textwidth]{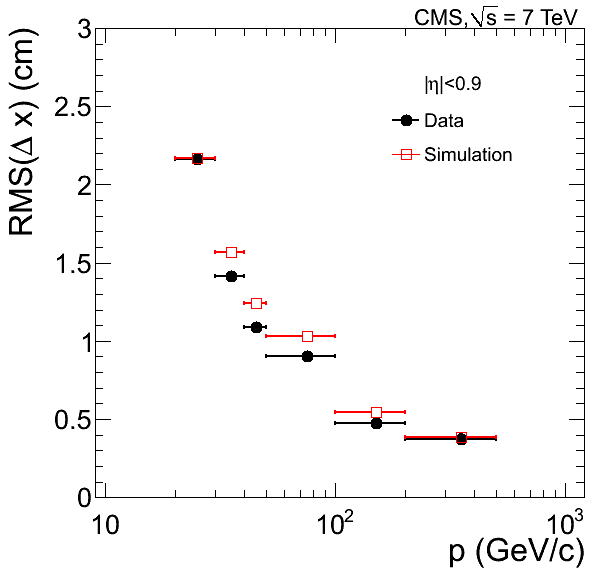}}\\
\subfigure[]{\includegraphics[width=0.45\textwidth]{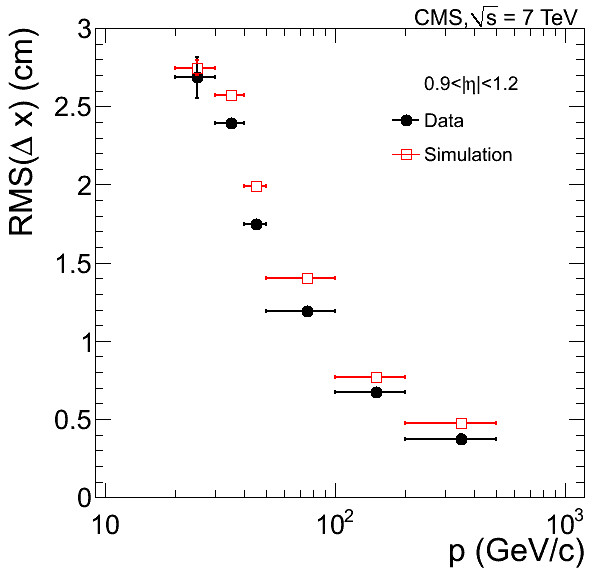}}
\subfigure[]{\includegraphics[width=0.45\textwidth]{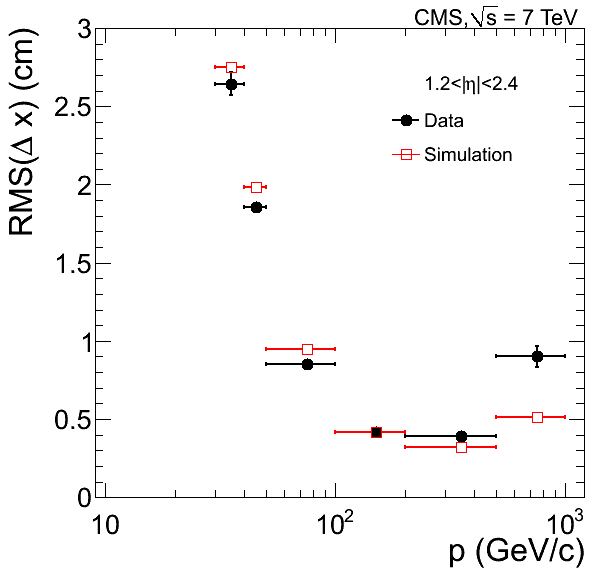}}
\caption{RMS width of residuals of the local $x$ position for
   the track-to-segment match in the first muon station
   (a) as a function of the muon pseudorapidity, with a requirement
   on momentum $p > 90\GeVc$, and (b)--(d)
   as a function of the muon momentum in different angular regions:
   (b) $|\eta|<0.9$; (c) $0.9<|\eta|<1.2$; (d) $1.2<|\eta|<2.4$.
   Data are compared with MC expectations.
}
\label{fig:sigmaVsPEta}
\end{figure}

The residuals for track-to-segment matches have also been studied as a function of the muon momentum and pseudorapidity.
The first muon station is the most important in the global track
reconstruction: it is where the track's sagitta, determined by the
magnetic field inside the solenoid, is largest, and its measurements are
the least affected by multiple-scattering and showering effects because
the material upstream of the first station
corresponds only to the inner detectors, the calorimeters, and the magnet cryostat, whereas
the stations downstream are also preceded by the steel sections of the magnet return yoke.
The average material thickness traversed by a muon reaching the first station is quite different 
depending on the angular region~\cite{PTDR1}.
In addition, the width of the position residual increases linearly with the propagation distance of the muon trajectory from the interaction region to the first muon chamber.
Figure~\ref{fig:sigmaVsPEta}(a) shows the RMS width of the residual of the
local $x$ position as a function of muon pseudorapidity.
A selection on the minimum momentum $p > 90\GeVc$ has been applied to remove the bias induced by the trigger threshold ($\pt > 15\GeVc$) in the endcap regions.
As expected from the $\eta$ dependence of the
distance between the inner tracker and the first muon station
and of the material thickness, the residual
width reaches the maximum values in the overlap region, $0.9<|\eta|<1.2$.
Figures~\ref{fig:sigmaVsPEta}(b), \ref{fig:sigmaVsPEta}(c), and
\ref{fig:sigmaVsPEta}(d) show the momentum dependence of the RMS width, 
separately for the barrel, overlap, and endcap regions.
The width decreases
with increasing momentum because of smaller multiple-scattering effects.
The shapes of the distributions are well reproduced by the Monte Carlo simulation, although the simulation predicts a somewhat larger width.
The only exception is the highest momentum bin in the endcap region (Fig.~\ref{fig:sigmaVsPEta}(d)),
where the distribution of position resolution in data has larger non-Gaussian
tails than predicted by simulation. 
This also leads to the discrepancy observed at the extreme $\eta$ bins in Fig.~\ref{fig:sigmaVsPEta}(a).

We have also examined the distributions of pulls of the local
positions and directions in both DT and CSC systems.  The widths of
the pulls were found to be close to unity and no large biases were
observed, thus demonstrating that the propagation works as expected
and that the uncertainties are well estimated.  The widths of the pull
distributions in the simulation are about 10\% larger than in data.
As demonstrated in the next section, such agreement between the
expected and the measured residuals and pulls is sufficient to obtain
a good description of the muon reconstruction and identification
efficiencies by the simulation.

High-momentum muons can give rise to electromagnetic showers in
the muon system. These may produce extended clusters of hits in the
muon chambers, which can degrade the quality of muon track
reconstruction. Hence an accurate simulation and 
reliable identification of such showers are needed.

\begin{figure}[hbtp] 
\begin{center}
    \includegraphics[width=0.35\textwidth, angle = 90]{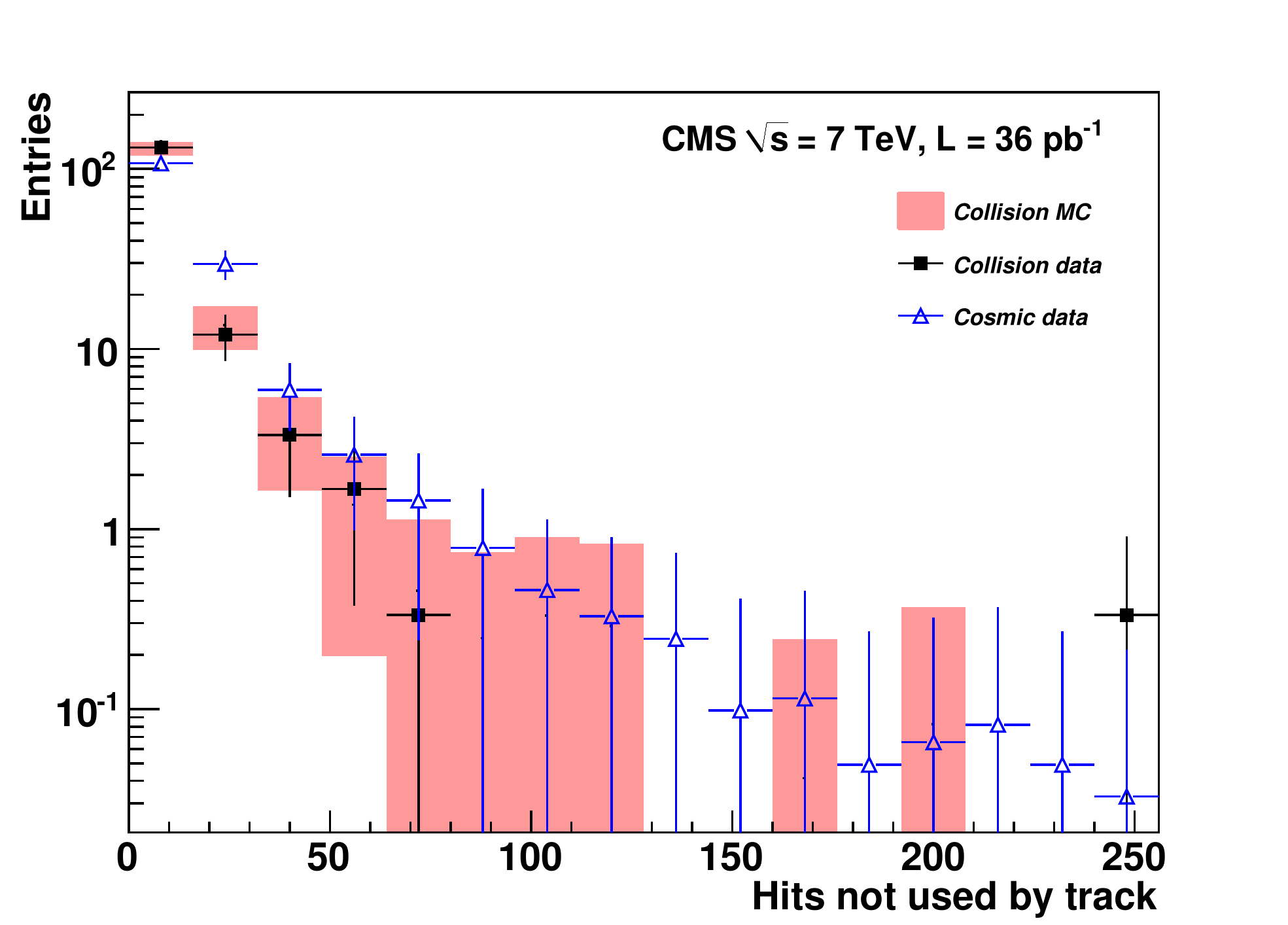}
    \includegraphics[width=0.35\textwidth, angle = 90]{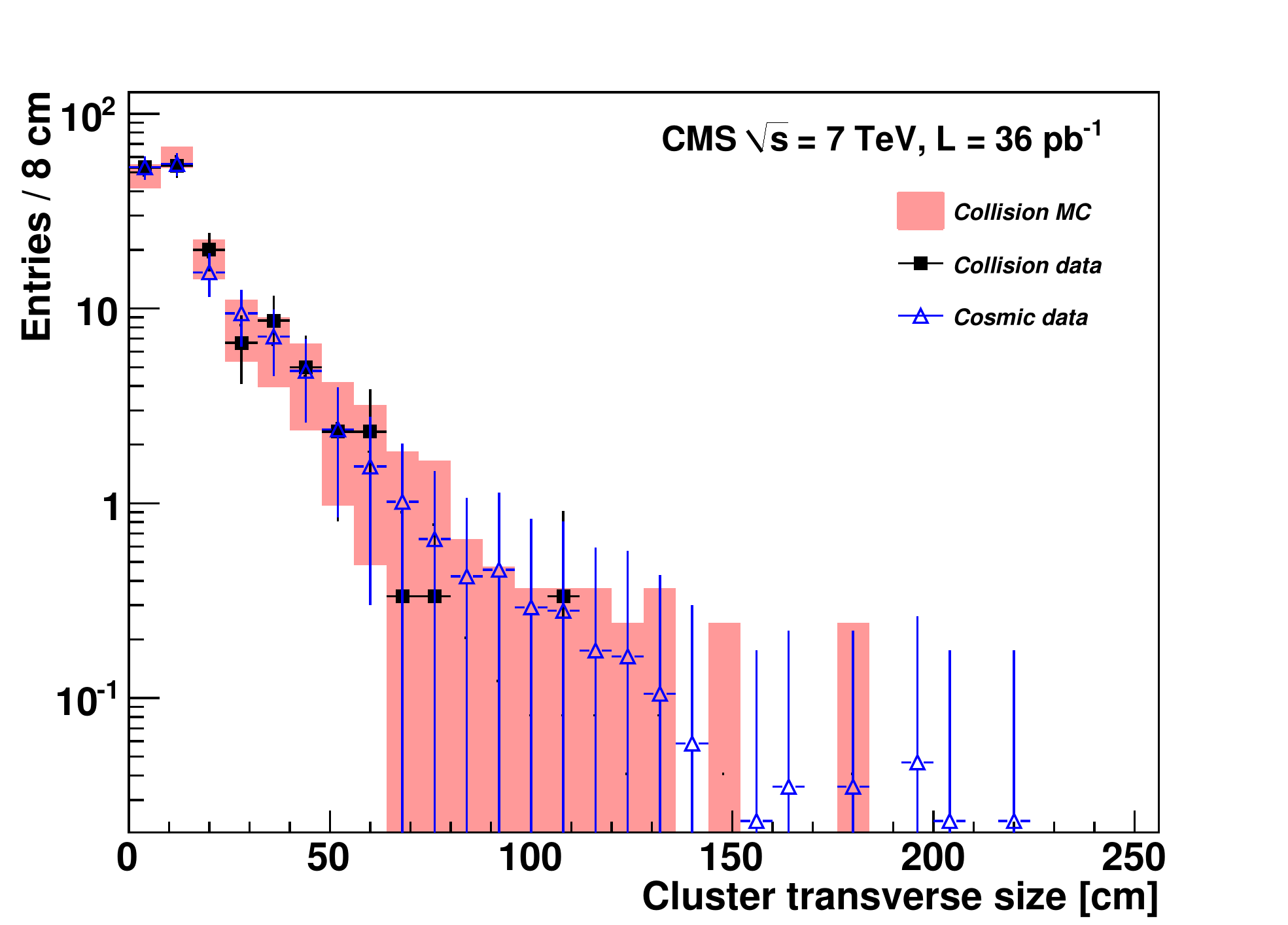}
    \caption{Comparison of data and simulation for variables characterizing
    electromagnetic showers, for muons with $p > 150 \GeVc$ in the barrel
    region: the number of reconstructed muon hits not used in
    a track fit (left); the transverse size of the cluster of hits around
    a track (right).  All distributions are normalized to the number of muons
    per muon station in the collision data sample.
    }
   \label{fig:em_uncorr_track}
\end{center}
\end{figure}

We have considered several variables that might indicate
the presence of electromagnetic showers in the muon system, and
examined how well they are reproduced by the simulation.
Figure~\ref{fig:em_uncorr_track} shows the distributions of two of them:
the number of hits reconstructed in the DT chamber crossed
by a track but not used in the track fit, and
the transverse size of the cluster of hits around a track. %($\Delta\phi_{T}$).
The cluster of hits is defined iteratively, starting from the impact
point of the extrapolated muon track and successively adding any hit if it lies
within $\Delta\phi < 0.05$~rad of the hit in the existing cluster with
largest radial distance from the impact point. The transverse size of
a cluster is defined as the maximum distance in the local $x$-$y$
plane between the impact point of the track and any hit in the
cluster.
The plots are made for two samples of
high-energy muons (with $p$ reconstructed by the tracker-only fit
above 150\GeVc): 1) collision muons, and 2) cosmic-ray muons
selected from collision data samples, with a topology similar to that of
collision muons.  Collision muons were selected by requiring
that an event has at least one primary vertex reconstructed close to the
nominal beam-spot position, and that the muon passes the
Tight Muon selection with additional
isolation and track-quality selections to reduce contamination from hadron
punch-through.
Cosmic-ray muons were selected by requiring events with at most two
tracks reconstructed in the inner tracker, one of which was also
reconstructed as a Global Muon.  The momentum spectra of selected
collision-muon and cosmic-muon data samples are quite similar, but
the cosmic-muon sample provides a larger number of high-energy muons.
Monte Carlo samples of Drell--Yan dimuon events
passing event selection criteria identical to those applied to the collision
muons were used for comparisons with the data. % in order to study
The vast majority of selected cosmic-ray muons 
are contained in the barrel region, so Fig.~\ref{fig:em_uncorr_track}
shows distributions for
the DT chambers only. 
We observe good general agreement between the data and MC simulation.

\section{Muon Reconstruction and Identification Efficiency}
\label{sec:muonideff}

The previous section focused on the comparison between data and simulation for various distributions of inclusive
samples of reconstructed muon candidates. In this section we study exclusive samples
of prompt muons, pions, kaons, and protons in data to determine the probability
that such a particle is reconstructed and identified as a muon.

Throughout this paper, efficiencies are defined in a relative manner, such that the total efficiency for the entire muon triggering, reconstruction, and identification chain
can be calculated as the product of the following individual factors:

\begin{equation}
\label{eqn:eff_factorization}
\epsilon_{\mu} = \epsilon_{\mathrm{track}}\cdot\epsilon_{\mathrm{rec+id}}\cdot\epsilon_{\mathrm{iso}}\cdot\epsilon_{\mathrm{trig}} \,.
\end{equation}

The efficiency to reconstruct a muon in the inner tracker $\epsilon_{\mathrm{track}}$ was measured separately~\cite{EWK-10-002,TRK-10-002}
and found to be 99\% or higher within the whole tracker acceptance, in good agreement with the expectation
from simulations.
Given the existence of a tracker track, the combined muon reconstruction and identification efficiencies of the different
selection algorithms $\epsilon_{\mathrm{rec+id}}$ can be measured using ``tag-and-probe'' techniques. In this section the tag-and-probe method is described, and measurements of the $\epsilon_{\mathrm{rec+id}}$ efficiencies are presented.
In Section~\ref{sec:isolation}, isolation efficiencies $\epsilon_{\mathrm{iso}}$ are calculated from a sample of identified muons.
In Section~\ref{sec:trigger}, trigger efficiencies $\epsilon_{\mathrm{trig}}$ are defined relative to muons identified offline and, unless otherwise mentioned, passing isolation criteria.

\subsection{Muon efficiency using the tag-and-probe method on dimuon resonances}\label{sec:muonideff_tnp}

\subsubsection{Method}

We evaluate the efficiencies for prompt muons by applying a tag-and-probe
technique to mu\-ons from $\jpsi$ and $\Z$ decays. Using this
technique it is possible to obtain almost unbiased estimates of the
efficiencies of the different stages
of muon trigger and offline reconstruction. Events are selected with strict
selection requirements on one muon (the ``tag'' muon) and with a more relaxed selection on the
other muon (the ``probe'' muon), such that the selection applied to the probe
muon does not bias the efficiency that one wants to measure. The fraction of
probe muons that passes the selection under study
gives an estimate of its efficiency.

In this section, muon efficiencies $\epsilon_{\mathrm{rec+id}}$ are measured with this technique. The probes are tracks reconstructed
using only the inner tracker, so there is no bias from the muon subdetectors.

In the case of the $\jpsi$ events, combinatorial backgrounds from other tracks in the event are generally high,
particularly at low \pt.
An effective way to suppress this background is to require that the candidate probe muon has
the signature of a minimum-ionizing particle (MIP) in the calorimeters.
In this way the background can be reduced by about a factor of three without using any information from the muon system.
The residual background in both $\jpsi$ and $\Z$ events
is subtracted by performing a simultaneous fit to the invariant mass spectra for passing and
failing probes with identical signal shape and appropriate background shapes; the efficiency is then computed from the
normalizations of the signal shapes in the two spectra.

The uncertainty on the fitted efficiency is determined from the likelihood function.
As normalizations of signal and background, efficiency of the background, and parameters
controlling the shapes of the signal and background are all
parameters of the fit, the uncertainty includes the contributions from
the background subtraction procedure. When the background is not negligible, as in the case of $\jpsi$, the uncertainty on the efficiency obtained by the fit is dominated by these contributions.

For the $\Z$ resonance, an unbiased sample of dimuon pairs can be collected efficiently using high-\pt single-muon triggers.
For $\jpsi$, specialized high-level triggers were implemented, as described in Section~\ref{sec:samples}.
The muon-plus-track trigger used for the $\mathrm{J}/\!\psi$ case does not bias the efficiencies related to the muon system,
but introduces a small positive bias in the efficiency for the Tight Muon selection, which includes quality requirements on the muon
tracker track. To measure this bias, the efficiencies for these quality requirements alone are extracted using
another special dimuon trigger that uses only the muon system for the reconstruction of one of the two muons.
The bias in efficiency, measured to be $(0.7\pm0.1)\%$ in the barrel and $(0.3\pm0.2)\%$
in the endcaps,
is well reproduced by the simulation and cancels out
in the ratio of efficiencies from data and from simulation.

Under certain kinematic configurations muons from $\jpsi$ decays can be close to each other in the muon system.
This can result in inefficiencies for some muon identification
algorithms. To obtain an unbiased measurement of single-muon efficiencies, a
separation criterion is applied to the tag-probe pairs: the extrapolated impact points of the two muon tracks on the surface of the first muon station
must have an angular separation $\Delta R = \sqrt{(\Delta \eta)^2 +(\Delta \phi)^2} > 0.5$.
The impact of this requirement on $\Zmm$ events is small: as the large opening angle at the production is preserved by the smaller bending in the
magnetic field ($\approx$0.1 rad for a $\pt$ of 25\GeVc), only 0.2\% of the $\Zmm$ events fail the separation criterion above.
A dedicated measurement of the dimuon efficiencies as a function of the separation between muons is described in
Section~\ref{sec:collimatedmus}.

\subsubsection{Results}
\begin{figure}[htb]
  \begin{center}

\includegraphics[width=0.32\textwidth]{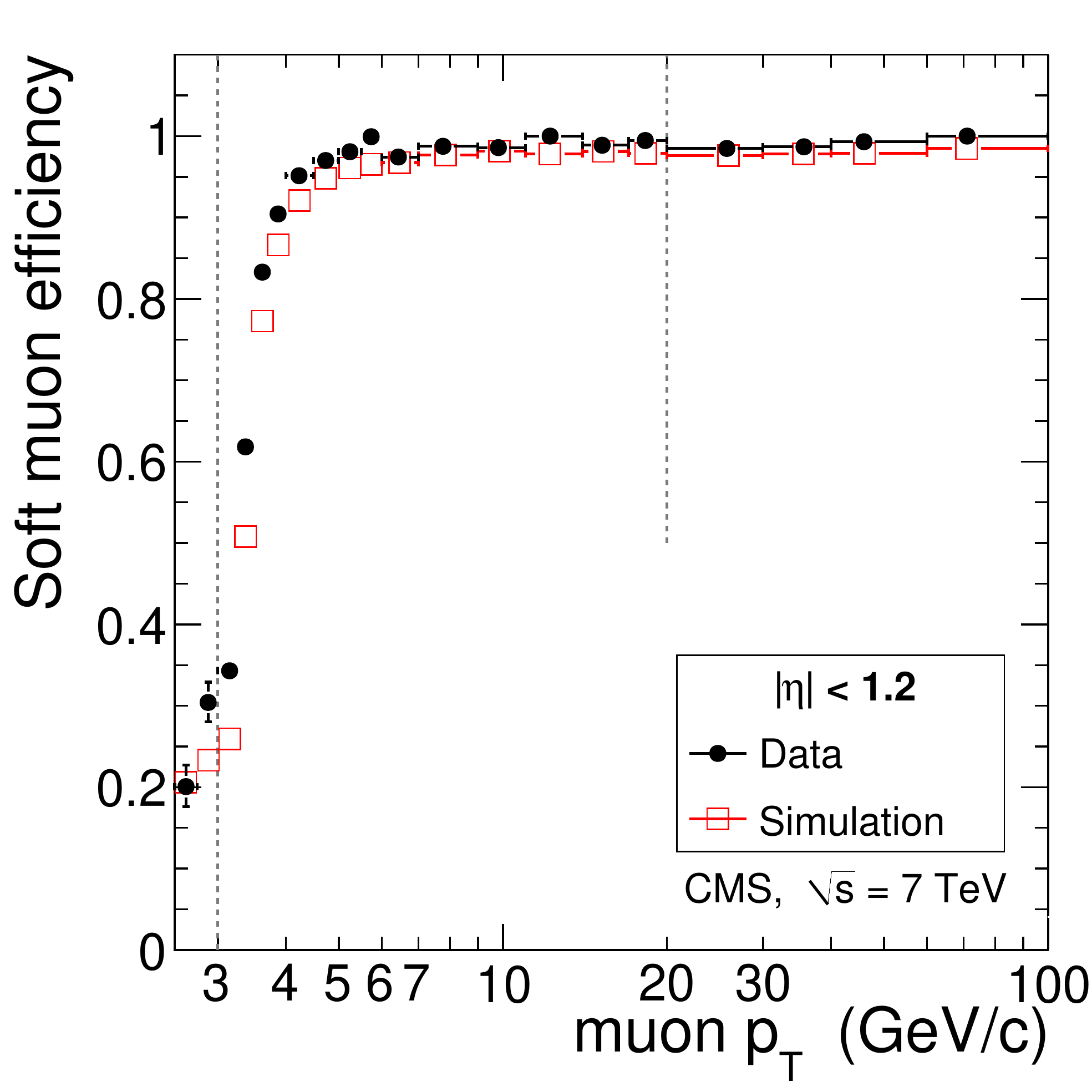}
\includegraphics[width=0.32\textwidth]{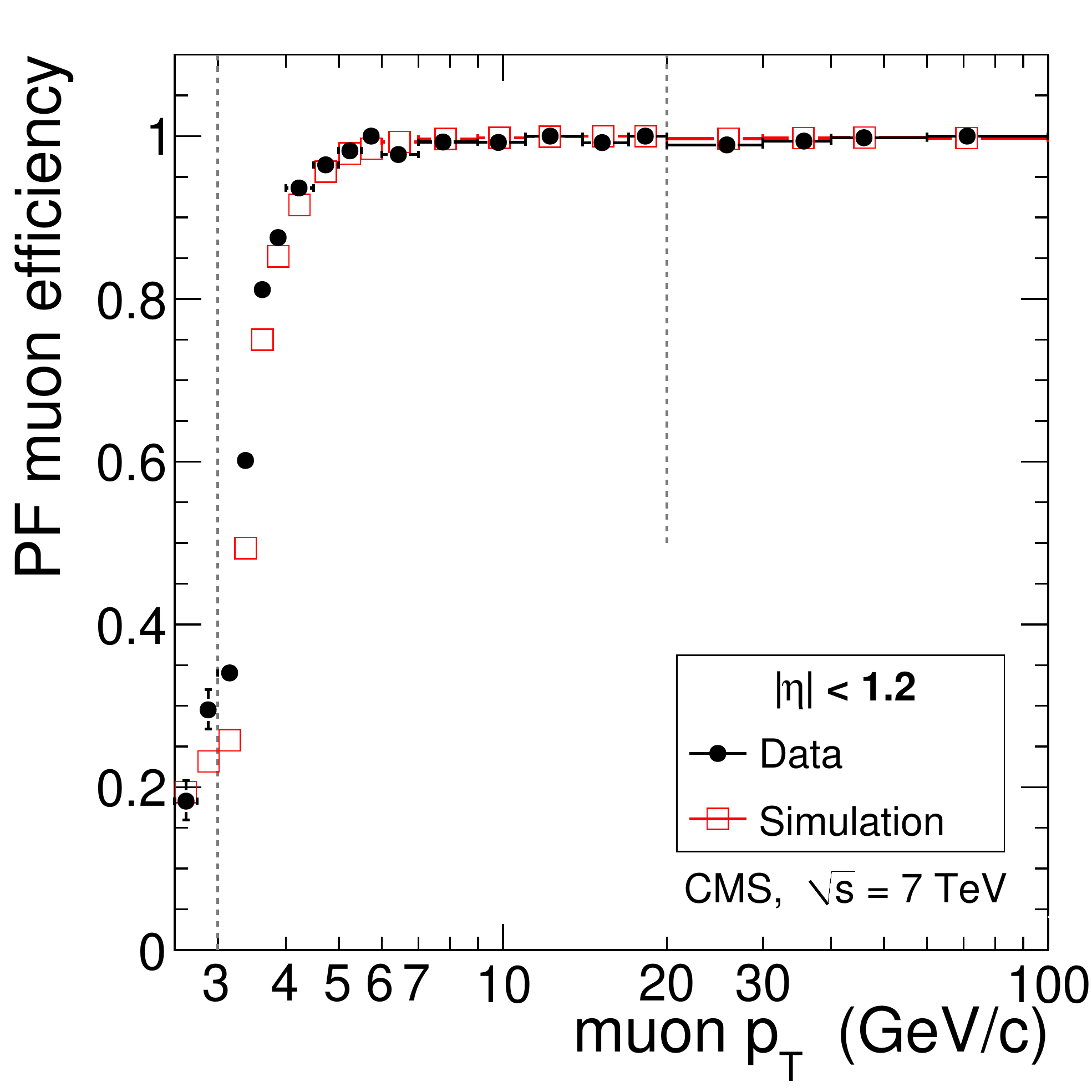}
\includegraphics[width=0.32\textwidth]{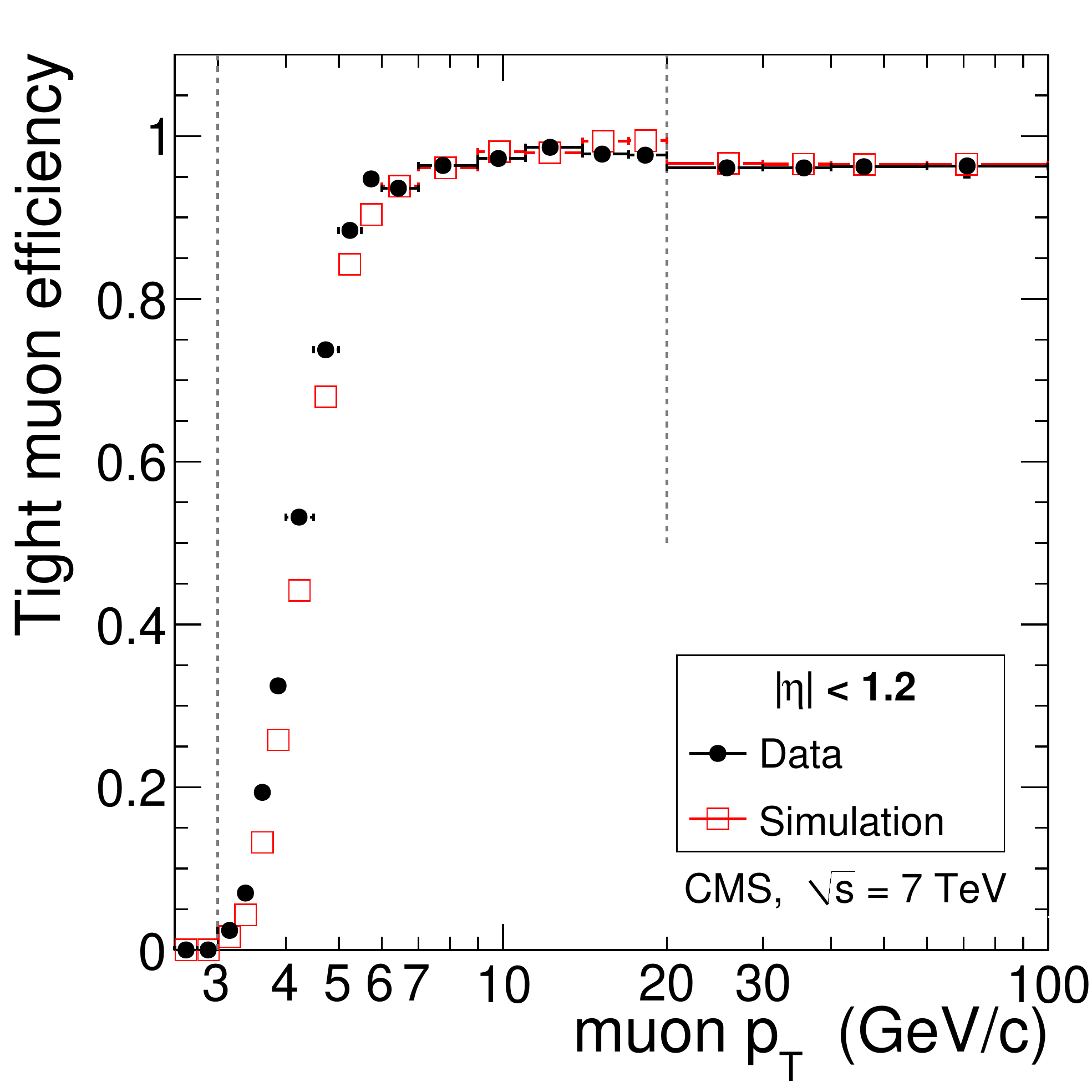}

\includegraphics[width=0.32\textwidth]{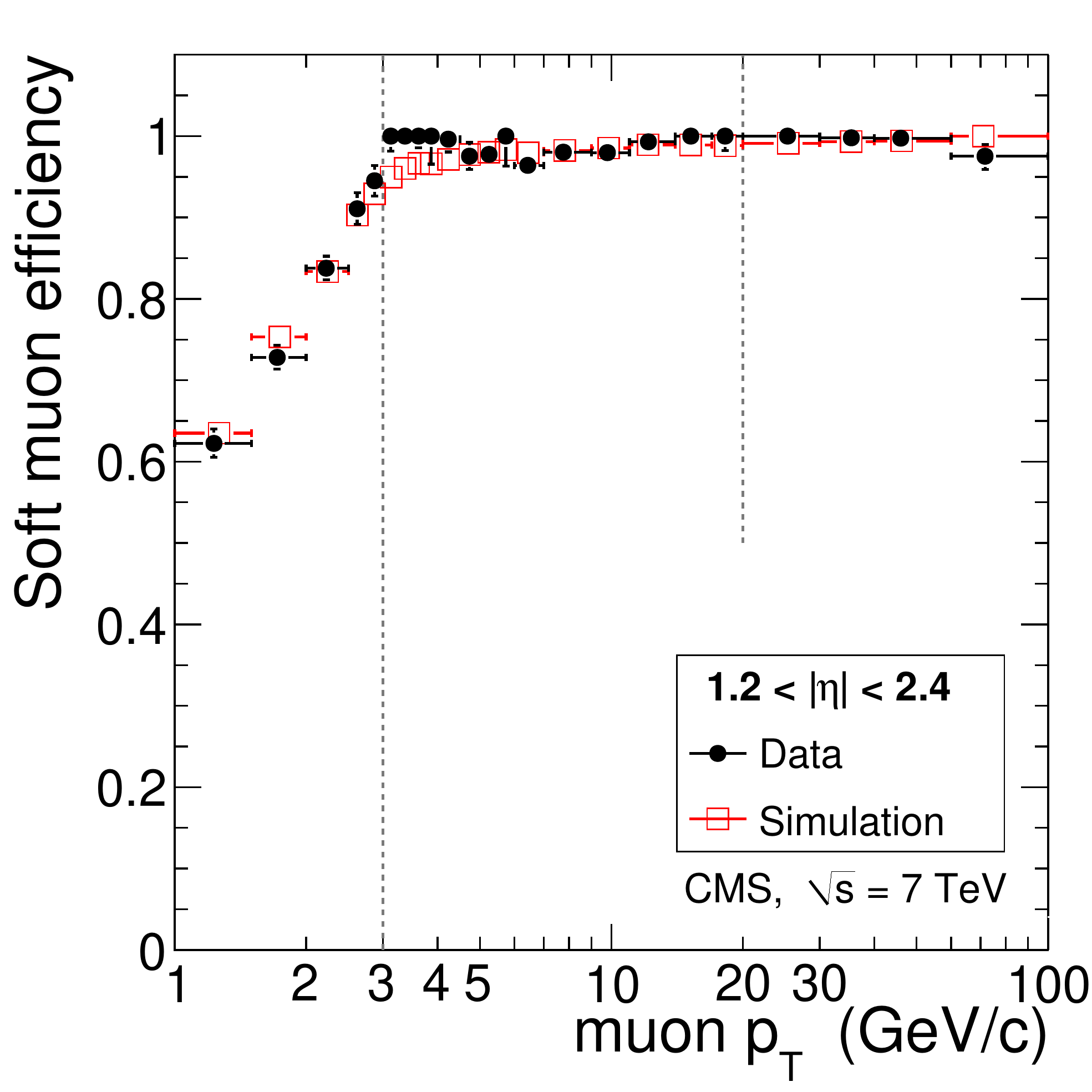}
\includegraphics[width=0.32\textwidth]{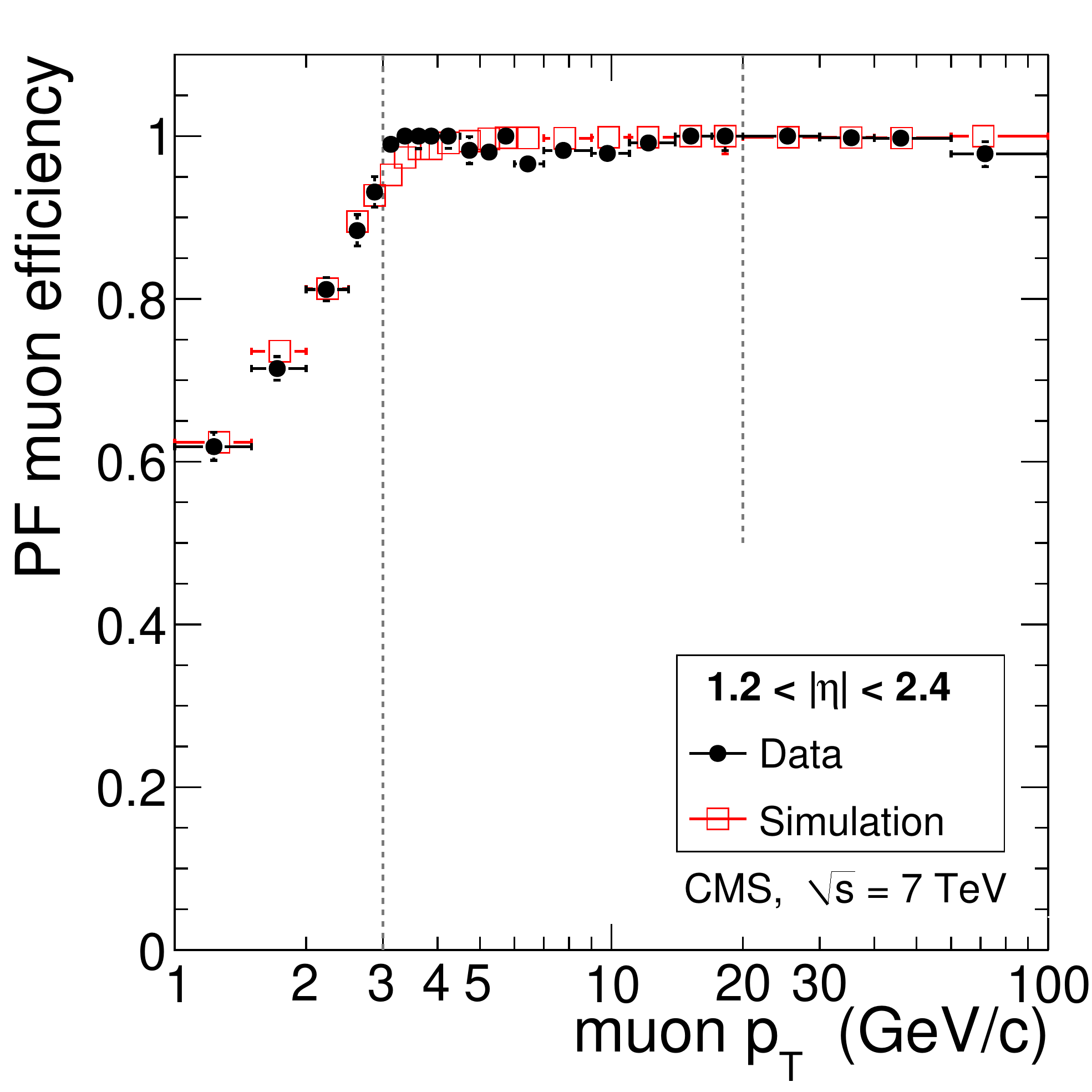}
\includegraphics[width=0.32\textwidth]{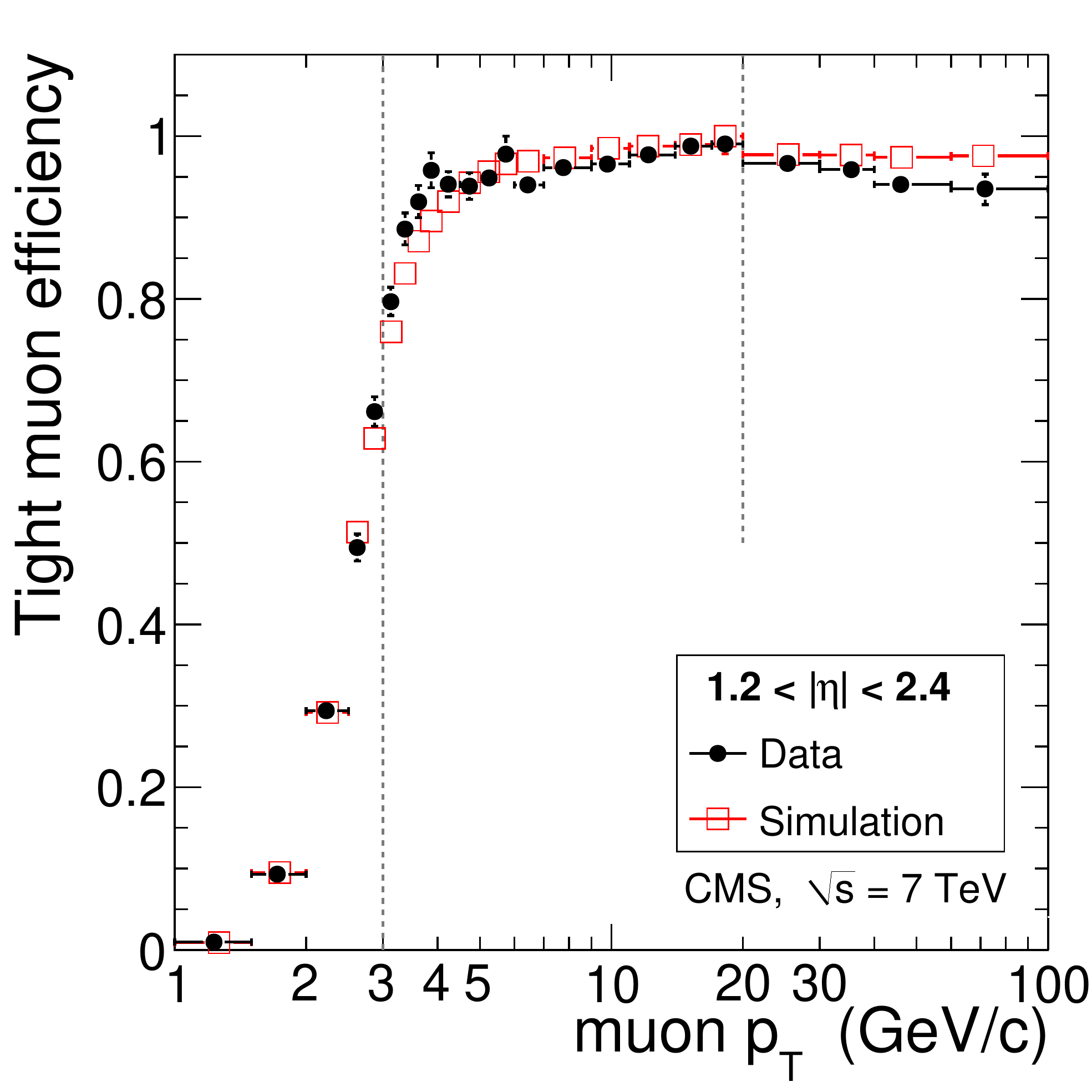}

    \caption{Tag-and-probe results for the muon efficiency $\epsilon_{\mathrm{rec+id}}$ in data compared to simulation.
    Given that a tracker track exists, the plots show the efficiency as a function of muon $\pt$ for
    Soft Muons (left), Particle-Flow Muons (middle), and Tight Muons (right) in the barrel and overlap regions (top), and in the endcaps (bottom).
    The measurement is made using $\mathrm{J}/\!\psi\to\mm$ events for $\pt < 20\GeVc$ and $\Zmm$ events for $\pt > 20\GeVc$.
    For $\pt < 3\GeVc$, to reduce the background, only tracks with MIP signature are considered.}
    \label{fig:TP1}
  \end{center}
\end{figure}

Figure~\ref{fig:TP1} shows the muon efficiency $\epsilon_{\mathrm{rec+id}}$ given that a tracker track exists, measured using
$\mathrm{J}/\!\psi\to\mm$ and $\Zmm$ events. The results obtained from the data collected in the 2010 LHC data-taking period are compared
with those from simulated events.

For comparisons with $\Zmm$ events, an unweighted sample of simulated events corresponding to an integrated luminosity of $\approx$330$\mathrm{pb}^{-1}$ is
used: the simulated samples are $\Zmm$, $\W+$jets, and muon-enriched QCD (see Section~\ref{sec:samples}).
For studies at the $\mathrm{J}/\!\psi$ peak, separate samples of prompt $\mathrm{J}/\!\psi\to\mm$ and $B\to\mathrm{J}/\!\psi+X\to\mm+X$ are used,
simulated as described in Section~\ref{sec:samples}.  All MC samples used
for the results in this section included simulation of pile-up.
Simulation of the background processes is not included for the $\mathrm{J}/\!\psi$ case, as it would be impractical to simulate a sufficient number of inclusive muon-plus-track events.
For studies of systematic uncertainties described below, samples
of background events have been generated according to the background invariant
mass spectra determined from fits to the $\JPsi\to\mm$ events in the data,
and added to the simulated signal events.

The tag-and-probe results in data and in simulation agree within the statistical uncertainties of the measurement almost everywhere.
The only significant discrepancy is in the barrel around the turn-on of the efficiency curves, where the efficiency in data is systematically
higher than in the simulation. This discrepancy arises from a small difference in the widths
of the track-to-segment pulls in data and in simulation discussed in
Section~\ref{sec:muonidvar}:
the efficiency of the track-to-segment matching is slightly higher in data, and in the region of rapidly rising efficiency the effect is amplified by the large variation of the efficiency in the bin.
The 1--2\% data-simulation difference in efficiency for Tight Muons in the endcaps is explained by the fact that several CSCs not operational during most of the 2010 data taking were simulated as fully efficient;
this has a negligible effect on efficiencies of the other muon selections
because they require a match with only a single muon segment.
Using a small sample of simulated events, we have verified that when
these chambers are properly accounted for in the
simulation, the efficiencies for Tight Muons in data and in simulation agree
to better than 1\%.

The efficiency for the Tight Selection measured on muons from $\Zmm$ is slightly lower than that measured on muons from
$\mathrm{J}/\!\psi\to\mm$. This difference is partly due to the bias introduced by the muon-plus-track
trigger on the track quality criteria described previously and partly due to the different kinematics of the probes. The effect is well
reproduced by the simulation.

For Soft Muons and Particle-Flow Muons the plateau of the efficiency is reached at $\pt$ $\approx$4\GeVc in the endcaps and $\approx$6\GeVc in the barrel,
while for Tight Muons it is reached at $\approx$10\GeVc in both regions.
The values of efficiencies at the plateau region obtained using $\mathrm{J}/\!\psi\to\mm$ and $\Zmm$ events in data and simulation are given in Table~\ref{tab:TPplateau}.
The efficiencies are high, and data and simulation are generally in good
agreement.
The plateau efficiency for Soft Muons is 1--2\% higher in data than in simulation, again due to a difference in the widths of the track-to-segment pulls.
The Particle-Flow and Tight Muon selections are much less affected by this difference because they use looser matching criteria between tracks and muon segments.
The efficiency at the plateau for the Particle-Flow Muon selection is very close to 100\% because the algorithm applies relaxed selection criteria to
the high-\pt muon candidates if they are isolated.

\begin{table}%[htpb]
    \begin{center}
        \topcaption{Muon efficiencies at the efficiency plateau for the different muon selections:
                 efficiency measured from data, and ratio between the measurements in data and simulation.
                 The first uncertainty quoted on the scale factor is the uncertainty on the efficiencies in data and
                 simulation from the fitting procedure, which includes the statistical uncertainty; the second is from the additional systematic
		 uncertainties described later in this section.}
        \begin{tabular}{|lc|cc|cc|} \hline
\multicolumn{2}{|l|}{Muon selection}  & \multicolumn{2}{c|}{$\mathrm{J}/\!\psi\to\mm$} &  \multicolumn{2}{c|}{$\Zmm$}  \\
                   & Region               & Eff. [\%]                  &  Data/Sim. ratio        &  Eff. [\%]        & Data/Sim. ratio            \\ \hline \hline & & & & & \\[-12pt]
Soft               & $0.0 < |\eta| < 1.2$ & $98.4_{-0.3}^{+0.3}$  &  $1.010 \pm 0.003 \pm 0.010$
                                          & $99.2_{-0.1}^{+0.1}$  &  $1.014 \pm 0.001 \pm 0.002$  \\[0.6mm]
                   & $1.2 < |\eta| < 2.4$ & $98.0_{-0.7}^{+0.7}$  &  $1.002 \pm 0.007 \pm 0.014$
                                          & $99.9_{-0.2}^{+0.1}$  &  $1.005 \pm 0.002 \pm 0.004$  \\[0.6mm] \hline & & & & & \\[-12pt]
Particle-          & $0.0 < |\eta| < 1.2$ & $98.8_{-0.3}^{+0.3}$  &  $0.993 \pm 0.003 \pm 0.010$
                                          & $99.7_{-0.1}^{+0.1}$  &  $0.999 \pm 0.001 \pm 0.002$  \\[0.6mm]
\multicolumn{1}{|r}{Flow }
                   & $1.2 < |\eta| < 2.4$ & $98.4_{-0.7}^{+0.7}$  &  $0.988 \pm 0.007 \pm 0.014$
                                          & $99.8_{-0.2}^{+0.1}$  &  $0.999 \pm 0.002 \pm 0.004$  \\[0.6mm] \hline & & & & & \\[-12pt]
Tight              & $0.0 < |\eta| < 1.2$ & $98.4_{-0.3}^{+0.3}$  &  $0.998 \pm 0.004 \pm 0.010$
                                          & $96.4_{-0.2}^{+0.2}$  &  $0.999 \pm 0.002 \pm 0.002$  \\[0.6mm]
                   & $1.2 < |\eta| < 2.4$ & $96.8_{-0.7}^{+0.7}$  &  $0.979 \pm 0.007 \pm 0.014$
                                          & $96.0_{-0.3}^{+0.3}$  &  $0.983 \pm 0.003 \pm 0.004$  \\[0.3mm] \hline       \end{tabular}
        \label{tab:TPplateau}
    \end{center}
\end{table}

\begin{figure}[htb]
  \begin{center}
\includegraphics[width=0.32\textwidth]{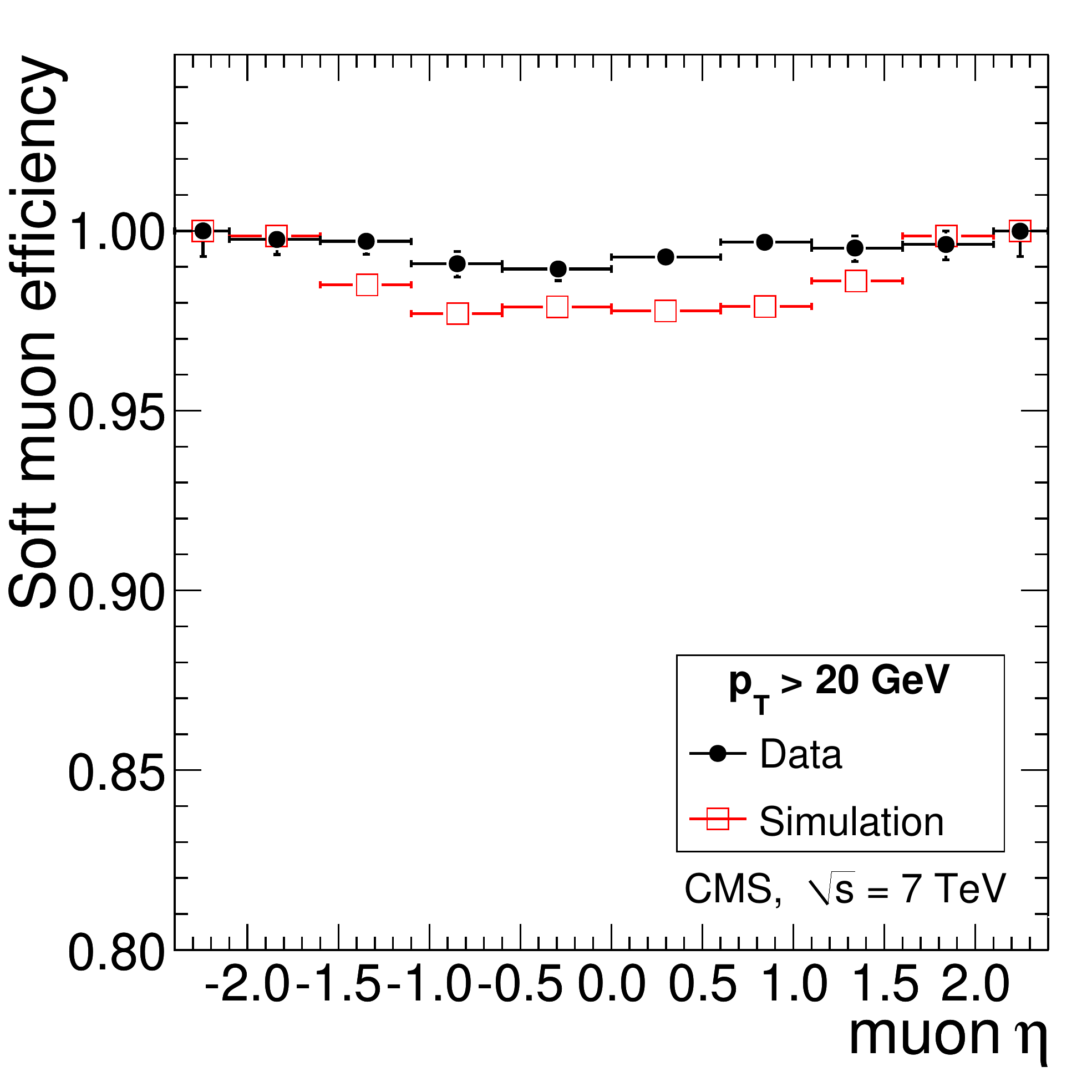}
\includegraphics[width=0.32\textwidth]{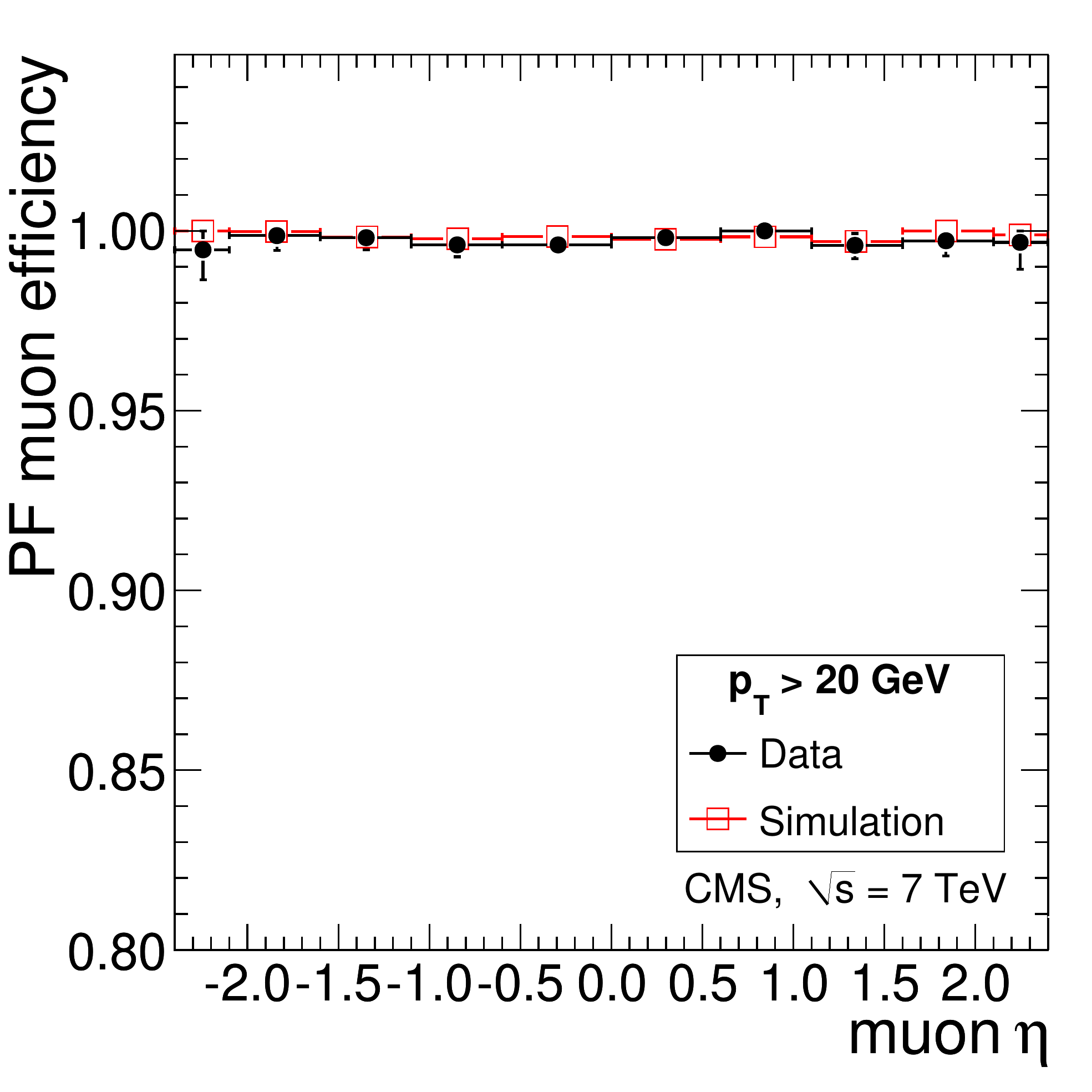}
\includegraphics[width=0.32\textwidth]{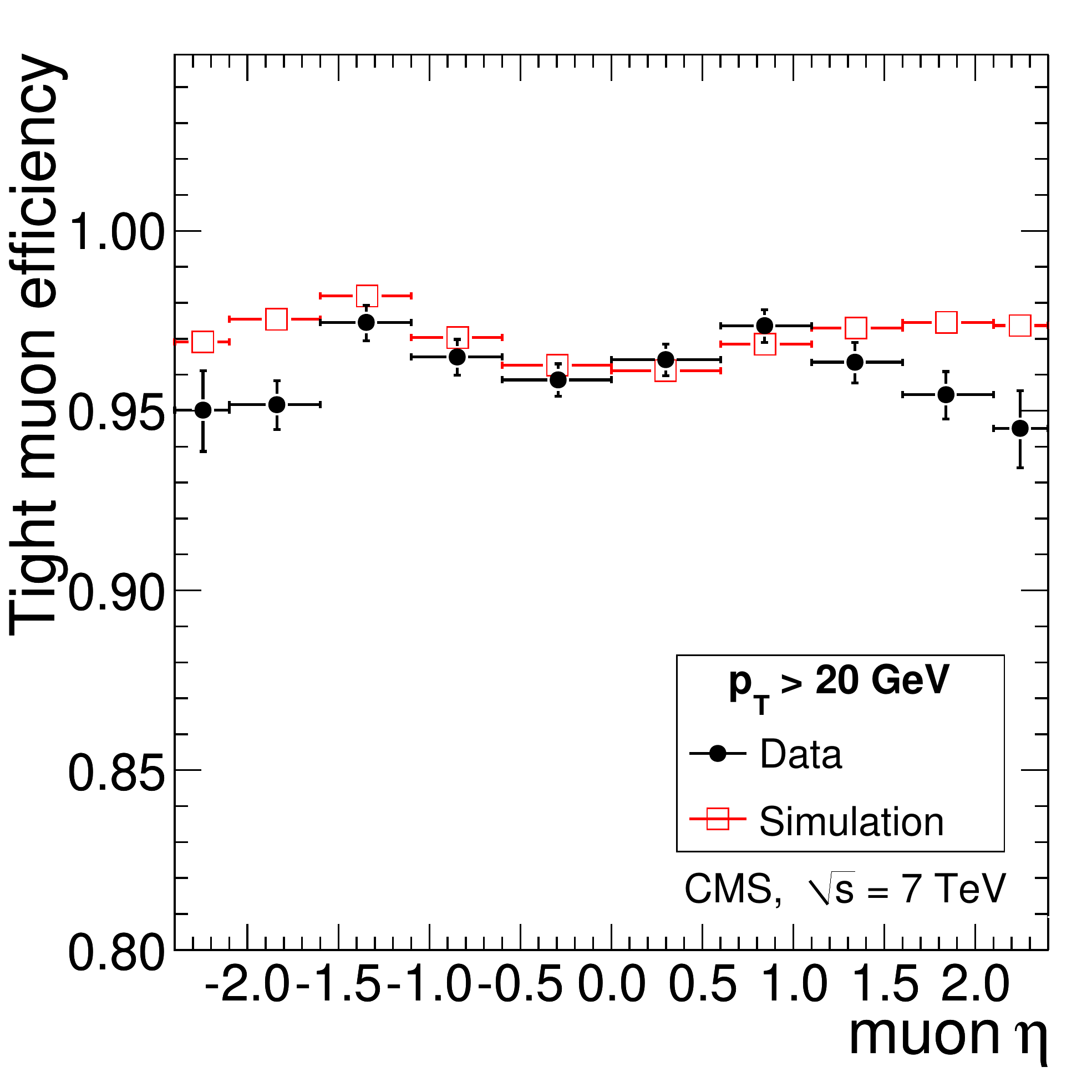}
    \caption{Muon efficiency $\epsilon_{\mathrm{rec+id}}$ in data and
    simulation as a function of muon pseudorapidity for Soft Muons (left),
    Particle-Flow Muons (middle), and Tight Muons (right).  The efficiencies
    were calculated relative to the tracker tracks with $\pt > 20\GeVc$
    by applying the tag-and-probe technique to $\Zmm$ events.}
    \label{fig:TP2}
  \end{center}
\end{figure}

The dependency of the plateau efficiency on the pseudorapidity is measured using $\Zmm$ events and is shown in Fig.~\ref{fig:TP2}. The data and
simulation agree to better than 2\%. % in the barrel for $\pt > 20\GeVc$.

To estimate the effect of pile-up on muon identification performance, the
efficiency at the plateau is measured for both $\mathrm{J}/\!\psi\to\mm$ and $\Zmm$ events as a function of the number of reconstructed primary vertices.
No loss of efficiency is observed for events containing up to six reconstructed primary vertices, the maximum
multiplicity for which a measurement could be made with a statistical uncertainty below 10\%.

Trigger efficiencies obtained by a similar tag-and-probe technique are
described in Section~\ref{sec:trigger}.

\subsubsection{Systematic uncertainties} \label{sec:muonideff_syst}

The various contributions to the possible systematic bias in the
data-to-simulation ratios of efficiencies calculated using the
tag-and-probe method are estimated using simulated and real data.

Bias in the measured efficiencies that could be introduced by
the tag-and-probe method and its implementation is studied by comparing
the efficiencies obtained by applying the tag-and-probe method to simulated
data containing $\mathrm{J}/\!\psi\to\mm$ ($\Zmm$) decays
and various background contributions with the ``true'' efficiencies
computed by simple counting of the passing and failing
probes in $\mathrm{J}/\!\psi\to\mm$ ($\Zmm$) MC events.
The difference in the efficiencies is less than 0.5\% for
muons from $\mathrm{J}/\!\psi\to\mm$. It is also less than 0.5\% for
most muons from $\Zmm$ ($20 < \pt < 60\GeVc $), and less than
1.5\% for the others. The differences are compatible with
zero within the statistical uncertainties; hence, no systematic
uncertainty is assigned as a result of this test.

For $\mathrm{J}/\!\psi$ events, the efficiencies are
recomputed with the tag-and-probe method using a simple Gaussian
instead of a Crystal Ball function~\cite{crystalball} to model the
resonance and with a quadratic polynomial instead of an exponential
to model the background. The differences in the efficiencies resulting
from this variation in the assumed signal shape are under 0.1\%. The
efficiencies obtained with a polynomial background shape are
systematically $\approx$1\% higher than those obtained using an
exponential background. The difference between the two results
is taken as a conservative estimate of systematic
uncertainty in the background modelling.
The same efficiencies have also been recomputed
without the requirement that the probe tracks have a MIP signature in the
calorimeters.  The results are fully compatible with those obtained with
a MIP requirement, but have larger uncertainties.
Simulation shows that in this low-$\pt$ range
tag-and-probe efficiencies estimated with a MIP requirement are
systematically higher, by 1--2\%, than without a MIP requirement,
due to small correlations between the energy deposition in the
calorimeters and the number of hits in the muon chambers.
This bias cancels out in the data/simulation ratio, so no corrections
accounting for it are made.

For $\Zmm$ events, the efficiencies are recomputed using only
isolated probe tracks, which reduces the background by a factor of
two. The results agree with those from all probes at the level of
0.1\%. As a conservative estimate of systematic uncertainty
on the plateau efficiencies resulting from the background estimation,
the scale of the largest difference between this estimate and that
from simulation is taken; this amounts to 0.2\% in the barrel
and 0.4\% in the endcaps.

For $\mathrm{J}/\!\psi$ events, the kinematic distributions of the
signal probes are extracted from the distributions of all probes
by using the SPlot technique~\cite{SPlot}. The distributions were
found to be in good agreement with those predicted by simulation, and
therefore no systematic uncertainty is assigned to the procedure
of averaging the efficiencies for different probes within each
$(\pt,\eta)$ bin.

Possible bias in the measurements of single-muon efficiencies due to
the presence of a second, tag muon in the event is studied by
changing the separation criteria from the angular separation $\Delta R
> 0.5$ to the tighter requirement that the distance between the coordinates
of the two muons in the innermost muon station be larger than 2 m
and by using only the pairs of muons that
bend away from each other in the magnetic field inside the
solenoid. The effect on the efficiencies measured using
$\mathrm{J}/\!\psi$ events is 1\% in the endcaps and 0.3\% in the
barrel; this difference is taken as an additional systematic uncertainty on
the data-to-simulation ratios of efficiencies.  The impact on the
efficiencies measured using $\Zmm$ events is negligible, and no
additional systematic uncertainty is assigned to them.

In the measurements of efficiencies using $\mathrm{J}/\!\psi\to\mm$
events, no attempt is made to separate promptly produced
$\mathrm{J}/\!\psi$ from those originating from the decay of b quarks.
Differences in efficiencies obtained using these two samples of muons
are studied in simulation and have been found to be less than 1\% and
compatible with the statistical uncertainties on each. In the
kinematic range over which the measurement is made, the fraction of
$\mathrm{J}/\!\psi$'s from the decays of b quarks is always below
50\%~\cite{BPH-10-002}, so the possible effects on the
data-to-simulation efficiency ratios are below 0.5\%.  No additional
systematic uncertainty is assigned.

The various contributions to systematic uncertainty were combined in
quadrature; the overall uncertainties at the efficiency plateau are
shown in Table~\ref{tab:TPplateau}.

\subsection{Reconstruction and identification efficiency for nearby muons}
\label{sec:collimatedmus}
If a muon has one or more other muons in its vicinity, their signals
in the muon system could overlap, resulting in identification
efficiencies lower than for single or well-separated muons.  For
example, such topologies are common for muon pairs produced in the decays
of low-mass resonances such as $\JPsi$.  Another example is
hypothetical highly collimated leptons, also referred to as ``lepton
jets'', predicted in different
models~\cite{bib-nima,bib-weiner,bib-tuckerweiner} proposed to explain
the excess of cosmic-ray leptons in recent astrophysical
observations~\cite{bib-PAMELA,bib-ATIC,bib-ATIC2}.  In this section,
we report the measurement of reconstruction and identification efficiency for such
nearby muons.

The muon identification performance for nearby muons is studied using the data collected
during the 2010 LHC data-taking period and compared to the expectations from
simulation, in which boosted muon pairs are generated
using \PYTHIA~\cite{PYTHIA}.
Two muon selections are considered: Tracker Muons with at least two
tightly matched %arbitrated
segments in the muon system, and Tight Muons.
Studies on simulated
events have shown that the purity of Tracker Muons with two or more well-matched muon segments is similar to that of Tight Muons.

In this study, the efficiency of identifying nearby muons as Tight or Tracker Muons is measured using a sample of dimuons from the decays of low-mass resonances:
$\mathrm{J}/\!\psi$, $\phi$, and $\rho/\omega$.  These resonances provide muon pairs
with kinematic and topological properties similar to those expected
for hypothetical collimated muons, notably a small angular separation
between the muons. % and therefore are also a good stand-in for collimated muons.

The sample used for this study consists of pairs of tracker tracks
each with \pt above 5\GeVc, associated with the same primary vertex,
and for which the invariant mass is in the vicinity of the invariant
mass of one of the above resonances.
Because the lower mass resonances ($\phi$ and $\rho/\omega$) have very
large combinatorial background, both tracks are also required to have
one loosely matched segment in the muon system.
To check for any bias introduced by this requirement a measurement was made with tracker tracks from $\mathrm{J}/\!\psi$ decays identified as possible muon candidates by using only the calorimeter information. This gives very similar results hence showing any bias is small.
The contribution from the residual background is evaluated from
fits to the side bands, by a procedure similar to that used in the
tag-and-probe method (Section~\ref{sec:muonideff_tnp}), and suitably
subtracted.
The efficiency is defined as the ratio of the number of dimuon candidates
passing both the above selection and the muon identification under study to
the number of dimuons passing the above selection.

\begin{figure}[thb]
  \begin{center}
   \includegraphics[width=0.45\textwidth]{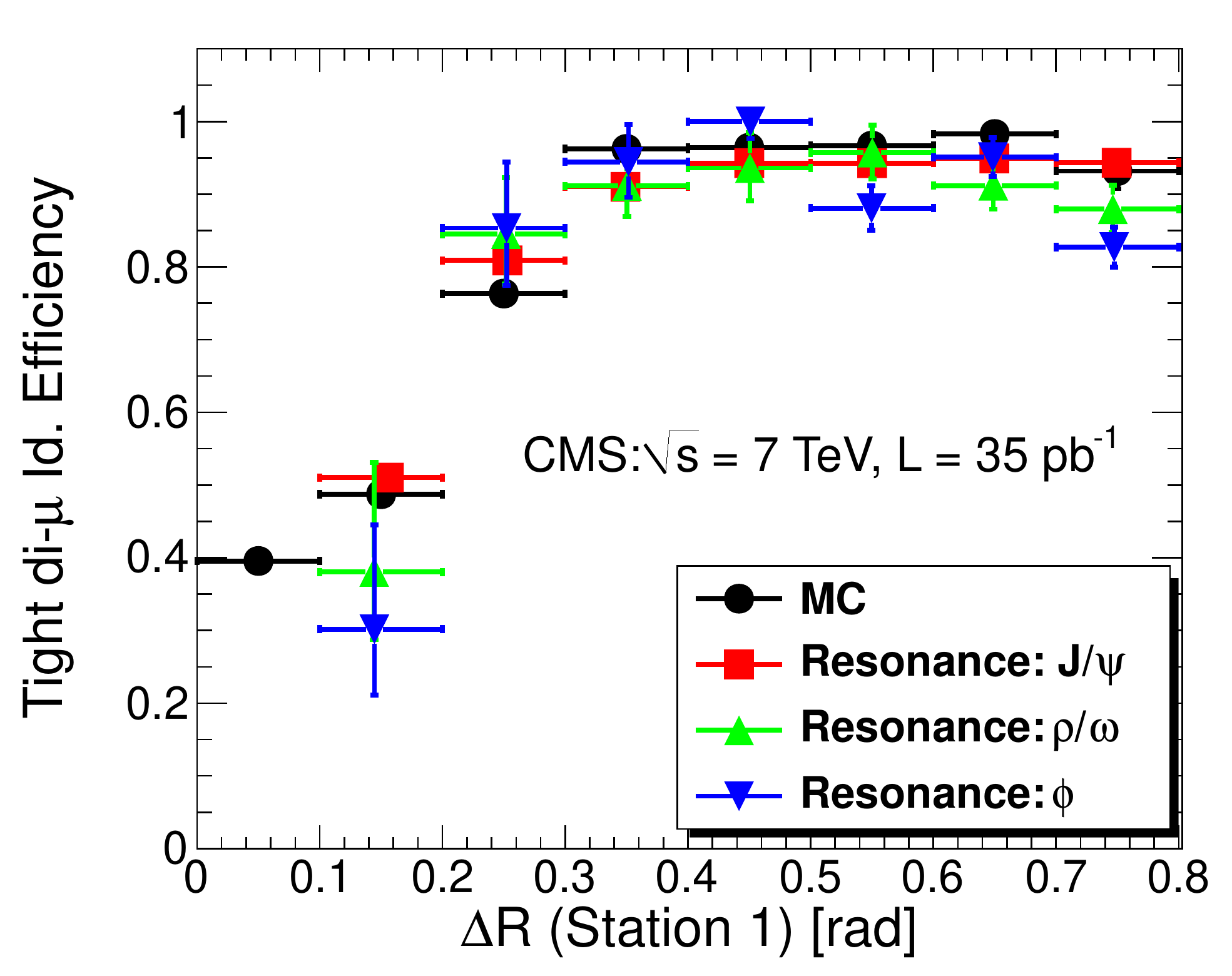}
   \includegraphics[width=0.45\textwidth]{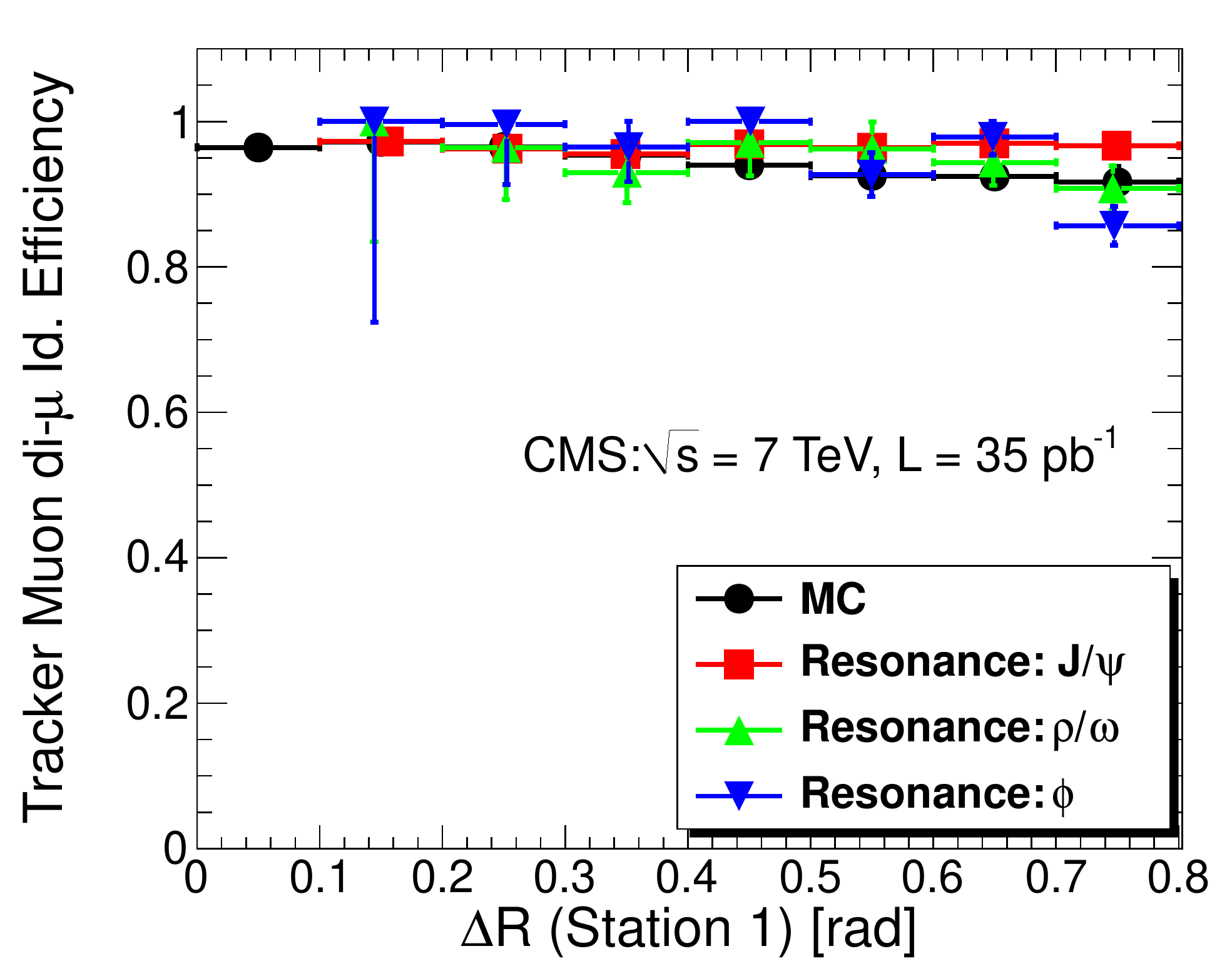}
    \caption{Efficiency for identifying both muons in the dimuon pair as Tight (left) and Tracker (right)
      Muons as a function of the angular separation of the two tracks computed at the surface of
      the first muon station. Measurements obtained using $\mathrm{J}/\!\psi$ (squares), $\phi$
      (inverted triangles), and $\rho$/$\omega$ (triangles) are compared with the expectations
      from the simulation (circles).}
    \label{fig:TP-CloseBy}
  \end{center}
\end{figure}

The efficiencies for identifying both muons in the dimuon pair as Tight or
Tracker Muons are shown in Fig.~\ref{fig:TP-CloseBy}.  The efficiencies
are plotted as a function of the angular separation of the two tracks,
$\Delta R =\sqrt{(\Delta \eta)^2 +(\Delta \phi)^2}$, computed at the
surface of the first muon station; the collimated muons are expected
to populate the range of $\Delta R \lesssim 0.7$.  For Tight Muons, a drop in
efficiency at small values of $\Delta R$ is observed; this
inefficiency is introduced by a cleaning procedure used at
the seeding stage of the global muon reconstruction to eliminate
muon seeds leading to duplicate muons.
The efficiency for Tracker Muons, however, remains
high at all $\Delta R$ values, demonstrating that Tracker Muons
are fully adequate for studies involving nearby muons.  For
both types of selections, the results obtained using different resonances
are in good agreement, demonstrating that the dependence of the efficiency
on $\Delta R$ is not affected by the decay kinematics and combinatorial
background.  The results of the measurements are also well
reproduced by the Monte Carlo simulation.

\subsection{Muon identification probability for particles other than muons}
\label{sec:ExclusiveRates}

One can obtain pure samples of kaons, pions, and protons from
resonances of particle decays such as $\mathrm{K}^0_\mathrm{S}
\rightarrow \pi^+ \pi^-$, $\Lambda \rightarrow \mathrm{p} \pi^-$ (and charge
conjugate), and $\phi \rightarrow \mathrm{K}^+ \mathrm{K}^-$. The
resonances are reconstructed using pairs of tracker tracks that are
associated with a common decay vertex, with a selection similar to
that described in Ref.~\cite{TRK-10-001}.  In $\Lambda$ decays, the
highest momentum track is assumed to be that of the proton. A data
sample collected with a jet trigger (minimum $\pt$ of 15\GeVc) is used,
and simulated QCD events, filtered using the same jet trigger, are
used for comparison.  The simulated events have been reweighted to
account for a small difference in the hadron momentum spectrum with
respect to the data sample.

We compute the fraction of events in which a hadron track is
identified as a Soft Muon, Particle-Flow Muon, or Tight Muon as a
function of several relevant track parameters.  Background subtraction
using resonance sidebands is performed to determine the muon
misidentification probability for the particles under study.  Invariant
mass spectra are fit with a sum of signal and background shapes, using
a double Gaussian for the signal and a power law for the background.  One fit
to the entire mass spectrum is made for each resonance to provide the
scale factor between the number of hadrons counted in the sideband
region and the background estimation in the signal region.  The scaled
number of hadrons in the sideband region is then subtracted from the
number counted in the signal region, in each bin of the distribution
of the hadron track parameter under study.  The same
background-subtraction procedure is then repeated only for hadrons
that share the tracker track with that of a muon.  By dividing the
sideband-subtracted number of muon-matched hadrons by the
sideband-subtracted number of hadrons before any matching to muons, we
obtain a misidentification probability for a given hadron type.  The
same method is applied to events in data and simulation.

\begin{figure}[tb]
  \begin{center}
        
    \includegraphics[angle=90,width=0.32\textwidth]{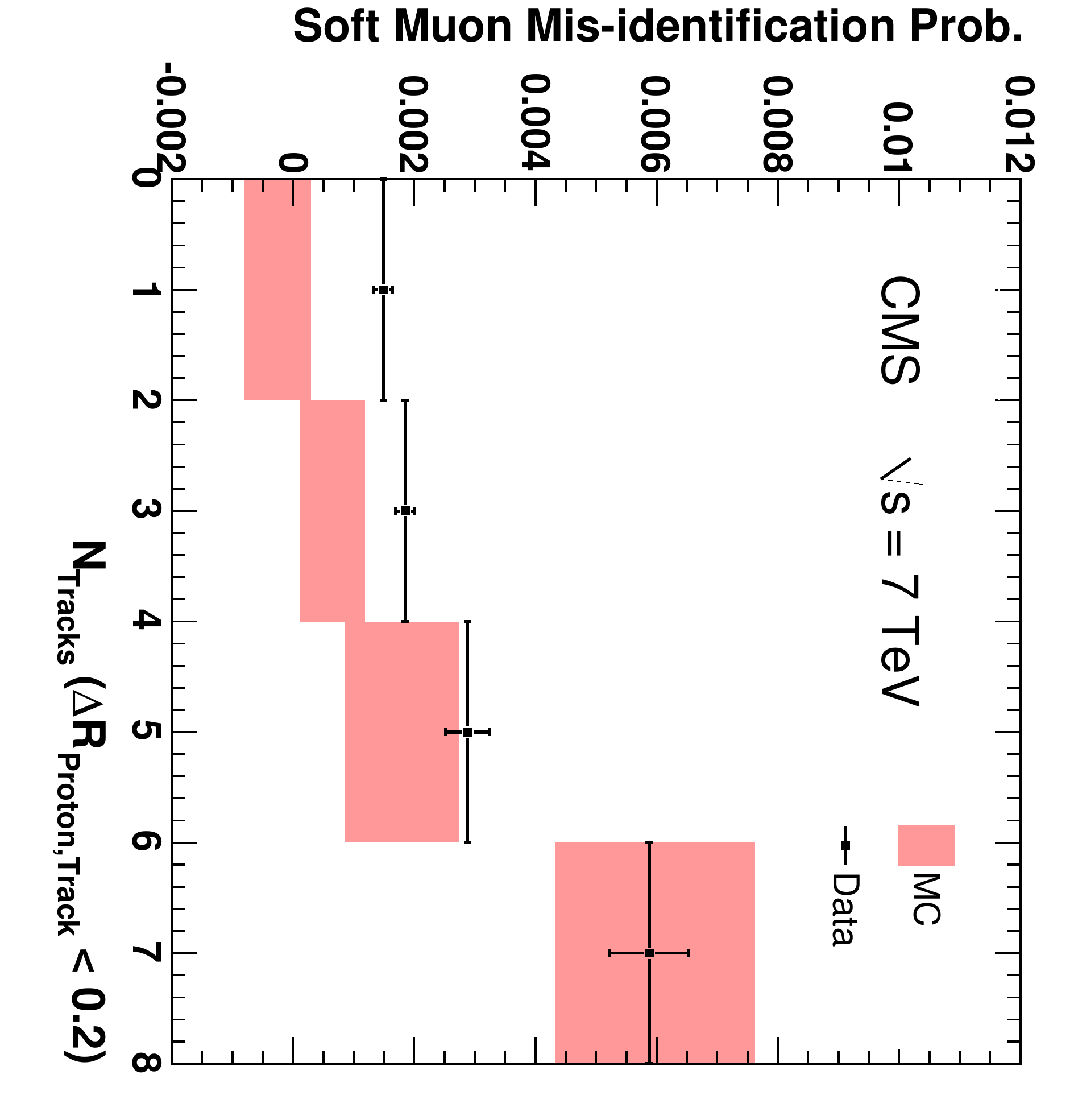} 
    \includegraphics[angle=90,width=0.32\textwidth]{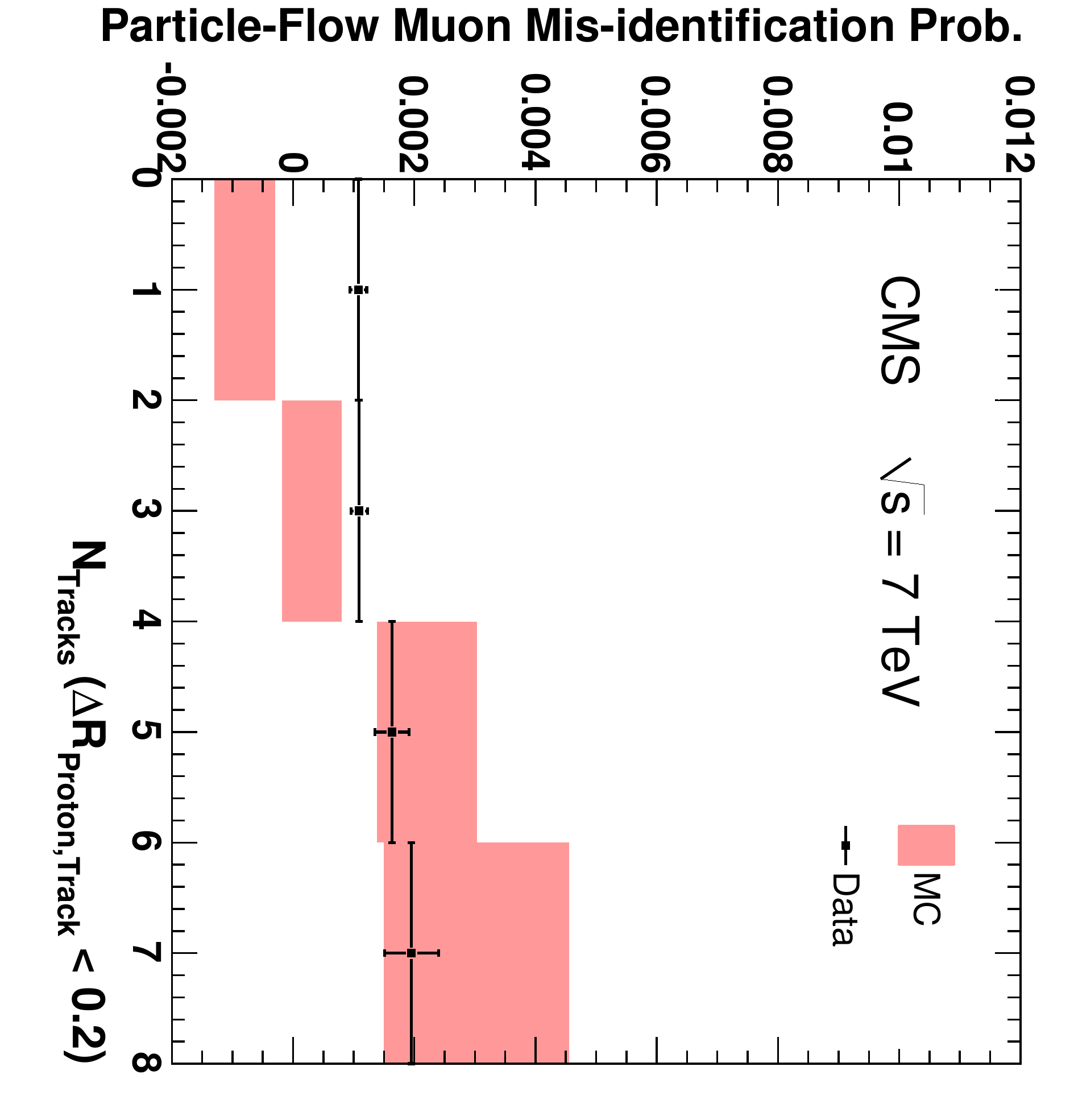} 
    \includegraphics[angle=90,width=0.32\textwidth]{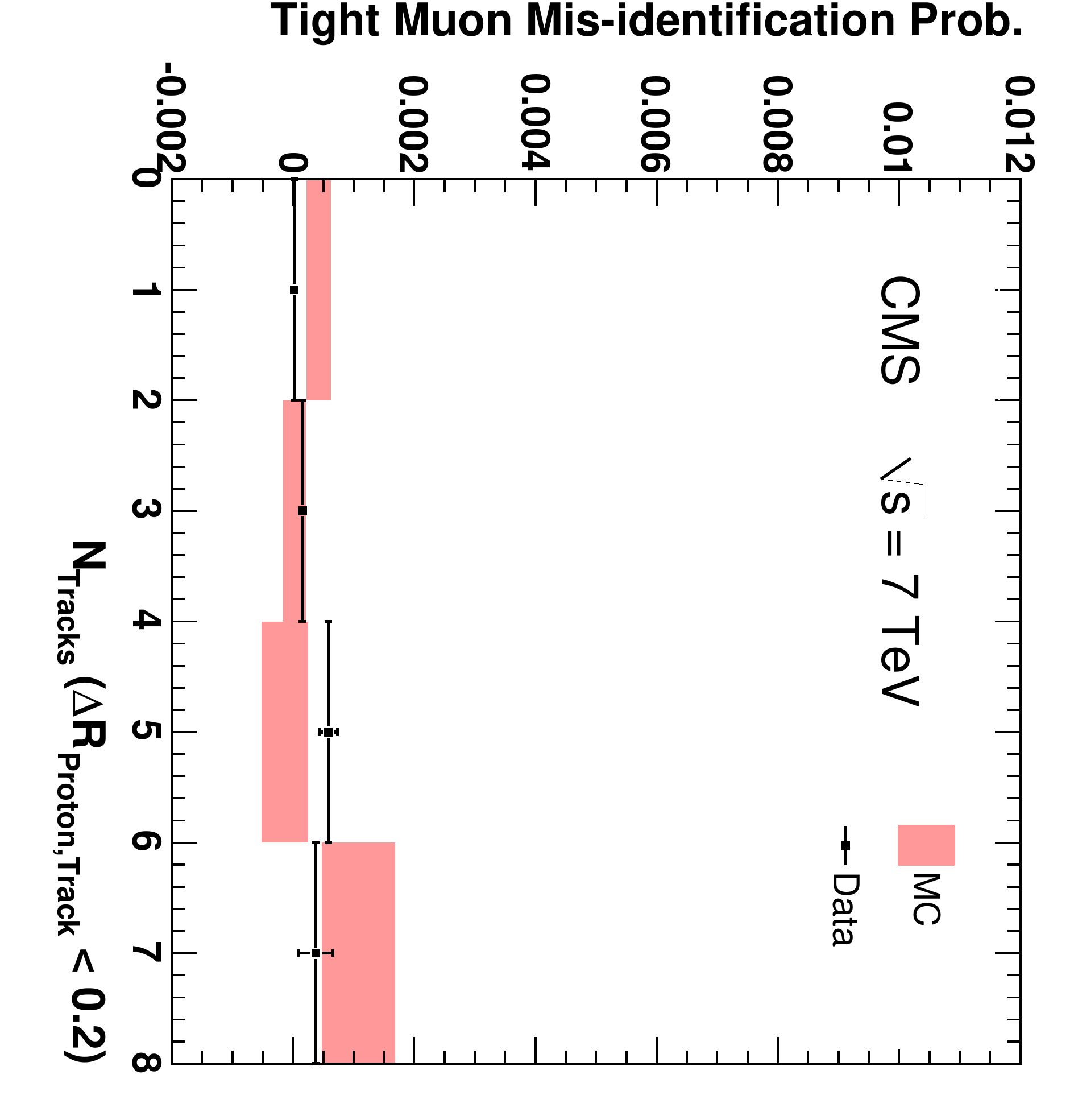}
    \caption{The fraction of protons that are misidentified as a Soft
    Muon (left), Particle-Flow Muon (centre), or Tight Muon (right) as
    a function of $N_{\mathrm{Tracks}}$, where $N_{\mathrm{Tracks}}$
    represents the number of tracks in the vicinity of the proton
    track (with ${\Delta}R = \sqrt{(\Delta\eta)^2 + (\Delta\phi)^2} <
    0.2$). Only protons with $p > 3\GeVc$ are included.  The first bin
    includes events with $N_{\mathrm{Tracks}} = 0$ and 1, the second
    bin includes those with $N_{\mathrm{Tracks}} = 2$ and 3, etc.  The
    uncertainties indicated by the error bars (data) and shaded boxes
    (\PYTHIA simulation) are statistical only.  Negative values arise
    from statistical fluctuations in the number of events in the signal
    and sideband regions.}
    \label{fig:Fakes_nTrk}
  \end{center}
\end{figure}

In addition to punch-through and decay in flight, there is a third mechanism by
which hadrons can be misidentified as prompt muons.  This mechanism is
random matching between the hadron track in the inner tracker and a track stub
in the muon system from one of the other tracks in the jet that may be due to a muon.  The
frequency of random matching is sensitive to the particular event
topology of the sample.  For example, in jet-triggered events, the
increase in the average number of tracks per event in comparison to
minimum-bias-triggered events leads to an increased probability of
random matching.  To illustrate the effect of
random matching, we present the proton-to-muon misidentification
probabilities as a function of $N_{\mathrm{Tracks}}$ in
Fig.~\ref{fig:Fakes_nTrk}, where $N_{\mathrm{Tracks}}$ is the number
of tracks in the vicinity of the proton track, within a cone of radius
${\Delta}R < 0.2$.  It is
clear for both data and simulation that the misidentification
probability increases with $N_{\mathrm{Tracks}}$ especially in the Soft Muon selection.
To remove much of the contribution due to random matching from the probability of misidentifying a hadron as a muon, we impose a requirement
of $N_{\mathrm{Tracks}} < 4$ for the rest of the results in this section.

\begin{figure}[tb]
  \begin{center} 
  
    \includegraphics[angle=90,width=0.32\textwidth]{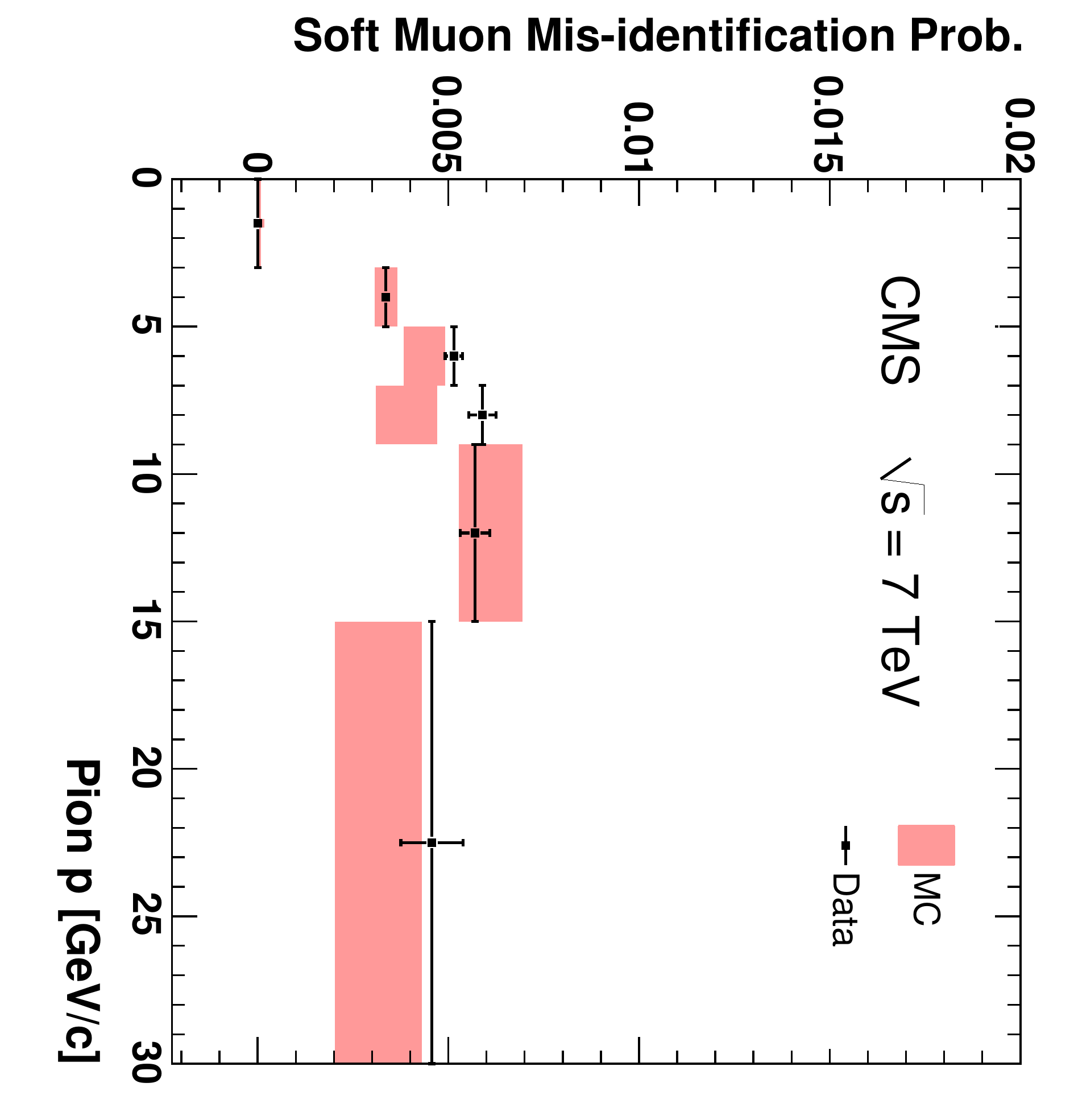}
    \includegraphics[angle=90,width=0.32\textwidth]{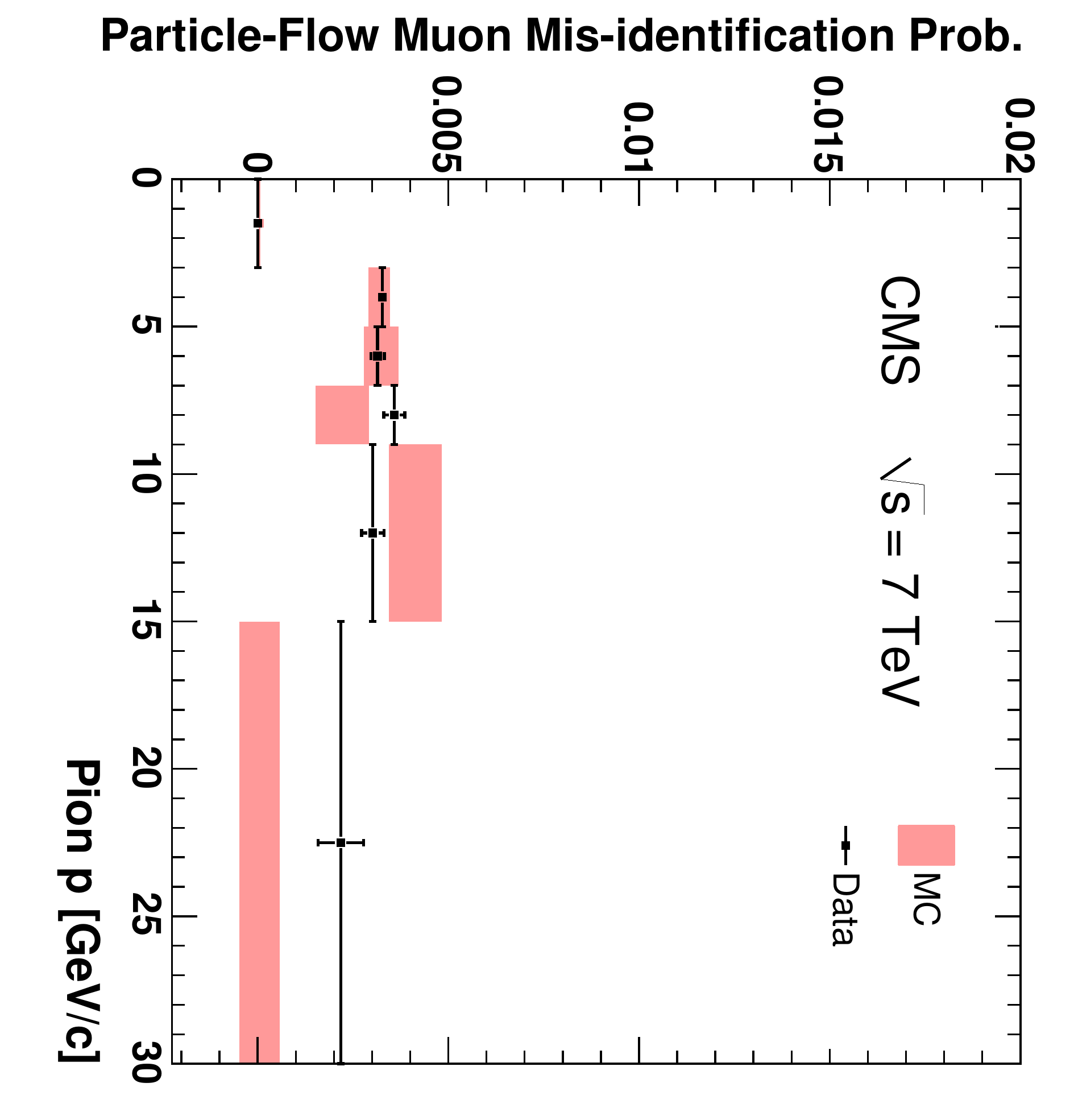} 
    \includegraphics[angle=90,width=0.32\textwidth]{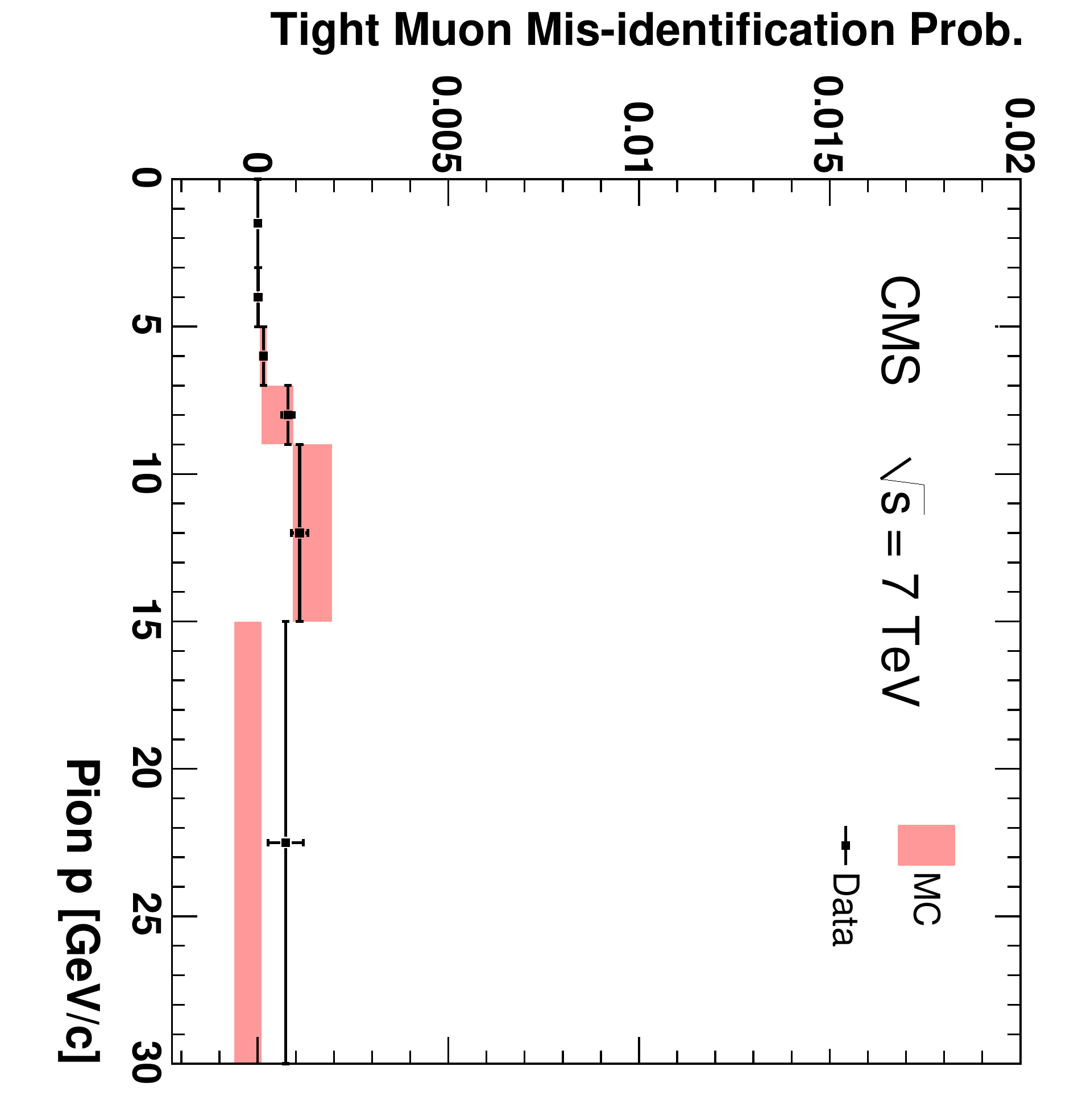} 
        
    \includegraphics[angle=90,width=0.32\textwidth]{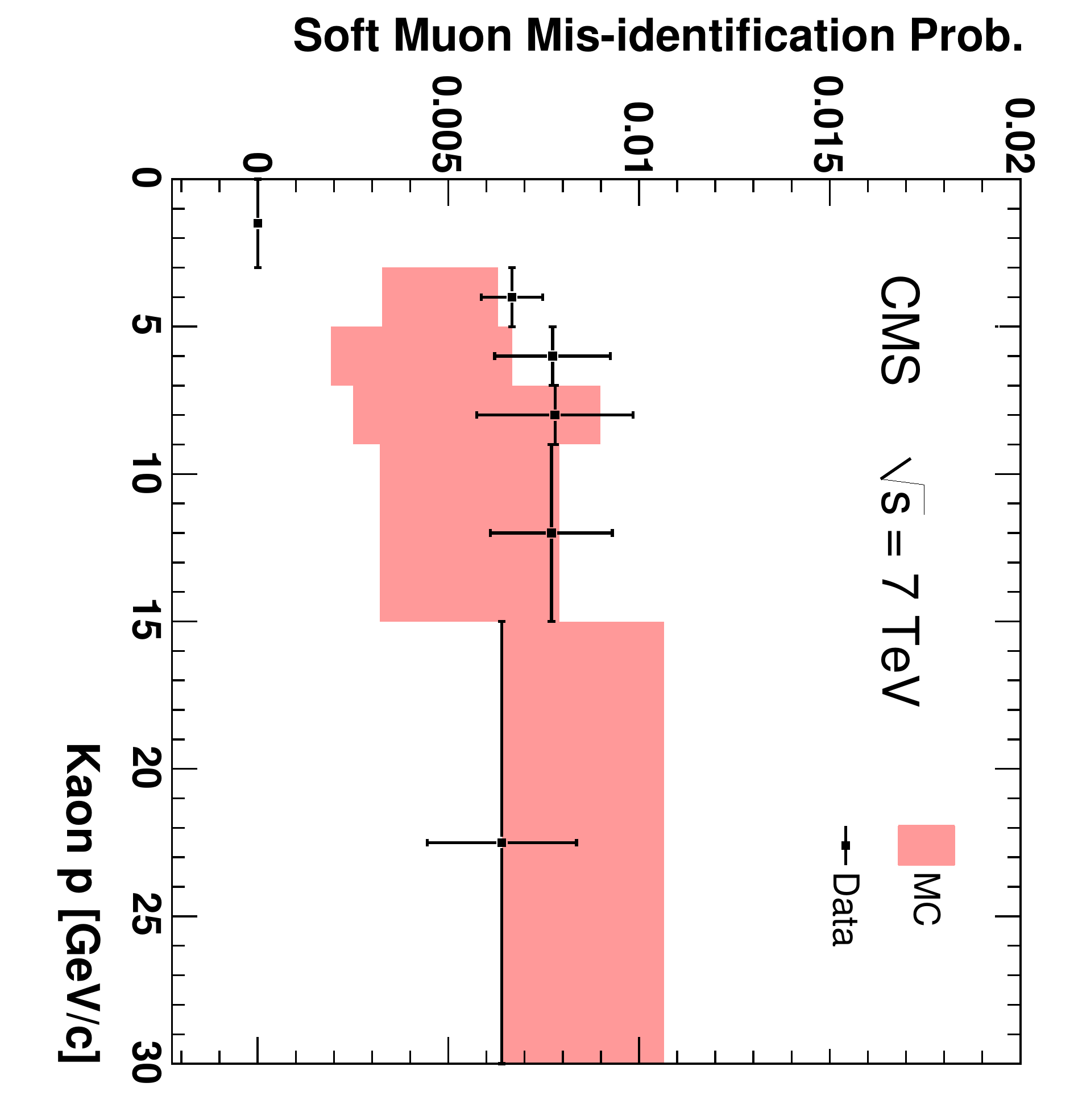} 
    \includegraphics[angle=90,width=0.32\textwidth]{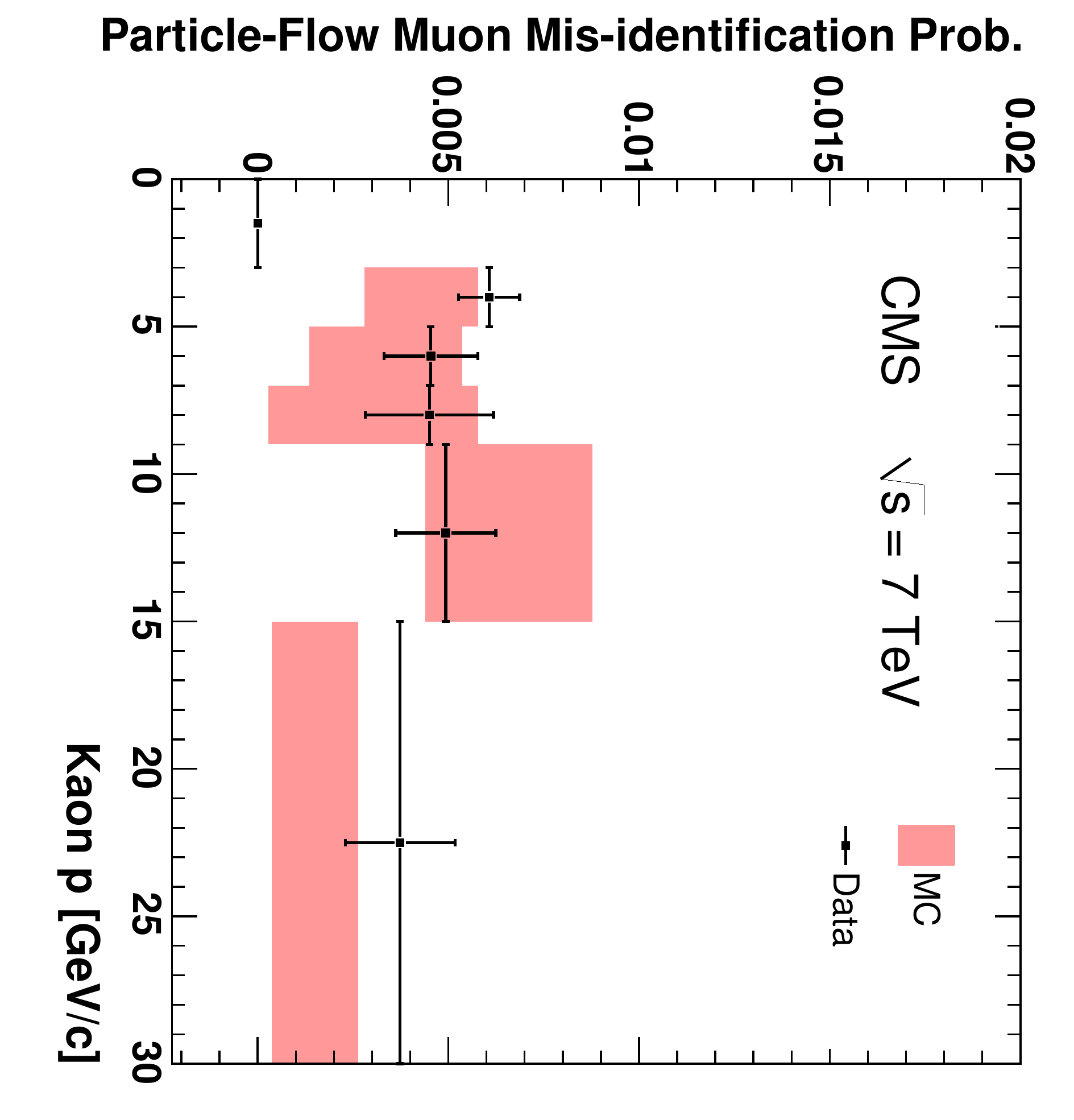} 
    \includegraphics[angle=90,width=0.32\textwidth]{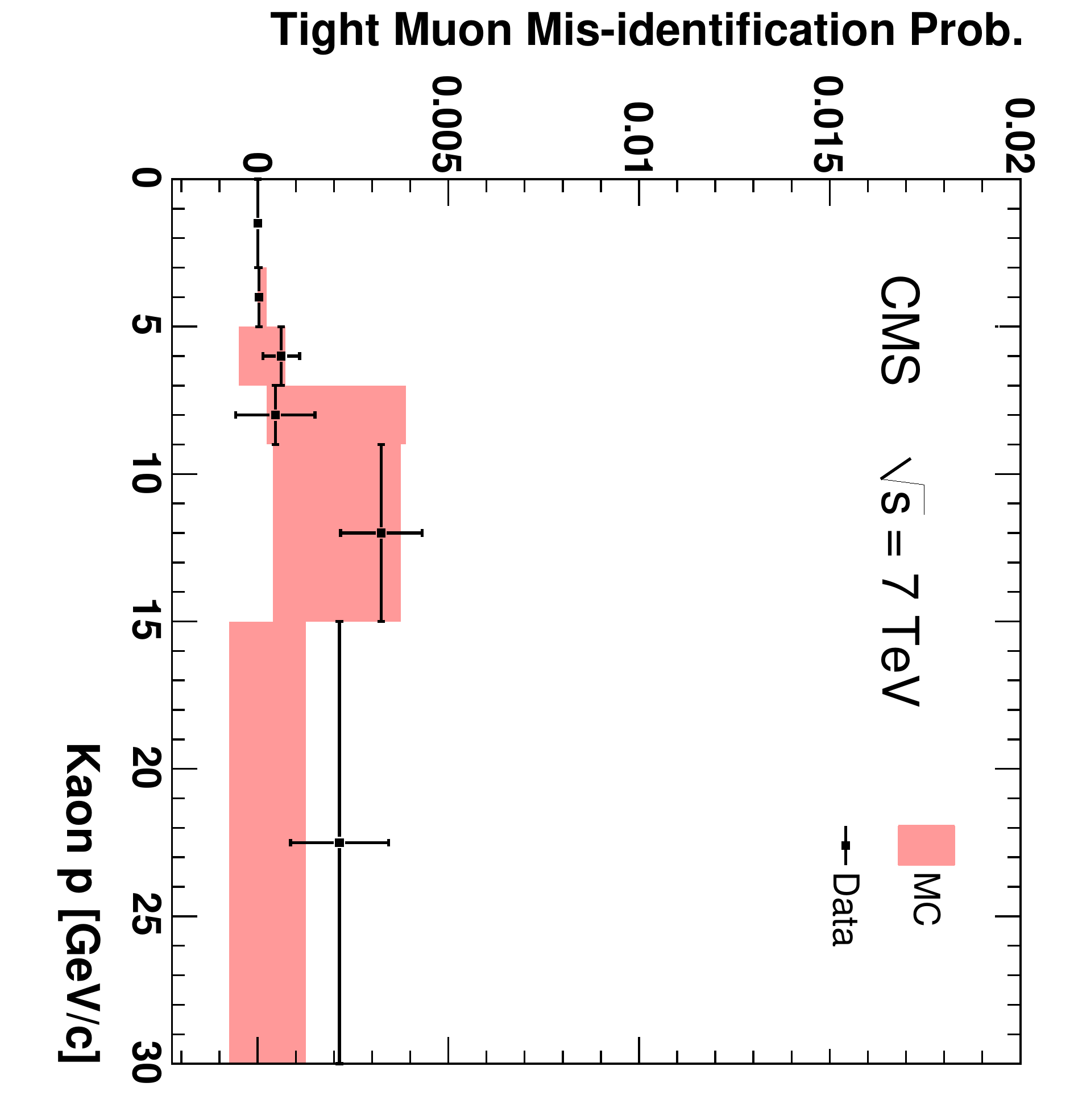} 
        
    \includegraphics[angle=90,width=0.32\textwidth]{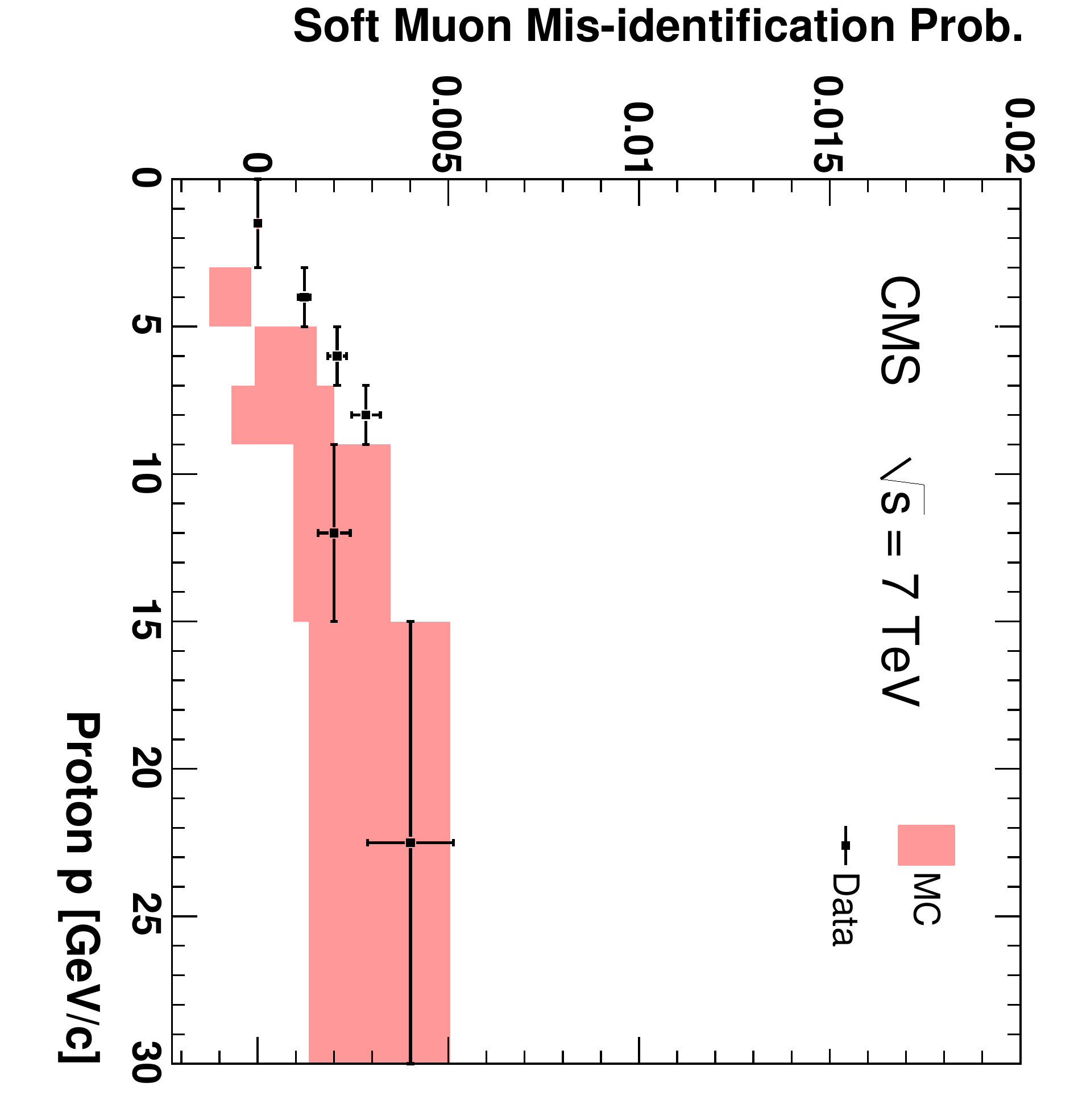} 
    \includegraphics[angle=90,width=0.32\textwidth]{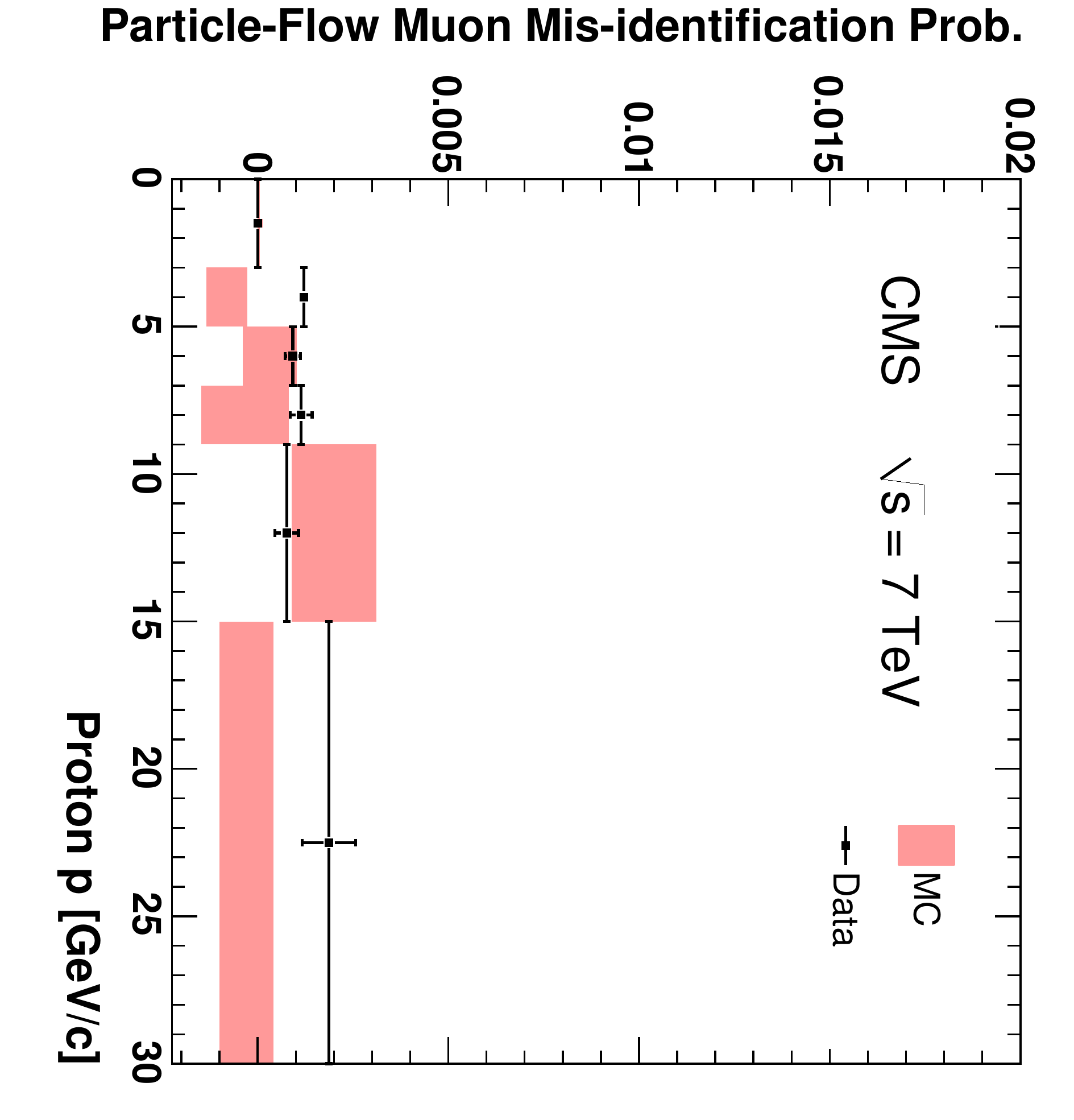} 
    \includegraphics[angle=90,width=0.32\textwidth]{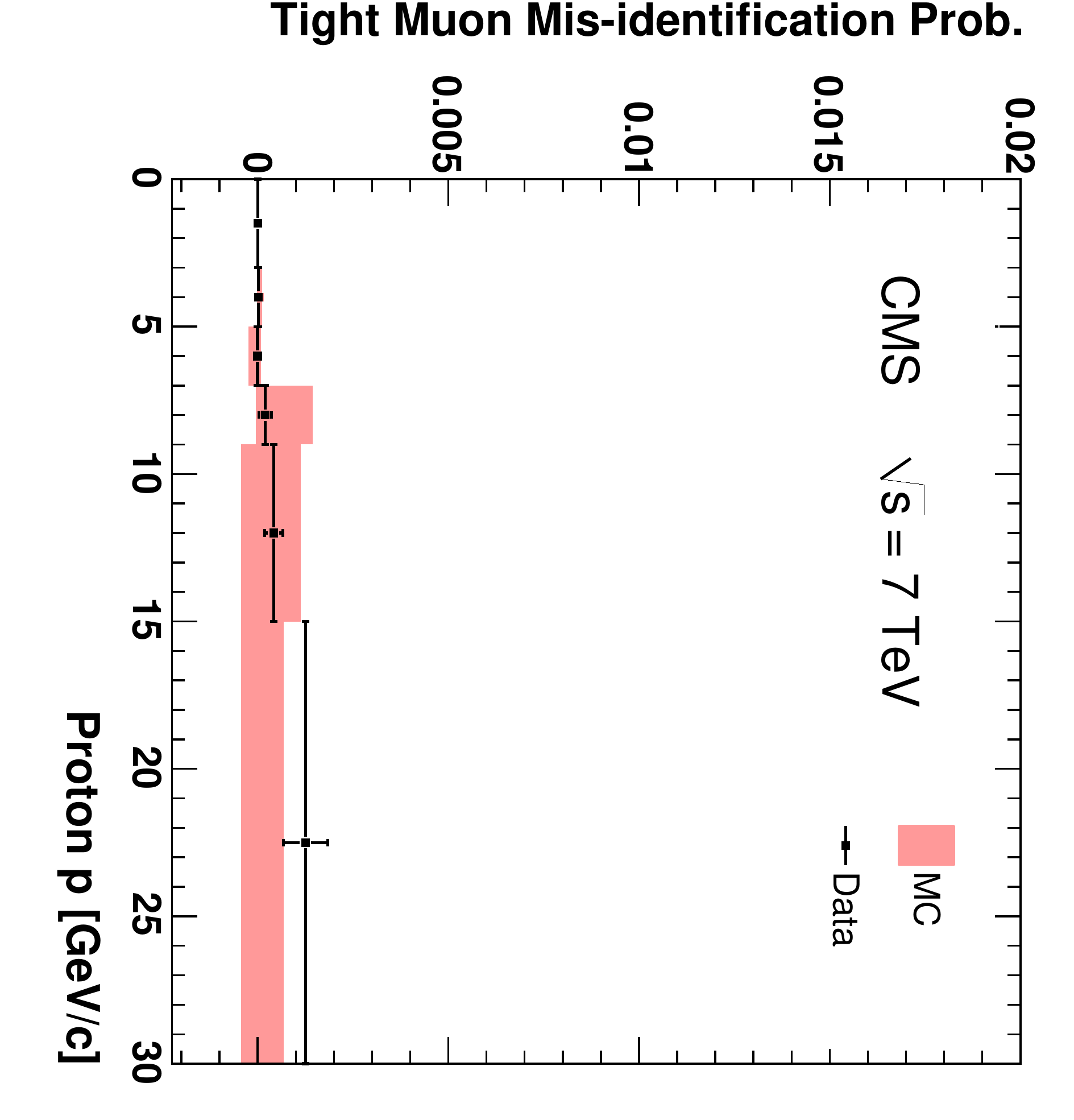}      
    \caption{The fractions of pions (top), kaons (centre), and protons
    (bottom) that are misidentified as Soft Muons (left),
    Particle-Flow Muons (centre), or Tight Muons (right) as a function
    of momentum.  Only particles with $N_{\mathrm{Tracks}} < 4$ are
    included.  The uncertainties indicated by the error bars (data)
    and shaded boxes (\PYTHIA simulation) are statistical only.}
    \label{fig:Fakes_p}
  \end{center}
\end{figure}

\begin{figure}[tb]
  \begin{center}
  
    \includegraphics[angle=90,width=0.32\textwidth]{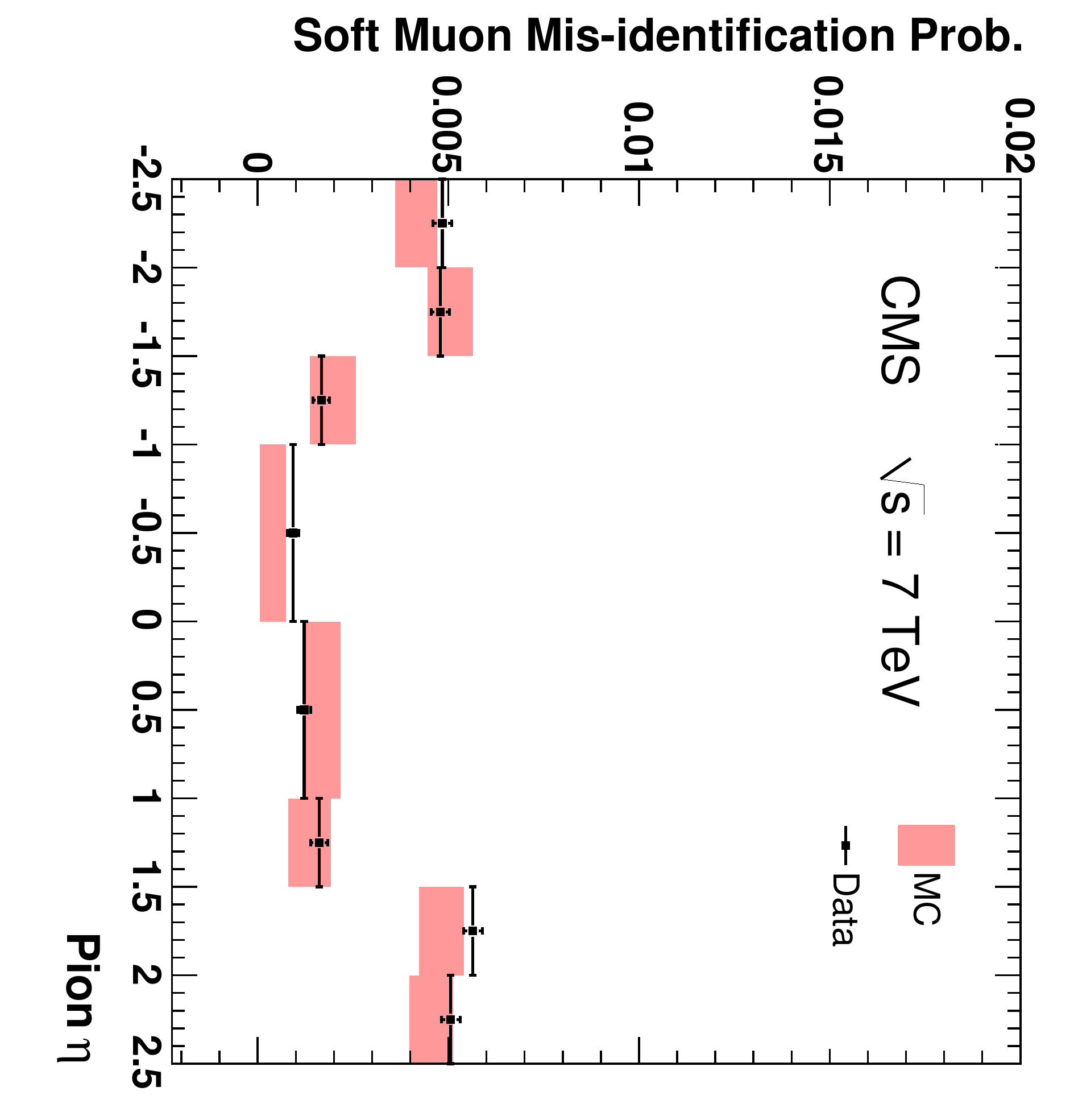}
    \includegraphics[angle=90,width=0.32\textwidth]{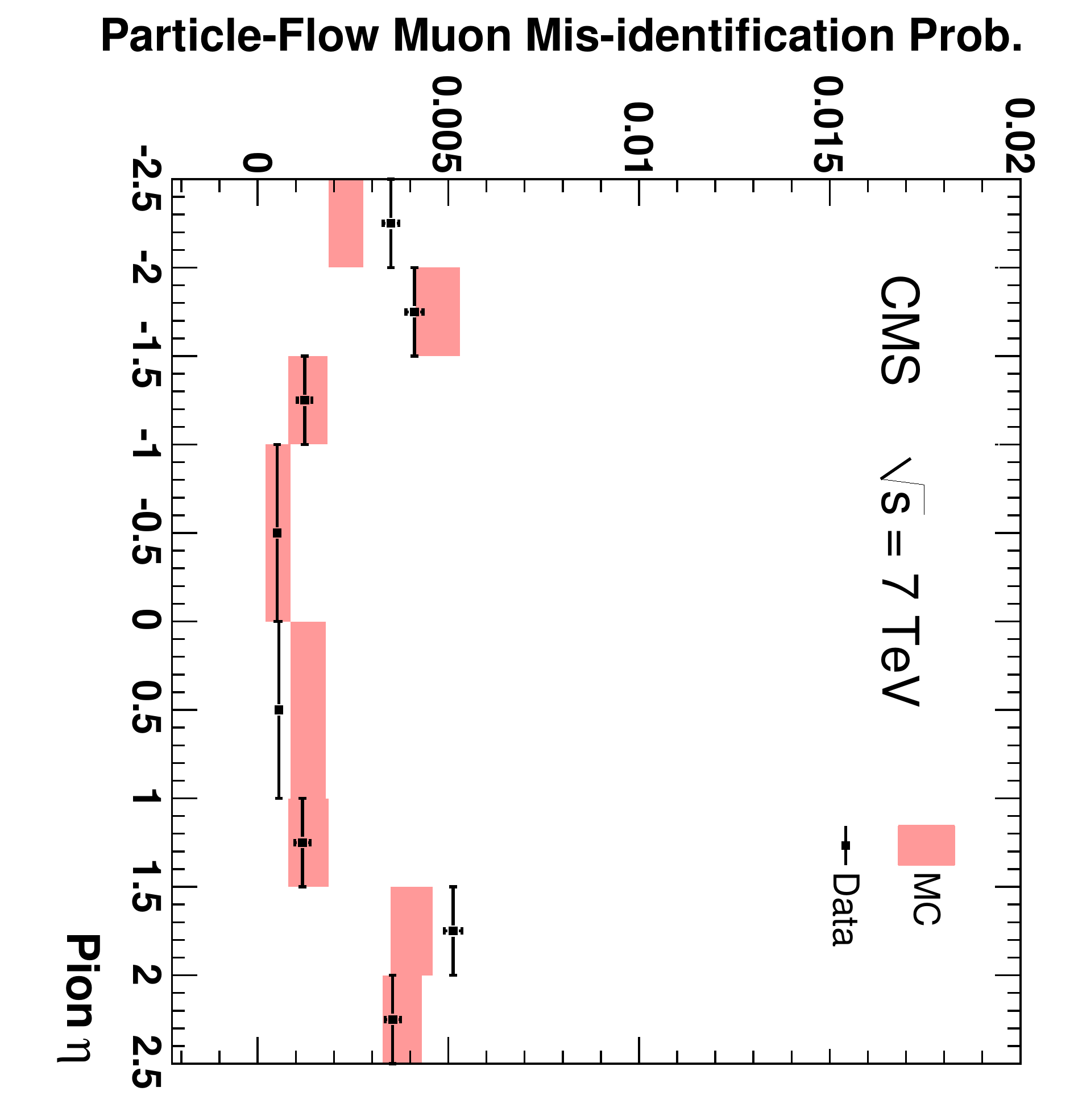} 
    \includegraphics[angle=90,width=0.32\textwidth]{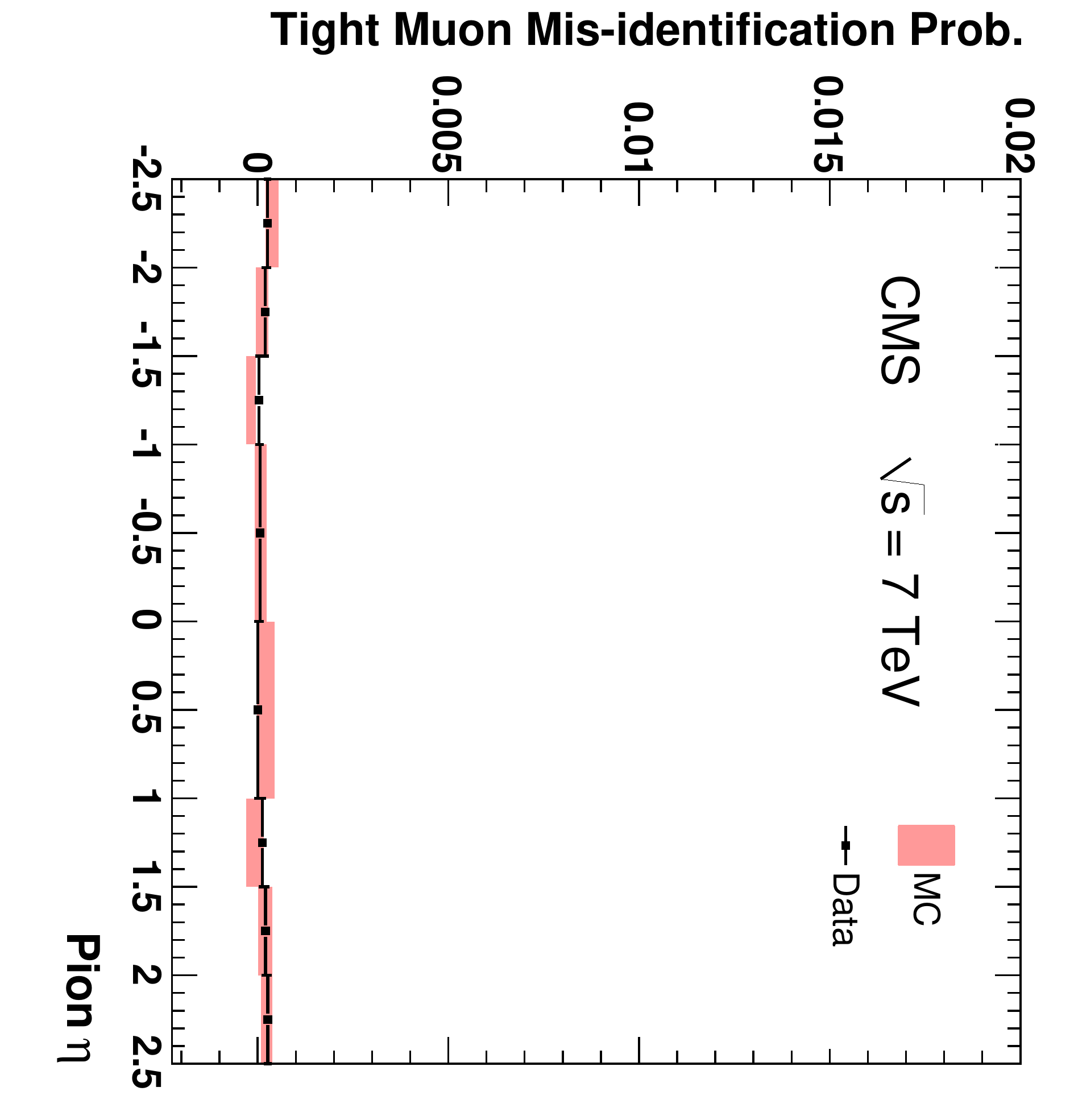} 
        
    \includegraphics[angle=90,width=0.32\textwidth]{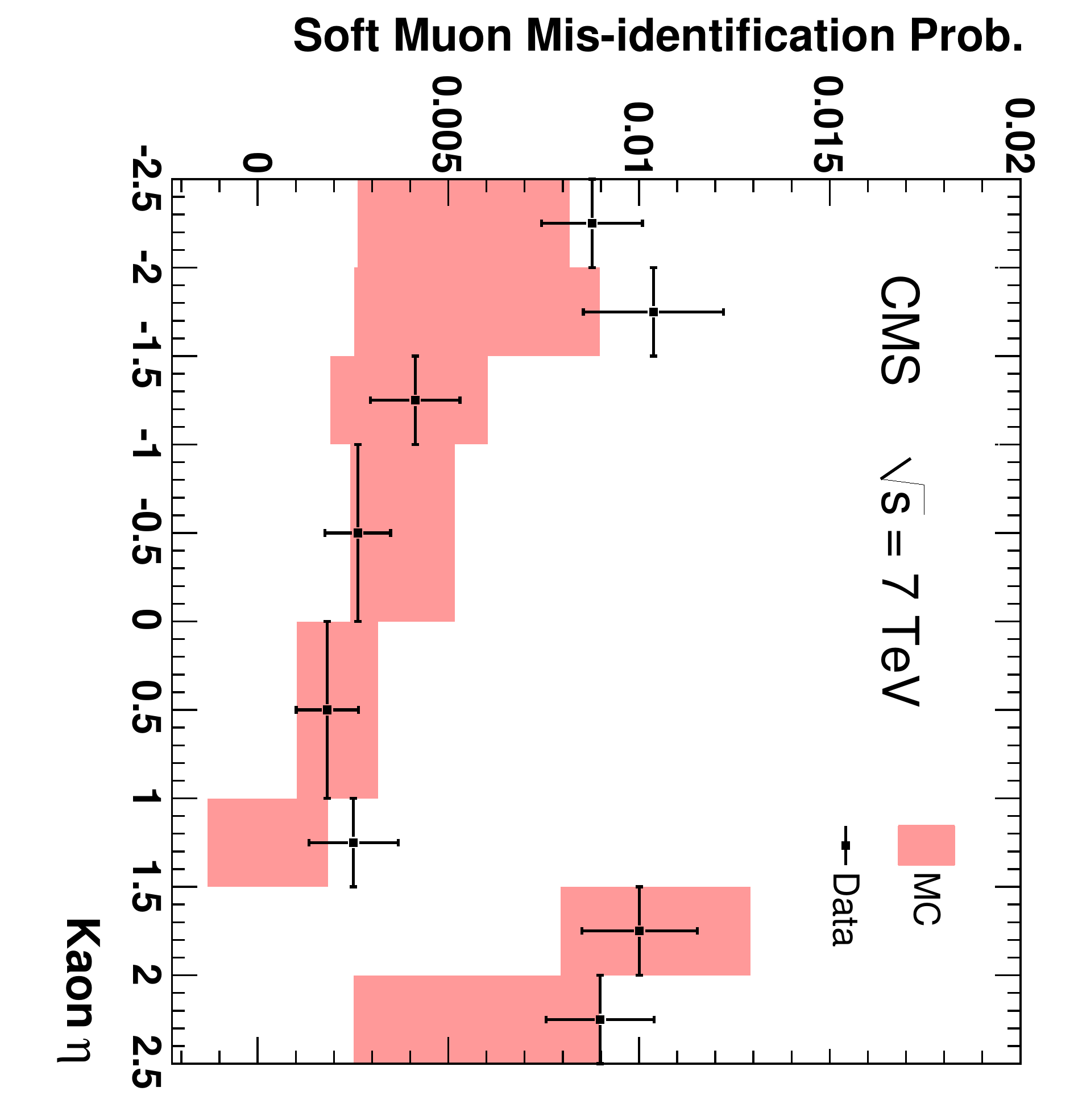} 
    \includegraphics[angle=90,width=0.32\textwidth]{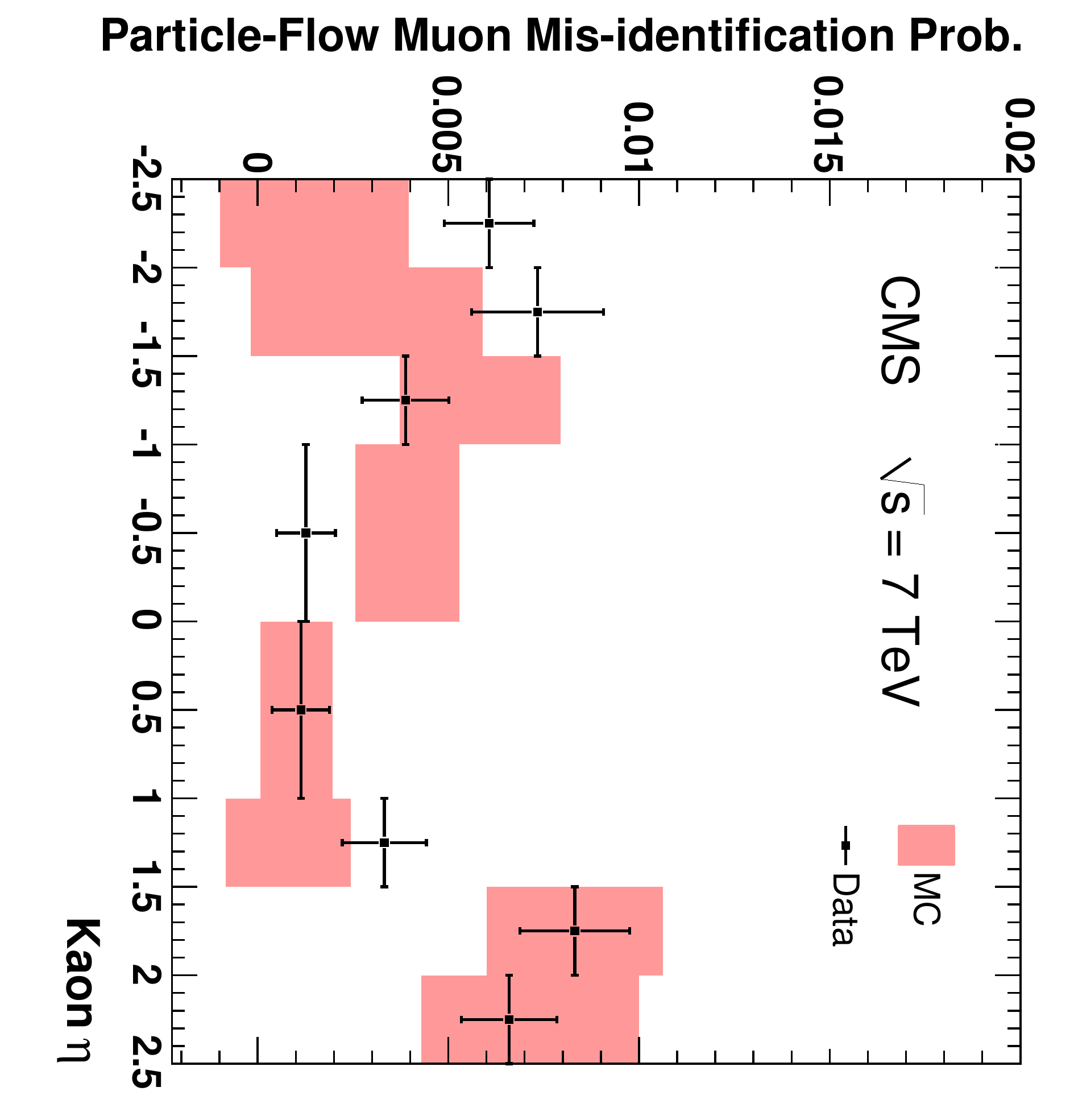} 
    \includegraphics[angle=90,width=0.32\textwidth]{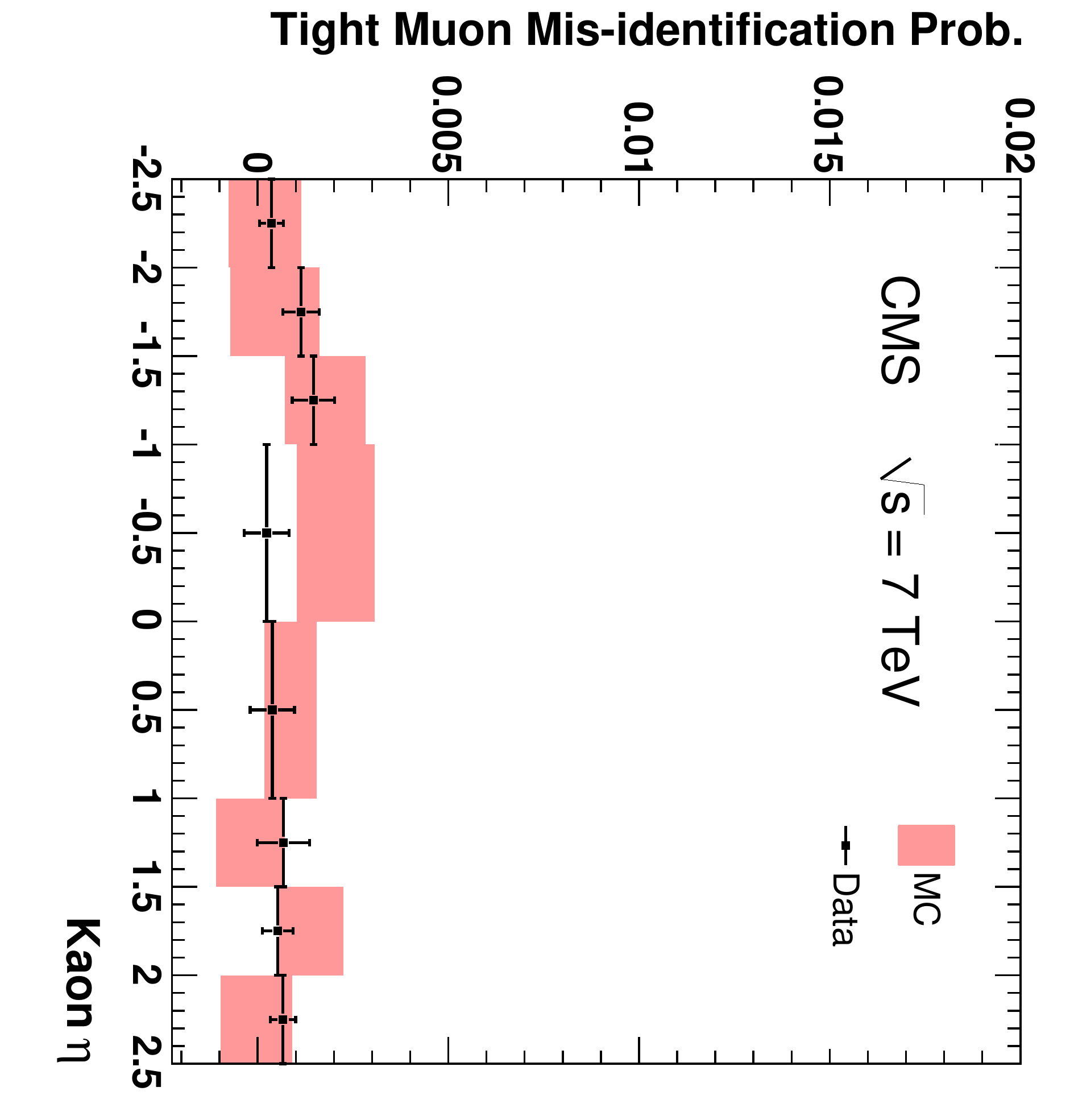} 
        
    \includegraphics[angle=90,width=0.32\textwidth]{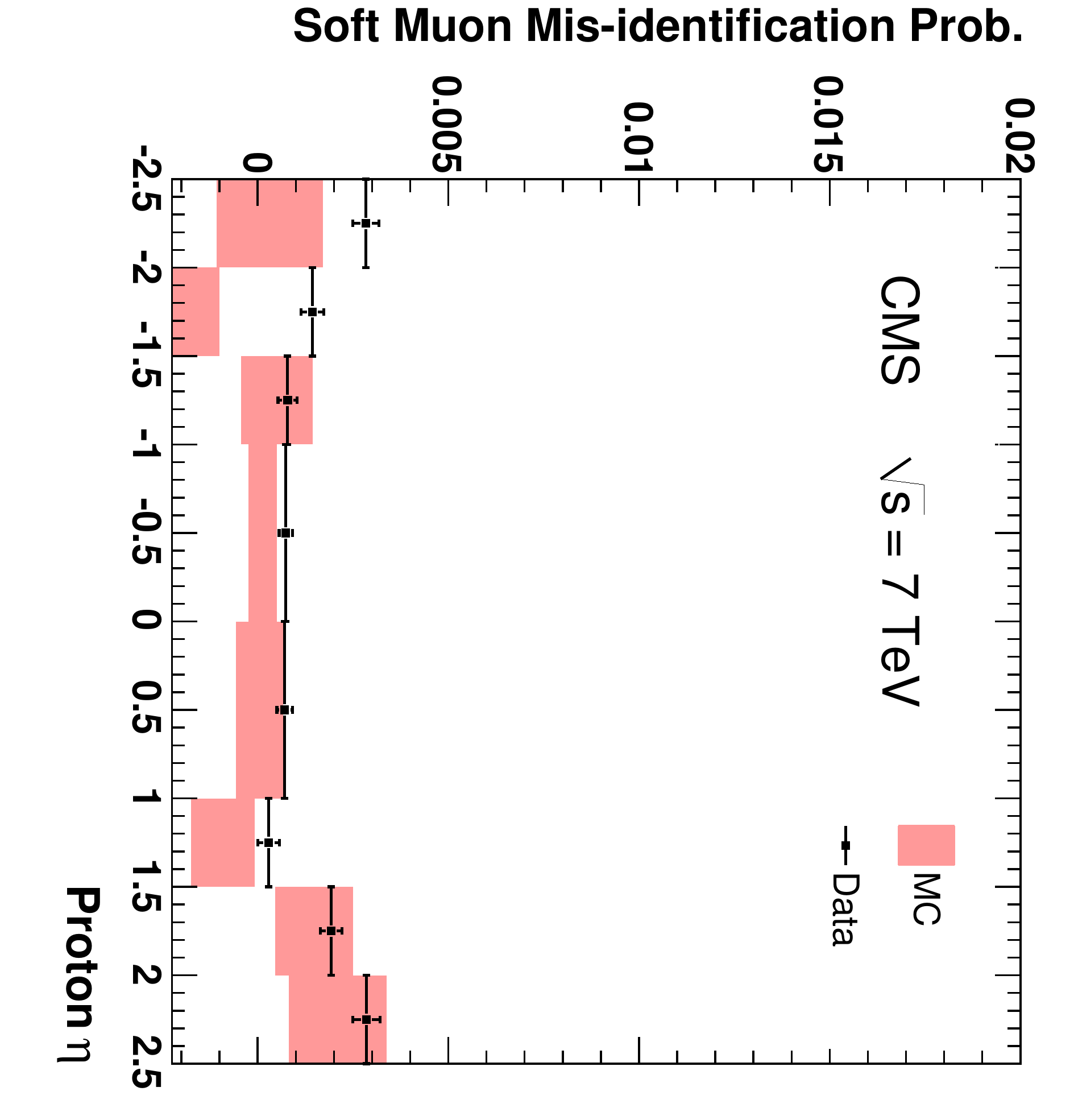} 
    \includegraphics[angle=90,width=0.32\textwidth]{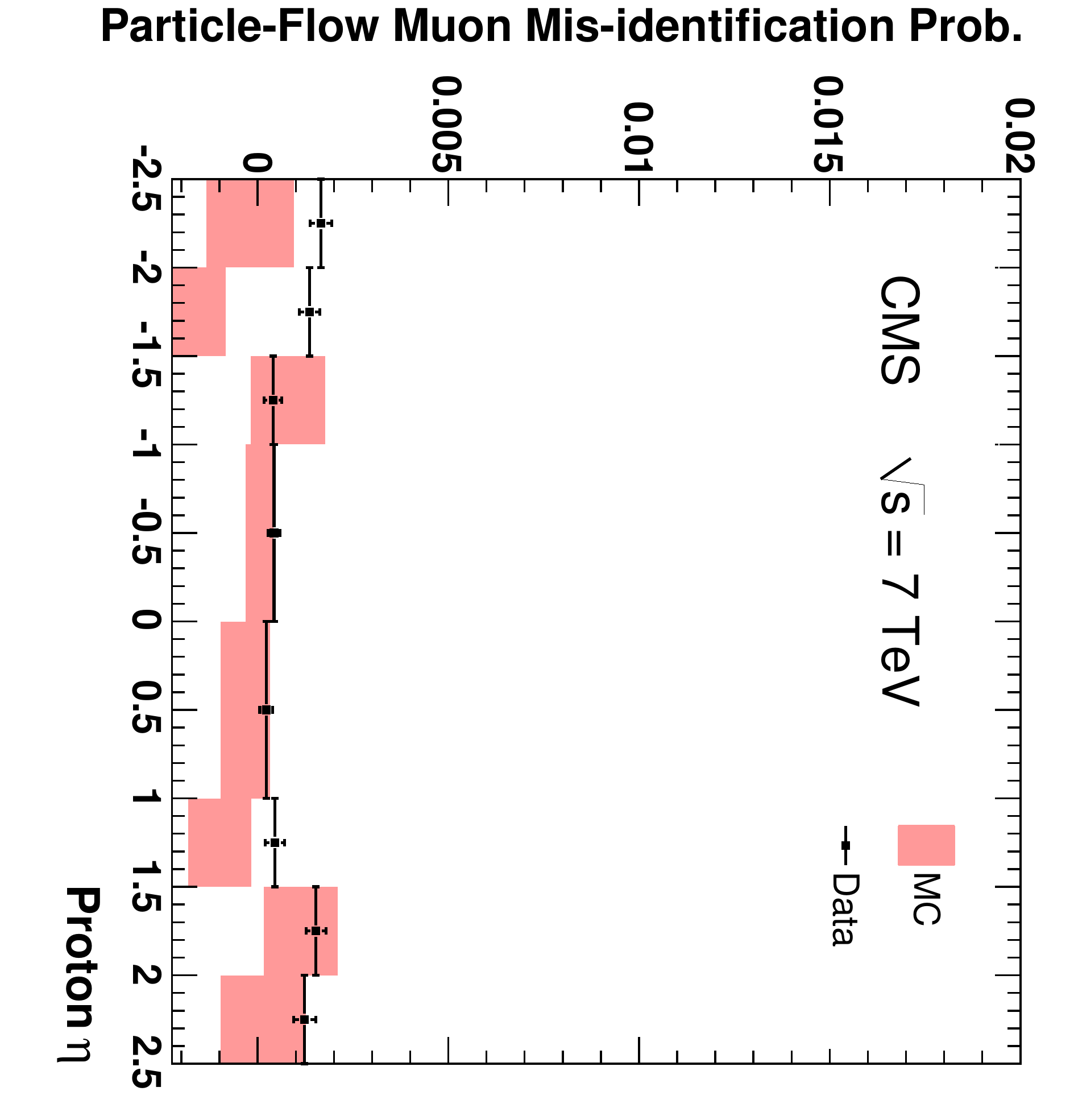} 
    \includegraphics[angle=90,width=0.32\textwidth]{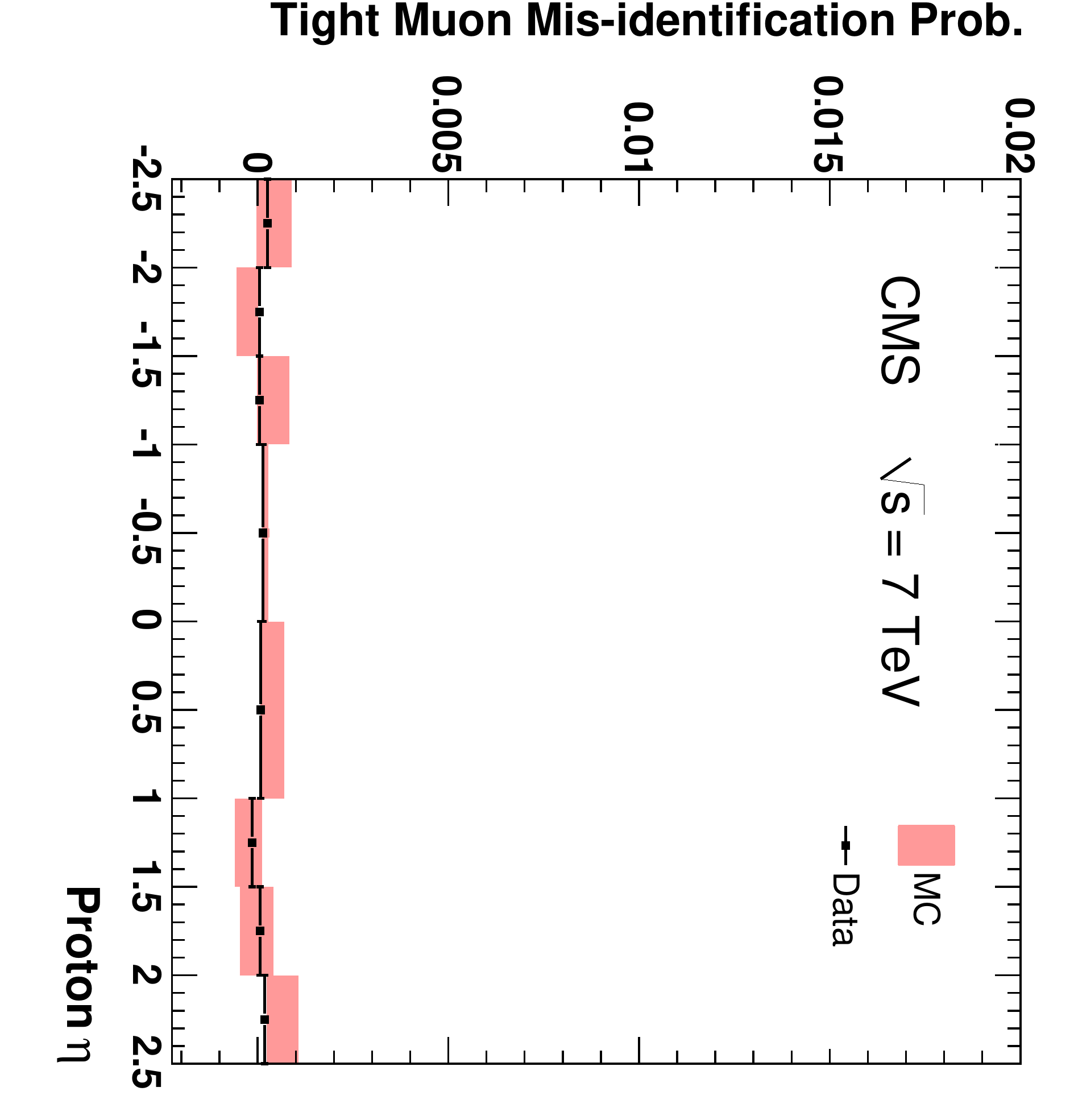}
    \caption{The fractions of pions (top), kaons (centre), and protons
    (bottom) that are misidentified as Soft Muons (left),
    Particle-Flow Muons (centre), or Tight Muons (right) as a function
    of pseudorapidity. Only particles with $p > 3\GeVc$ and
    $N_{\mathrm{Tracks}} < 4$ are included.  The uncertainties
    indicated by the error bars (data) and shaded boxes (\PYTHIA
    simulation) are statistical only.}
    \label{fig:Fakes_eta}
  \end{center}
\end{figure}

The resulting muon misidentification probabilities are shown in
Figs.~\ref{fig:Fakes_p} and \ref{fig:Fakes_eta} as a function of
particle momentum and pseudorapidity, respectively.
The shapes of the distributions, well reproduced by simulation, are due
to a combination of acceptance (a minimum momentum is required to
reach the muon system), the amount of material before the muon system,
and the distance available for pions and kaons to decay before
reaching the calorimeters.  For pions and kaons, the
misidentification probabilities are below 1\% for all
muon selections and decrease at $p \gtrsim$ 10--15\GeVc due
to fewer of the hadrons decaying to muons within the detector volume.
For protons, the probability to be identified as a muon slowly
increases with momentum but remains low in the accessible momentum
range, which confirms that punch-through is small and that at low
momenta the main reason for misidentification of pions and kaons
is decays in flight, in agreement with the
predictions from simulation discussed in Section~\ref{sec:kinematics}.
As expected, the misidentification probabilities are found to be
independent of the azimuthal angle and the decay length of the mother
particle within the statistical uncertainty.

Overall, the probability to misidentify a hadron as a muon is the
largest for Soft Muons, decreases slightly for Particle-Flow Muons,
and drops significantly for Tight Muons.  As shown in
Section~\ref{sec:muonideff_tnp}, the lower misidentification probability
for Tight Muon selection comes at the cost of a few percent lower
muon identification efficiency.  It is this trade-off between
misidentification probability and efficiency that motivates using
different muon selections for different analyses.

\section{Muon Momentum Scale and Resolution}
\label{sec:momentumscale}

The measurement of the muon transverse momentum is 
highly sensitive to the alignment of the tracker
and of the muon chambers, to the composition of material and its distribution inside the tracking
volume, and to the knowledge of the magnetic field inside and outside the solenoid volume.

The relative bias $\Delta(\pt)/\pt$ in reconstructed muon transverse momentum with respect to its true value
that could be caused by imperfect knowledge of the magnetic field is generally constant as a function of momentum.
Similarly, inaccuracies in the modelling of the energy loss (dependent
on the material distribution) produce relative biases that are essentially independent of the muon momentum.
On the other hand, alignment effects produce relative biases that generally increase linearly with momentum.

The momentum scale and resolution of muons are studied using different approaches
in different $\pt$ ranges.
At low and intermediate $\pt$ ($\lesssim$100\GeVc), the mass constraint of
dimuon decays from the $\jpsi$ and $\Z$ resonances is used to calibrate the
momentum scale and measure the momentum resolution.
In the high-$\pt$ range ($\gtrsim$100\GeVc), the muon momentum scale and resolution can be measured using
cosmic-ray muons (with the exception of the high-$|\eta|$ region).

The lower $\pt$ range of the muon spectrum, $\pt \lesssim 10 \GeVc$, has been studied in Ref.~\cite{TRK-10-004}.
In this region of $\pt$, alignment effects are less important, and biases in the reconstructed momentum mostly arise
from uncertainty in the modelling of the detector material and in the description of the magnetic field used when
reconstructing the track.
Results obtained using \jpsi{} events show that the overall relative bias in the tracker measurement of the muon $\pt$ 
in this momentum range is $\approx$0.1\%. 
The muon $\pt$ resolution $\sigma(\pt)/\pt$ was found to be between 0.8\% and 3\% depending on $\eta$ and in good agreement with the simulation.

In the intermediate-$\pt$ range, two approaches to study the muon $\pt$ measurement have been developed.
The first, referred to as MuScleFit (Muon momentum Scale calibration Fit),
produces an absolute measurement of momentum scale and resolution
by using a reference model of the generated
$\Z$ lineshape convoluted with a Gaussian function.
The second, called SIDRA (SImulation DRiven Analysis), compares the data 
with the full simulation of the $\Z$ decay to two muons in the
CMS detector and provides a way to directly modify the simulation to better match the data.
As these two methods have different approaches to the same problem, 
the difference between the results provides a useful crosscheck and gives an estimate of a systematic uncertainty in the measurement.  The results obtained with these methods are reported in
Section~\ref{sec:momentumScaleAtMediumPt}.

At high $\pt$ (Section~\ref{sec:resscalefromcosmics}),
the resolution is determined by comparing cosmic-muon
tracks reconstructed independently in the upper and lower halves of the detector,
while the scale bias
is evaluated by using what is called the "cosmics endpoint method''.

\subsection{\texorpdfstring{Measurements at intermediate $\pt$}{Measurements at intermediate pT}}
\label{sec:momentumScaleAtMediumPt}
As previously mentioned, a sample of muons produced in the decays of $\Z$
bosons is well suited for measuring the muon
momentum scale and resolution in the intermediate range of transverse
momentum, $20 < \pt < 100\GeVc$.
Muons from $\Z$-boson decays are identified using the Tight Muon selection. 
They are compared to the simulated sample of $\Zmm$ and Drell--Yan dimuon events
reconstructed using the realistic misalignment scenario (see
Section~\ref{sec:samples}) and passing the same set of muon selection
criteria.
Tight Muon selection is chosen since it is used in all electro-weak precision measurements in CMS. However using a different selection has no significant impact on the results presented in this section.

Measurements of the muon momentum scale and resolution, reported
in Sections~\ref{sec:muscl} and \ref{sec:mures},
refer to the $\pt$ of
the tracker-only fit as it is the default muon momentum assignment in
the range of $\pt < 200\GeVc$ (see Section~\ref{sec:reco}).
Studies of the resolution of the $\pt$ measured by the muon system only, 
provided by the standalone-muon track fit, are described in 
Section~\ref{sec:staMuonResolData}.

\subsubsection{Muon momentum scale}
\label{sec:muscl}
\begin{figure}
 \centering
 \includegraphics[width=0.32\textwidth]{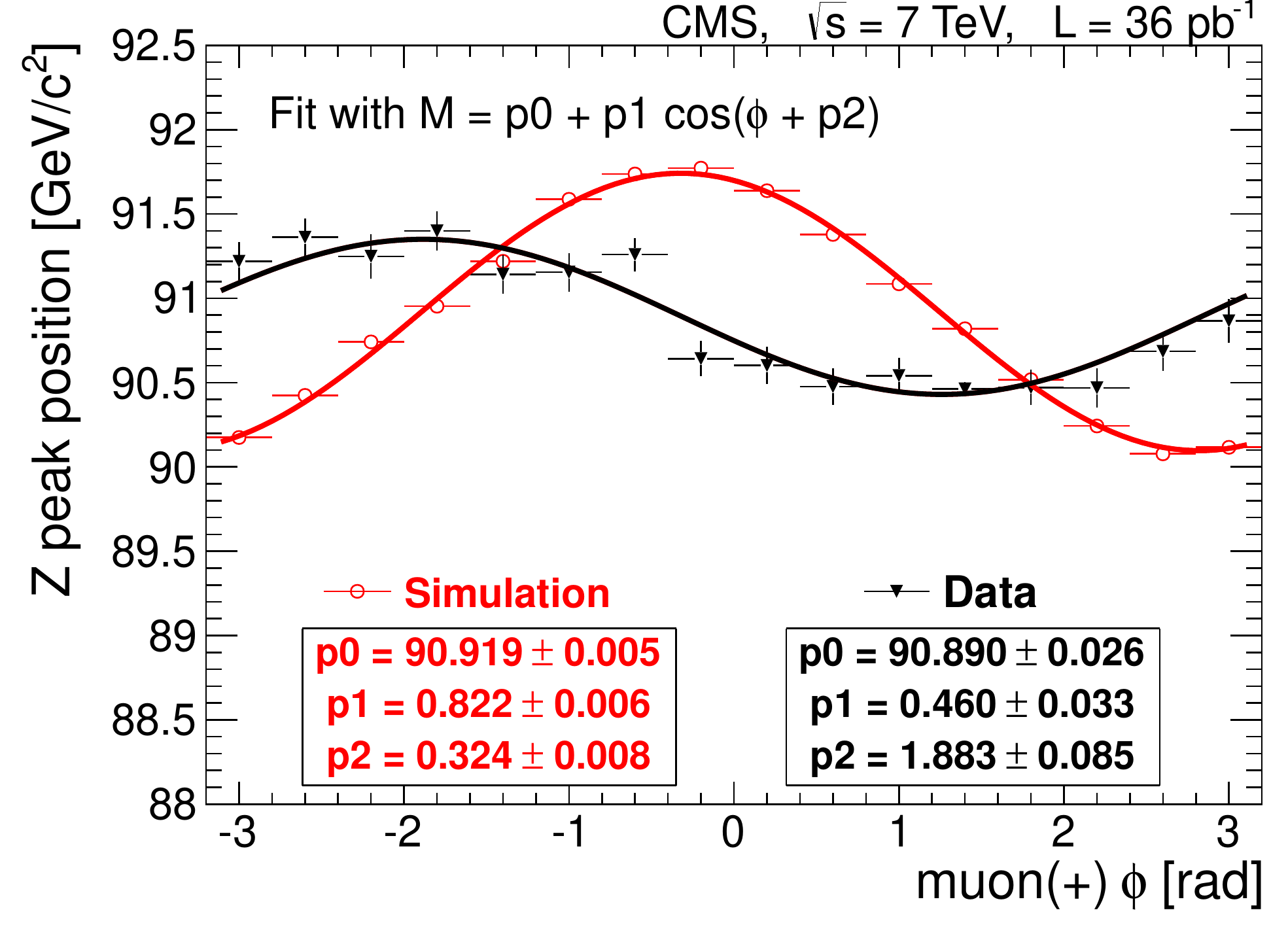}
 \includegraphics[width=0.32\textwidth]{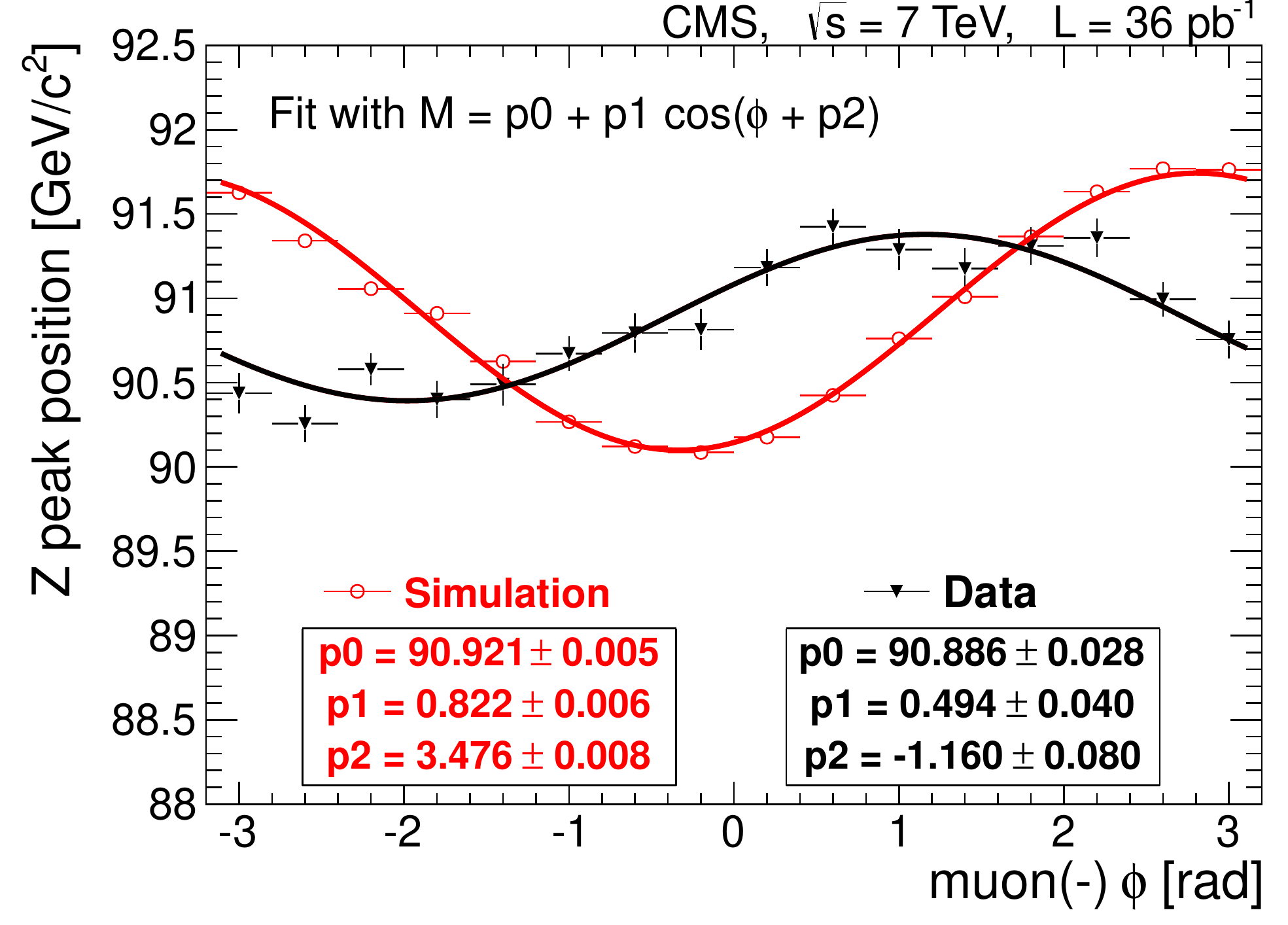}
 \includegraphics[width=0.32\textwidth]{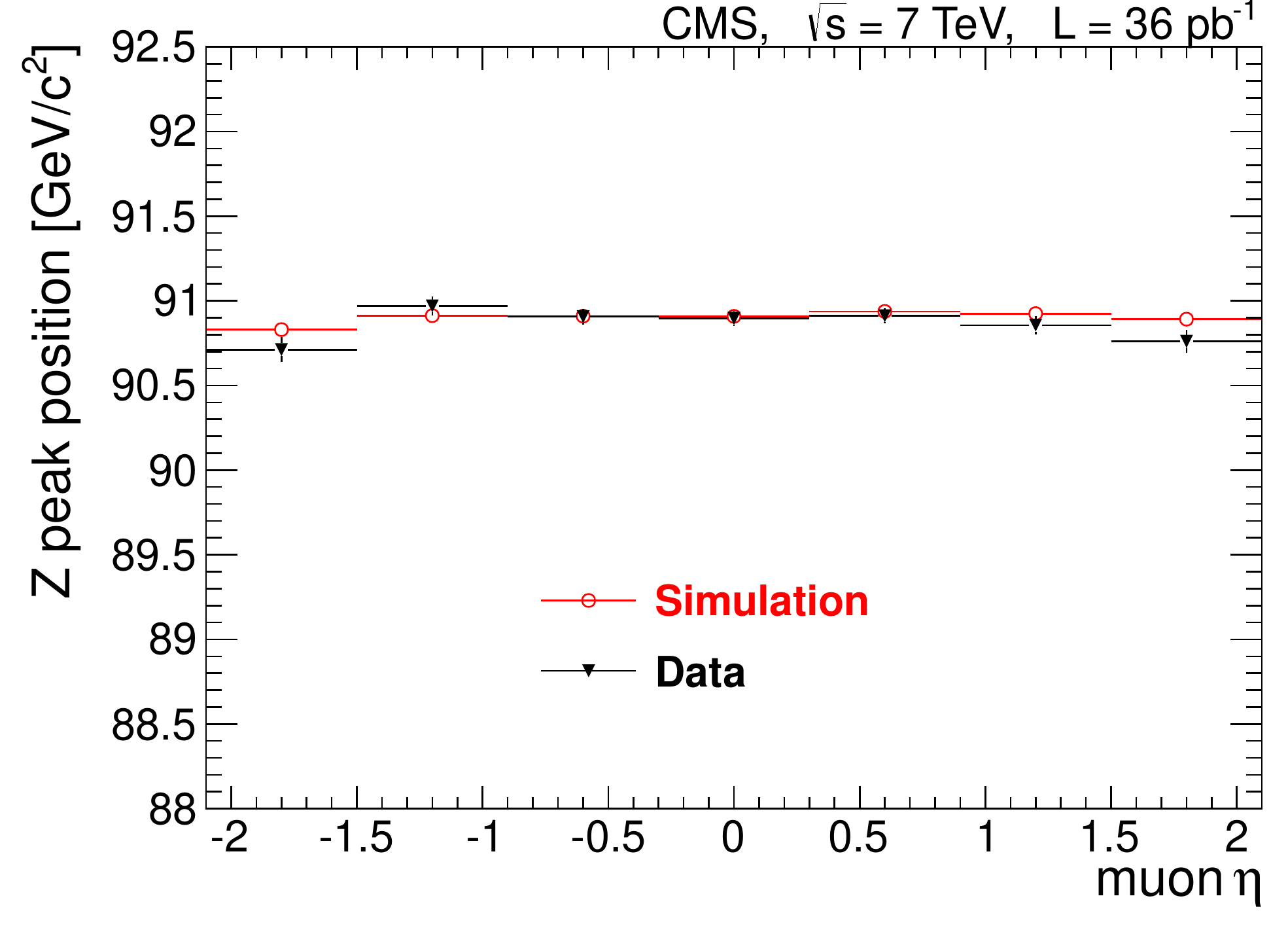}
 \caption{The position of the $\Z$ peak reconstructed by the MuScleFit method
  as a function of muon $\phi$ for positively charged muons (left),
  $\phi$ for negatively charged muons (middle), and $\eta$ for muons
  of both charges (right). Results obtained from data (black triangles)
  are compared with those from
  simulation (red circles). The $\phi$ plots also show the results
  of sinusoidal fits; the values of the fit parameters are
  given in the text boxes below the labels.}
 \label{fig:massBias}
\end{figure}

In the MuScleFit approach, the biases in the reconstructed muon $\pt$ are determined from the position of the 
$\Z$ mass peak as a function of muon kinematic variables.
Figure \ref{fig:massBias} shows the position of the peak determined from a
fit to a Voigtian (the convolution of a Lorentzian and a Gaussian) in bins
of muon $\phi$, shown
separately for positively and negatively charged muons, and muon $\eta$, for both data and
simulation\footnote{Note that the peak position returned by the fit is not expected to perfectly match the
PDG value of the $\Z$ mass. The generator-level $\Z$ lineshape
is not symmetric around the peak and has a higher tail at lower mass values. When it is convoluted
with the detector resolution effects (which can be approximated with a Gaussian), the peak shifts towards
lower values. A simplified test performed by convolving the reference lineshape with a Gaussian with $\sigma = 1.5\GeVcc$
gives a peak position of 90.8\GeVcc. 
}. 
We observe a
sinusoidal bias as a function of $\phi$, antisymmetric for muons of opposite
charges.  The maximum shift of the position of the mass peak is about 0.5\%.
This is significantly larger than the relative bias observed at low $\pt$ using the
\jpsi{} events, indicating that the relative bias is $\pt$ dependent and that its
main source at intermediate $\pt$ is the residual tracker misalignment.
As explained in Section~\ref{sec:samples}, the misalignment scenario used in the simulation was obtained by the alignment procedure similar to that used in the data but was
based only on the sample of cosmic-ray muons available at the time of the MC production.
This results in a different phase and slightly larger amplitude of the bias in MC simulation.
In addition, a small $\eta$-dependent bias is also present. This bias
exhibits no dependence on muon charge, has a
parabolic shape, and is reasonably well reproduced by the Monte Carlo
simulation.

\begin{figure}
 \centering
 \includegraphics[width=0.32\textwidth]{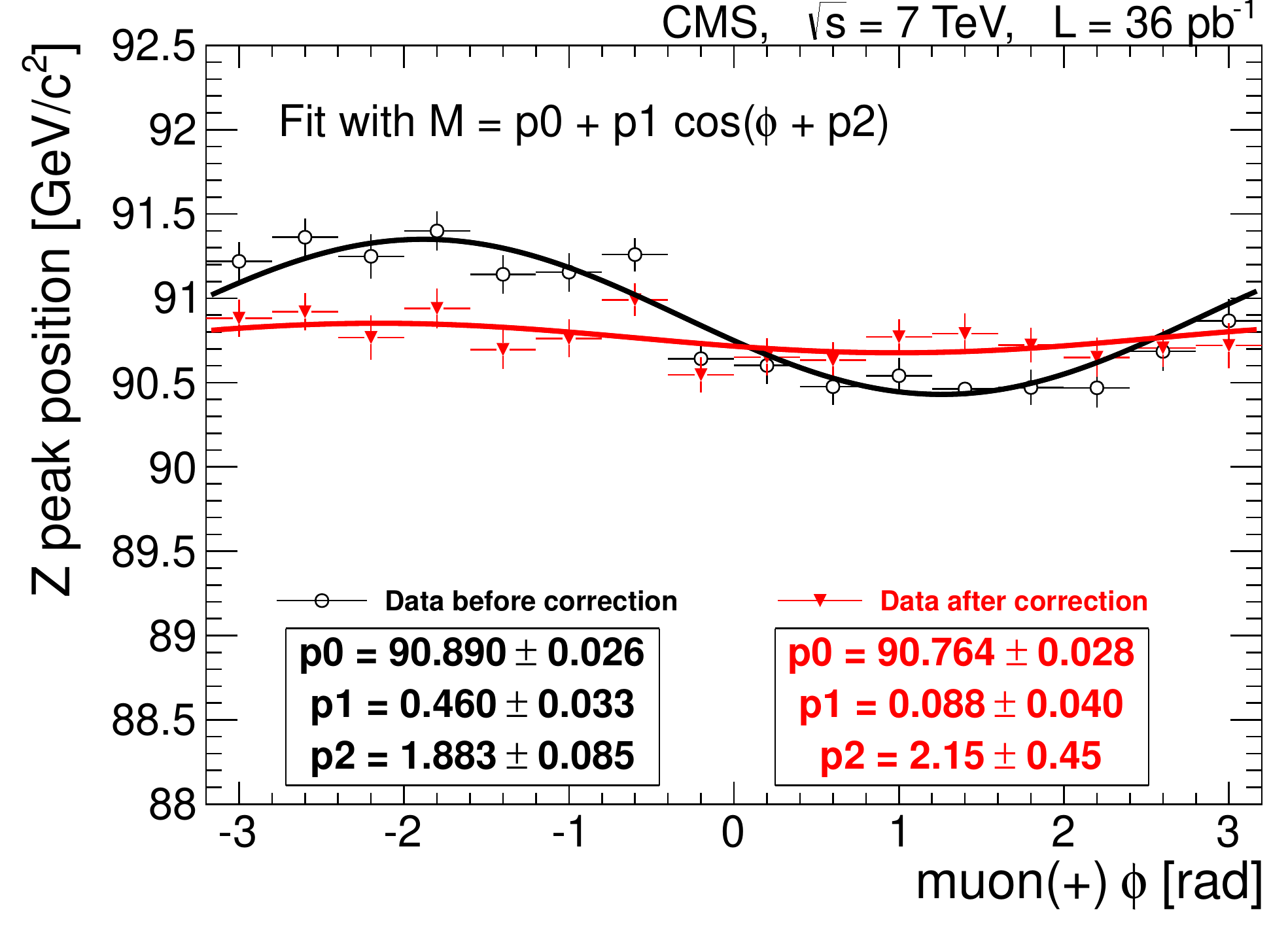}
 \includegraphics[width=0.32\textwidth]{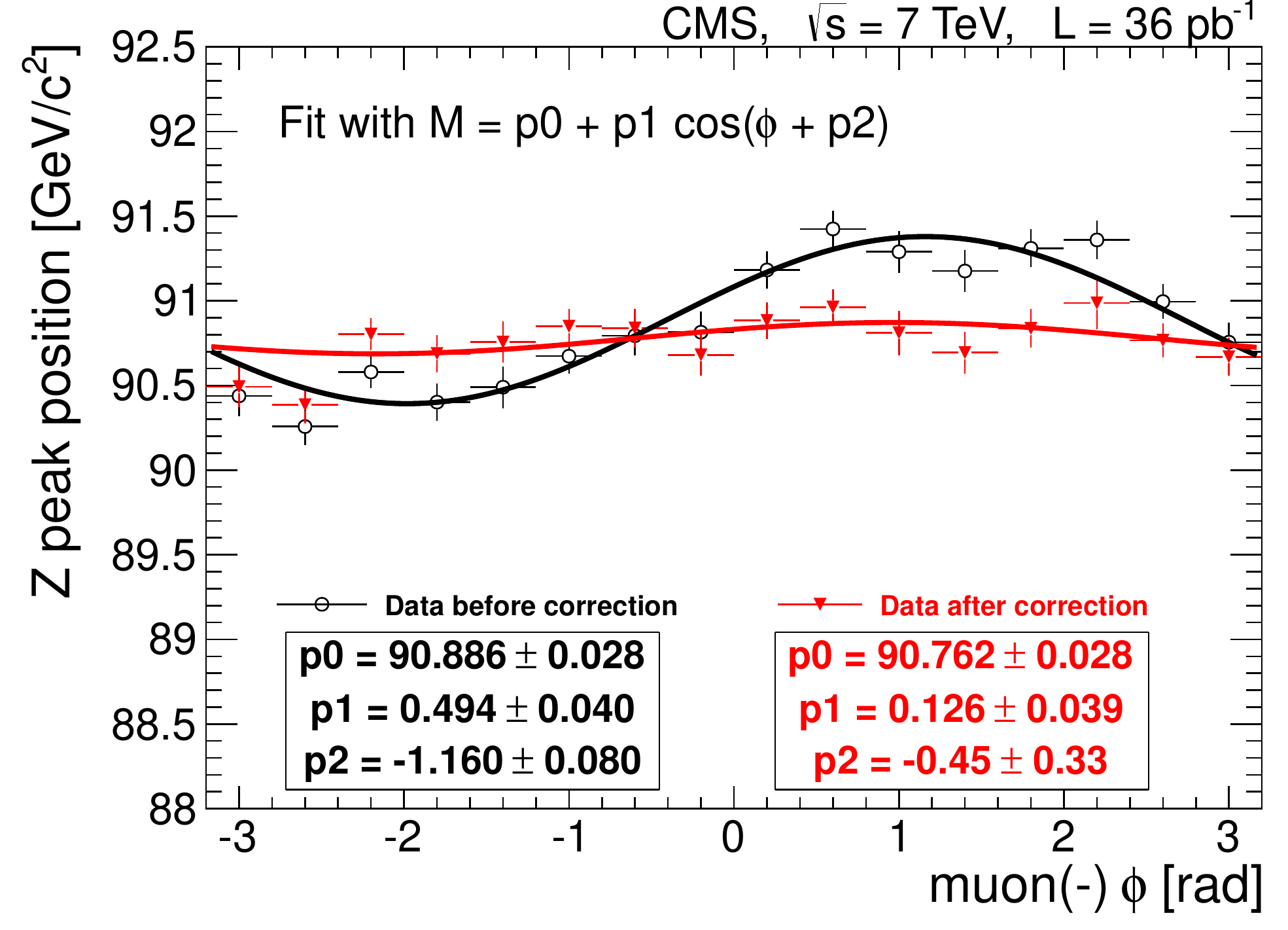}
 \includegraphics[width=0.32\textwidth]{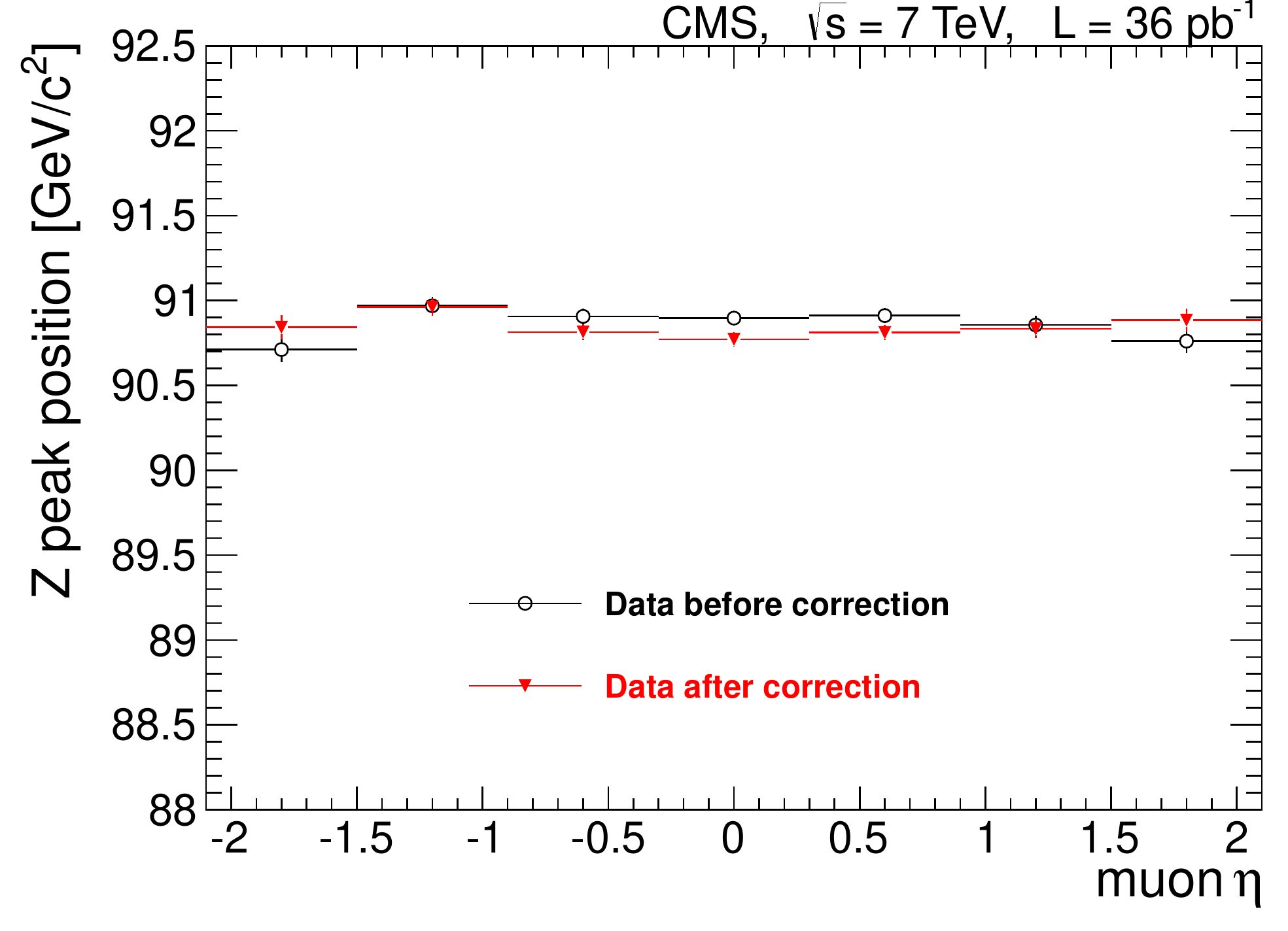}
 \caption{The position of the $\Z$ peak in data as a function of muon $\phi$
          for positively charged muons (left), $\phi$ for
          negatively charged muons (middle), and $\eta$ for muons of both
          charges (right) before (black circles) and after (red triangles)
          the MuScleFit calibration.  The $\phi$ plots also show the results
          of sinusoidal fits; the values of the fit parameters are
          given in the text boxes below the labels.}
 \label{fig:calibrationResultOnData}
\end{figure}

MuScleFit uses an unbinned likelihood fit with a reference model to correct the momentum scale.
Given the shape of the $\Z$-peak position biases, the following ansatz function is used for the calibration of the momentum scale:

\begin{equation} 
 \pt' = \pt(1 + b\cdot \pt + c\cdot\eta^2 + q\cdot d\cdot \pt \cdot\sin(\phi + e)) \ ,
\label{eq:pdfMUSCLE}
\end{equation}

where $q$ is $+1$ for $\mu^+$ and $-1$ for $\mu^-$, and $b$, $c$, $d$, and $e$
are the fit parameters.
The model taken as calibration reference is a lineshape of the $\Z$ decaying to dimuon pairs
as described in Ref.~\cite{Dittmaier} and generated with a high granularity in 1001 bins between 71.2 and 111.2\GeVcc.
Figure~\ref{fig:calibrationResultOnData} shows the results of this calibration procedure on data.
The calibrated position of the mass peak is consistent with being flat
within the statistical uncertainties, demonstrating that
the biases are successfully removed.
When averaged over $\phi$, the correction is small,
and the position of the dimuon invariant-mass peak remains
practically unchanged.
The same calibration procedure also successfully eliminates
momentum scale biases present in the simulation.

The strategy implemented in the SIDRA method consists of modifying the
reconstructed inva\-ri\-ant-mass spectrum of simulated $\Zmm$ events
by additional shifts 
and resolution distortions to make it agree with the invariant-mass 
distribution observed in data. This means that unlike the MuScleFit method, the SIDRA method 
calibrates only relative biases between data and simulation.
This approach assumes that the resolution
in data is slightly worse than in simulation and is well suited for the
present study.  The $\Zmm$ candidates in data and in
simulation are binned according to their reconstructed parameters, and the
difference between the two distributions is minimized by using a binned
maximum likelihood fit.
At each minimization step, the reconstructed transverse momentum $\pt$ 
of the simulated muons, $p_{\rm T,\,sim}$, is modified 
as follows:

\begin{equation}
      \frac{1}{p'_{\rm T,\,sim}} = \frac{1}{p_{\rm T,\,sim}} + \delta_{\kappa_{\rm T}}(q,\phi,\eta) + 
\sigma_{\kappa_{\rm T}}(q,\phi,\eta)~\mathrm{Gauss}(0,1)\,,
\label{eq:pdfSIDRA}
\end{equation}

where $\delta_{\kappa_{\rm T}}$ and $\sigma_{\kappa_{\rm T}}$ are
parameters controlling the scale shifts and resolution distortions,
respectively, and $\mathrm{Gauss}(0,1)$ denotes a sampling according
to a Gaussian of zero mean and unit variance.  The fit parameters
depend in general on the muon charge $q$, 
its azimuthal angle $\phi$, and its pseudorapidity $\eta$.
This ansatz assumes that the differences between the data and the simulation 
are due to misalignment, with relative effects increasing with $\pt$.
This assumption is justified by the excellent agreement between 
data and simulation for low-mass dimuon resonances~\cite{TRK-10-004}.
Similarly to the MuScleFit case, 
several exploratory studies suggest dependencies of the type
\begin{eqnarray}
      \delta_{\kappa_{\rm T}}(q,\phi,\eta) & = & A + B\eta^2 + q\,C\,\sin(\phi-\phi_0)\,;\\
      \sigma_{\kappa_{\rm T}}(q,\phi,\eta) & = & A' + B'\eta^2\,,
\end{eqnarray}
where $A$, $B$, $C$, $\phi_0$, $A'$, and $B'$ are the parameters to be determined in 
the fit. The dependence on charge influences the choice of the 
binning for the fit: since the proposed ansatz function has a charge-dependent 
term as a function of $\phi$, we employ a two-dimensional grid, binning
events according to the reconstructed dimuon invariant mass and the
azimuthal angle of one of the two muons.

\begin{figure}[thb]
\begin{center}
  \includegraphics[width=0.7\textwidth]{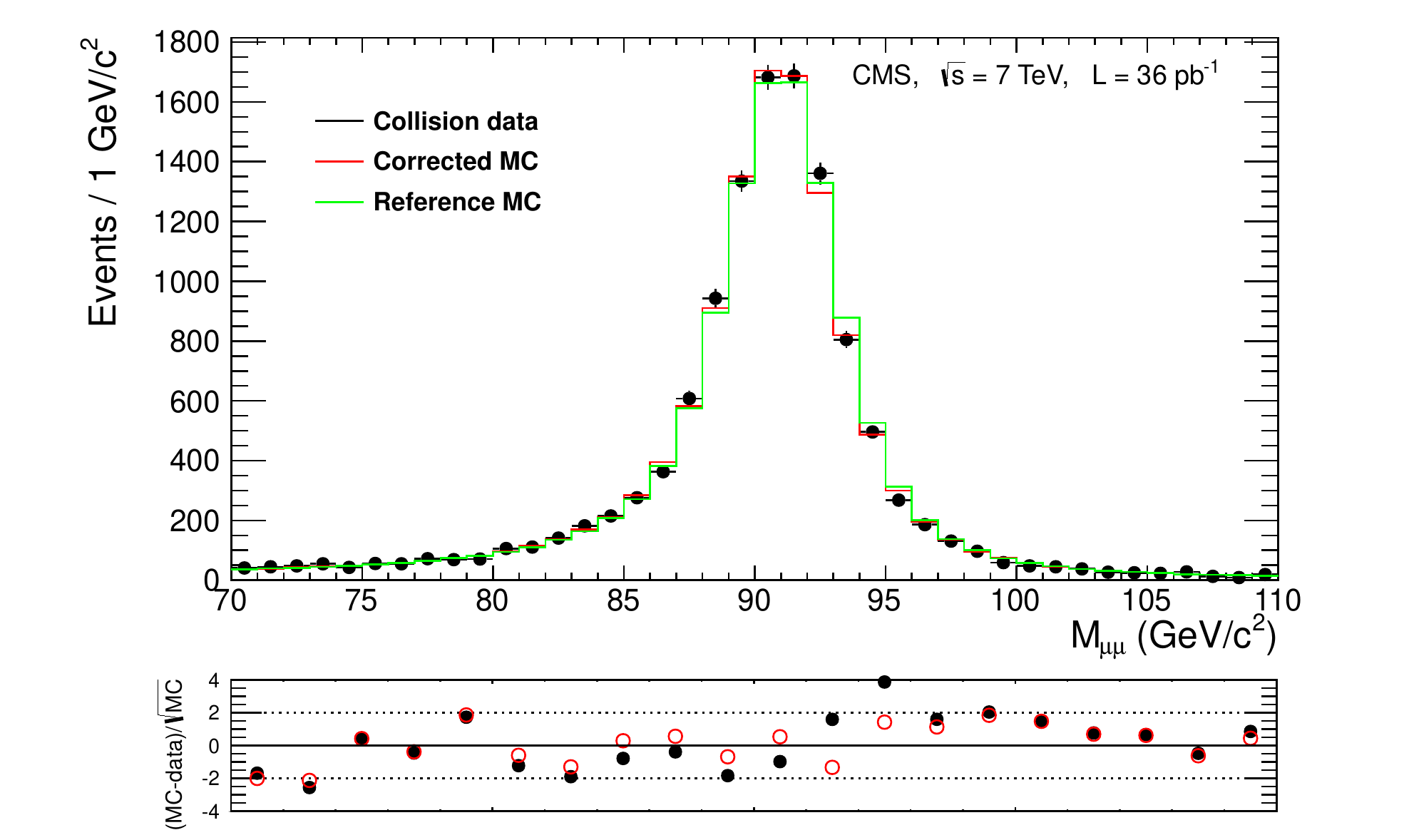}
  \caption{Top: distributions of the dimuon invariant mass for the
  selected $\Zmm$ candidates in data (points with error bars) and
  in simulation without (``reference MC'') and with (``corrected MC'')
  corrections from SIDRA applied. 
  Bottom: bin-by-bin difference (rebinned for clarity)
  between the simulation and the data, divided by the expected statistical
  uncertainty, for MC samples without (filled black circles) and with
  (open red circles) the SIDRA corrections.  The uncertainties are statistical
  only.
}
 \label{fig:sidraZ_after_correction}
 \end{center}
\end{figure}

\begin{figure}[thb]
\begin{center}
  \includegraphics[width=0.48\textwidth]{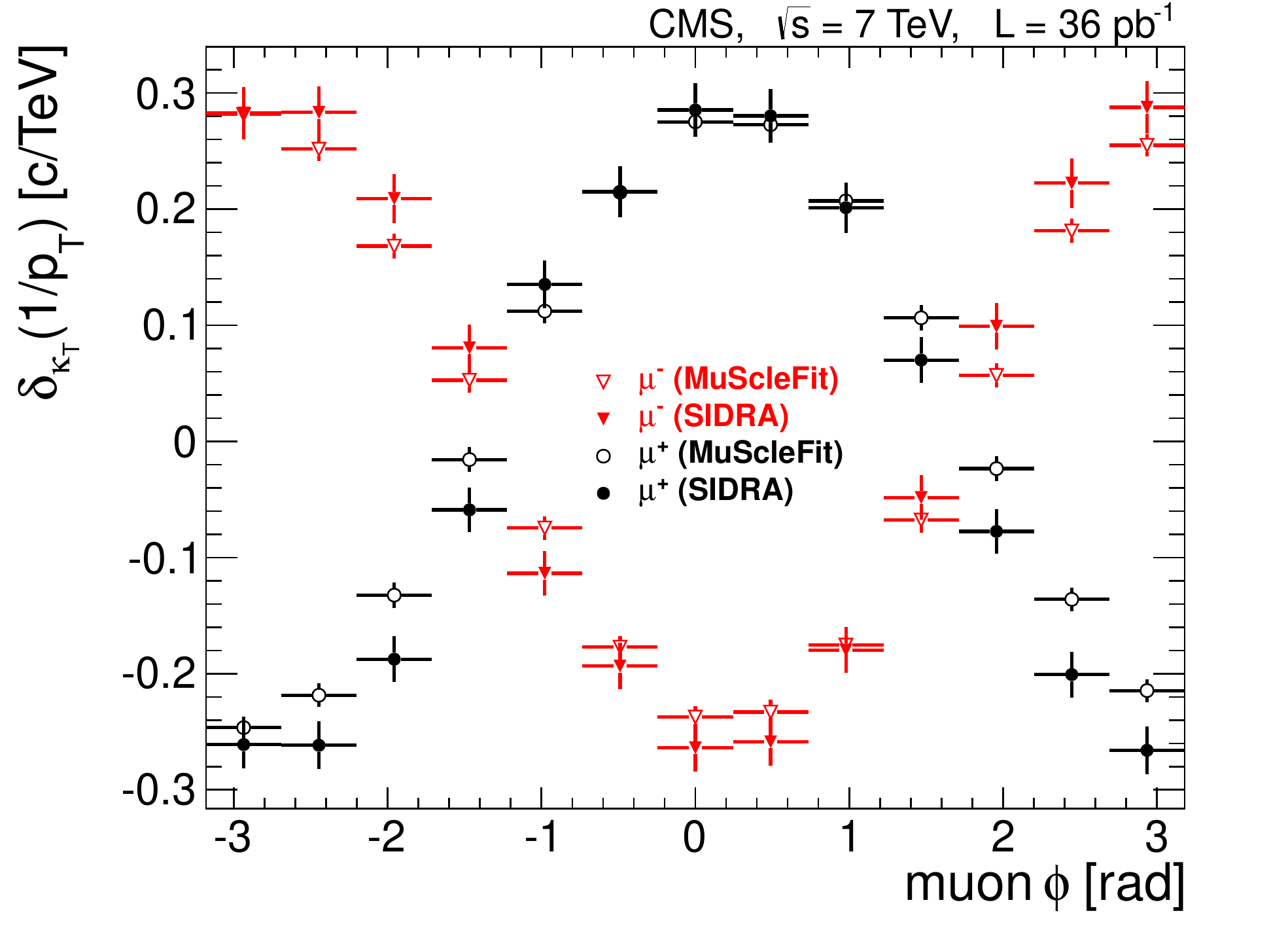}
  \includegraphics[width=0.48\textwidth]{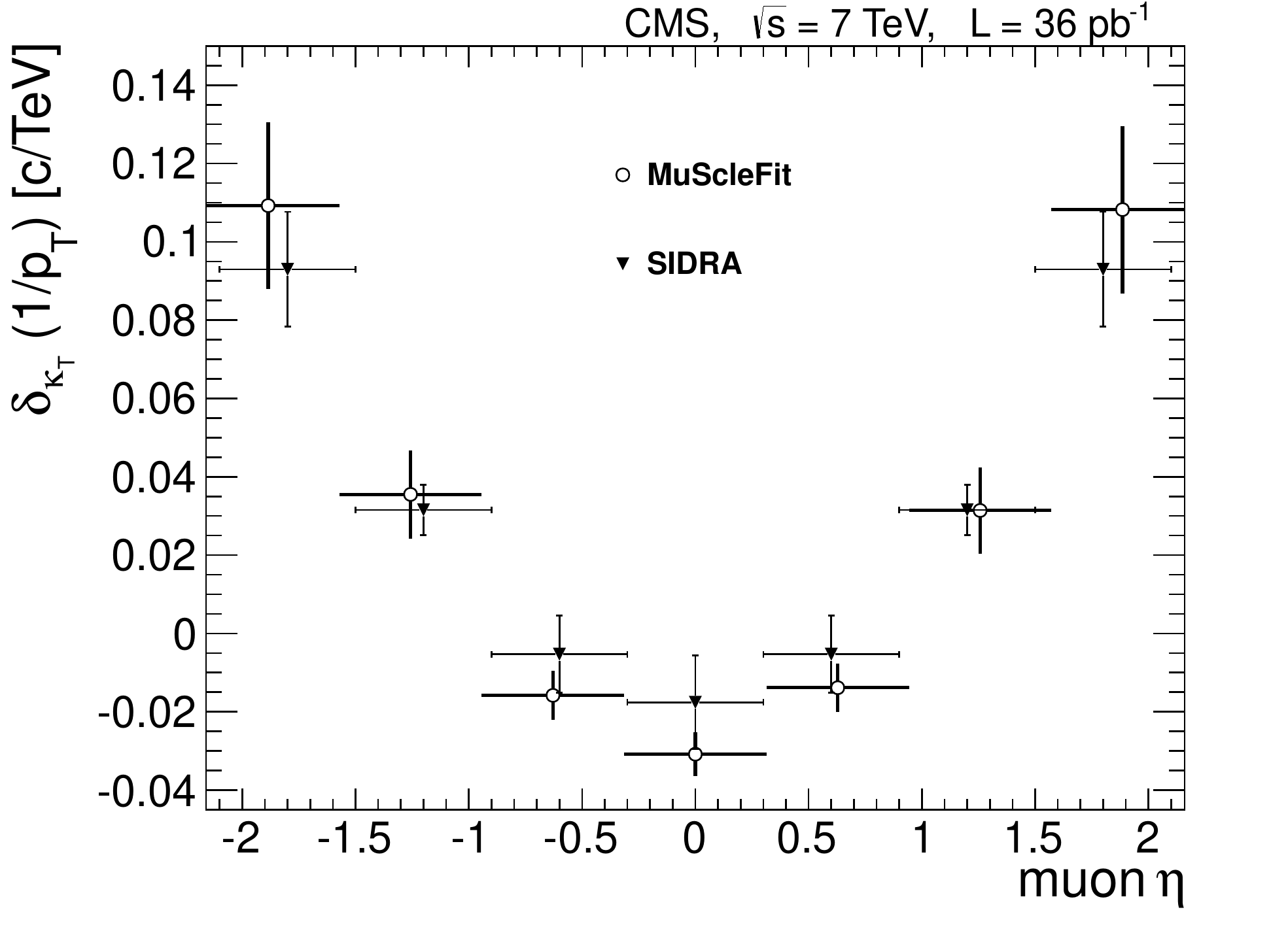}
  \caption{Comparison of the differences between muon momentum scale
    in data and that in the simulation
    obtained with the SIDRA and MuScleFit methods
    as a function of the azimuthal angle for positive and negative muons (left)
    and pseudorapidity (right). Only statistical uncertainties
    are shown.}
 \label{fig:Z_shifts}
 \end{center}
\end{figure}

The results of the application of the SIDRA method to the dimuon 
invariant-mass spectrum in 2010 data are shown in
Fig.~\ref{fig:sidraZ_after_correction}.  Applying SIDRA corrections
to the simulation improves agreement with the data.
The scale shifts $\delta_{\kappa_{\rm T}}$ as a function of $\phi$ and $\eta$ are shown in 
Fig.~\ref{fig:Z_shifts}, superimposed with the corresponding shifts
obtained with the MuScleFit method.  As also shown in Fig.~\ref{fig:massBias},
the phases and amplitudes of the $\phi$-dependent biases present in data
and in simulation are different: for $\pt \approx M_{\Z}/2$,
characteristic of this study, the amplitude of the 
sinusoidal correction to be applied to the simulation in order
to obtain the best match with the data
is $\approx$1.5\%; differences between the corrections 
from the two methods do not exceed 0.3\%.
When examined as a function of $\eta$,
the scale shifts between data and simulation are consistent with zero in the
barrel region and increase with
$|\eta|$. For $\pt = M_{\Z}/2$ and $|\eta| > 2$,
they are $\approx$0.5\% with a difference between the
two methods of $\approx$0.1\%.
When integrated over $\phi$ and $\eta$, the overall difference in muon momentum scale $\delta_{\kappa_{\rm T}}$
between data and simulation is found to be 0.016 $\pm$ 0.012 (stat.)~$c$/TeV
for SIDRA and 0.020 $\pm$ 0.006 (stat.)~$c$/TeV for MuScleFit.

We use the difference between the MuScleFit and SIDRA relative
simulation-to-data corrections
as the systematic uncertainty in the
measurements of the absolute momentum scale bias obtained with
MuScleFit.  Another source of systematic uncertainties are the
theoretical uncertainties in the reference models used by both SIDRA
and MuScleFit, such as
uncertainties in parton distribution functions, initial state
radiation and higher-order QCD effects, and weak and QED interference effects.
Dedicated simulation studies show that these uncertainties can produce
shifts in the $\Z$ mass peak position of at most 0.1\%, which corresponds
to an uncertainty in $\pt$ scale of order $0.1\% \cdot \pt/[91\GeVc]$ if
positive and negative muons are equally affected.
Summing both types of systematic uncertainties in quadrature, the amplitude
of the $\phi$-dependent scale correction at $\pt = M_{\Z}/2$ is found
to be
0.266 $\pm$ 0.010 (stat.) $\pm$ 0.046 (syst.)\GeVc.
When integrating over $\phi$, the scale correction for the same $\pt$ varies from 
$\Delta(\pt) = -0.130 \pm 0.022$ (stat.) $\pm$ 0.046 (syst.)\GeVc for $\eta = 0$ to
$\Delta(\pt) = 0.234 \pm 0.048$ (stat.) $\pm$ 0.046 (syst.)\GeVc for $|\eta| = 2.1$.
The future versions of the alignment workflow will include a
$\Z$-mass constraint as an integral part of the alignment procedure,
which will strongly reduce the $\phi$ and $\eta$ dependence of the
momentum bias.
Averaged over the whole acceptance, the relative bias in the muon momentum
scale is measured with a precision of better than 0.2\% and is found to be
consistent with zero up to $\pt$ values of 100\GeVc.

\subsubsection{Muon momentum resolution}
\label{sec:mures}
The techniques used to calibrate the muon momentum scale can also be used
to measure
the muon transverse momentum resolution.
In SIDRA, an extra smearing in data with respect to the resolution
predicted by the simulation is obtained from the $\sigma_{\kappa_{\rm T}}$
parameter, as explained in Section~\ref{sec:muscl}.  The full resolution in
data can then be evaluated by applying the scale shifts and extra smearing
to the reconstructed $\pt$ of muons in simulation and comparing the obtained
$\pt$ values with the ``true'' $\pt$.  The MuScleFit
procedure takes into account the correlations between the two muons, but neglects
contributions to the mass resolution from the $\phi$ and $\eta$ resolutions, as well as covariance terms.
These contributions were found to be small with respect to other terms at the typical $\pt$ range of muons from $\Z$; nevertheless,
the systematic uncertainty due to these approximations is estimated from the simulation and accounted for in the final
uncertainty of the result.
Resolution and scale biases are fit in successive iterations to
minimize correlation effects.
The assumed functional form to fit the momentum resolution in data
contains a term linear in $\pt$, and separate terms for positive and
negative $\eta$, each parabolic in $\eta$ with a common minimum at
$\eta = 0$. In the simulation many more events are available, and the
$\eta$ component contains a third term, symmetric about $\eta=0$.

\begin{figure}
\begin{center}
  \includegraphics[width=0.6\textwidth]{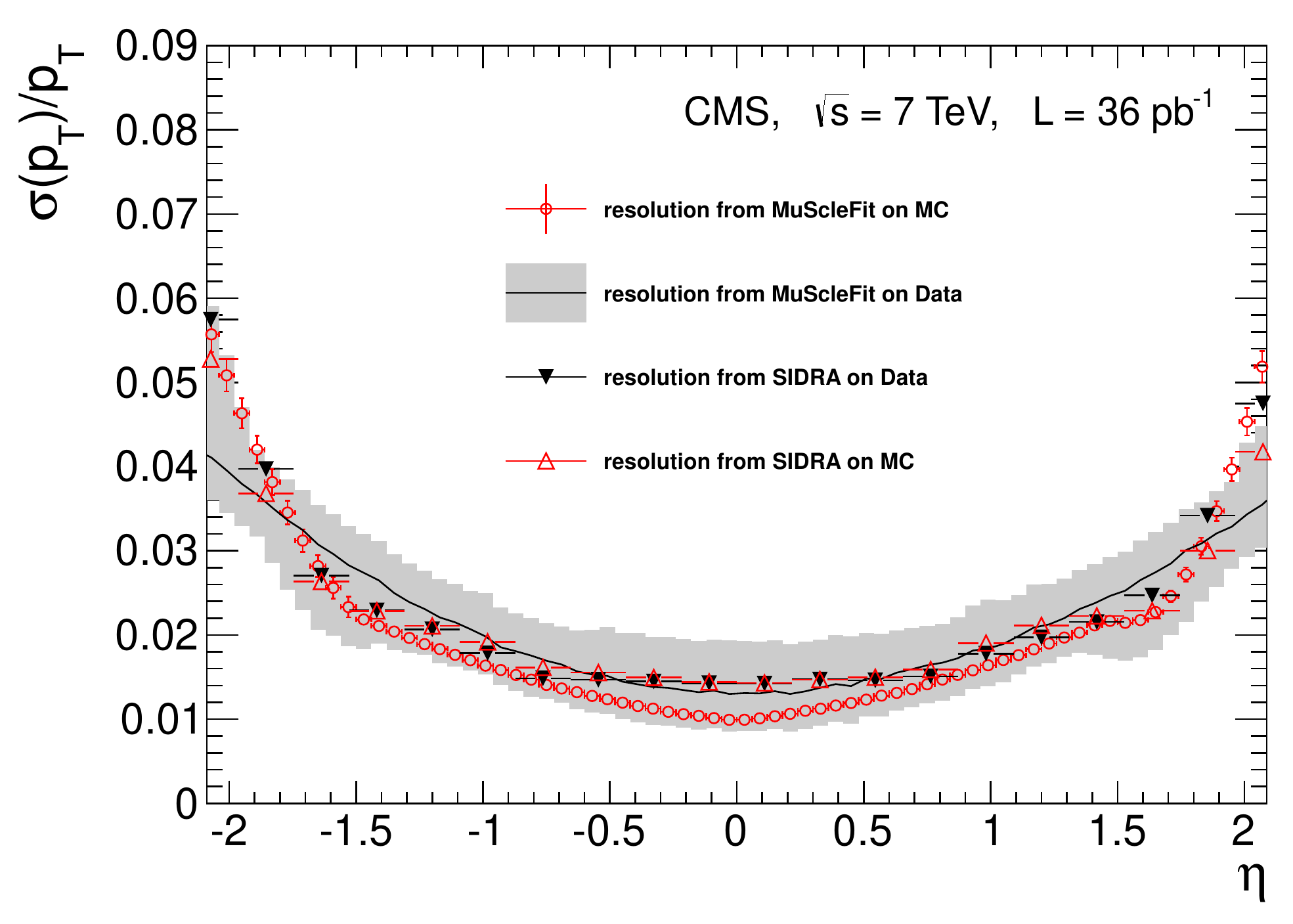}
  \caption{Relative transverse momentum resolution $\sigma(\pt)/\pt$ in
   data and simulation measured by applying the MuScleFit and SIDRA
   methods to muons produced in the decays of $\Z$ bosons and passing the
   Tight Muon selection.
   The thin line shows the result of MuScleFit on data, with the grey
   band representing the overall (statistical and systematic) 1$\sigma$
   uncertainty of the measurement. The 
   circles are the result of MuScleFit on simulation.
   The downward-pointing and upward-pointing triangles are the results
   from SIDRA obtained on data and simulation, respectively; the resolution
   in simulation was evaluated by comparing the reconstructed and ``true''
   $\pt$ once the reconstructed $\pt$ was corrected for $\phi$-dependent
   biases.  The uncertainties for SIDRA are
   statistical only and are smaller than the marker size.}
 \label{fig:sidraZ_eta_resolution}
 \end{center}
\end{figure}

The calibration of the momentum scale described in Section~\ref{sec:muscl}
improves the muon momentum resolution by $\approx$2\%.
Figure~\ref{fig:sidraZ_eta_resolution} shows the MuScleFit and SIDRA measurements of muon transverse momentum resolution versus $\eta$ after
correcting for biases in the momentum scale, for both data and simulation. 
The grey band enveloping MuScleFit results in data shows statistical and systematic uncertainties summed in quadrature.
The uncertainties from the choice of a particular function for the resolution shape and the approximations in the
method are estimated by comparing the result of the fit using the same function in simulation with the true MC resolution.
The bin-by-bin difference between the two results is taken as the
systematic uncertainty. The relative difference is on average 6\%
with an RMS of 4\%.
Another source of systematic uncertainty included in the band is
theoretical uncertainties in the reference models discussed in Section~\ref{sec:muscl}.  They
can produce an extra smearing of the $\Z$ mass distribution of at most
0.5\%, which can be interpreted as 
an uncertainty of $0.5\%/\sqrt{2}$ for muons with $\pt\approx M_{\Z}/2$.
The statistical and systematic uncertainties are of similar
magnitude; the overall (statistical and systematic combined in quadrature)
1$\sigma$ uncertainty of the measurement varies from 20\% to 40\% of the
resolution in the studied acceptance range.

As can be seen in Fig.~\ref{fig:sidraZ_eta_resolution},
the results obtained with the two methods agree within the uncertainties:
the largest difference in the barrel is
($\sigma(\pt)/\pt)_{\rm MuScleFit} - (\sigma(\pt)/\pt)_{\rm SIDRA} = 0.003 \pm 0.003 ({\rm stat.} \oplus {\rm syst.})$, % ($\sim 22\%$)
while in the endcaps it is $-0.018 \pm 0.013 (\rm{stat.} \oplus \rm{syst.})$. % ($\sim 32\%$).
The relative $\pt$ resolution in the intermediate $\pt$ range obtained using MuScleFit is found to be
in the range from 1.3\% to 2.0\% for muons in the barrel and up to $\approx$6\%
for muons in the endcaps, in good agreement with the results obtained
from simulation.
The $\sigma(\pt)/\pt$ averaged over $\phi$ and $\eta$ varies in $\pt$ from
$(1.8 \pm 0.3 (\rm{stat.}))\%$ at $\pt = 30 \GeVc$ to $(2.3 \pm 0.3 (\rm{stat.}))\%$ at
$\pt = 50 \GeVc$, again in good agreement with the expectations from
simulation.

\subsubsection{Momentum resolution of standalone muons}
\label{sec:staMuonResolData}

The momentum resolution for standalone muons is estimated using
the $\pt$ of the tracker track as reference: 

\begin{equation}
    R_{\mathrm{sta}}(1/\pt) = \left((1/\pt)^{\mathrm{sta}}-(1/\pt)^{\mathrm{trk}}\right)\left/\left(1/\pt\right)^{\mathrm{trk}}\right.\,.
  \label{eq:staResolData}
\end{equation}

As the resolution of the 
tracker tracks at low and intermediate $\pt$ is expected to be about an
order of magnitude better than the resolution of the standalone-muon tracks,
Eq.~(\ref{eq:staResolData}) provides a good estimate of the resolution
for standalone muons.
The relative difference between the resolutions measured with respect to the
tracker-track $\pt$ and the true $\pt$ was evaluated from
simulation and found to be smaller than 1\%
in the barrel and smaller than 5\% in the endcaps. 

The sample of muons used for this study is selected from events passing
the single-muon trigger with a minimum $\pt$ threshold of 5\GeVc by
applying the standard requirements for Tight Muons
(see Section~\ref{sec:reco}). 
Selected muons were subdivided into subsamples according to the
$\pt$ and $\eta$ of the tracker track, and a Gaussian fit to
the distribution of $R_{\mathrm{sta}}$ (Eq.~(\ref{eq:staResolData}))
is performed for each subsample, with the fit restricted
to a range $\pm 1$ RMS about the sample mean.
The resolution thus measured using data is compared with results obtained from
simulation. Simulated muons are selected from MC samples of QCD, 
$\Wmn$, and $\Zmm$ events using the 
same trigger and selection criteria as applied to data.

\begin{figure}[ht!]
  \begin{center}
  \includegraphics[width=0.477\textwidth]{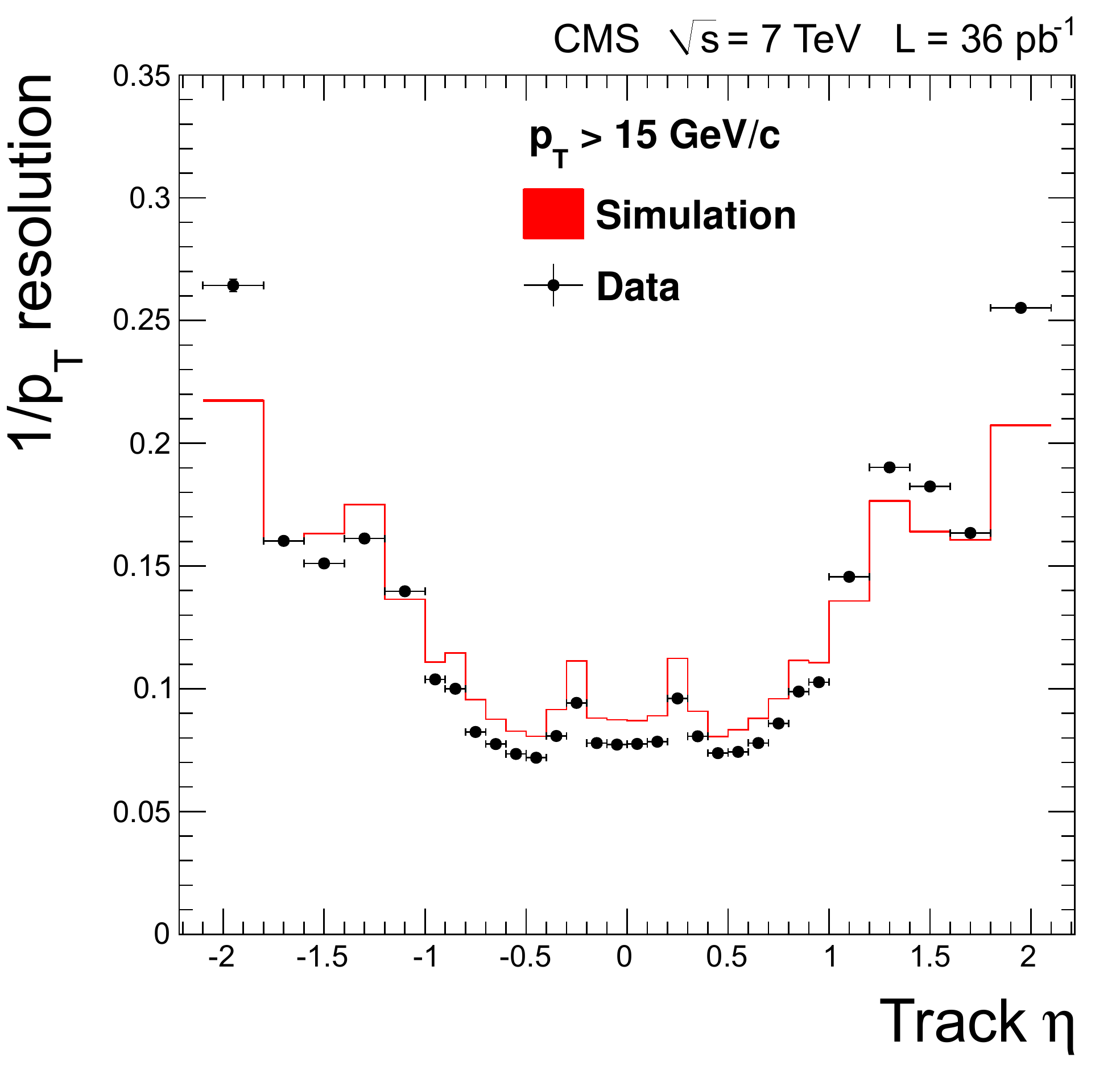}
 \end{center}
 \vspace*{-0.3cm}
 \caption{The $1/\pt$ resolution of standalone-muon tracks with respect
  to tracker tracks, as a function of $\eta$ of the tracker track
  for muons with $\pt>15\GeVc$.
  Resolution is estimated as the $\sigma$ of a
  Gaussian fit to $R_{\mathrm{sta}}(1/\pt)$ defined in
  Eq.~(\ref{eq:staResolData}).} 
  \label{fig:staResol}
\end{figure}

We report results for standalone muons with a beam-spot constraint
applied in the track fit.
Figure~\ref{fig:staResol} shows the widths of the Gaussian fits
to the distributions of $R_{\mathrm{sta}}$ as a function of $\eta$
for muons with $\pt > 15\GeVc$.
The resolution in the barrel remains better than 10\%
up to $\pt = 100\GeVc$.  The overall shape of the $\eta$ dependence is
reproduced by the MC simulation to within 10--15\%.
Considering that standalone muons are usually not used in physics analyses
on their own and mainly serve as a component of the global-muon reconstruction,
the performance of which is well described by the simulation, this
difference between resolutions for standalone muons in data and
simulation is acceptable. A 
slight asymmetry between resolutions in the negative and positive
endcaps is due to small differences in the precision of the alignment
of the muon chambers.
The bias in the momentum scale of standalone muons, given by the mean
values of the fits to the $R_{\mathrm{sta}}$ distributions, does not
exceed 1\% in the barrel region and 5\% in the region of $0.9 < |\eta| < 2.1$.

\subsection{\texorpdfstring{Measurements at high $\pt$}{Measurements at high pT}}
\label{sec:resscalefromcosmics}

High-$\pt$ muons are
an important signature in many searches for new physics, so it is crucial
that the performance of their reconstruction, which has some significant
differences to that of lower-$\pt$ muons (such as an increased role
of the muon system in momentum measurement and larger impact of showering)
is well understood. While there are few high-$\pt$ muons from LHC collisions in the 2010 dataset,
cosmic-ray events provide a source of muons with a momentum spectrum
extending to quite high $\pt$; cosmic-ray muons having momentum up to
a few \TeVc have been observed in CMS. Here we present results
using cosmic-ray events collected during periods in 2010 when the LHC
was not delivering collisions.

\subsubsection{Muon momentum resolution from cosmic-ray muon data}

Cosmic-ray muons that traverse the entire CMS detector
can be used to evaluate the momentum resolution by comparing
the momenta reconstructed independently in the upper and lower
halves of the muon system, a procedure that was first applied to
cosmic-ray muons collected in
2008~\cite{CMS_CFT_09_014}.  The angular distribution of cosmic-ray muons that
traverse the detector is strongly peaked in the vertical direction.
To select events in which the muons are most similar to those produced
in collision events and used in physics analyses, we only use pairs of
tracks that pass close to the interaction point, enforced in practice
by requiring each track of the pair to contain hits from at least eight
different layers of the silicon strip tracker and at least one hit from
the pixel detector.
The results presented in this section are for the barrel region only,
because few cosmic muons cross both endcaps and pass near the centre of CMS.

To estimate the muon $q/\pt$ resolution, we define the relative
residual,

\begin{equation}
\label{eqn:relative_residual}
R(q/\pt) = \frac{{(q/\pt)}^\text{upper} - {(q/\pt)}^\text{lower}}{\sqrt{2} {(q/\pt)}^\text{lower}}\,,
\end{equation}

where ``upper'' and ``lower'' refer to the results of the fits in the
two halves of CMS, and the factor of $\sqrt{2}$ accounts for the fact
that the two fits are independent.  We also define the normalized
residual (pull) as

\begin{equation}
\label{eqn:cosmics_pull}
P(q/\pt) = \frac{{(q/\pt)}^\text{upper} - {(q/\pt)}^\text{lower}}{\sqrt{\sigma_{(q/\pt)^\text{upper}}^2 + \sigma_{(q/\pt)^\text{lower}}^2}}\,,
\end{equation}

to examine the behavior of the uncertainties from the track fit. To study the
various effects as a function of $\pt$, we examine both the truncated
sample RMS and a Gaussian fit to the distributions of $R(q/\pt)$ and
$P(q/\pt)$, with the fit range in each $\pt$ bin under consideration
restricted to the region of $\pm 1.5$ RMS
centred on the sample mean. The $\pt$ used in the binning is that of the lower
tracker-only track.

\begin{figure}[thbp]
  \begin{center}
      \subfigure[]{\includegraphics[width=0.49\textwidth]{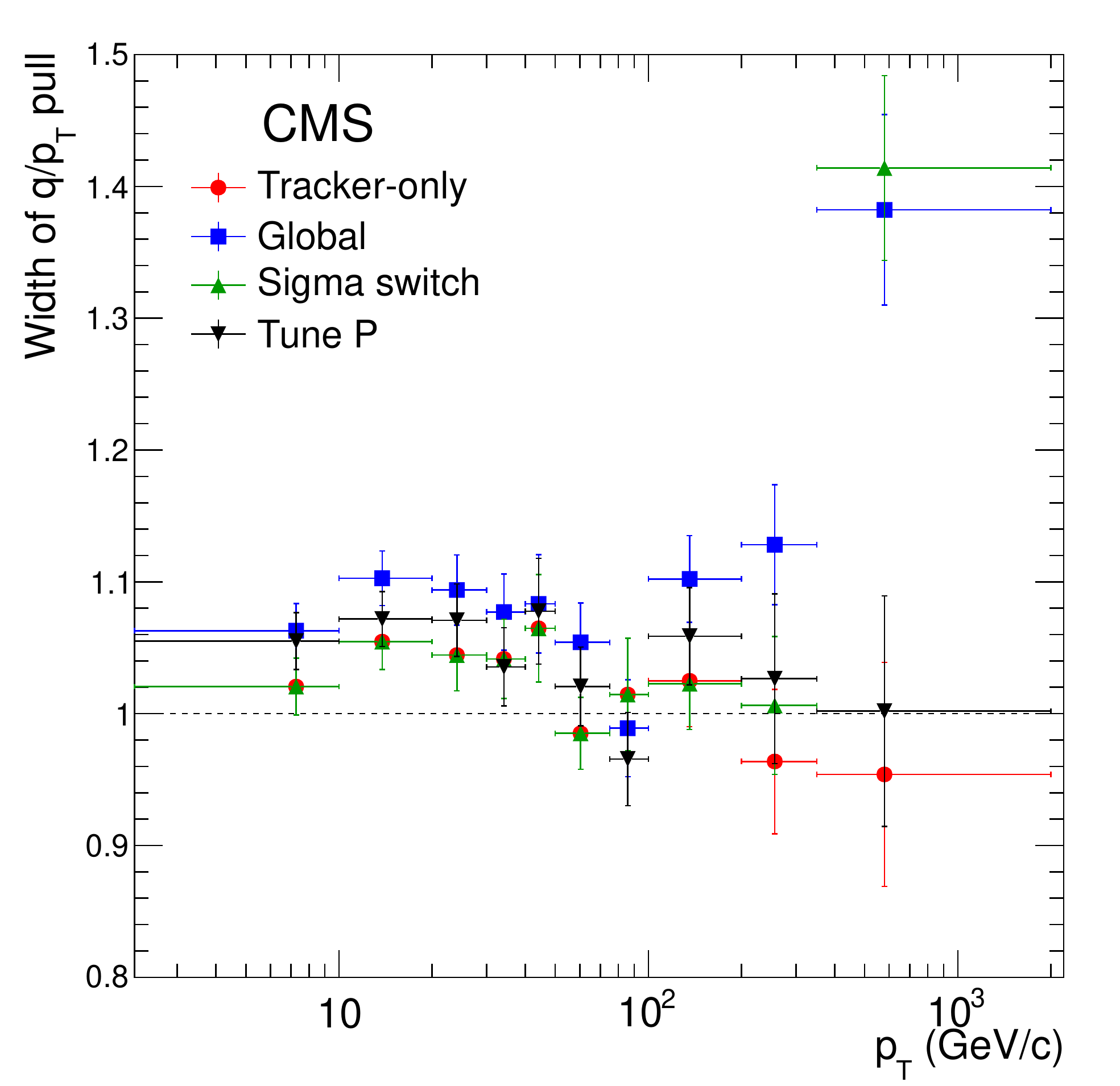}}
      \put(-175,135){\footnotesize $|\eta| < 0.9$}
      \subfigure[]{\includegraphics[width=0.49\textwidth]{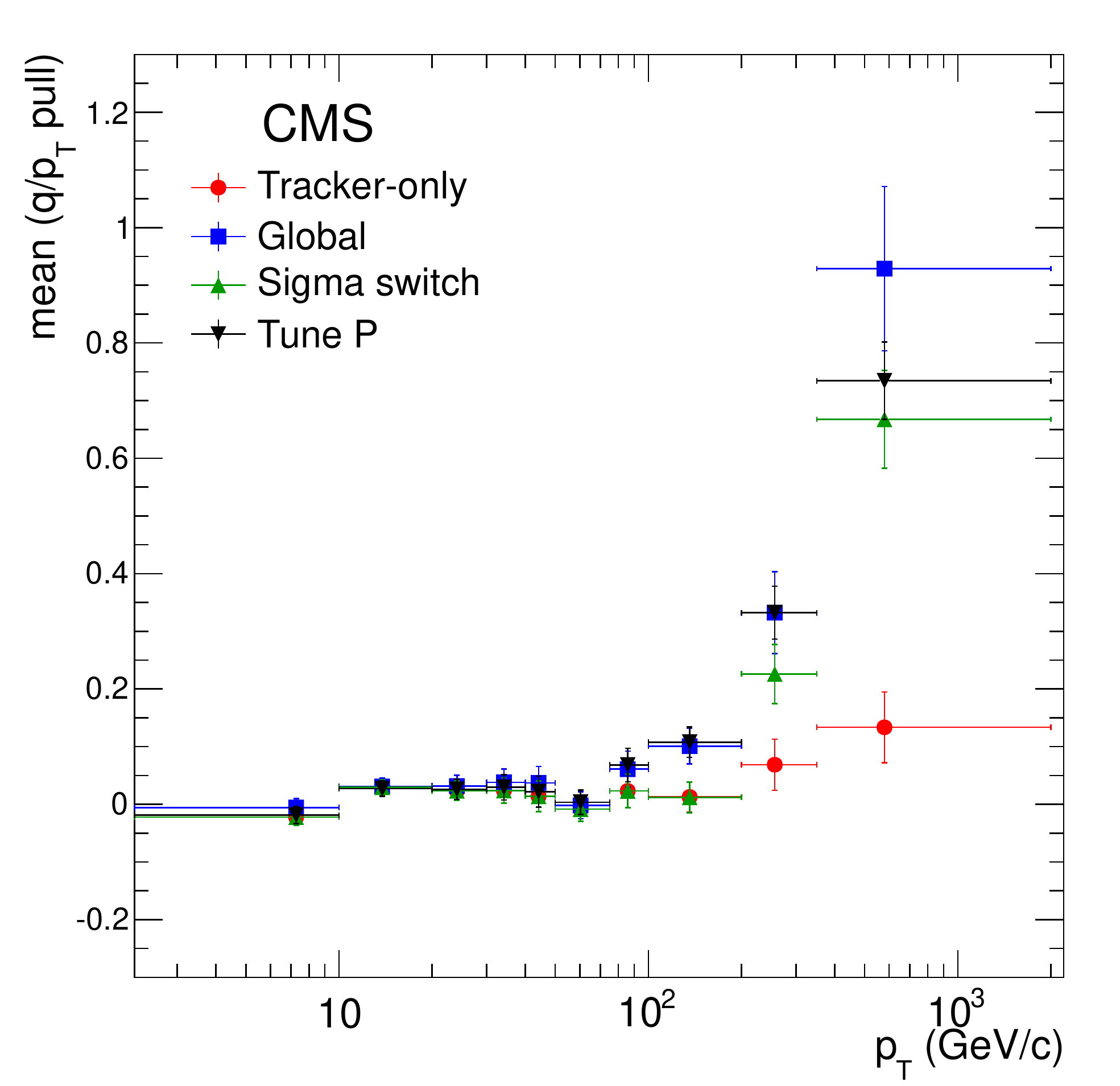}}
      \put(-175,135){\footnotesize $|\eta| < 0.9$}
      \caption{(a) Widths of Gaussian fits to the distributions of the
        muon $q/\pt$ pulls (as defined by Eq.~(\ref{eqn:cosmics_pull}))
        for the tracker-only and global fits, and for the output of the
        sigma-switch and Tune P algorithms, as a function of the $\pt$ of the
        muon; (b) the means of the same distributions.}
      \label{fig:cosmics_pulls}
  \end{center}
\end{figure}

Figure~\ref{fig:cosmics_pulls}(a) shows the Gaussian
widths of the pulls as a function of $\pt$ for the tracker-only and
global fits, and for the sigma-switch and Tune P algorithms (see
Section~\ref{sec:reco}).  While the cores of the pull distributions are well
described by Gaussians, the tails are
non-Gaussian, so these widths are slightly smaller than the sample
RMS values, which are not shown here. The widths of the pulls for the
tracker-only fit are within $\pm$10\% of unity over the entire $\pt$
range studied, indicating that the track uncertainties are well estimated.
The widths of the pulls for the global fit are close to
unity in the range of $\pt \lesssim 200 \GeVc$, where the tracker
information dominates in the fit, and start to become larger
at $\pt \approx 200 \GeVc$.
One important factor contributing to this is that the muon alignment position
uncertainties, which account for the precision with which the positions
of different detector components are known, are set to zero in the track
fits.  As a consequence, the total uncertainties are underestimated at high
$\pt$. As Tune P uses less information from the muon system than the global
fit, the pulls are in between those of tracker-only and global fits.

Figure~\ref{fig:cosmics_pulls}(b) shows the sample
means for the distributions of $P(q/\pt)$ for the same four muon
reconstruction algorithms.  The means are consistent
with zero up to $\pt$ values of 100\GeVc, above which fits with muon
information begin to show a sizeable bias. The biases seen
in the means of the residual distributions $R(q/\pt)$, which are not
shown here, are small, indicating that
the source of the bias in the means of $P(q/\pt)$ is most likely to be
the underestimated uncertainties in the fit.
This bias should become smaller as the muon system alignment
and its uncertainties are further refined.

\begin{figure}[hbtp]
  \begin{center}
      \subfigure[]{\includegraphics[width=0.49\textwidth]{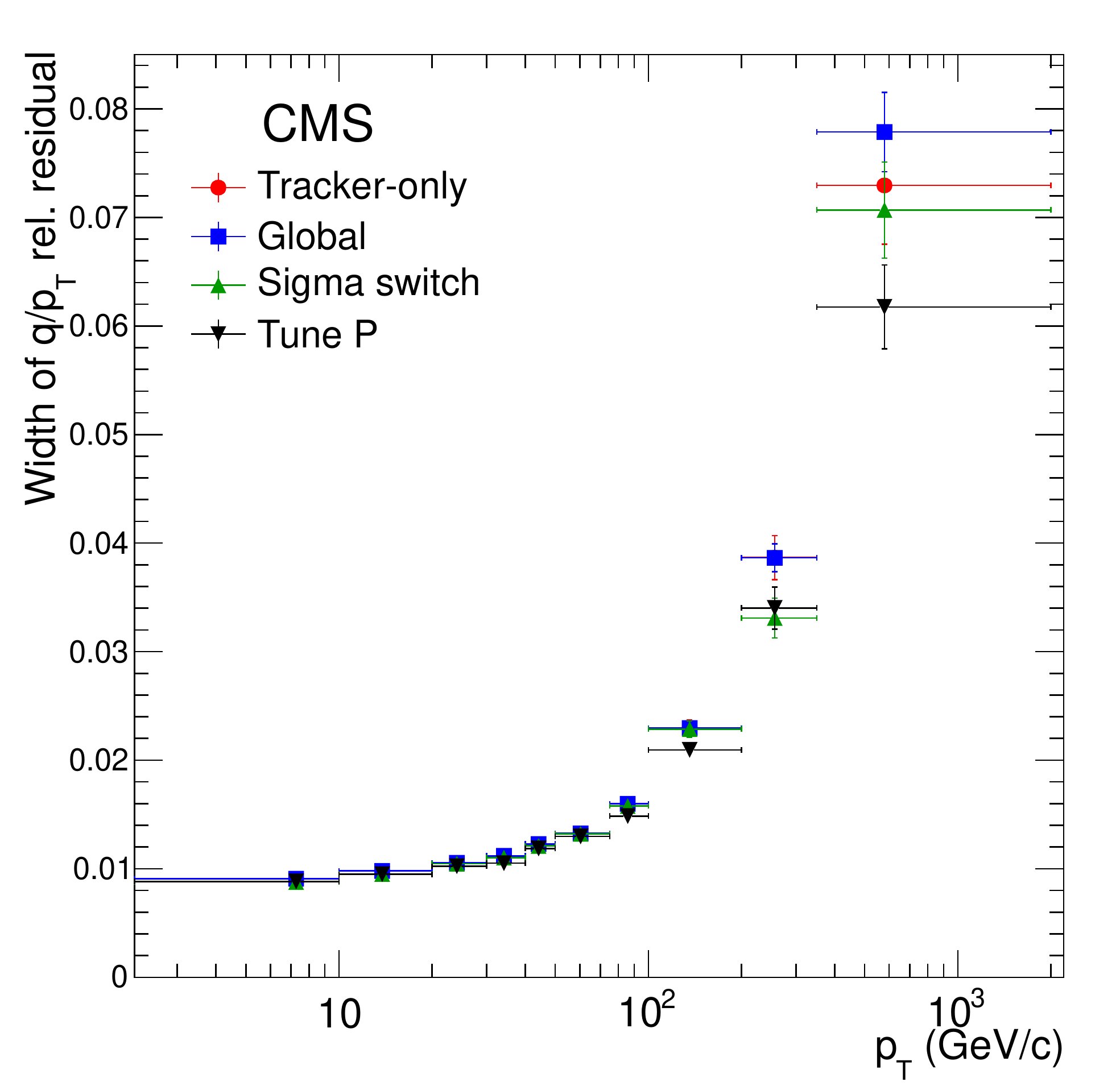}}
      \put(-175,135){\footnotesize $|\eta| < 0.9$}
      \subfigure[]{\includegraphics[width=0.49\textwidth]{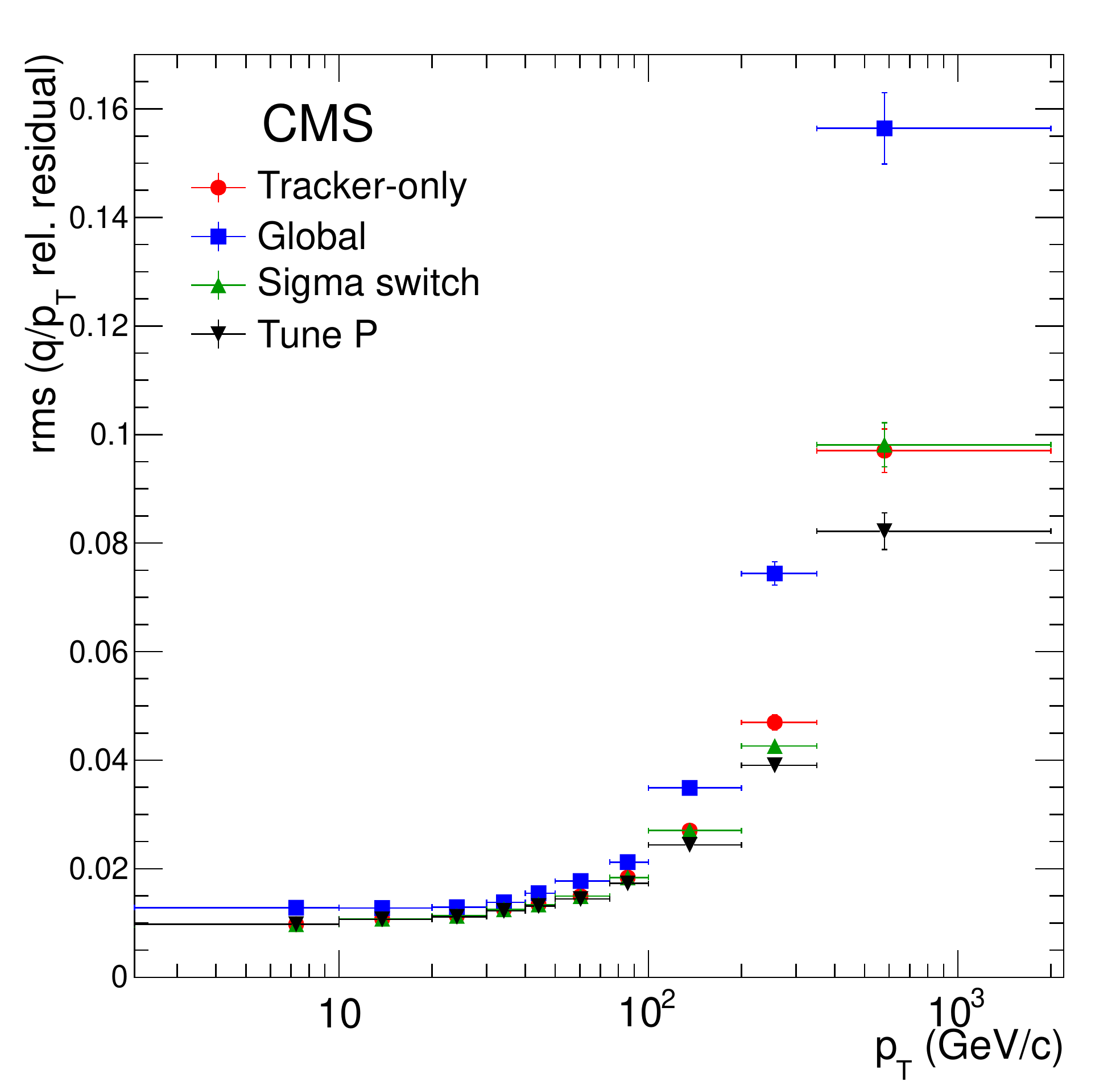}}
      \put(-175,135){\footnotesize $|\eta| < 0.9$}
      \caption{(a) Widths of
        Gaussian fits of the distributions of
        the muon $q/\pt$ relative residuals (as defined by
        Eq.~(\ref{eqn:relative_residual})) for the tracker-only and global
        fits, and for the output of the sigma-switch and Tune P algorithms,
        as a function of the $\pt$ of the muon; (b) sample RMS values
        (truncated at $\pm 1$) of the same distributions. }
      \label{fig:cosmics_resolution}
  \end{center}
\end{figure}

Figures~\ref{fig:cosmics_resolution}(a) and (b) show the relative
resolution as measured by the Gaussian width and the truncated sample RMS,
respectively, for the tracker-only and global fits, and for the sigma-switch
and Tune P algorithms as a function of the $\pt$ of the muon.
Table~\ref{tab:highptres} summarizes the performance of these algorithms for
muons with transverse momentum in the range $350 < \pt < 2000\GeVc$.
The Gaussian width gives a
measure of the core resolution, while the truncated sample RMS
includes the effects of the tails of the distribution; both can be
separately important for considerations of momentum resolution,
possibly depending on the details of the physics analysis being
considered.

\begin{table}[tbh!]
\centering
\topcaption{Comparison of the fitted Gaussian width $\sigma$, sample RMS (truncated at $\pm 1$),
  and counts in the tails of the $R(q/\pt)$ distribution for 294 muons with
  measured transverse momentum in the range $350 < \pt < 2000\GeVc$.
}
\begin{tabular}{|l|r|r|c|c|}
\hline
             Fit/selector &       Fitted $\sigma$, \% &                   RMS, \% &           $R(q/\pt) < -1$ &            $R(q/\pt) > 1$ \\ \hline \hline
             Tracker-only &             7.3 $\pm$ 0.3 &             9.7 $\pm$ 0.4 &                         0 &                         0 \\
                   Global &             7.8 $\pm$ 0.3 &            15.6 $\pm$ 0.7 &                         4 &                         9 \\
             Sigma switch &             7.1 $\pm$ 0.3 &             9.8 $\pm$ 0.4 &                         0 &                         0 \\
                   Tune P &             6.2 $\pm$ 0.3 &             8.2 $\pm$ 0.3 &                         0 &                         0 \\
\hline
\end{tabular}
\label{tab:highptres}
\end{table}

Muon momentum resolution in the
$\pt$ region below approximately 200\GeVc is dominated by the track
measurement in the inner tracker, so one already has the best
performance by using the tracker-only fit for
low-to-intermediate-$\pt$ muons. At high $\pt$, the extended path length
through the magnetic field
between the tracker and the muon system leads to improved resolution,
provided that the pattern recognition carefully selects hits in the
muon system, avoiding hits due to showers
and hits in any muon chamber whose position is poorly known.
In Fig.~\ref{fig:cosmics_resolution} these two effects are visible:
at high $\pt$, the global fit has larger resolution tails
and worse core resolution than the tracker-only fit, while the
more selective Tune P performs better than tracker-only and sigma switch in
both respects. These results are in agreement with previous studies and
expectations~\cite{CMS_CFT_09_014}.

\subsubsection{Momentum scale from the cosmics endpoint method}

The flux of cosmic-ray muons falls steeply as momentum increases.
If there is a $q/\pt$ bias present in the reconstruction at high
momentum, the shape of the $q/\pt$ spectrum for cosmic-ray muons
will be distorted, with the location of the minimum shifted from zero;
the $q/\pt$ bias of high-momentum tracks can be extracted from the
location of the minimum.  The ``cosmic endpoint'' method used to
estimate the bias is described in Ref.~\cite{MUO-10-001} and briefly
summarized below.

We perform a binned comparison between the
$q/\pt$ distributions of the data and of simulated events,
introducing various artificial biases into the spectrum of simulated
events and calculating the $\chi^2$ as the sum of the squared
differences between each bin of the data histogram and the corresponding bin of
the new simulated histogram, divided by the value in the simulated bin.
To ensure that the extraction of the bias is not affected by possible
data-simulation differences in the flux ratio of positive to negative
cosmic-ray muons, the $q/\pt$ distribution in the simulation is normalized
to that in the data separately in the $q/\pt < 0$ and $q/\pt > 0$ regions.
We then fit the distribution of the $\chi^2$ values as a function of
the introduced bias with a %n eighth-order
polynomial and take the location of its minimum
as the estimate of the bias in data.  The one standard deviation
statistical uncertainty is computed as the $\Delta(q/\pt)$ corresponding
to an increase in the function value by $\Delta\chi^2=1$ from its minimum.

We report results for the tracker-only reconstruction to compare
directly with the measurements performed at intermediate $\pt$
(Section~\ref{sec:momentumScaleAtMediumPt}).
Using cosmic-ray muons reconstructed with $\pt > 200\GeVc$ and passing
the selection criteria as described in the previous section, and introducing
in the simulation a bias $\kappa$ of the form

\begin{equation}
q/\pt \rightarrow q/\pt + \kappa
\end{equation}

motivated by the arguments used for the choice of the functional forms
presented in Eqs.~(\ref{eq:pdfMUSCLE}) and (\ref{eq:pdfSIDRA}),
the bias in data is found
to be $-0.20 \pm 0.12\ {\rm (stat.)}$~$c/$TeV.

To test the robustness of this procedure we generate ensembles of
pseudo-experiments using simulated cosmic-ray muons
for five equally spaced values of $\kappa$ from $-0.2$ to $0.2~c/$TeV.
For each pseudo-experiment, we repeat the procedure outlined above and
calculate a pull as the difference between
the estimated bias and the input bias divided by the
uncertainty.  The distributions of these pulls are consistent with
a Gaussian of zero mean and unit width.
Considering all the pull distributions tested,
we assign the largest deviation from zero mean, 0.02 $c$/TeV,
as a systematic uncertainty on the estimate of the bias.
We also consider the following sources of systematic uncertainty and
find their impact to be negligible: the resolution model in the
simulation, the functional form and the fit range used in finding the
$\chi^{2}$ minimum, and uncertainty in the charge ratio of cosmic-ray muons.

To compare this measurement to that obtained using muons from
$\Z$ decays (Section~\ref{sec:momentumScaleAtMediumPt}), we divide the
cosmic-ray muon sample into three bins in $\phi$ and repeat the
above procedure. The results using these $\phi$ bins and for the whole
studied $\phi$ range are shown in
Table~\ref{fig:cosmics_endpoint_v_phi}.
The results for the momentum bias at high $\pt$ obtained with the endpoint
method are compatible with those observed at intermediate $\pt$.  The
accuracy of the measurement is currently statistically limited by the available
sample of high-$\pt$ cosmic-ray muons and will improve
as more cosmic muons are collected.

\begin{table}[tbh!]
\centering
\topcaption{The average $q/\pt$ bias obtained with the cosmic endpoint method
in three bins of the azimuthal angle $\phi$ and for the whole studied
$\phi$ range.  The first uncertainty quoted is statistical; the second
is systematic.  Also shown are results obtained with
the MuScleFit method (Section~\ref{sec:momentumScaleAtMediumPt}) by
rescaling $\pt$ of each of the muons from $\Z$ decays according to
Eq.~(\ref{eq:pdfMUSCLE}) and computing the average difference between
the rescaled and the original $q/\pt$ values in bins of $\phi$.  The
distribution of muons from $\Z$ decays was reweighted to have the same $\phi$
distribution as that of cosmic-ray muons.  The uncertainty quoted is
statistical only.}
\begin{tabular}{|c|c|c|} \hline
$\phi$ range, rad & Endpoint method, $c$/TeV & MuScleFit method, $c$/TeV \\ \hline \hline & & \\[-12pt]
$-\pi< \phi <-2.1 $ & $-0.14 \pm 0.09 \pm 0.02$ & $-0.10 \pm 0.01$ \\
$-2.1< \phi <-1.05$ & $-0.31 \pm 0.09 \pm 0.03$ & $-0.11 \pm 0.01$ \\
$-1.05< \phi <0$    & $-0.09 \pm 0.07 \pm 0.01$ & $-0.06 \pm 0.01$ \\ \hline & & \\[-12pt]
$-\pi< \phi <0$     & $-0.20 \pm 0.12 \pm 0.02$ & $-0.10 \pm 0.01$ \\ \hline
\end{tabular}
\label{fig:cosmics_endpoint_v_phi}
\end{table}

\section{Background from Cosmic-Ray and Beam-Halo Muons}
\label{sec:cosmics}

\subsection{Cosmic-ray muons}

As shown in the previous section and in Ref.~\cite{CMS_CFT_09_014},
cosmic-ray muons are very valuable for studying the performance
of the muon system and muon reconstruction tools.  At the same time, they
are a source of several types of background in physics analyses:
\begin{itemize}
\item A cosmic-ray muon passing close to the interaction point can be reconstructed as a collision muon, or as
a pair of oppositely charged muons in the upper and lower halves of CMS.
\item A muon that is not reconstructed in the tracker (either because it is out-of-time or passes
too far from the interaction point) can still be reconstructed as a standalone muon in the muon system and accidentally matched
to a tracker track, forming a mismeasured global muon.
\item A cosmic-ray muon can deposit energy in the calorimeters but avoid detection in the tracking detectors, which would result in, \eg, mismeasured missing
transverse energy.
\end{itemize}

To discriminate between cosmic-ray and collision muons, several
quantities can be used, related to either properties of the
whole event (vertex quality and track multiplicity) or
properties of a muon candidate (transverse impact parameter $|d_{\rm xy}|$,
timing information, and collinearity of two tracks reconstructed in the
upper and lower halves of the detector and associated with the same muon).
Their performance was studied
using two different data samples: a sample of muons from collisions with
a small (of the order of 1\%) contamination from cosmic-ray muons, selected
by requiring events with a high-quality reconstructed primary vertex, and a sample of cosmic-ray muons with a contamination from collision events smaller than 0.1\%, selected by requiring
events without a well-reconstructed primary vertex and with at most two tracks reconstructed in the inner tracker.
For comparisons with the data,
Monte Carlo samples of $\Zmm$ and Drell--Yan dimuon events were used as representative of muons from collisions for
the variables under study. Only events with at least one Global Muon reconstructed
with $\pt > 10 \GeVc$ and passing loose quality criteria were selected.
All distributions were normalized to the number of collision muons in data.

\begin{figure}[thb]
 \centering
 \subfigure[]{\includegraphics[height=0.49\textwidth,angle=90]{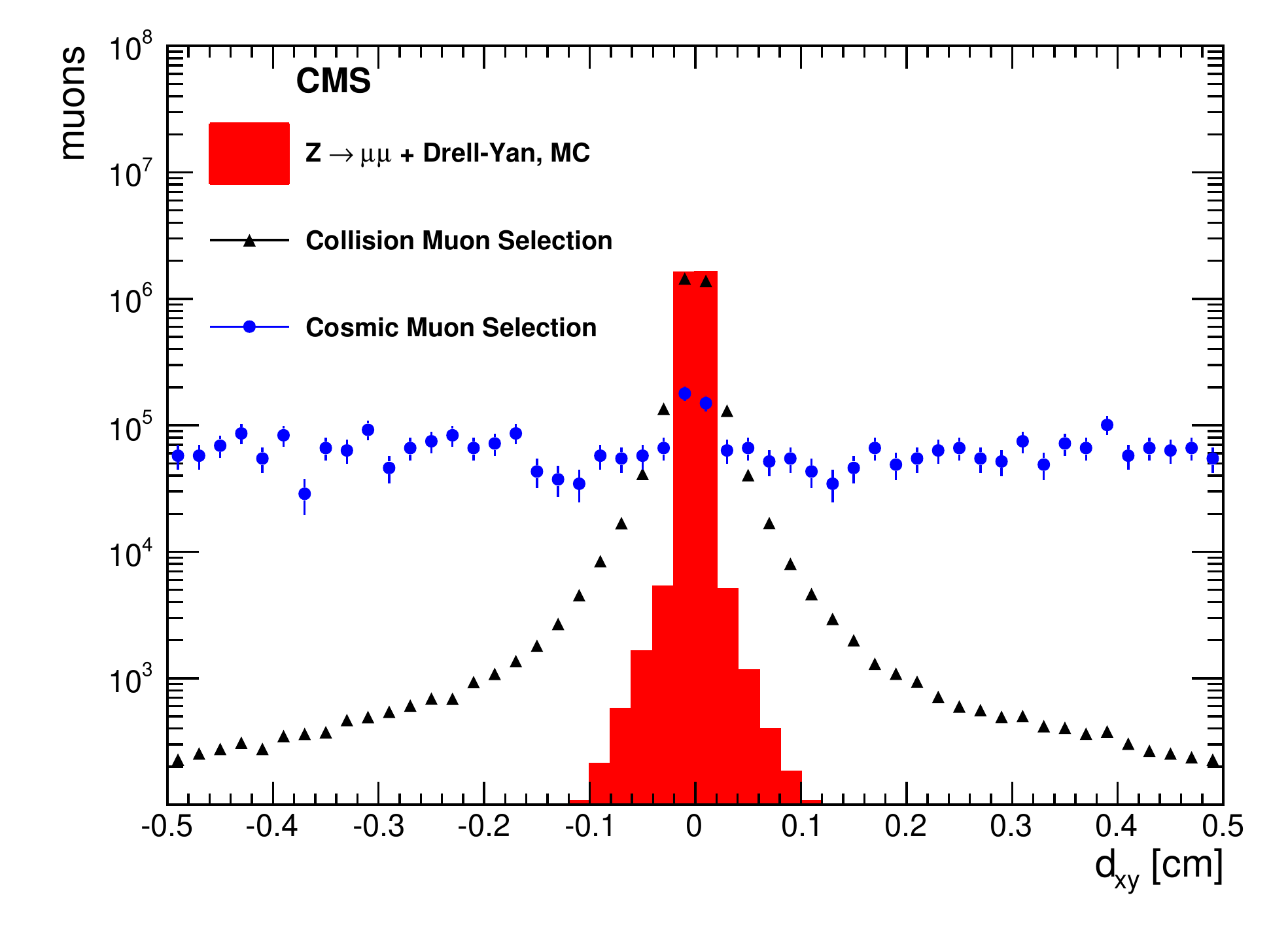}}
 \subfigure[]{\includegraphics[height=0.49\textwidth,angle=90]{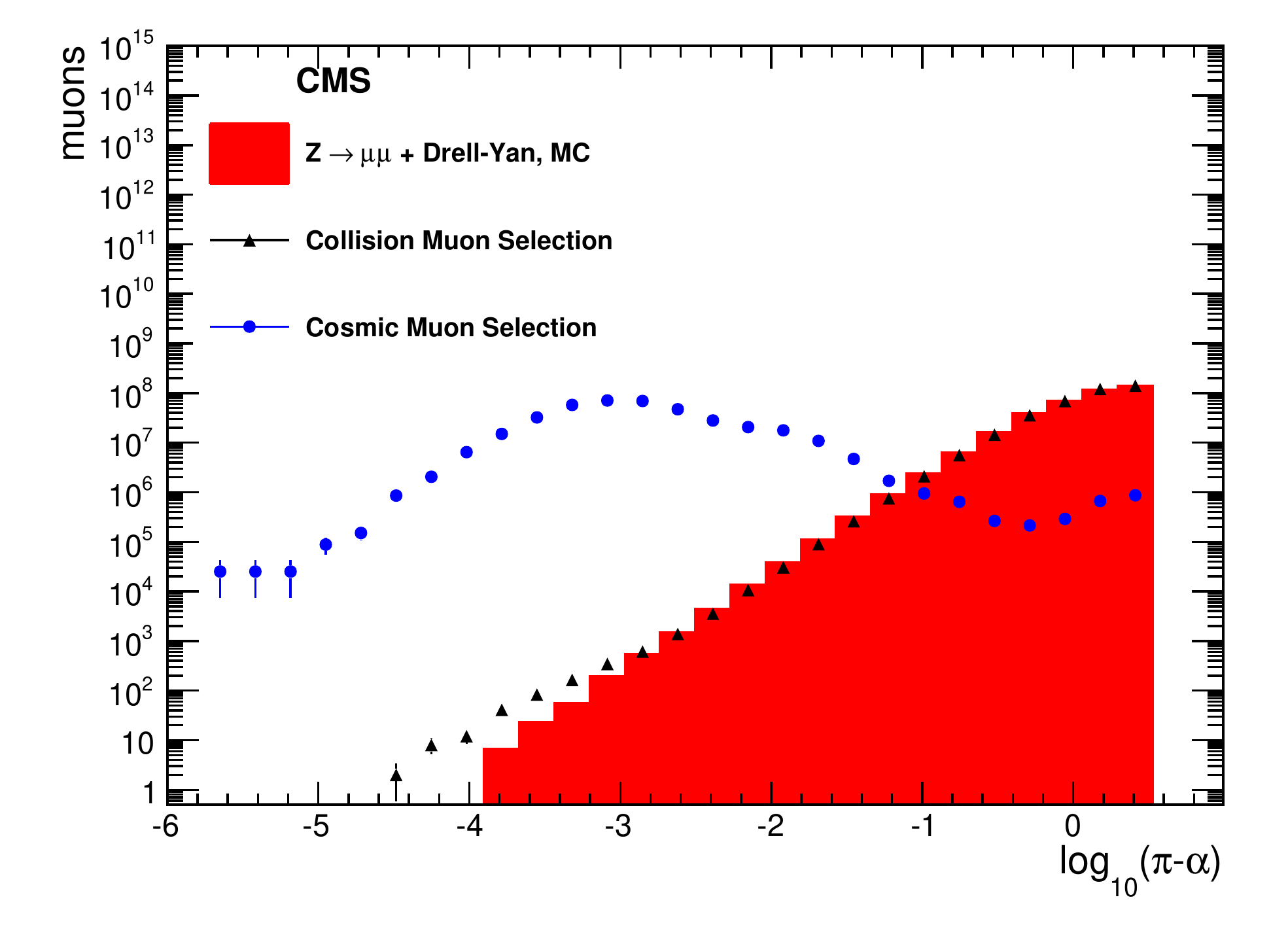}}\\
 \subfigure[]{\includegraphics[height=0.49\textwidth,angle=90]{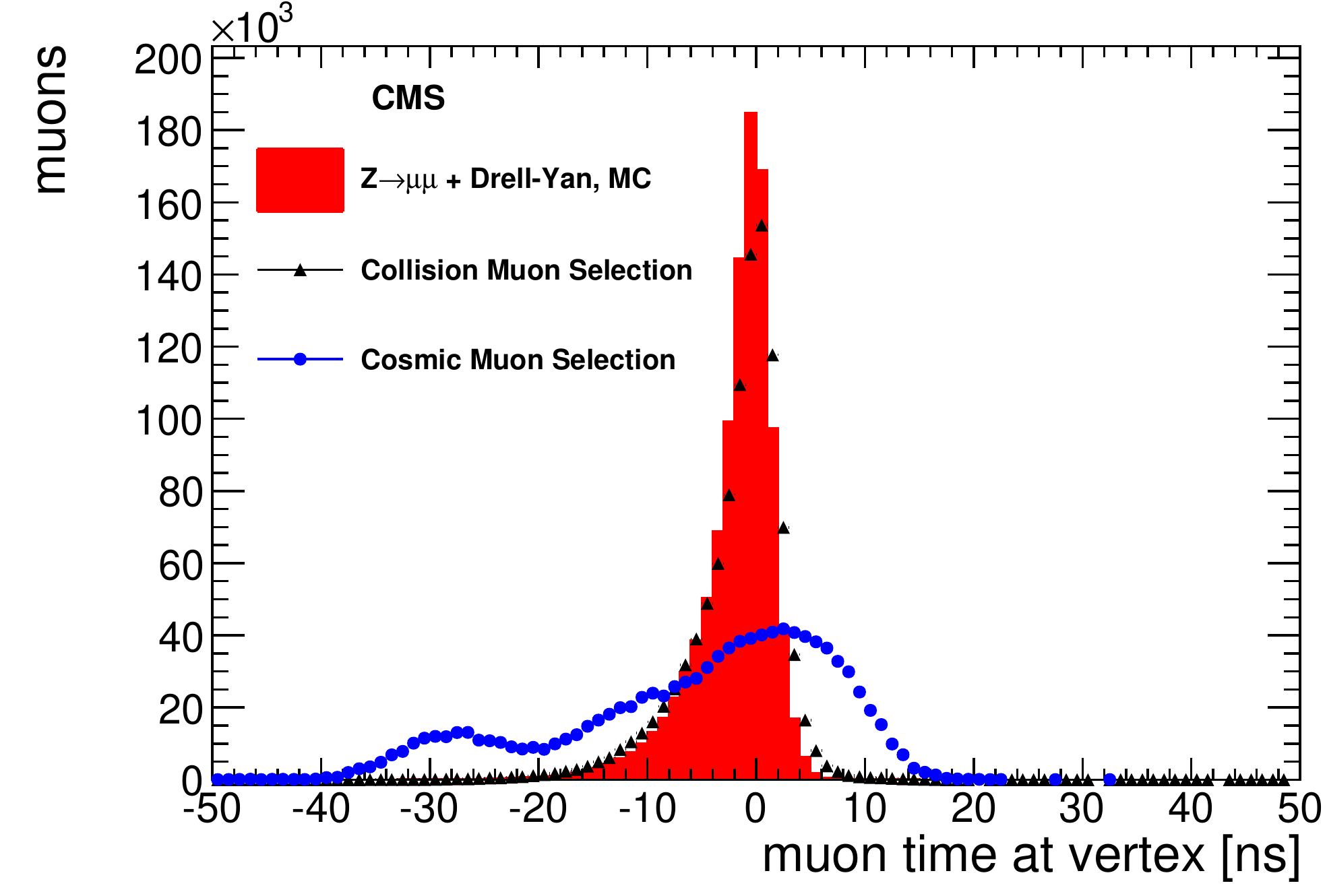}}
 \subfigure[]{\includegraphics[height=0.49\textwidth,angle=90]{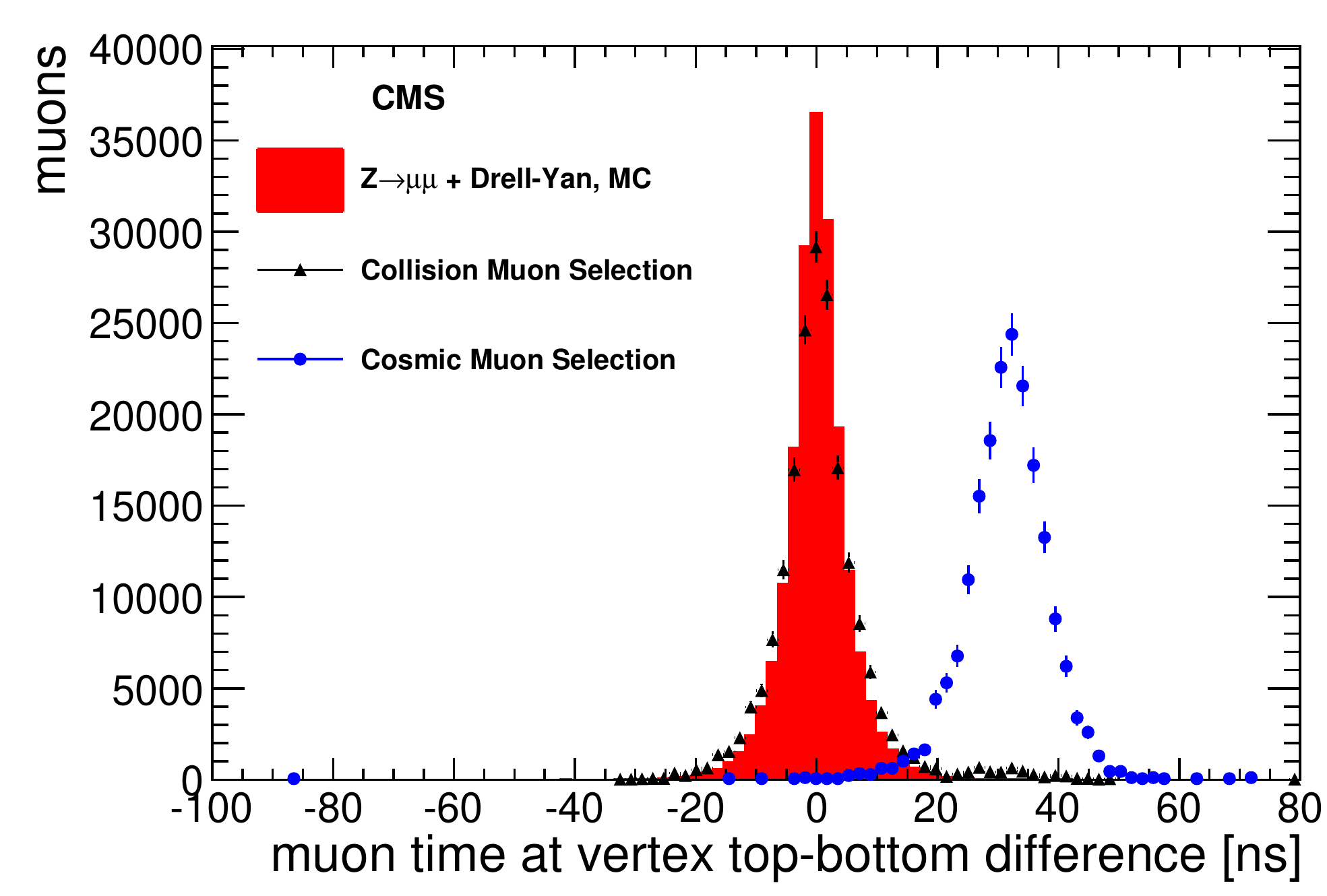}}
 \caption{Distributions of variables used for identification of cosmic-ray
 muons, shown for collision and cosmic-ray muon data samples described in the text, and for
 $\Z$+Drell--Yan MC samples: (a) muon transverse impact parameter with respect to the nominal
 beam-spot position; (b) $\log_{10}(\pi-\alpha)$ (see text); (c) muon timing;
 (d) timing difference between upper and lower muon legs.}
 \label{fig:cosmicVars}
\end{figure}

A few examples of variables that can be used to distinguish cosmic
muons from collision muons are shown in Fig.~\ref{fig:cosmicVars} for
cosmic-ray muons, muons from collision data, and muons in simulated
$\Z$ and Drell--Yan events.  Figure~\ref{fig:cosmicVars}(a) shows the
distributions of the transverse impact parameter calculated relative to
the beam-spot position $d_{\rm xy}$.  The $d_{\rm xy}$ distribution
for cosmic-ray muons is flat, while for collision muons it is strongly
peaked at zero. The typical requirement used in muon analyses to
suppress the cosmic-muon background is $|d_{\rm xy}| <0.2$~cm.
For illustration, the $\Z$+Drell--Yan MC prediction is shown as well, exhibiting
an even narrower transverse impact parameter peak than the inclusive
collision muon sample (cf. Fig.~\ref{fig:ID1})
and demonstrating that a tighter and more effective $d_{\rm xy}$ requirement
can be applied in specific cases.

If a cosmic muon passes close enough to the centre of the tracker,
it is expected to be reconstructed
as two tracks (``legs'') in the upper and lower halves of CMS with
almost equal $\pt$ and opposite directions.  The distinct topology
of cosmic muons is well represented by the opening angle $\alpha$,
defined as the largest angle between the inner-tracker track of a
muon and any other track in the event.
The distributions of $\log_{10}(\pi-\alpha)$ for cosmic-ray
and collision muons are shown in Fig.~\ref{fig:cosmicVars}(b).  As expected,
for most cosmic muons a track opposite in direction could be found,
which is not the case for collision muons or muons from simulated
$\Z$+Drell--Yan samples.  The excess in the number
of collision muons observed at the negative tail of the distribution
over that predicted is due to a small admixture of cosmic-ray
muons present in the data.

Muon timing information can also be used to discriminate between
cosmic-ray muons and mu\-ons from collisions.
The timing distributions observed in data and MC samples are shown in
Fig.~\ref{fig:cosmicVars}(c).
The muon timing is defined as the time at which a
muon would pass the interaction point relative to the time of the
bunch crossing.  It is calculated from timing
information available in the muon system under the assumption
that a particle is moving at the speed of light from the centre of the CMS
detector outward.
Therefore, for collision muons
one should expect a distribution peaked around zero with a width determined by the
resolution of the time measurement.
For cosmic-ray muons the actual arrival time distribution is flat but the reconstructed distribution is also
centred around zero. This is due to the trigger
and inner-tracker reconstruction being most efficient for in-time particles.
The additional peak visible at negative time
is due to the upper reconstructed legs in events where the lower leg is the source of the trigger.
In events where both legs are reconstructed, the timing difference between the upper and lower legs
is a good discriminator between dimuons from collisions and cosmic-ray muons,
as can be seen in Fig.~\ref{fig:cosmicVars}(d).

Cosmic-ray muon backgrounds must be removed from all physics
analyses using muons.  For the vast majority of analyses, a simple
impact-parameter requirement $|d_{\rm xy}| < 0.2$~cm, which suppresses
the contamination from cosmic muons in a sample of Tight Muons with
$\pt > 10$\GeVc to below 10$^{-4}$, is sufficient.  Some
analyses, in particular searches for new phenomena involving
very energetic muons, need to employ additional cosmic-ray background
rejection criteria because the proportion of cosmic muons to collision
muons increases as the muon momentum increases.
For example, in the search for extra charged gauge bosons
(\ensuremath{\mathrm{W'}}) decaying to a muon and a
neutrino~\cite{wprime},
the standard impact-parameter cutoff at 0.2~cm was replaced with
a tight cutoff of 0.02~cm.
In the search for
extra neutral gauge bosons ($\cPZpr$) decaying to muon
pairs~\cite{zprime}, the impact-parameter requirement was complemented
with a selection on the opening angle, ($\pi - \alpha$) $>$ 0.02 rad.
As the amount of accumulated data increases,
the reach of physics analyses is extending to higher values of dimuon mass
and muon momentum; since the spectrum of muons predicted
by the Standard Model falls more rapidly with increasing $\pt$ than
the cosmic-muon spectrum, more sophisticated methods
may become necessary
to suppress cosmic-ray muon background to the required level in those kinematic regions.
A cosmic-muon identification algorithm has been developed that
combines several quantities sensitive to cosmic muons in order to
maximize the cosmic-ray muon selection efficiency and minimize
misidentification of muons from pp collisions as cosmic-ray muons.
This algorithm quantifies the compatibility
of any given muon with the hypothesis that it is a cosmic-ray muon;
the quantities used are the ones described
earlier in this section plus a few others characterizing the event
activity and whether there are hits in the opposite (upper or lower)
half of CMS which could be associated to a muon track by the dedicated
cosmic-muon reconstruction algorithm.

Table \ref{tab:perfTable1} shows the performance of cosmic-ray muon identification with some typical
individual selections and that of the cosmic-muon identification algorithm.  The
efficiency is defined as the fraction of events identified as cosmic-ray
muons by each algorithm in the sample of cosmic-ray
muons selected from collision data, as discussed above.
The misidentification fraction is the
fraction of simulated $\Z$+Drell--Yan events identified as cosmic-ray muons.
The first row shows the performance of the
back-to-back identification, which requires that the
opening angle $\alpha > (\pi - 0.1)$~rad and that
$|p_\mathrm{T1} - p_\mathrm{T2}|/\sqrt{p_\mathrm{T1} \cdot p_\mathrm{T2}} < 0.1$.
The second row shows the effect of applying an impact-parameter requirement:
muons with $|d_{\rm xy}|>0.2$~cm are flagged as cosmic-ray muons.
This requirement is quite efficient in identifying cosmic-ray muons,
while maintaining a low misidentification probability for collision muons.
In the third and fourth rows, we show the performance of the cosmic-muon
identification algorithm. This algorithm allows the analyzer to tune the
identification efficiency and misidentification fraction according to the needs of the
analysis.  The tight selection minimizes the misidentification at the
cost of some loss of efficiency in cosmic-ray muon rejection.
The loose selection maximizes cosmic-ray muon rejection efficiency at
the cost of some higher misidentification.

\begin{table}%[htpb]
\begin{center}
\topcaption{The efficiency of cosmic-ray muon identification and misidentification
fractions for some relevant individual quantities and the loose and tight
versions of the cosmic-muon identification algorithm.  Uncertainties are
statistical only.}
\begin{tabular}{|c|c|c|}
\hline
 & Efficiency ($\%$) & Misidentification ($\%$)\\ \hline
\hline
Back-to-back      & $89.97 \pm 0.13$  & $0.153  \pm 0.002$ \\
\hline
$|d_{\rm xy}|>0.2$ cm   & $99.05 \pm 0.04$ & $0.0045 \pm 0.0003$ \\
\hline
Cosmic ID tight & $97.58 \pm 0.07$ & $0.0001 \pm 0.0001$ \\
\hline
Cosmic ID loose & $99.52 \pm 0.03$ & $0.153  \pm 0.002$ \\
\hline
\end{tabular}
\label{tab:perfTable1}
\end{center}
\end{table}

\subsection{Beam-halo muons}
Accelerator-induced backgrounds (``beam halo'') contribute
to a variety of physics analyses. The CMS detector can identify
beam-halo muons that overlap with collision events and might otherwise
be considered as part of the collision itself. The CSC system has
extensive geometrical coverage in the plane transverse to the beamline
and hence can detect halo muons traveling parallel to the beamline as
well as muons originating from collisions.  An algorithm has been
developed to identify beam-halo muons using information from the CSCs
available at both the trigger and the reconstruction level.  The
following three categories of information have proved effective: 1) a
dedicated Level-1 beam-halo trigger (based on patterns of hits in the
CSC system consistent with a charged particle traveling parallel to
the beamline); 2) CSC trigger primitives (patterns of hits within an
individual CSC consistent with a traversing charged particle) that are early
in time relative to the actual pp collision; and 3) a reconstructed
standalone-muon track with a trajectory parallel to the
beamline. Requiring any one of these conditions leads to excellent
beam-halo muon identification efficiency. We refer to this as
\emph{Loose Halo ID}. Requiring any two of these conditions ensures
fewer collision muons are spuriously accepted as halo, but at the
expense of slightly lower efficiency for true halo muons. This is
referred to as \emph{Tight Halo ID}.

The performance of the beam-halo identification algorithms was studied with 2010 collision data and Monte Carlo samples.   Simulated beam-halo muons that
passed through the barrel or endcap calorimeters were found to satisfy the requirements of the Loose Halo ID 96\% of the time, while satisfying the requirements of the Tight Halo ID
65\% of the time.  To measure these efficiencies in collision data we used $\MET$-triggered events.  Beam-halo muons can induce a large missing energy signal by
traversing the barrel or endcap calorimeters at constant $\phi$, and
thus $\MET$-triggered data provide an enriched sample of beam-halo muons.
A pure sample of beam-halo muons was selected
by requiring no reconstructed collision muon in an event and $\MET > 50\GeV$
opposite in $\phi$ to at least one reconstructed hit in the CSCs.
The fraction of muons in the selected sample identified as beam halo
was found to be 89\% for the Loose and 73\% for the Tight Halo ID.
Given the number of simplifying assumptions involved in the simulation of
beam-halo events, the agreement between the data and the simulation is
considered to be satisfactory.

The fraction of simulated minimum-bias events in which a particle produced
in a proton--proton collision was misidentified as a beam-halo muon was
$\approx$$5 \times 10^{-5}$ and $\approx$10$^{-7}$ for the Loose and Tight Halo ID
algorithms, respectively.   When evaluated from minimum-bias collision data, these misidentification probabilities were found to be somewhat larger but still small, with
values of $\approx$$2 \times 10^{-4}$ and $\approx$$8 \times 10^{-7}$, respectively.

The probability for a beam-halo muon crossing the CSCs to be misidentified
as a collision muon was also evaluated in data and found to be
$\approx$$10^{-4}$ for Soft Muons with $\pt < 5\GeVc$ and $\lesssim 5 \times 10^{-5}$ for Soft Muons with higher $\pt$ and for Tight Muons.
Another important number is the fraction of collision
events contaminated by beam-halo muons.  In 2010, it
ranged from $\approx$$10^{-5}$ to $\approx$$10^{-3}$, depending on the LHC fill,
for muon-triggered events~\cite{JME_10_009}; the fraction could be
higher by an order of magnitude for hadronic-triggered events, due to the beam-halo muons' role in triggering the event via energy deposited in the calorimeters.  Several searches for new physics in
CMS~\cite{StoppedGluinos,SUSYRA2} employed the beam-halo identification algorithms described above to veto events with beam-halo contamination.

\section{Isolation}
\label{sec:isolation}

The requirement that a muon is an isolated particle in the event,
meaning that the energy flow in its vicinity is below a certain
threshold, can effectively discriminate muons from the decays of
$\W$ and $\Z$ bosons from those produced in heavy-flavor decays and hadron
decays in flight.  Three different isolation algorithms have been studied: 
\begin{itemize}
\item {\em Tracker relative isolation ($\IRelTrk$).} This algorithm calculates
the scalar sum of the $\pt$ of all
tracker tracks reconstructed
in a cone of radius $\Delta R
\equiv \sqrt{(\Delta \varphi)^2+(\Delta \eta)^2} < 0.3$
centred on the muon track direction.
The $\pt$ of the muon track itself is not included in the sum.
For the muon to be considered isolated, the ratio of the $\pt$ sum
to the muon track $\pt$ is required to be below a certain threshold.
Track directions and values of $\pt$ are
computed at the point of closest approach to the
nominal centre of the detector. 
\item {\em Tracker-plus-calorimeters (combined) relative isolation ($\IRelComb$).}
The discriminating variable is similar to $\IRelTrk$, but the numerator
of the ratio also includes the sum of energies measured in ECAL and HCAL
towers found within a cone of radius $\Delta R < 0.3$ centred on the
muon track direction.  The energy deposits associated with the muon
track itself are not included in the sum.
\item {\em Particle-flow relative isolation ($\IPF$).} The discriminating
variable is the sum of the $\pt$ of all charged
hadrons, the transverse energies $\ET$ of all photons, and $\ET$ of
all neutral hadrons reconstructed by the particle-flow
algorithm~\cite{PFT-09-001} within a cone of radius $\Delta R < 0.4$
centred on the muon track direction, divided by the muon track~$\pt$. 
\end{itemize}
The optimization of the cone sizes was performed independently for the 
different algorithms, resulting in the different $\Delta R$ values mentioned
above.  Each of these algorithms has features that suit the requirements
of different analyses. For example, the tracker-plus-calori\-meters relative  
isolation with a threshold of 0.15 is the algorithm chosen for the
measurement of the $\W$ and $\Z$ cross sections~\cite{EWK-10-002}. 
The search for heavy resonances decaying into muon pairs~\cite{zprime} 
instead used the tracker relative isolation given that high-energy
muons are expected to deposit a significant amount of energy in the
calorimeters. % and therefore
Photons emitted in final-state radiation can lead to energy deposited in the
ECAL only. An isolation algorithm that 
uses energy deposits in the inner tracker and the HCAL only
is not considered in this paper. 
Analyses involving $\tau$ leptons~\cite{EWK-10-013, HIG-10-002}
use particle-flow isolation for the
muons from the $\tau$ decays, since particle flow is used for the 
global event reconstruction in these analyses.

The efficiency of the various isolation algorithms was measured in
data, using a sample of muons from $\Z$ decays.  Two methods were used:
tag-and-probe and the Lepton Kinematic Template (LKT) method. 
The tag-and-probe method is described
in Section~\ref{sec:muonideff_tnp}.
The LKT method~\cite{MUO-10-002} is an extension of the
random-cone~\cite{CMS_NOTE_2006-033} method. 
It relies on the assumption that
the kinematics of muons from decays of $\W$ or $\Z$ bosons
produced in the hard parton scattering is unrelated to accompanying
interactions of the other partons in the colliding protons
(the underlying event), which are responsible for the energy flow around
the muons.  An isolation variable can then be computed relative to
any specific direction in an event with underlying event activity   
similar to that of a signal event.
Events containing a $\Z$ decaying into a pair of muons, with the reconstructed
muon tracks and activity associated with them discarded,
were used as approximations to underlying events.
Directions were drawn from template kinematic
distributions of muons obtained from simulation.
As every event can be re-used multiple times, the LKT method
provides an important cross-check for the results obtained with the
tag-and-probe technique, which are statistically limited in some
kinematic regions.

An almost pure sample of $\Zmm$ events was obtained by
selecting events with a pair of oppositely charged muons that 
form an invariant mass in a mass region around the nominal $\Z$ mass
(between 70 and 110\GeVcc for the tag-and-probe method and within
10\GeVcc of the nominal $\Z$ mass for the LKT method)
and with each muon satisfying the Tight
Muon identification criteria.  For the tag-and-probe method, the tag muon
was also required to be isolated with $\IRelComb < 0.15$.
A sample of simulated $\Zmm$ and Drell--Yan dimuon events passing the same
selection criteria was used for comparisons with the data, with
simulated minimum-bias collision events overlaid to
reproduce the observed pile-up (PU) distribution.

\begin{figure}[htbp]
  \centering
\includegraphics[height=0.45\textwidth,angle=0]{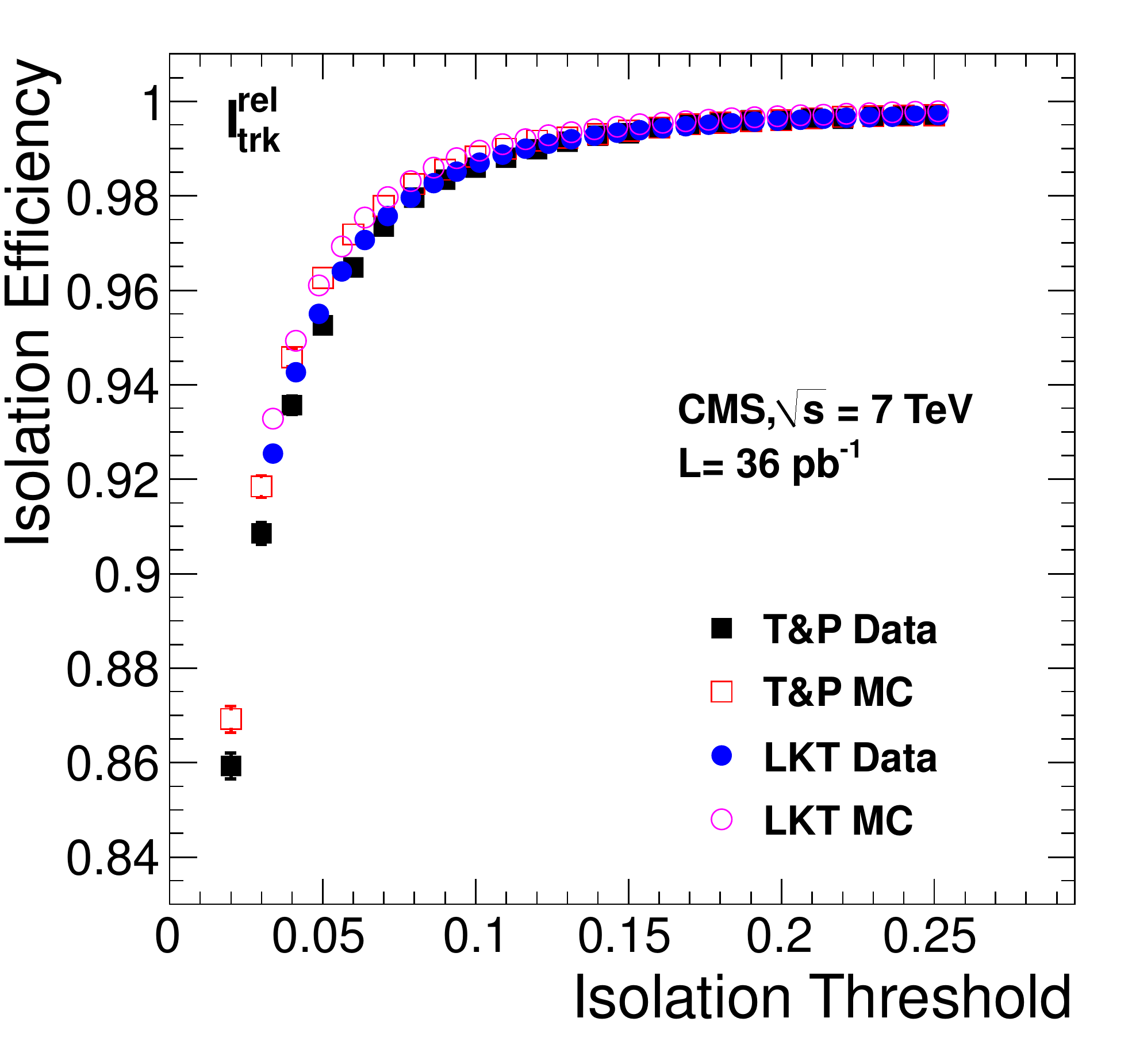}
\includegraphics[height=0.45\textwidth,angle=0]{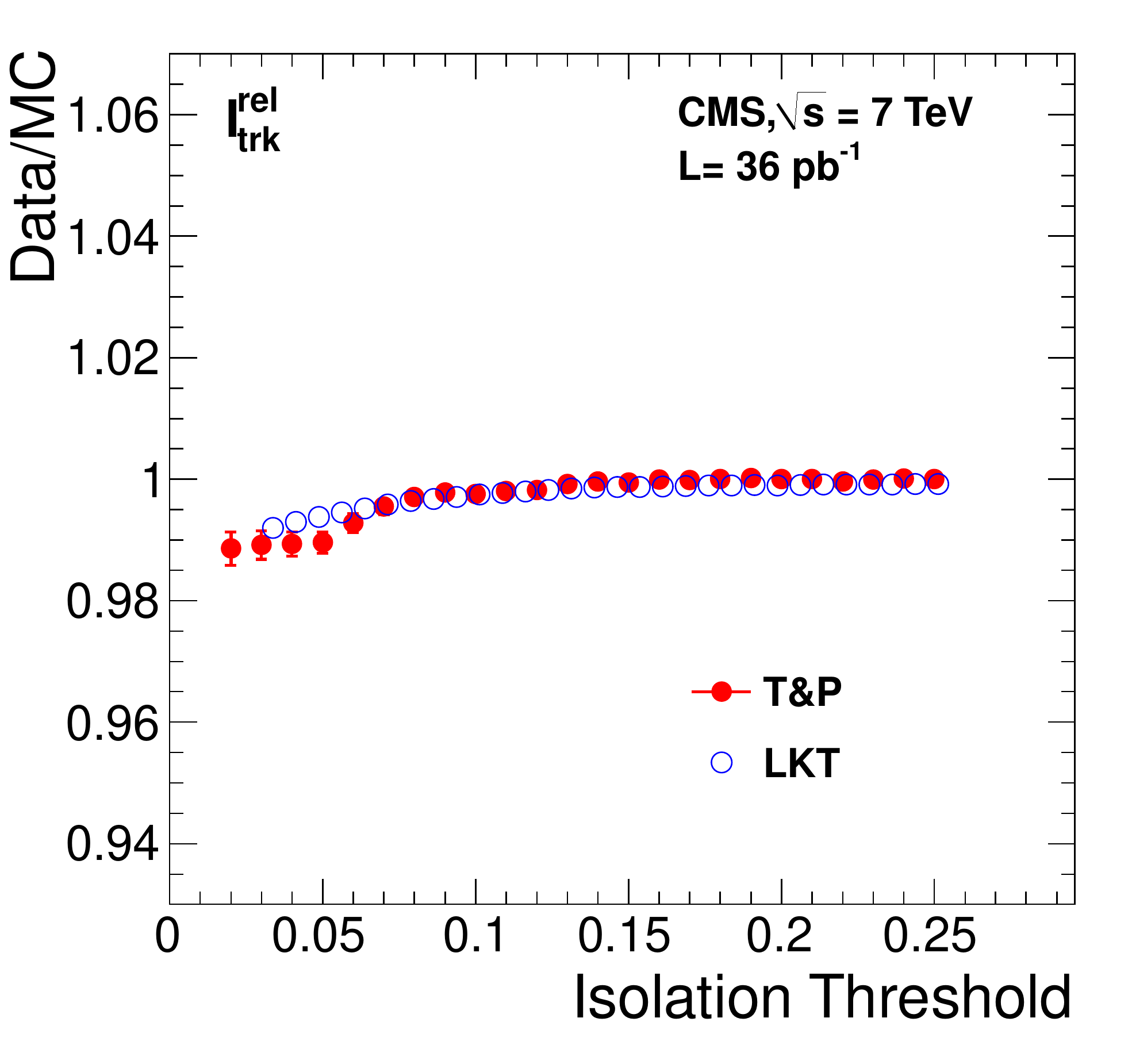}
\includegraphics[height=0.45\textwidth,angle=0]{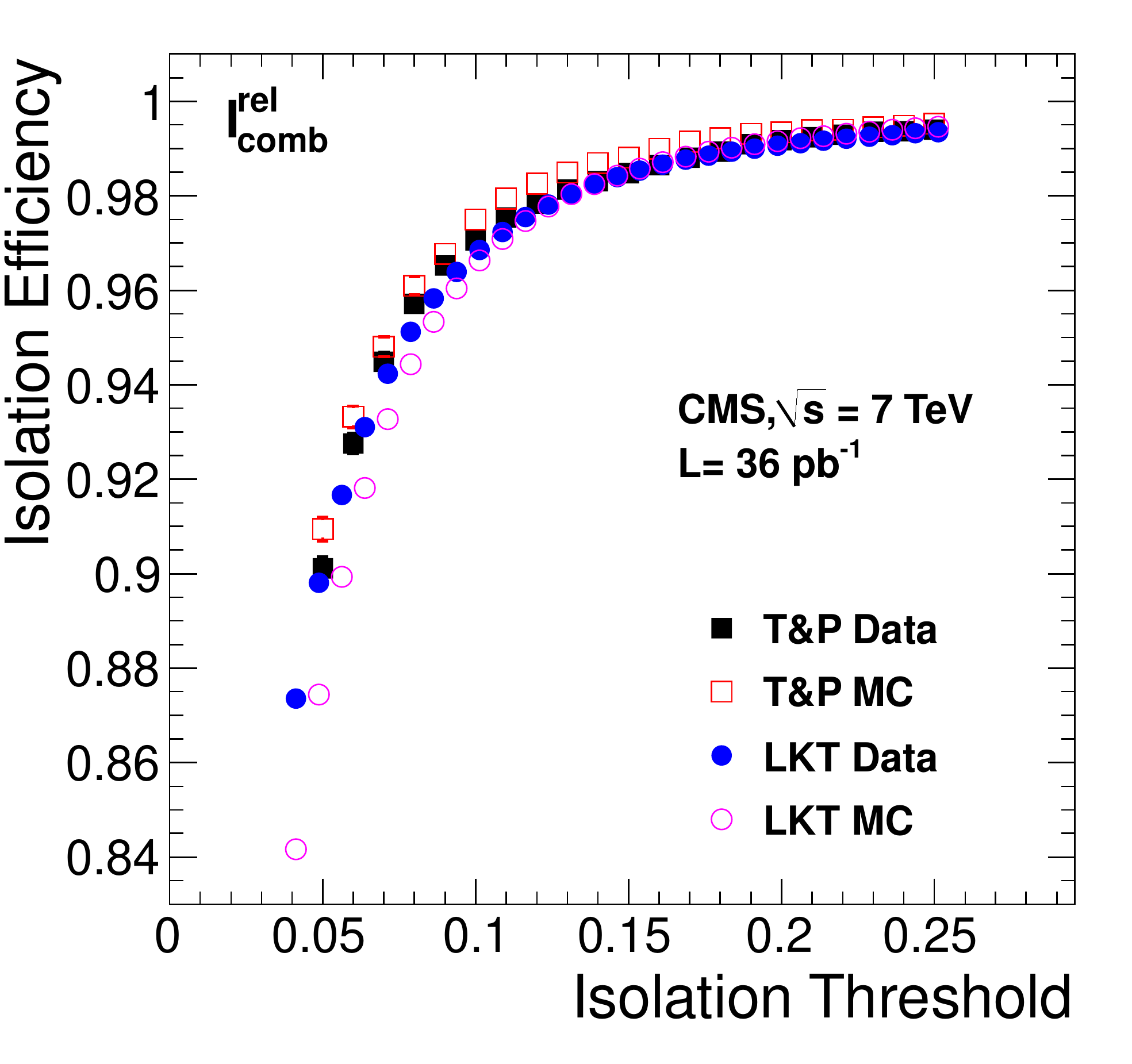}
\includegraphics[height=0.45\textwidth,angle=0]{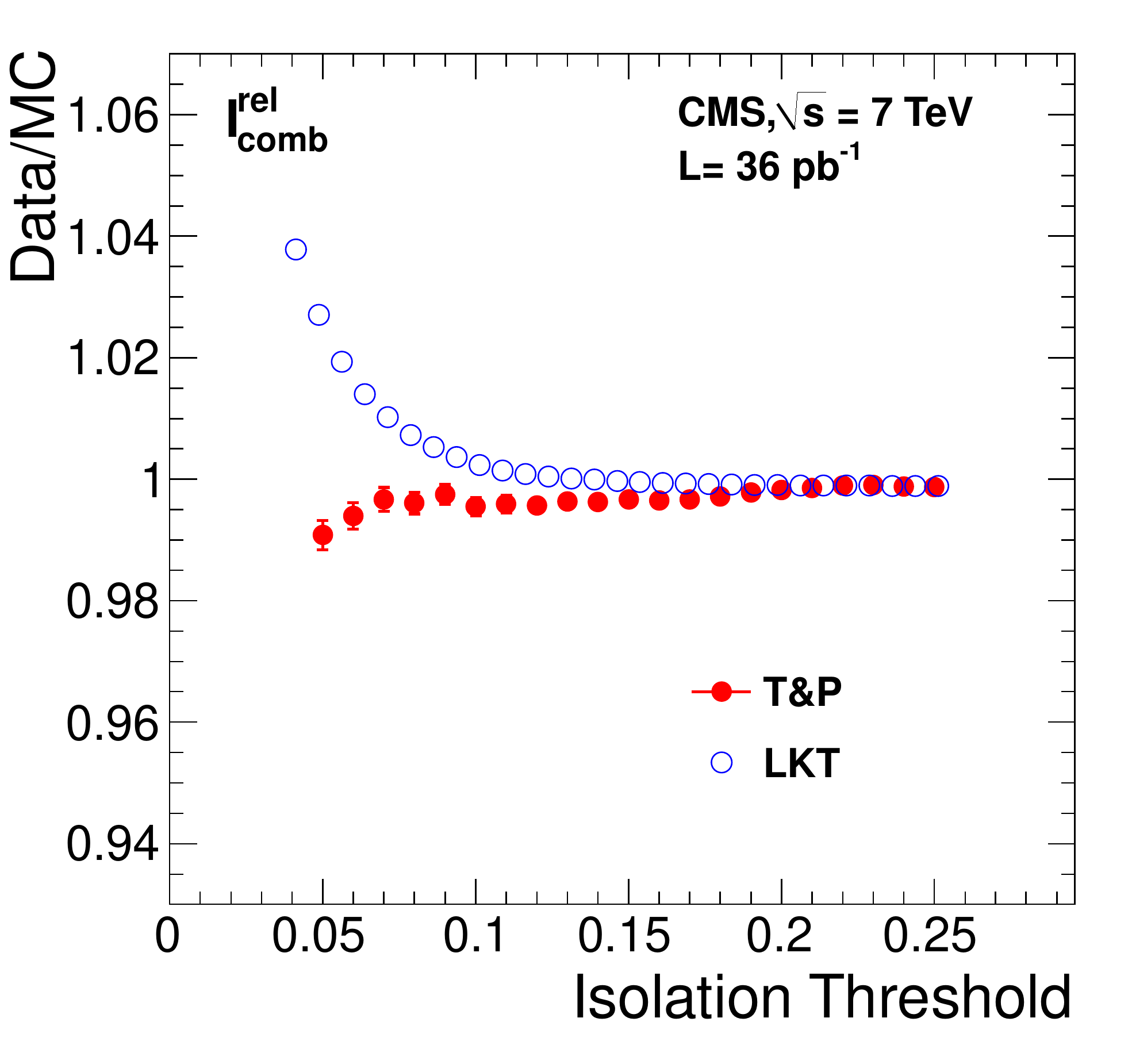}
\includegraphics[height=0.45\textwidth,angle=0]{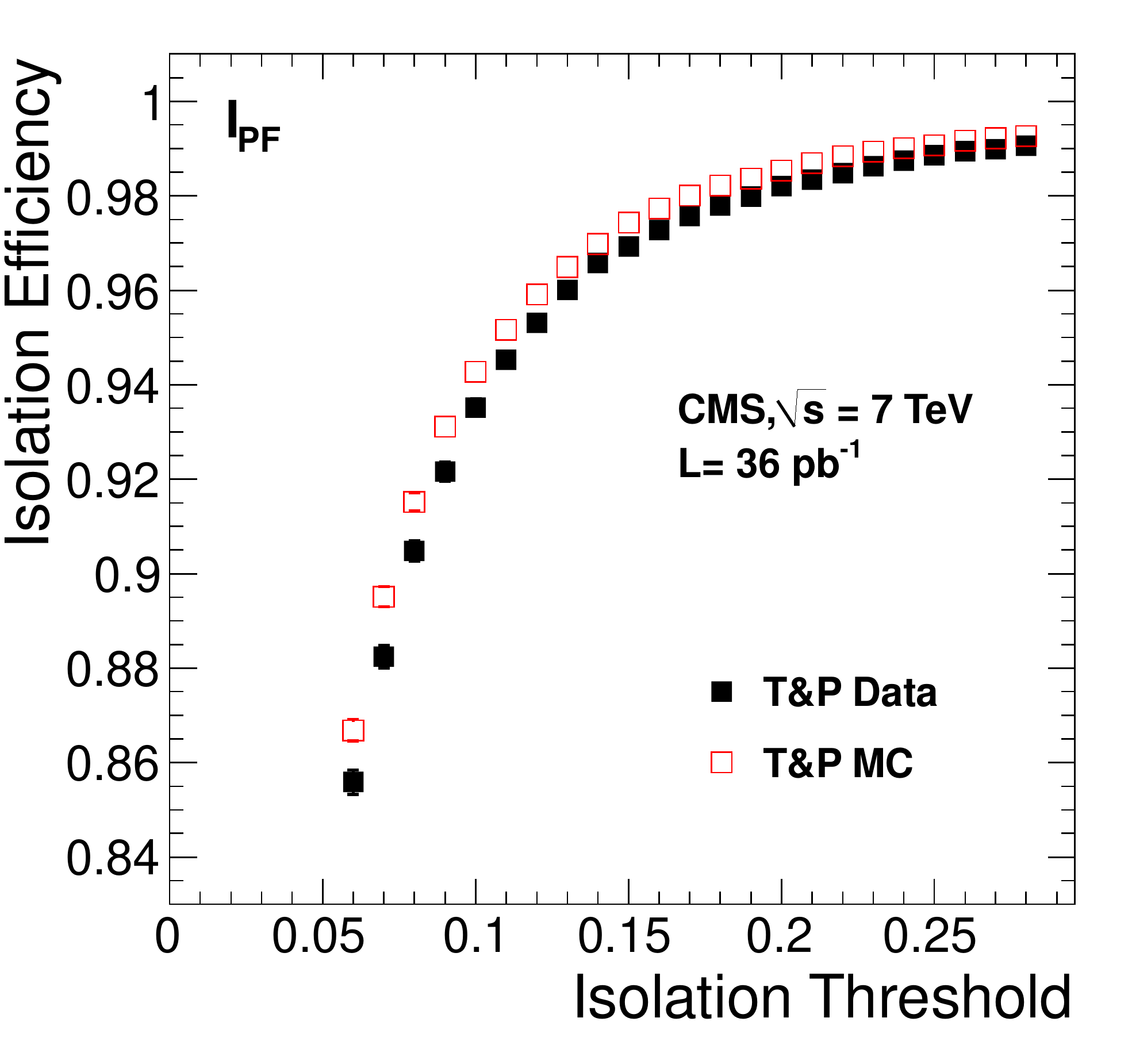}
\includegraphics[height=0.45\textwidth,angle=0]{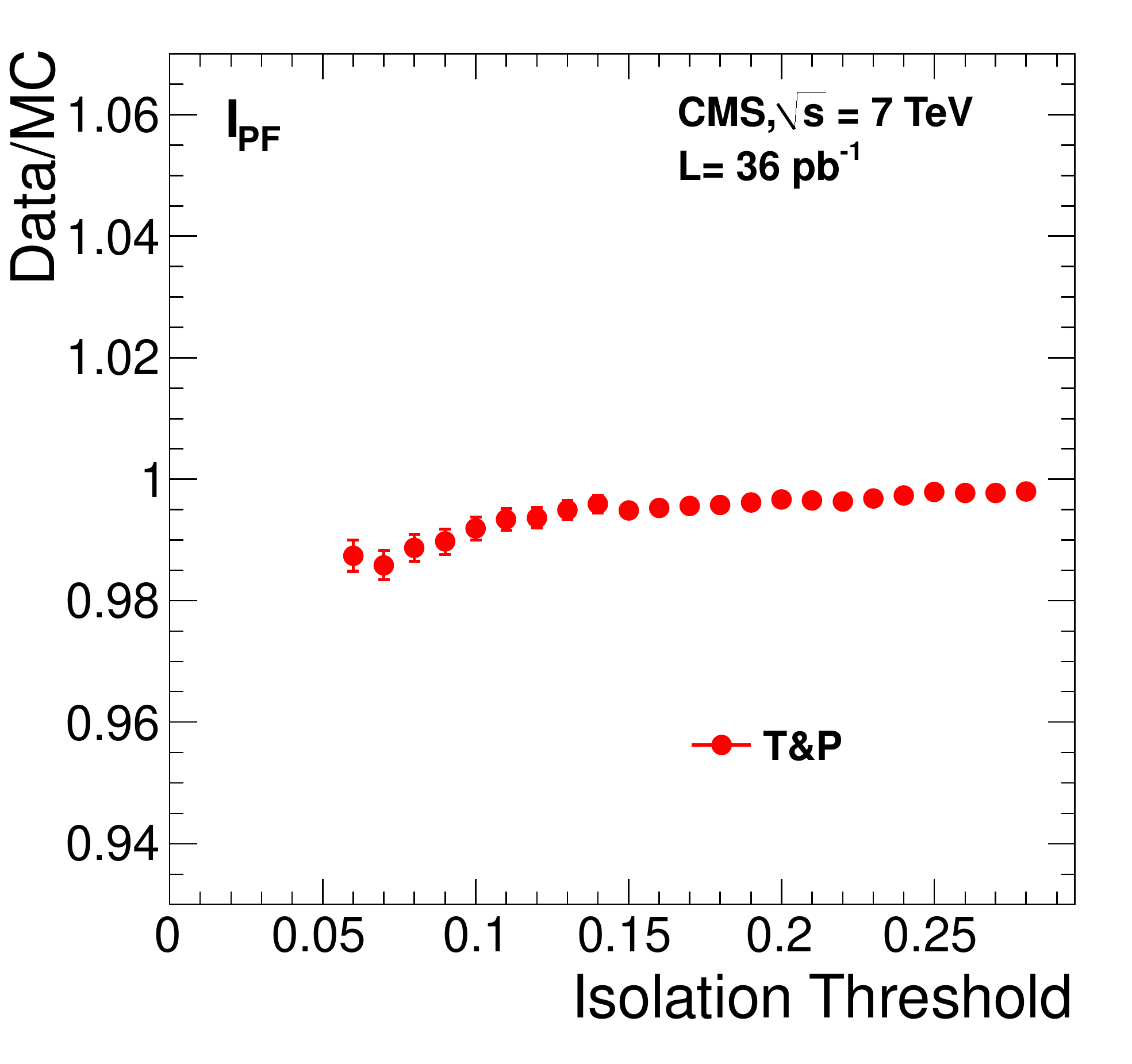}
  \caption{
Left: efficiencies of various isolation algorithms for muons with
$20<\pt<50\GeVc$ from $\Z$ decays as a function of the isolation threshold.
Results are shown for both data and simulation using the tag-and-probe
(``T\&P'') and Lepton Kinematic Template (``LKT'')
methods; the LKT method is not used for the
particle-flow algorithm.  Right: data to simulation ratios.  Plots are shown
for tracker relative ($\IRelTrk$, top), tracker-plus-calorimeters
relative ($\IRelComb$, middle), and particle-flow relative ($\IPF$, bottom)
isolation algorithms. The MC samples include simulation of pile-up events.} 
  \label{fig:Isolation_1}
\end{figure}

Figure~\ref{fig:Isolation_1} shows the efficiency of the various
isolation algorithms evaluated on muons with $20<\pt<50\GeVc$ 
from $\Z$ decays as a function of the threshold on the corresponding isolation
variable. Results obtained with the tag-and-probe (for all three isolation
algorithms) and the LKT (for $\IRelTrk$ and $\IRelComb$) methods
are shown for both data and simulation.
Efficiencies in data and MC simulation are generally in good agreement, with
the difference between the two not exceeding 1.5\%. A single exception
is the results from the LKT method 
for low values of the $\IRelComb$
threshold, where efficiencies from MC simulation are lower than those from 
data by up to 4\%.  The origin of this discrepancy is under study.
The ratios of efficiencies obtained on data and on
simulation are also reported in Fig.~\ref{fig:Isolation_1} and can be
considered as scale factors to correct MC results.
The agreement between the tag-and-probe and LKT results on data 
is within 1\%.

\begin{figure}[htb!]
  \centering
\includegraphics[height=0.45\textwidth,angle=0]{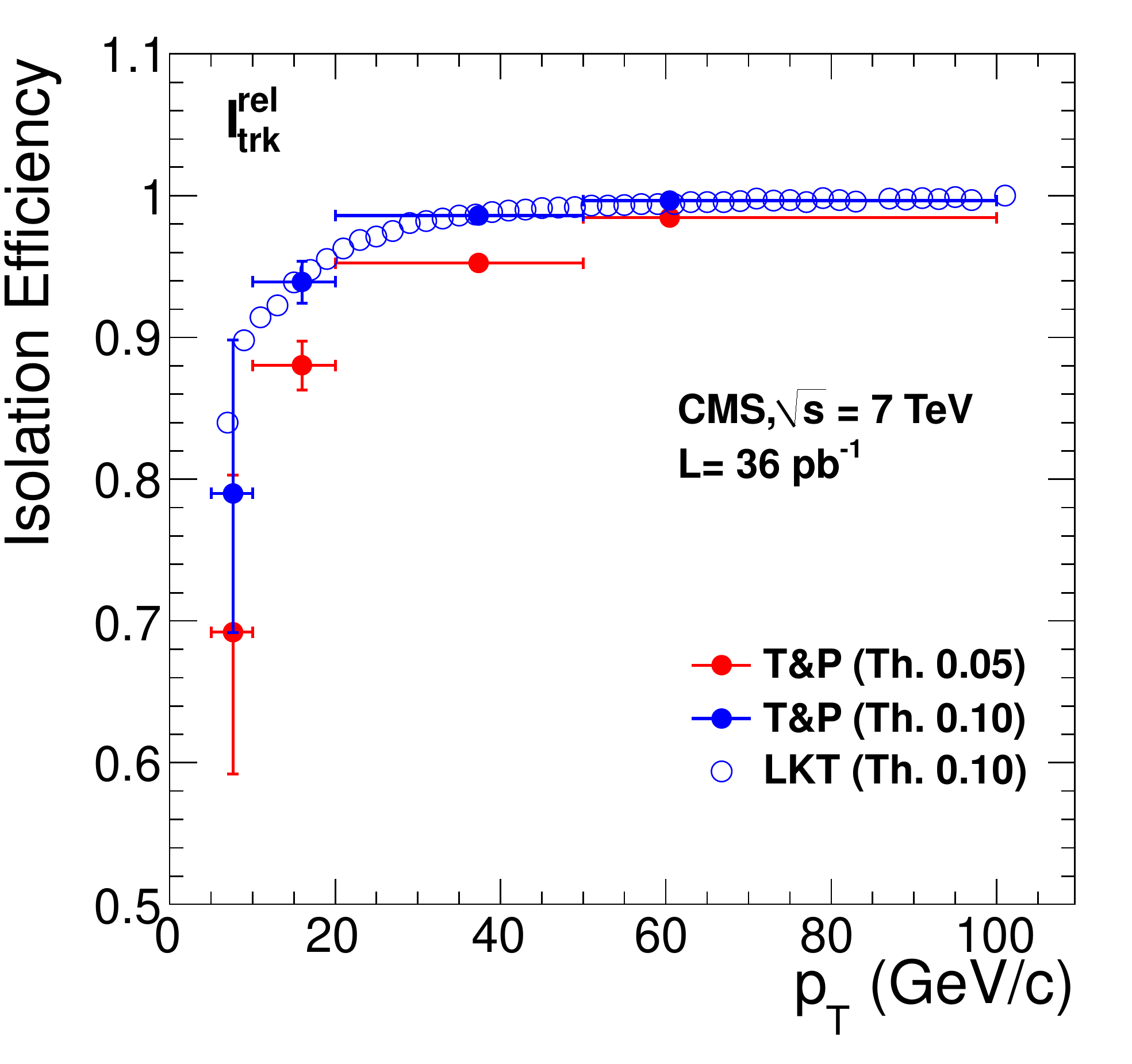}
\includegraphics[height=0.45\textwidth,angle=0]{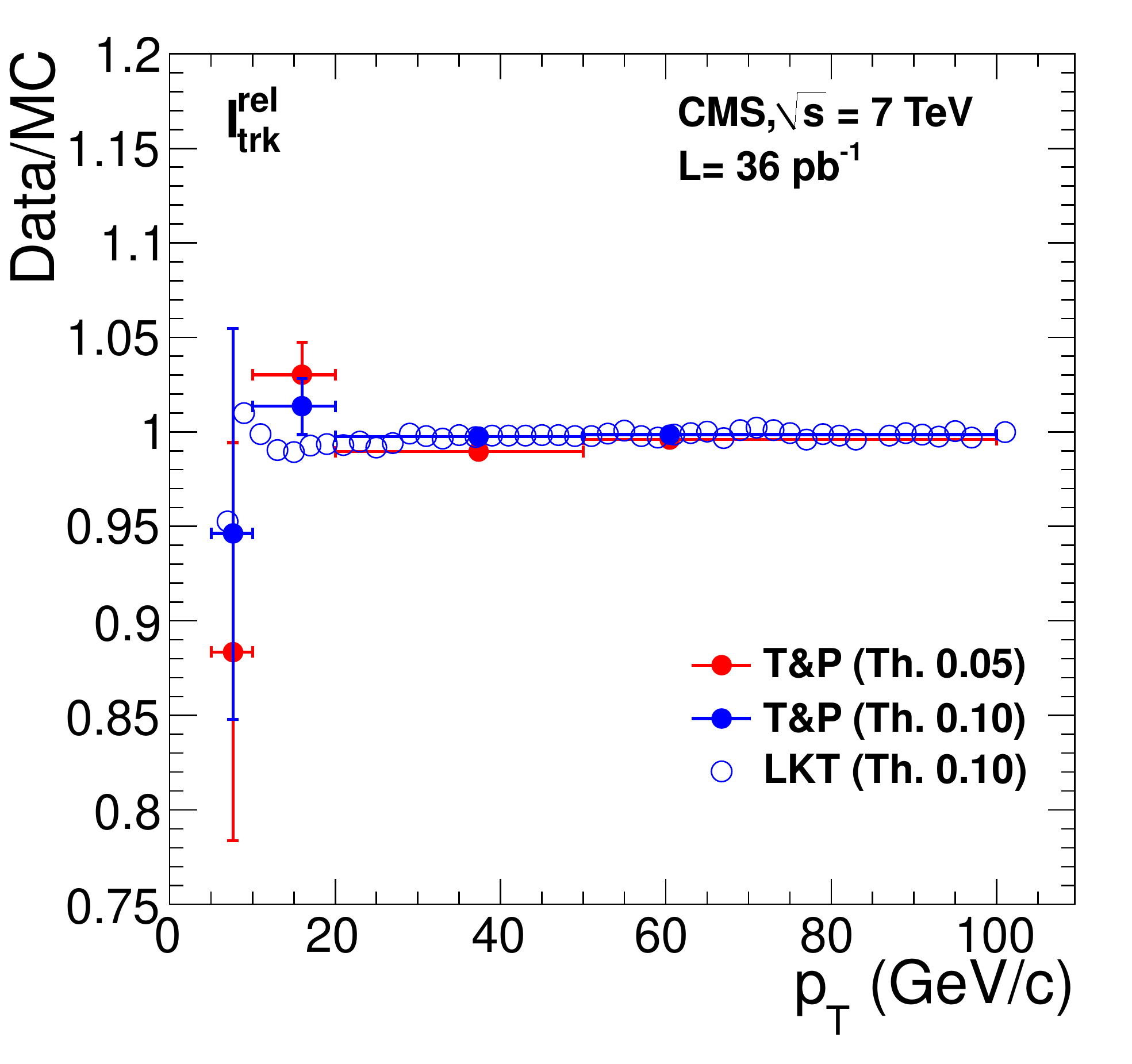}
  \caption{
Left: efficiency of tracker relative isolation $\IRelTrk$
for muons from $\Z$ decays as a function of muon $\pt$, measured
with data using the tag-and-probe (``T\&P'') method (in bins of
[5, 10], [10, 20], [20, 50], and [50, 100] $\GeVc$) and the LKT method
(finer binning).  Results corresponding to the threshold values of
0.05 and 0.1 are shown. Right: ratio between data and MC efficiencies. The
MC samples include simulation of pile-up events.
}
  \label{fig:Isolation_21}
\end{figure}

\begin{figure}[htb!]
  \centering
\includegraphics[height=0.45\textwidth,angle=0]{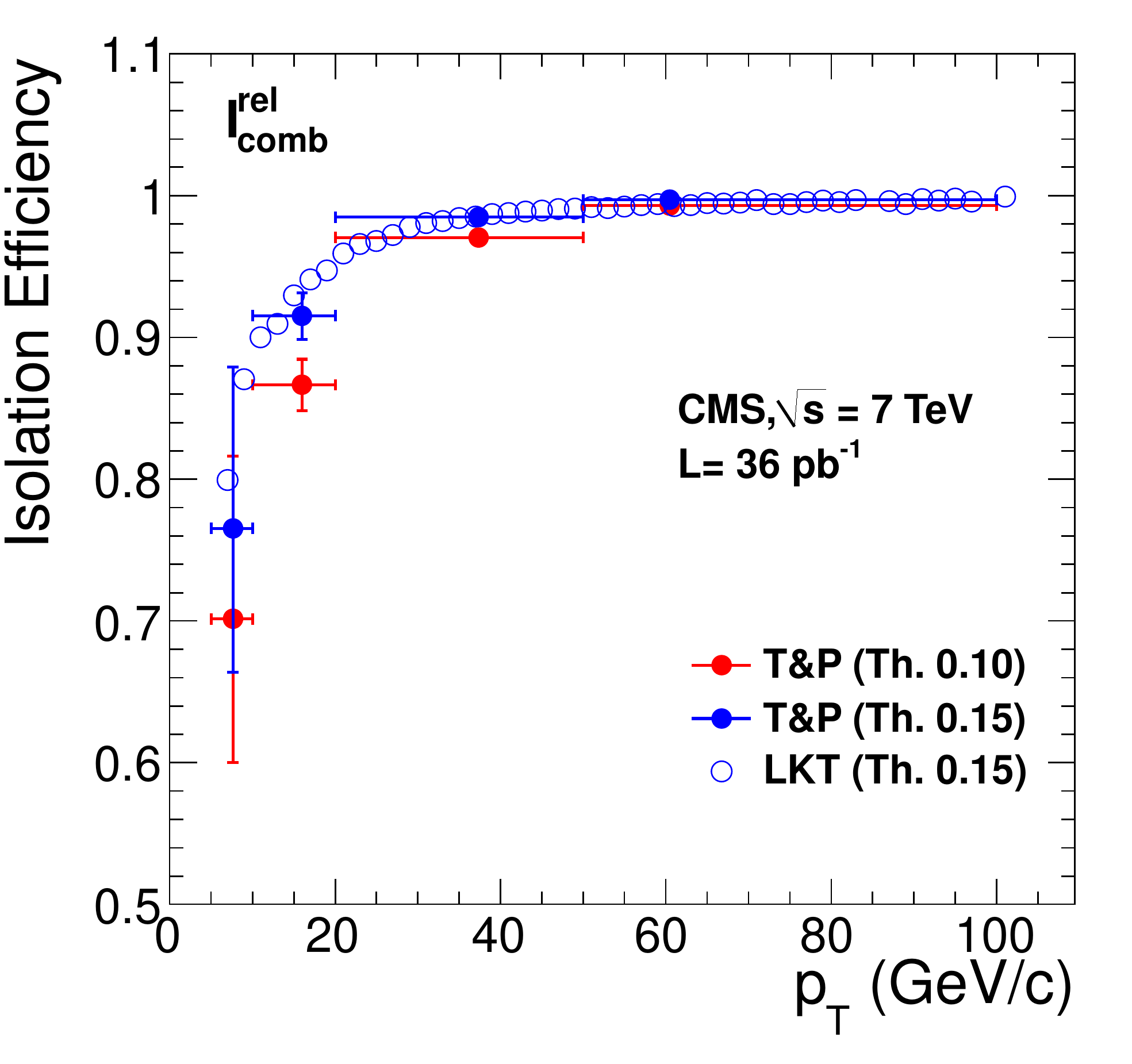}
\includegraphics[height=0.45\textwidth,angle=0]{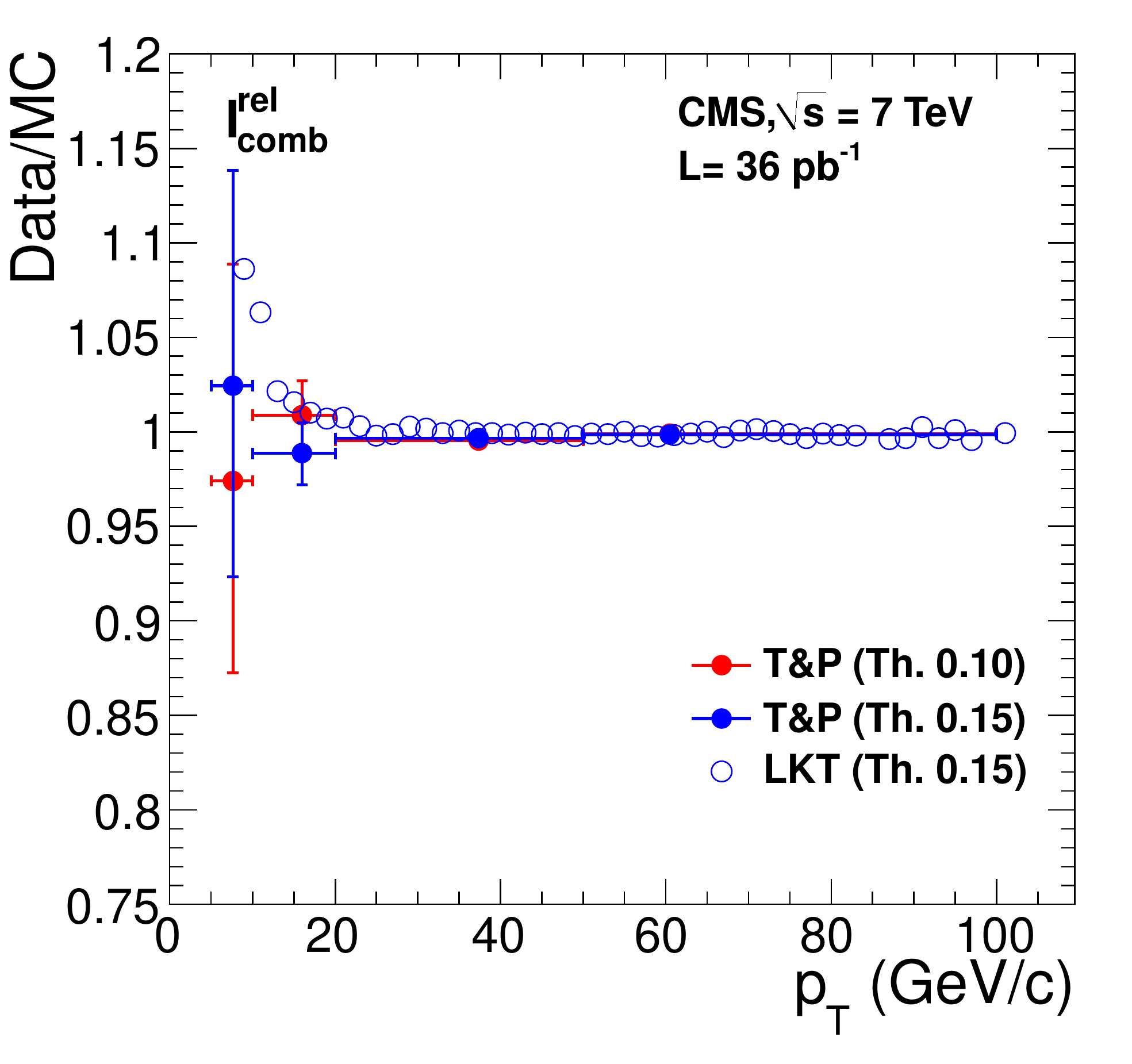}
  \caption{
Left: efficiency of tracker-plus-calorimeters relative isolation $\IRelComb$
for muons from $\Z$ decays as a function of muon $\pt$, measured
with data using the tag-and-probe (``T\&P'') method (in bins of
[5, 10], [10, 20], [20, 50], and [50, 100] $\GeVc$) and the LKT method (finer
binning).  Results corresponding to the threshold values of 0.10 and
0.15 are shown. Right: ratio between data and MC efficiencies. The MC samples
include simulation of pile-up events.  
}
  \label{fig:Isolation_22}
\end{figure}

\begin{figure}[htb!]
  \centering
\includegraphics[height=0.45\textwidth,angle=0]{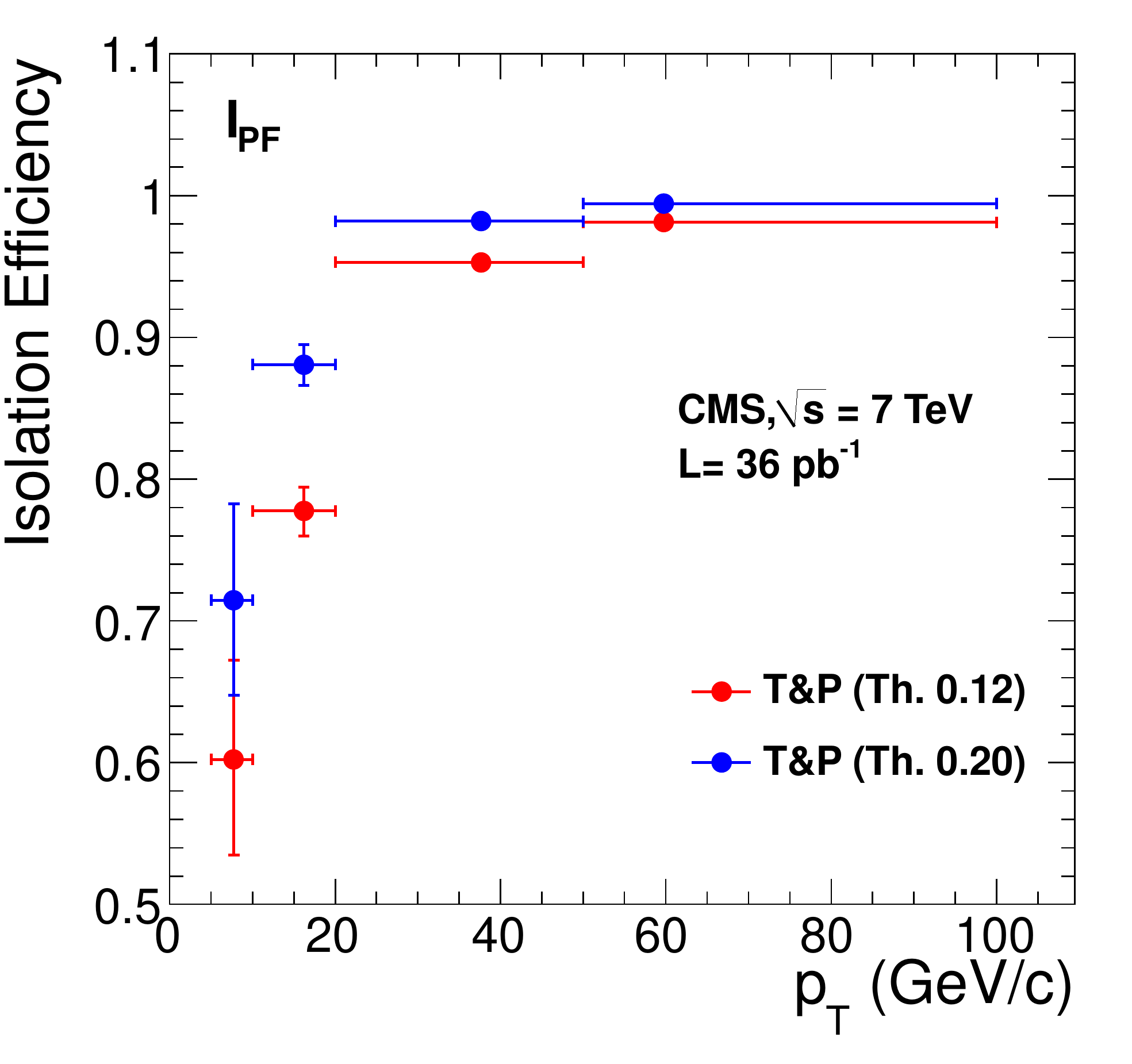}
\includegraphics[height=0.45\textwidth,angle=0]{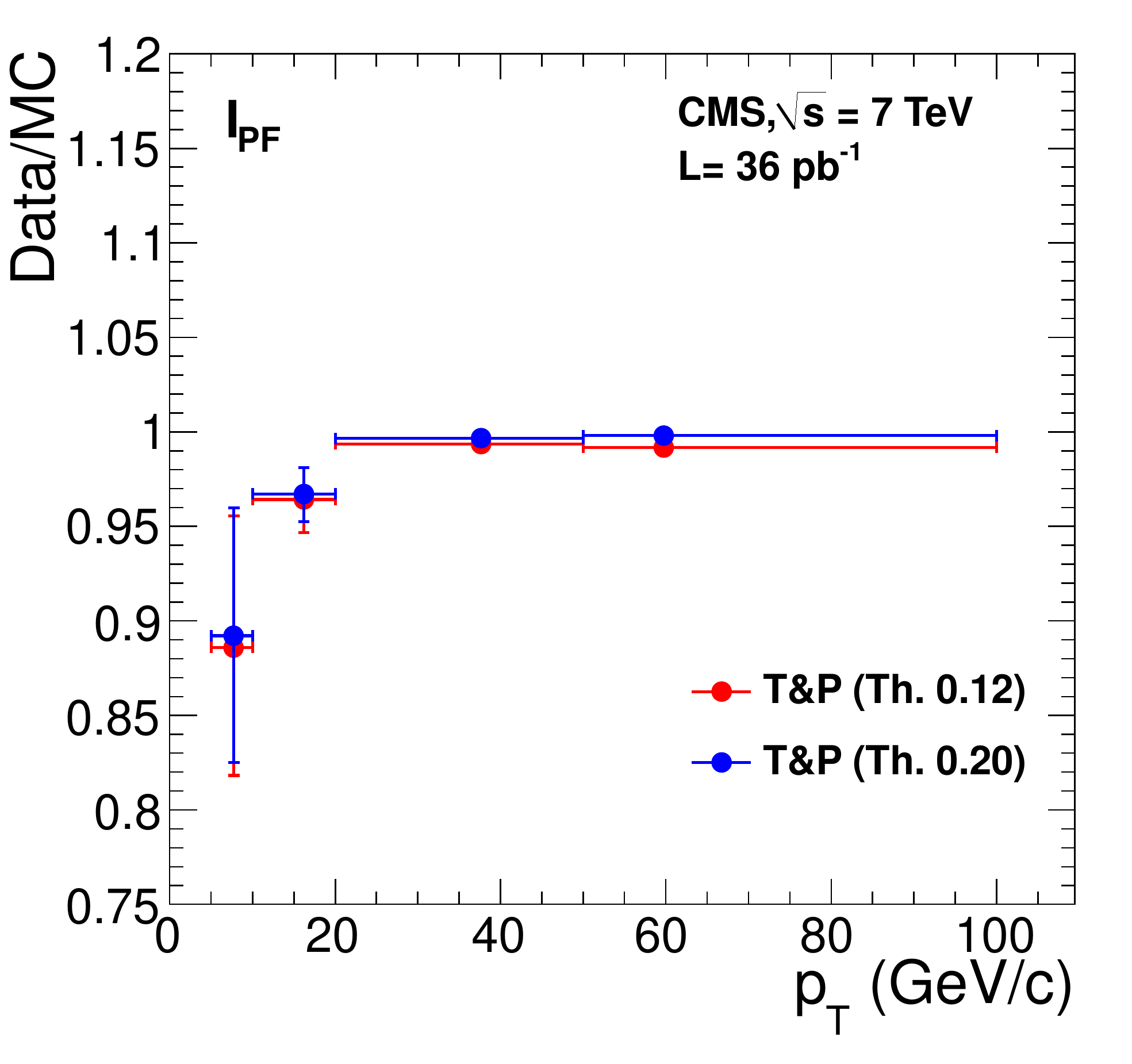}
  \caption{
Left: efficiency of particle-flow relative isolation $\IPF$
for muons from $\Z$ decays as a function of muon $\pt$, measured
with data using the tag-and-probe (``T\&P'') method.
Results corresponding to the threshold values of
0.12 and 0.20 are shown. Right: ratio between data and MC efficiencies. The
MC samples include simulation of pile-up events. 
}
  \label{fig:Isolation_23}
\end{figure}

The efficiency 
of the various isolation algorithms for muons from
$\Z$ decays is shown as a function of the muon $\pt$ and for
two threshold values in Fig.~\ref{fig:Isolation_21} for $\IRelTrk$,
Fig.~\ref{fig:Isolation_22} for $\IRelComb$,
and in Fig.~\ref{fig:Isolation_23} for $\IPF$.
In each of the three figures the ratio between data and MC results
is also reported. The two threshold values studied for each isolation
algorithm are those most commonly used in physics analyses and
correspond approximately to the end of the rapid rise of the efficiency curve
and to the beginning of the efficiency
plateau shown in Fig.~\ref{fig:Isolation_1}.
The results obtained with the tag-and-probe and LKT methods agree within
the statistical uncertainties down to the lowest
tested muon $\pt$ (5\GeVc). 
The LKT results, which have very small statistical
uncertainties, indicate that for muon $\pt$ as low as 5\GeVc, the
agreement between data and MC efficiencies 
for $\IRelTrk$ ($\IRelComb$)
is within 5\% (10\%), while for
$\pt$ greater than 15\GeVc the agreement is
within 1\%.   

Another important characteristic of the performance of an isolation
algorithm is its power to reject muons from typical background events.
To estimate the rejection power, a sample enhanced with events from QCD
processes is used. These events are selected by requiring the following:
one offline reconstructed muon with $20<\pt<50\GeVc$
satisfying the Tight Muon requirements; no other Tight Muon reconstructed
in the event; at least one particle-flow jet
reconstructed with $\ET$ greater than 30\GeV; the azimuthal angle
$\Delta\phi_{\mu,\MET}$ between the muon and the missing transverse
energy directions
smaller than 1.5 rad; and the transverse mass, defined as
$M_{\rm T} = \sqrt{2\pt\MET(1-\cos(\Delta\phi_{\mu,\MET}))}$,
smaller than 20\GeVcc. The last two
requirements reduce the contamination from $\W$ events
significantly.
A sample of simulated events from QCD processes, $\W$ decays, $\Zmm$,
Drell--Yan, and $t\bar{t}$ production, mixed according to the corresponding
cross sections, is used to study the expected
composition of the data sample resulting from the above selection and for
comparing results obtained from data and MC simulations that
include pile-up observed in data. Before isolation requirements
are applied, events from QCD
processes represent 99.6\% of the MC sample. The next most important
contributions come from $\W$ (0.3\%) and $\Z$+Drell--Yan events (0.1\%).

\begin{figure}[htb!]
  \centering
\includegraphics[height=0.45\textwidth,angle=0]{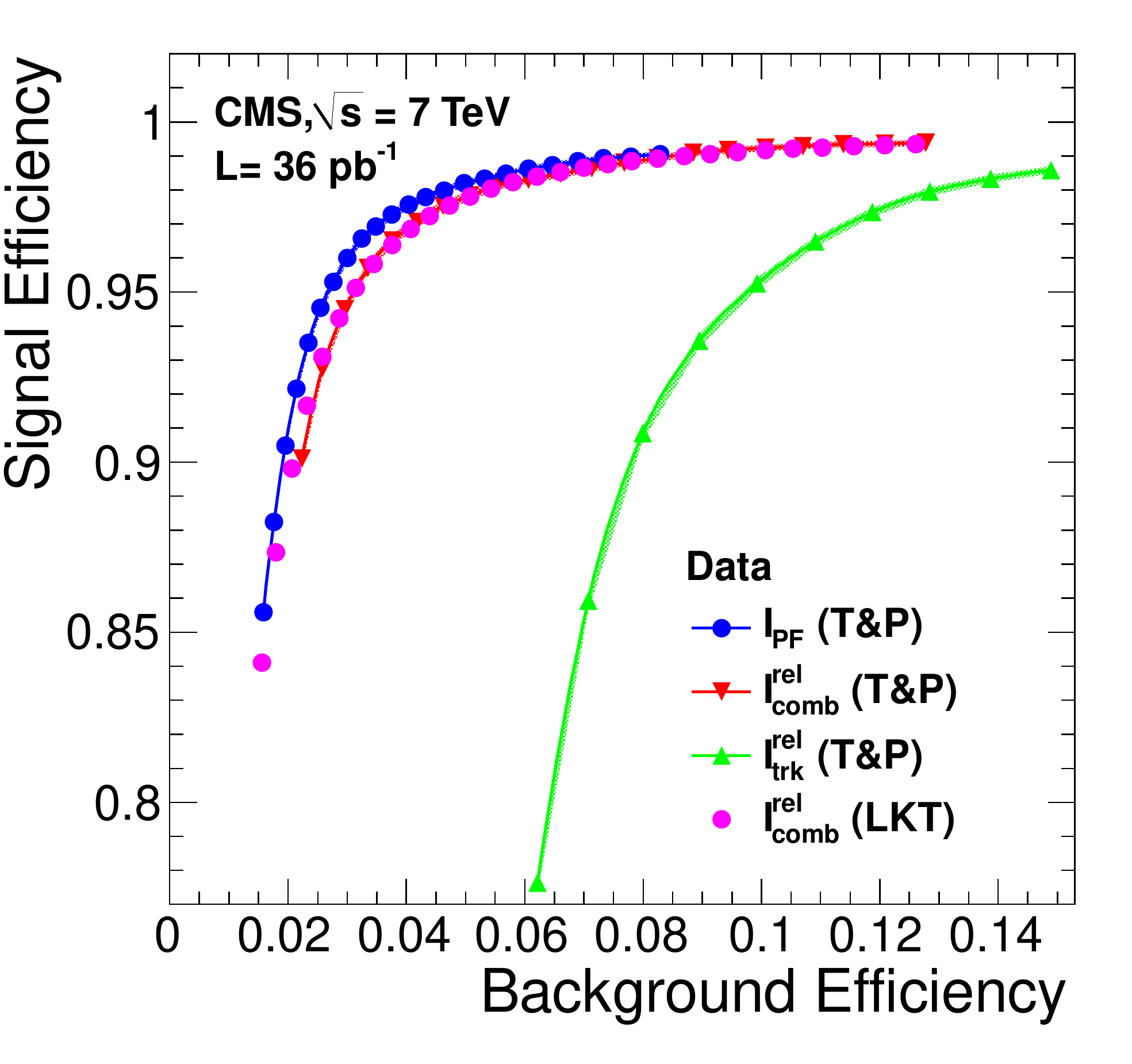}
  \caption{
Isolation efficiency for muons from $\Z$ decays versus isolation
efficiency for muons in the QCD-enhanced
dataset described in the text, for tracker relative, %(top left),
tracker-plus-calorimeters relative, %(top right),
and particle-flow relative %(bottom left)
isolation algorithms. 
Muons are required to have $\pt$ in the range between 20 and 50\GeVc. 
The background rejection is limited by the 0.4\% contamination from
truly isolated muons.}
  \label{fig:Isolation_3}
\end{figure}

Figure~\ref{fig:Isolation_3} shows the
efficiency of the various isolation algorithms for muons from $\Z$
decays versus the efficiency of the same algorithm for muons from the
QCD-enhanced dataset. In both cases muons
are required to have $20<\pt<50\GeVc$. Results %from data and MC
obtained with the tag-and-probe method are reported; the LKT results
agree with the tag-and-probe results within 1\% for both data and
simulation, and are shown only for the $\IRelComb$ algorithm for comparison.
As can be seen in the figure,
the best performance for the considered datasets and $\pt$ range 
is given by $\IPF$, followed closely by $\IRelComb$.
For a given signal efficiency, the background rejection evaluated using
the MC sample (not shown) is slightly better than that in the data, by
up to 0.5\% for $\IRelTrk$ and up to 1\% for $\IRelComb$ and $\IPF$.
It should be noted that the reported background
efficiency includes the absolute 0.4\% contamination from truly isolated
muons from $\W/\Z$ decays in the background sample, as described above. 

\begin{figure}[htb!]
  \centering
\includegraphics[height=0.45\textwidth,angle=0]{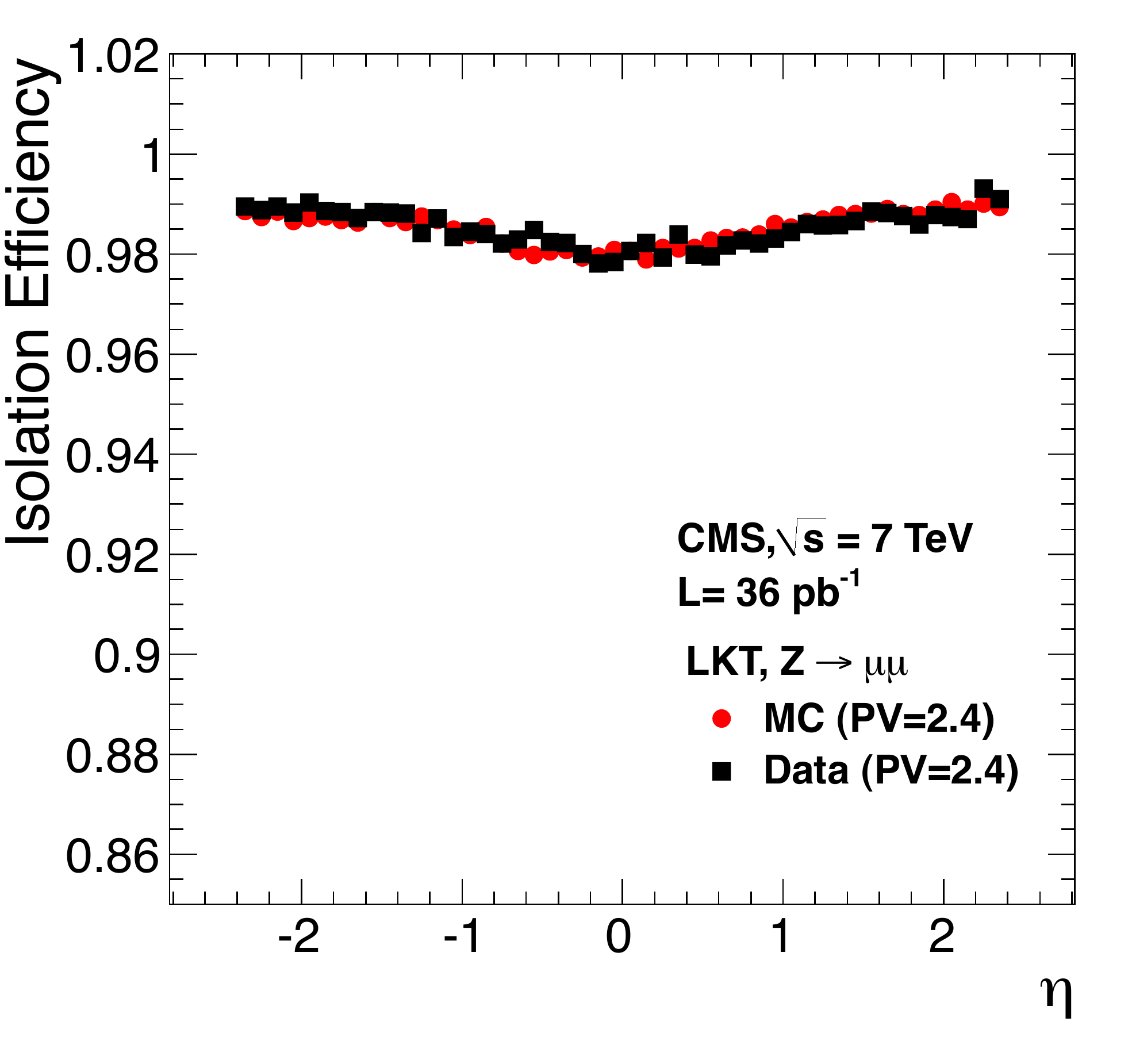}
  \caption{
Muon isolation efficiency versus $\eta$.
Efficiency for muons with $\pt$ between 20 and 50\GeVc from $\Z$ decays 
observed in 2010 data, in which the average number of reconstructed primary
vertices (PV) is 2.4, is compared with the results obtained using
MC samples with the same average number of primary vertices.
The results were obtained with
the LKT method using the $\IRelComb$
algorithm with a threshold of 0.15.}
  \label{fig:Isolation_4}
\end{figure}

The efficiency of any isolation algorithm features a dependence on the
muon pseudorapidity, and this dependence is expected to become more
pronounced as the number of pile-up collisions increases. 
Figure~\ref{fig:Isolation_4} shows the efficiency 
for muons with $20<\pt<50\GeVc$
from $\Z$ decays observed in data compared with the results
obtained using MC samples having the same distribution
of the number of primary vertices as in data, with an average of 2.4.
The results are obtained
with the LKT method using the $\IRelComb$ algorithm with a
threshold of 0.15. The particular $\eta$ dependence reflects the
distribution of transverse energy flow expected for minimum-bias events.
Methods to mitigate the impact of pile-up on the performance of
isolation algorithms have been developed for higher luminosity running. One such technique
is based on the measurement, event-by-event, of the average transverse
momentum per unit area $\rho$ added to the event by minimum-bias pile-up
collisions~\cite{Cacciari:2007fd}. Another technique uses
reconstructed tracks and primary vertices to compute a correction
factor $\beta$ to be applied to all or part of the numerator of the isolation
variables described above. In both cases, the $\rho$ and
$\beta$ variables allow the energy in the isolation cone
due to particles produced in pile-up collisions to be estimated.

\begin{figure}[htb!]
  \centering
\includegraphics[height=0.45\textwidth,angle=0]{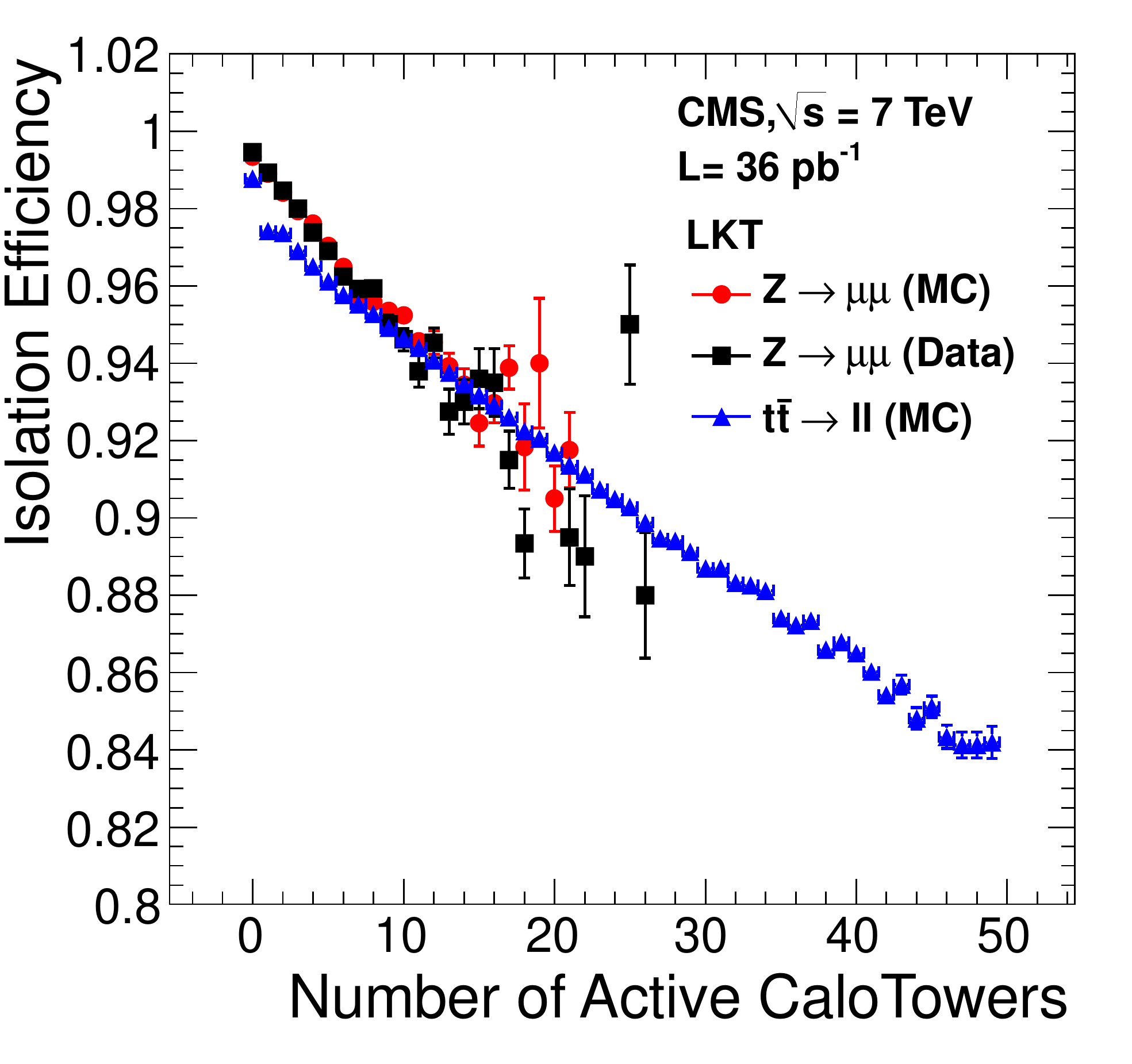}
  \caption{
Muon isolation efficiency 
versus number of calorimeter towers in the region of $|\eta| < 2.7$ 
having transverse energy larger than 1\GeV.  Results for muons with
$20<\pt<50\GeVc$ from $\Z$ decays obtained in data and in simulation
are compared with the efficiency for prompt muons in a simulated sample
of top pairs decaying leptonically.
All results were obtained with the LKT method using the $\IRelComb$
algorithm with a threshold of 0.15. 
The MC samples include simulation of pile-up events.}
  \label{fig:Isolation_5}
\end{figure}

An additional advantage of the LKT method is that it can be used to estimate
the isolation efficiency for muons produced in any signal process. 
The basic assumption is made that the kinematics of a
muon produced in the leptonic decay of a $\W$, $\Z$, or a heavier
new particle is unrelated to the activity of the rest of
the event.  
Under this assumption, the isolation efficiency for
muons from such a signal can be inferred from the
corresponding efficiency measured using muons from $\Z$, by extrapolating
from underlying event activity of $\Z$ events to that of the signal.
The number of ECAL and HCAL calorimeter towers with transverse energy
above threshold can be used as a measure of the event activity. 
The power of this technique can be appreciated in 
Fig.~\ref{fig:Isolation_5}, where
the isolation efficiency for prompt muons from simulated $\ttbar$
events forced to decay leptonically
is plotted
versus the number of calorimeter towers with $|\eta| < 2.7$ and
$\ET > 1\GeV$.  The isolation efficiency for muons from $\Z$
decays (data and MC simulation) is shown superimposed. 
The efficiencies shown are those of the $\IRelComb$ algorithm
computed by the LKT method.
Compared to $\ttbar$, %and supersymmetric particle production
$\Z$ events are characterized by relatively low
occupancies.  In the region where the $\Z$ and $\ttbar$ efficiencies
overlap, however, the differences between them do not exceed 2\%,
and both sets of points tend to align along
an approximately straight line.  The parameters of this line can be 
obtained by fitting the points from the $\Z$ dataset; the parameters are then
used to predict, with an accuracy of a few percent, the isolation efficiency
for muons in events with large underlying activity, which
cannot be directly measured with the tag-and-probe technique.
The pile-up contribution can then be accounted for and corrected by means
of the techniques using the $\rho$ and $\beta$ variables described above.

\section{Muon Trigger}
\label{sec:trigger}

The CMS trigger system consists of two basic stages: the hardware-based
\Lone trigger~\cite{L1TDR} and the software-based high-level
trigger (HLT)~\cite{HLTTDR}.

The \Lone muon trigger uses signals from all three CMS muon detector systems: DT, CSC, and RPC.
It has a latency of 3.2~$\mu$s and reduces the rate of inclusive muon candidate events read-out from detector front-end
electronics to a few kHz by applying selections on the estimated muon $\pt$ and
quality.% depending on which muon stations participated in the $\pt$ measurement.

In the muon HLT, first a \Lone trigger object is used as a seed to
reconstruct a standalone-muon track in the muon system, leading to an improved $\pt$ estimate. At this point,
$\pt$ threshold filters are applied to the standalone (also called \Ltwo) muon.
Then seeds in the inner tracker are generated in the region around
the extrapolated \Ltwo muon, and tracker tracks are reconstructed.
If a successful match is made between a tracker track and the \Ltwo muon, a global fit combining
tracker and muon hits is performed, yielding a \Lthree muon track on which the final $\pt$ requirements
are applied. In this way, the rate of recorded inclusive muon events is reduced to a few tens of Hz.
The average processing time of the HLT reconstruction is about 50 ms.

In this section we report on trigger efficiency measurements and rejection rates,
and compare them to predictions from the simulation.  We study the trigger efficiency using two
complementary methods: one using tag-and-probe and one using single muons
reconstructed offline.
The efficiencies are shown separately for the barrel %($|\eta|<0.9$)
and for the overlap-endcap
regions since differences in the
trigger response have been observed between them.

\subsection{Trigger efficiency using the tag-and-probe method on dimuon resonances}
\label{sec:Trigger_TnP}

In this subsection, we report the measurements of the trigger efficiency
for prompt muons performed by applying
the tag-and-probe method described in Section~\ref{sec:muonideff_tnp}
to muons from the
decays of $\jpsi$ and $\Z$ resonances.
With this method, the trigger efficiency can be measured with respect to
any offline muon selection.
In the following, we use Soft Muons and Tight Muons as probes to check how
efficient the trigger is in selecting muons that can be reconstructed
offline. In the case of the $\Z$, the probe muons are also required
to be isolated by requiring $\IRelComb$ to be smaller than 0.15.
Measured efficiencies are compared to those evaluated using the
MC samples described in Section~\ref{sec:muonideff_tnp}.

The muon trigger efficiency was studied separately in two regions
of muon transverse momentum: below and above 20\GeVc.  In the
region of $\pt > 20\GeVc$, the efficiency was measured using a
sample of $\Zmm$ events collected with single-muon triggers. In the
region of $\pt < 20\GeVc$, the $\mathrm{J}/\!\psi\to\mm$ events collected
by the specialized muon-plus-track triggers described in Section~\ref{sec:samples}
were used.
Several instances of the muon-plus-track trigger have been deployed, with
different thresholds on the $\pt$ of the tracker track;
those with the lower thresholds became prescaled as the instantaneous
luminosity increased during 2010.  As a result, the amount of data used for the
measurements presented here varies with the $\pt$ region:
it corresponds to an integrated luminosity of 0.9~pb$^{-1}$ for
$\pt < 3 \GeVc$, 0.6~pb$^{-1}$ for $3 < \pt < 5 \GeVc$, and
5.3~pb$^{-1}$ for $5 < \pt < 20 \GeVc$.  The sample of $\Zmm$ events
used for the measurements in the range of $\pt$ above $20 \GeVc$
corresponds to an integrated luminosity of 31~pb$^{-1}$.%check again with the correct run range

To evaluate the trigger efficiency, trigger objects must be
matched to the muons reconstructed offline.  \Lone muon-trigger candidates
are matched to offline muons by position, extrapolating a muon tracker track
to the muon system.
The HLT muons are matched to the muons reconstructed offline
by direction at the vertex.
The HLT-only efficiencies are computed matching the probe
with the \Lone candidate and requiring it to be also matched with the
HLT candidate. The combined efficiencies of \Lone and HLT are obtained simply
requiring the probe to be matched with the HLT candidate.

\subsubsection{Trigger efficiency for Soft Muons}
\label{sec:triggerSoft}
Trigger efficiencies for Soft Muons in the range of $\pt < 20 \GeVc$
were measured using muons from the decays of $\jpsi$ particles.

\begin{figure}[thb]
  \begin{center}
    \includegraphics[width=0.32\textwidth]{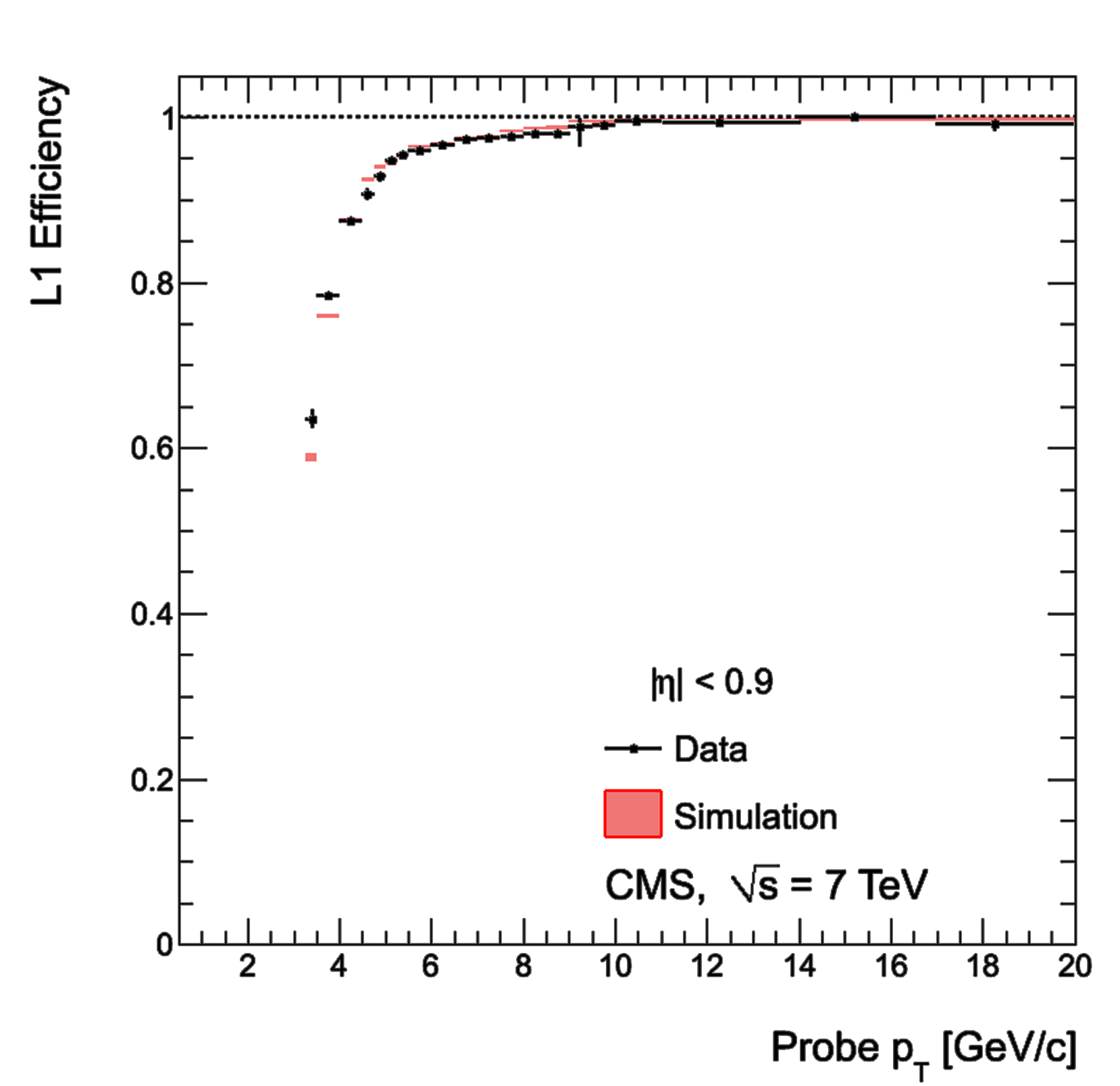}
    \includegraphics[width=0.32\textwidth]{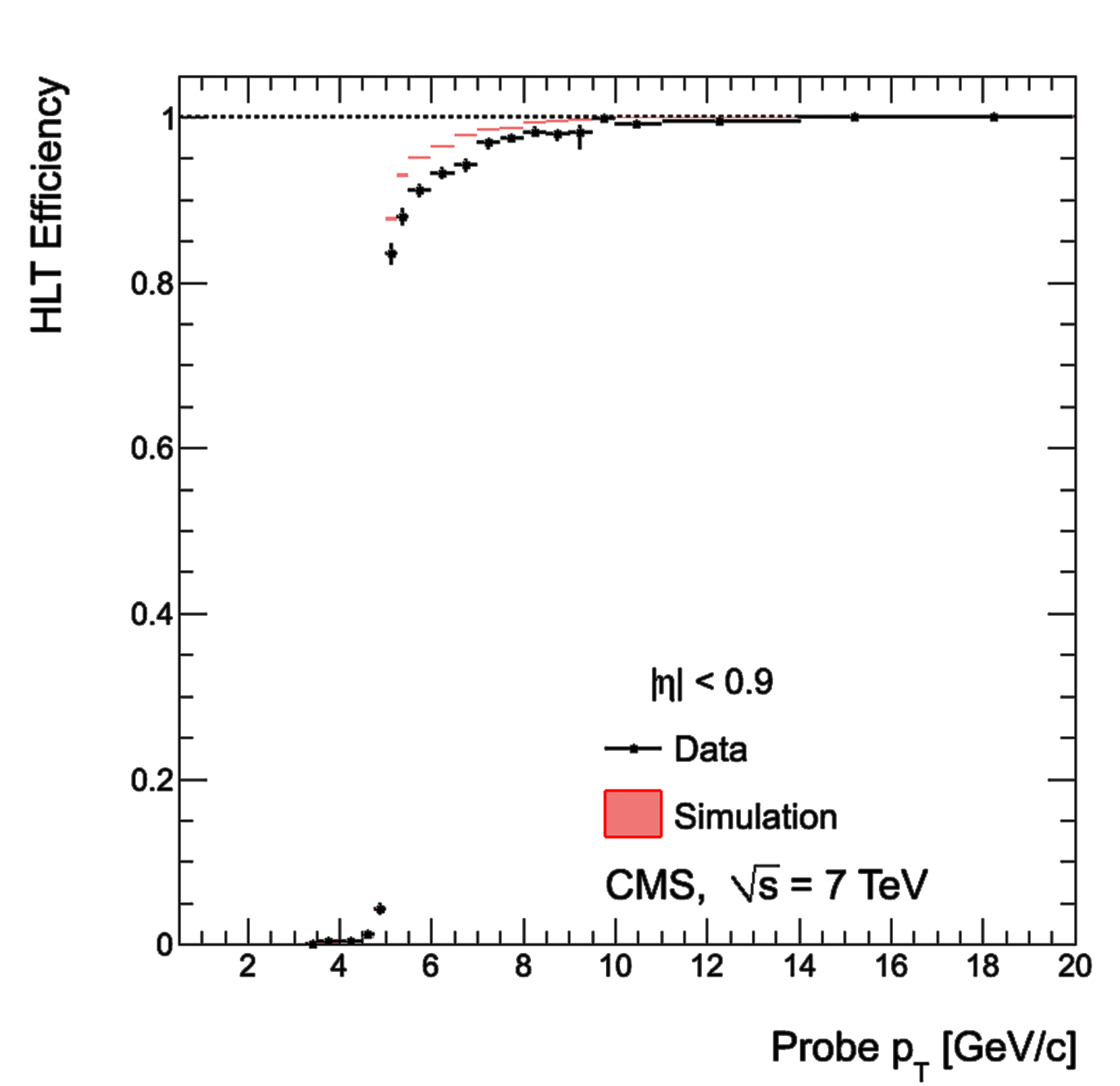}
    \includegraphics[width=0.32\textwidth]{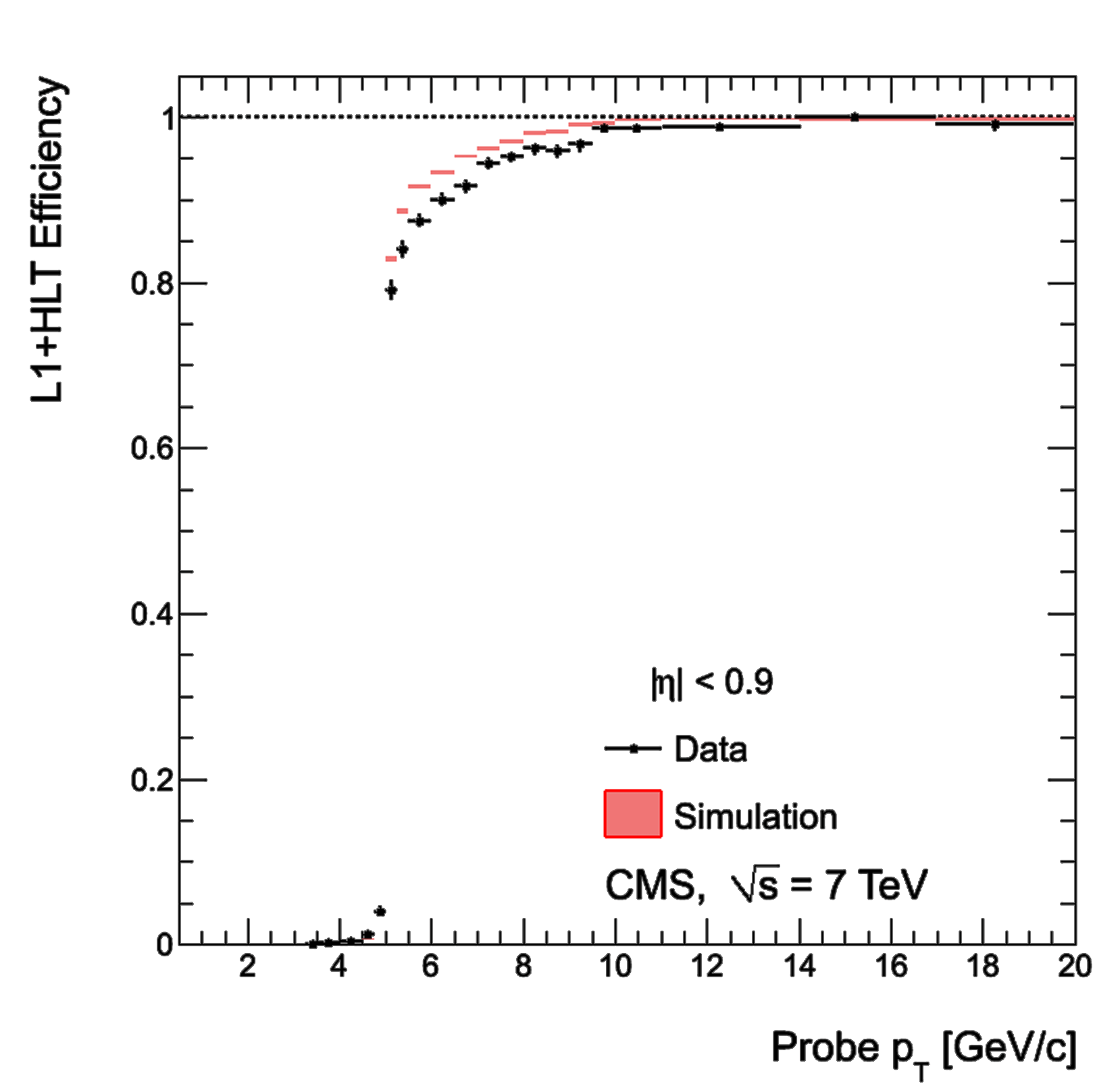}
    \includegraphics[width=0.32\textwidth]{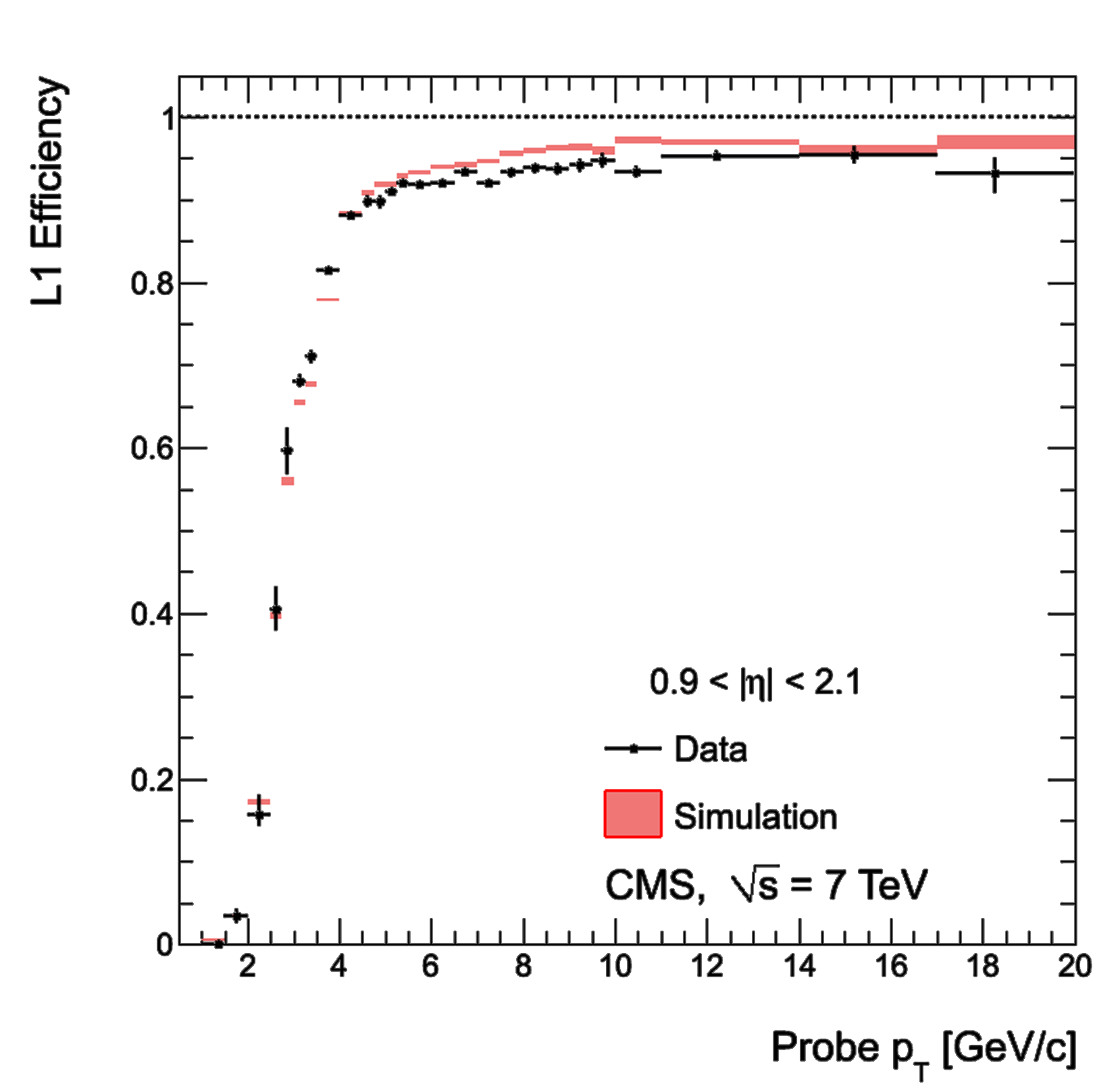}
    \includegraphics[width=0.32\textwidth]{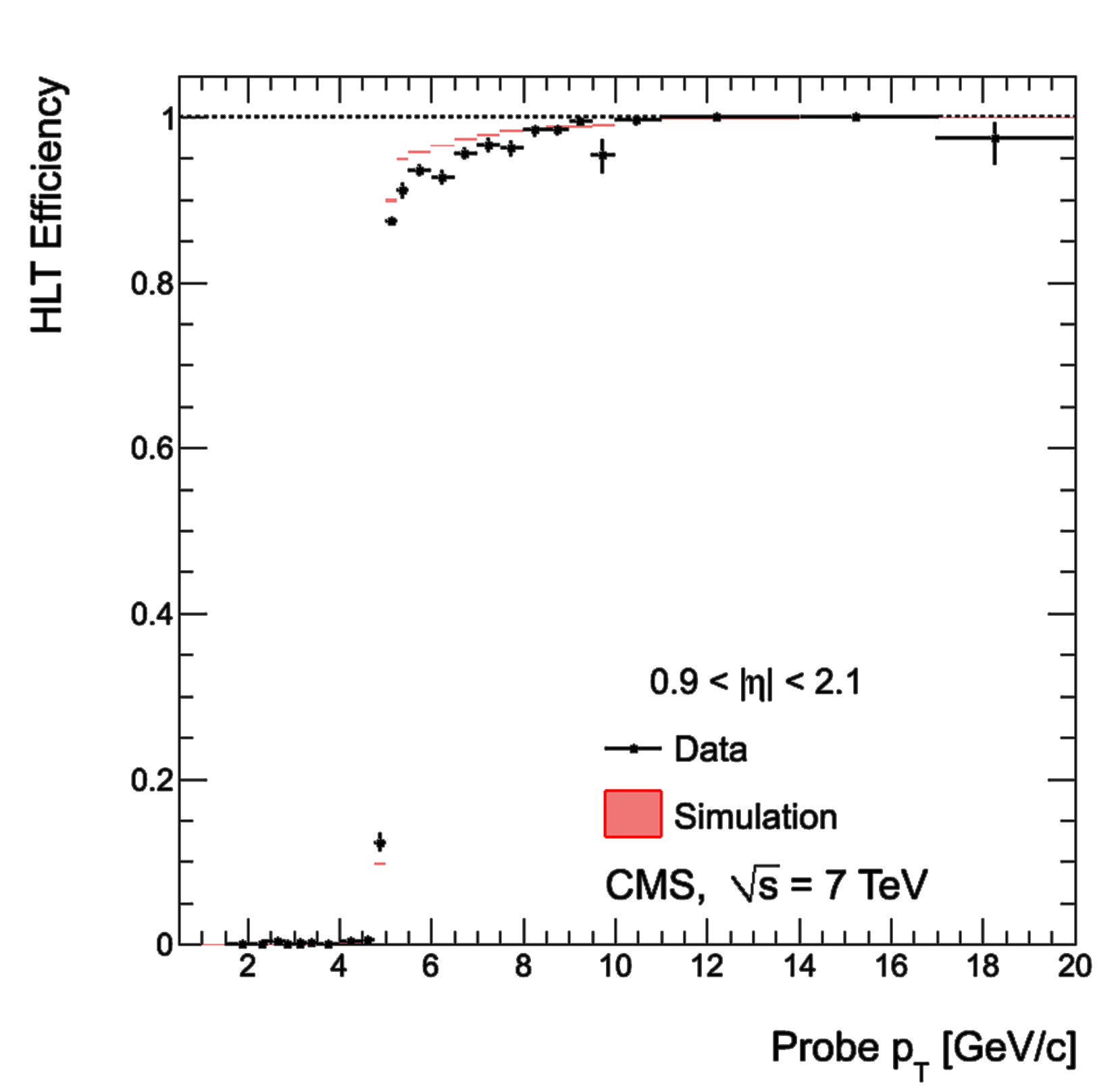}
    \includegraphics[width=0.32\textwidth]{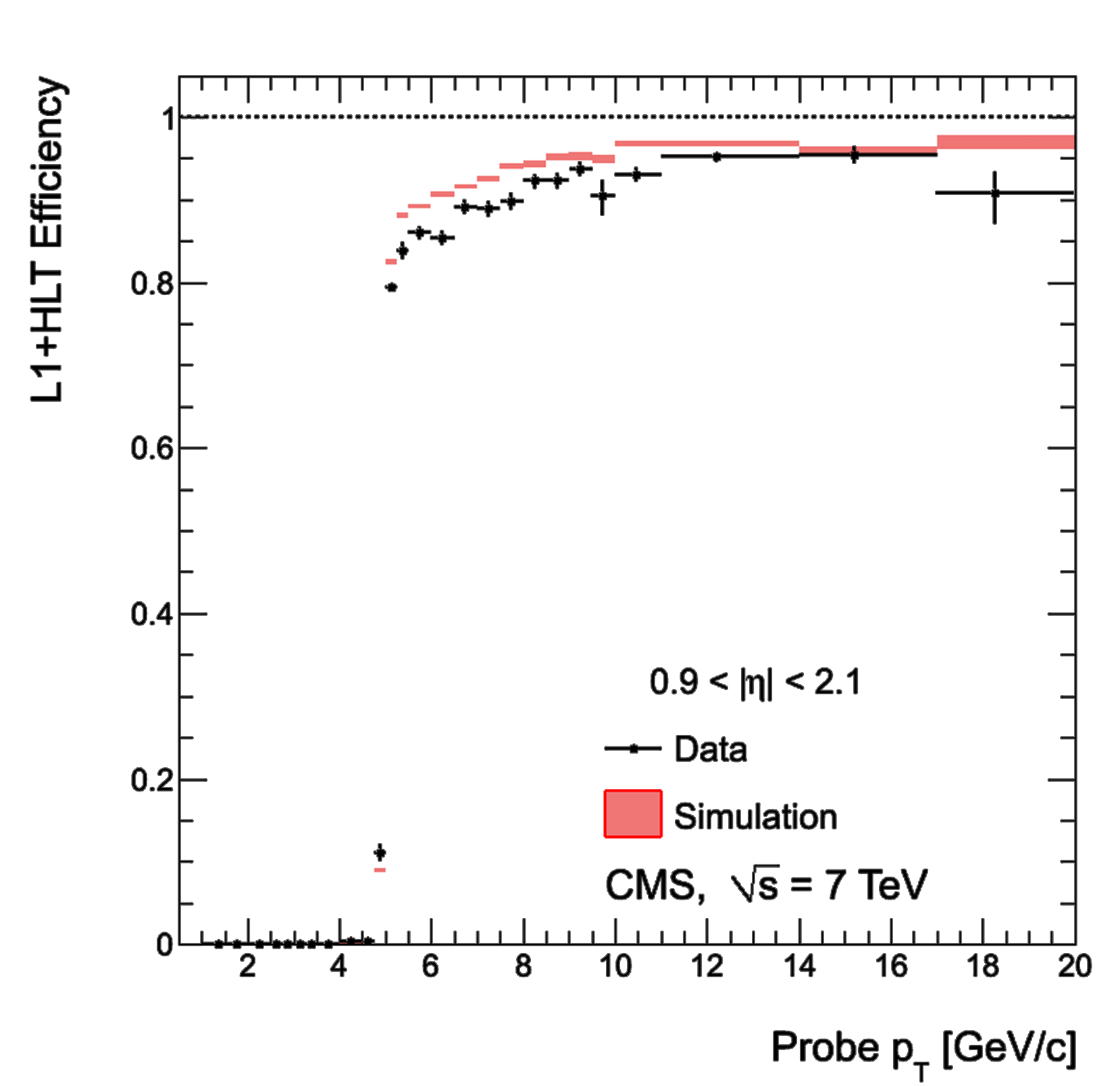}
    \caption{Single-muon trigger efficiencies for Soft Muons as a
      function of the Soft Muon $\pt$ in the barrel (top) and the
      overlap-endcap (bottom) regions: the efficiency of the \Lone trigger
      with $\pt$ threshold at 3\GeVc (left), the efficiency of the HLT
      with $\pt$ threshold of 5\GeVc with respect to \Lone
      (middle), and the combined efficiency of \Lone and HLT (right).  The
      efficiencies obtained using $\mathrm{J}/\!\psi\to\mm$ events
      (points with error bars) are compared with predictions from the MC
      simulation ($\pm 1\sigma$ bands); the uncertainties are statistical
      only.}
    \label{fig:JpsiSoft_pt}
  \end{center}
\end{figure}

\begin{figure}[thb]
  \begin{center}
    \includegraphics[width=0.32\textwidth]{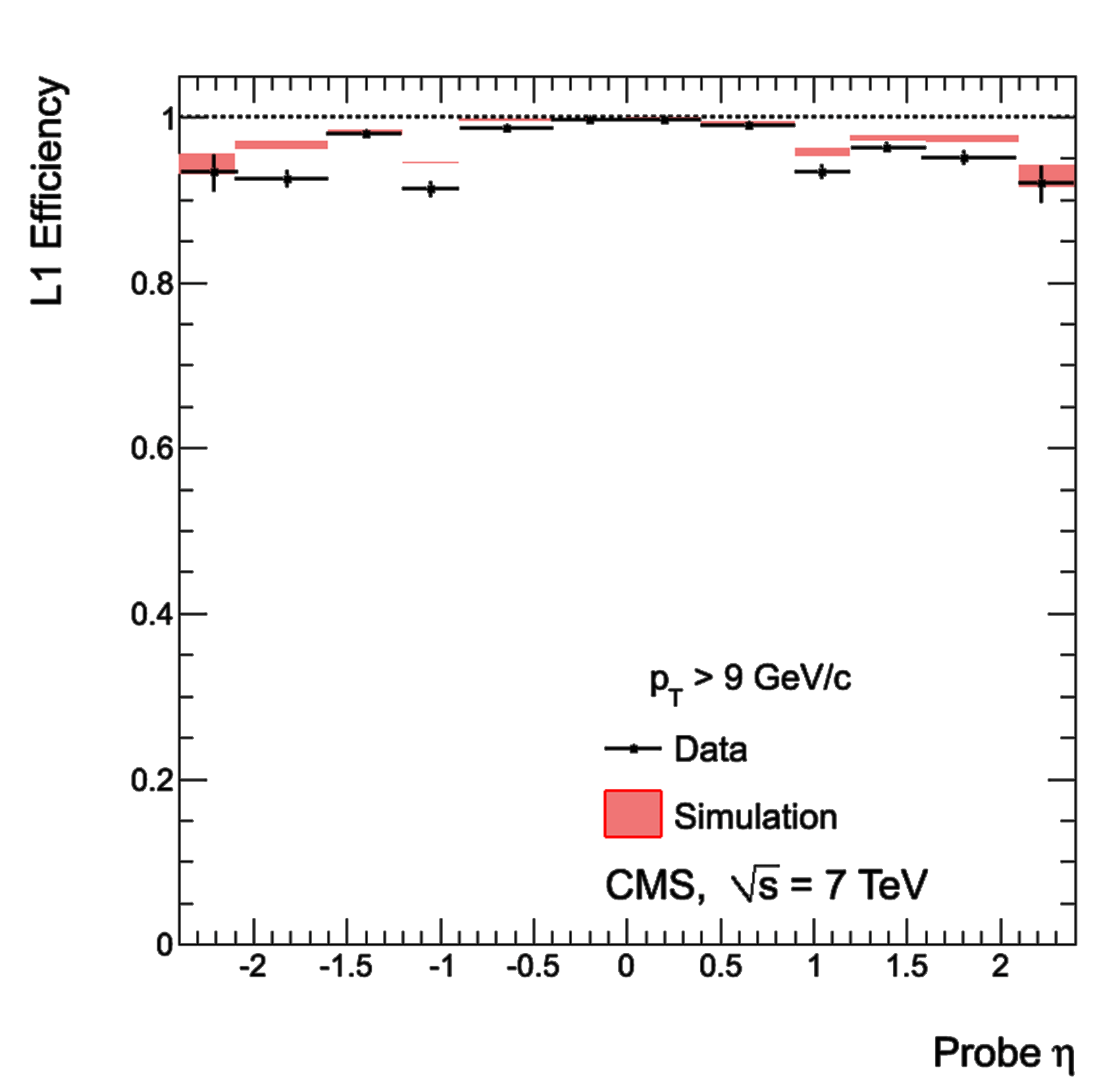}
    \includegraphics[width=0.32\textwidth]{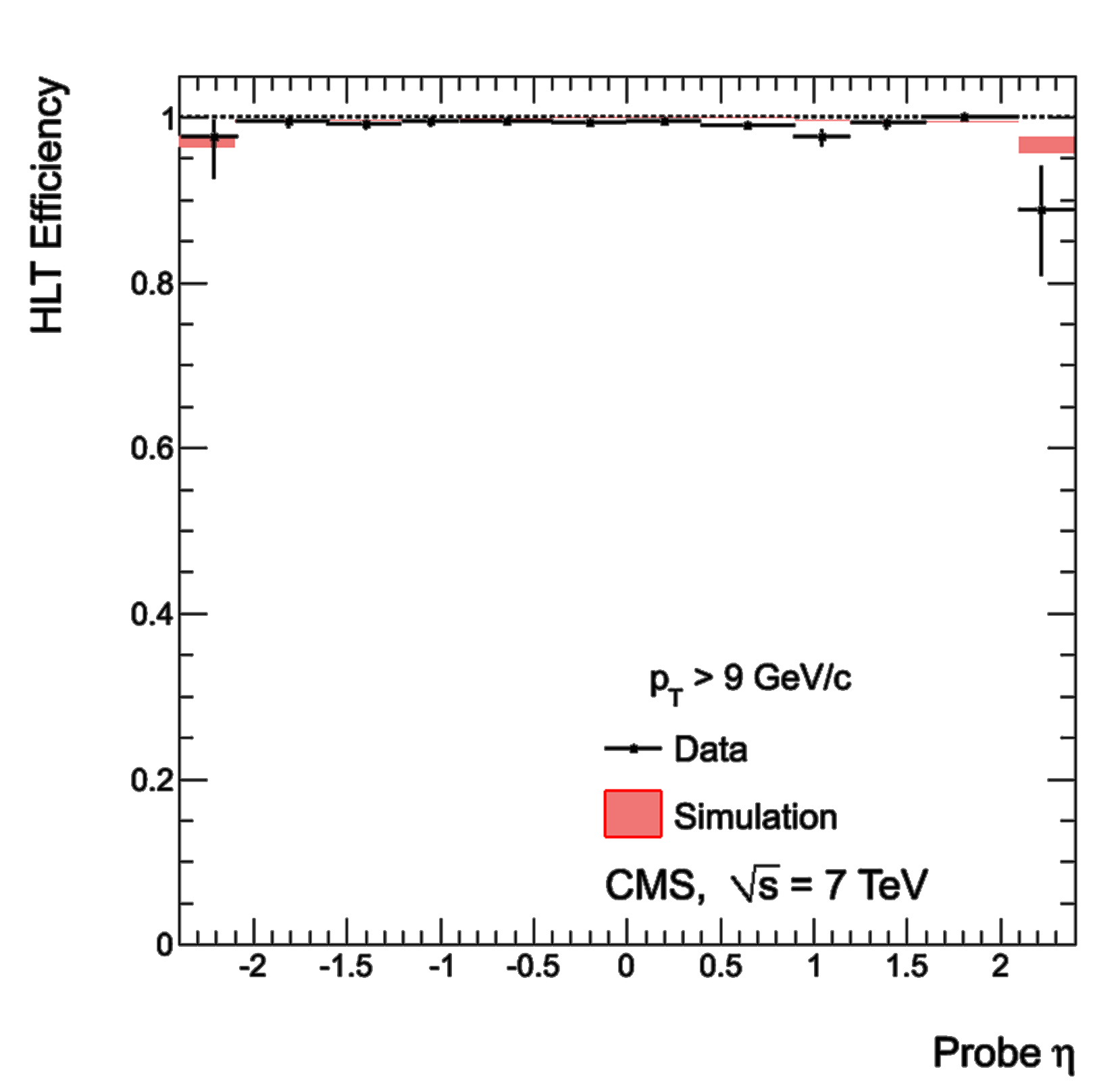}
    \includegraphics[width=0.32\textwidth]{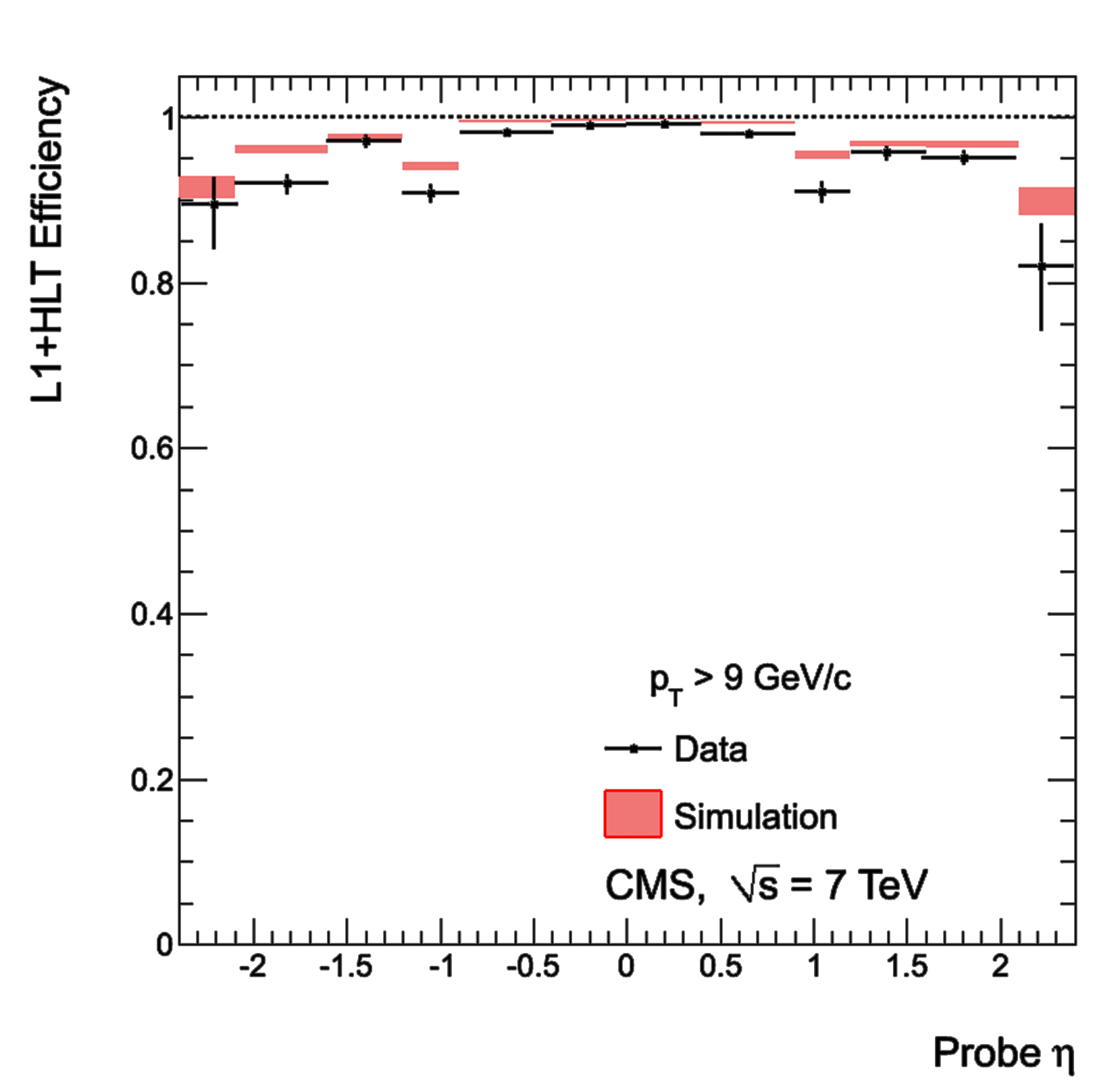}
    \caption{Single-muon trigger efficiencies for Soft Muons as a
      function of the Soft Muon $\eta$, for muons with $\pt$ in the
      range of $9 < \pt < 20 \GeVc$: the efficiency of the \Lone
      trigger with $\pt$ threshold at 3\GeVc (left), the efficiency
      of the HLT with $\pt$ threshold of 5\GeVc with respect to \Lone
      (middle), and the combined efficiency of \Lone and HLT (right).
      The efficiencies obtained using $\mathrm{J}/\!\psi\to\mm$
      events (points with error bars) are compared with predictions from
      the MC simulation ($\pm 1\sigma$ bands); the uncertainties are
      statistical only.}
    \label{fig:JpsiSoft_eta}
  \end{center}
\end{figure}

The single-muon trigger efficiencies for Soft Muons are shown in
Fig.~\ref{fig:JpsiSoft_pt} as a function of muon $\pt$, separately for
the \Lone trigger with $\pt$ threshold at 3\GeVc, the HLT with $\pt$
threshold at 5\GeVc, and for the full \Lone--HLT online selection.  The
efficiencies at the plateau, for muons with $\pt$ in the range of $9 <
\pt < 20 \GeVc$, are reported as a function of muon $\eta$ in
Fig.~\ref{fig:JpsiSoft_eta} and summarized in
Table~\ref{tab:triggerSumJpsiMuons}.

\begin{table}[thb]
\begin{center}
\topcaption{\Lone, HLT, and overall single-muon trigger efficiencies
  for Soft Muons in the efficiency plateau ($9 < \pt < 20 \GeVc$) for
  different pseudorapidity regions.  The first column shows
  efficiencies measured from data; the second column shows the ratio
  between the measurements in data and simulation.  The uncertainties are
  statistical only.}
\begin{tabular}{|lc|cc|} \hline
\multicolumn{2}{|l|}{Trigger Level} & \multicolumn{2}{c|}{Tag-and-Probe $\jpsimm$} \\
          & Region           & Eff. [\%]      & Data/MC           \\ \hline \hline
\Lone     & $|\eta|<2.1$     & 97.1 $\pm$ 0.2 & 0.990 $\pm$ 0.002 \\
          & $|\eta|<0.9$     & 99.2 $\pm$ 0.1 & 0.995 $\pm$ 0.001 \\
          & $0.9<|\eta|<2.1$ & 94.5 $\pm$ 0.3 & 0.978 $\pm$ 0.004 \\ \hline
HLT       & $|\eta|<2.1$     & 99.1 $\pm$ 0.2 & 0.995 $\pm$ 0.002 \\
          & $|\eta|<0.9$     & 99.2 $\pm$ 0.2 & 0.993 $\pm$ 0.002 \\
          & $0.9<|\eta|<2.1$ & 99.0 $\pm$ 0.3 & 0.997 $\pm$ 0.004 \\ \hline
Level-1+HLT & $|\eta|<2.1$   & 96.2 $\pm$ 0.2 & 0.985 $\pm$ 0.003 \\
          & $|\eta|<0.9$     & 98.5 $\pm$ 0.2 & 0.989 $\pm$ 0.002 \\
          & $0.9<|\eta|<2.1$ & 93.6 $\pm$ 0.5 & 0.975 $\pm$ 0.005 \\ \hline
\end{tabular}
\label{tab:triggerSumJpsiMuons}
\end{center}
\end{table}

All efficiency curves show a sharp ``turn-on'' at the trigger $\pt$
threshold, well described by the simulation.  As expected, the turn-on
is sharper for the HLT, due to an improved $\pt$ resolution.  The
plateau efficiencies are about 99\% for the \Lone trigger in the
barrel region and for the HLT in the whole studied pseudorapidity
range.  The \Lone efficiency in the overlap-endcap region is slightly
lower, at about 95\%.  Most of the 5\% efficiency loss is due to stringent
quality criteria used in the selection logic of \Lone muon triggers during
2010 data taking; these criteria were further optimized during the
2010--11 winter technical stop of the LHC.  Measured efficiencies are
generally in good agreement with those expected, with the ratios
between the two (see Table~\ref{tab:triggerSumJpsiMuons}) being within
1\% of unity.  The only exception is a slightly larger, about 2\%,
difference between data and simulation in the overlap-endcap region
due to a few non-operational CSCs not accounted for in the simulation
(see Section~\ref{sec:muonideff_tnp}).

The very forward region, $2.1 < |\eta| < 2.4$, is characterized by
higher rates of low-$\pt$ muons and poorer momentum resolution.
At the nominal LHC luminosity, this region is
intended to be used to improve trigger efficiency for events with multiple muons
but not to be included in triggering on single muons.
In 2010, however, the luminosity and hence the rates of
low-$\pt$ muons were sufficiently low to allow the single-muon trigger to be
extended to include the entire acceptance of muon detectors, up to
$|\eta|=2.4$.  In this region of $|\eta|=2.1$--2.4, the 3:1 ganging of
48 CSC cathode strips into 16 readout channels in the first muon
station~\cite{MUO-11-001} leads to an ambiguity in $\pt$ assignment at
\Lone trigger.  The algorithm to resolve this ambiguity was configured
with the goal to reduce the rate of muons with overestimated $\pt$ to
a minimum.  As a consequence, the efficiency of the \Lone triggers in
the very forward region was high for triggers with low $\pt$
thresholds (about 95\% for the \Lone trigger with $\pt$ threshold at
3\GeVc, see Fig.~\ref{fig:JpsiSoft_eta}), but substantially lower for
triggers with thresholds above 3\GeVc (see next section).

\subsubsection{Trigger efficiency for Tight Muons}
Trigger efficiencies for Tight Muons were measured by applying the
tag-and-probe method to
$\mathrm{J}/\!\psi\to\mm$ events in the region of $\pt$ below 20\GeVc
and $\Zmm$ events in the region of $\pt$ above 20\GeVc.

\begin{figure}[thb]
  \begin{center}
    \includegraphics[width=0.32\textwidth]{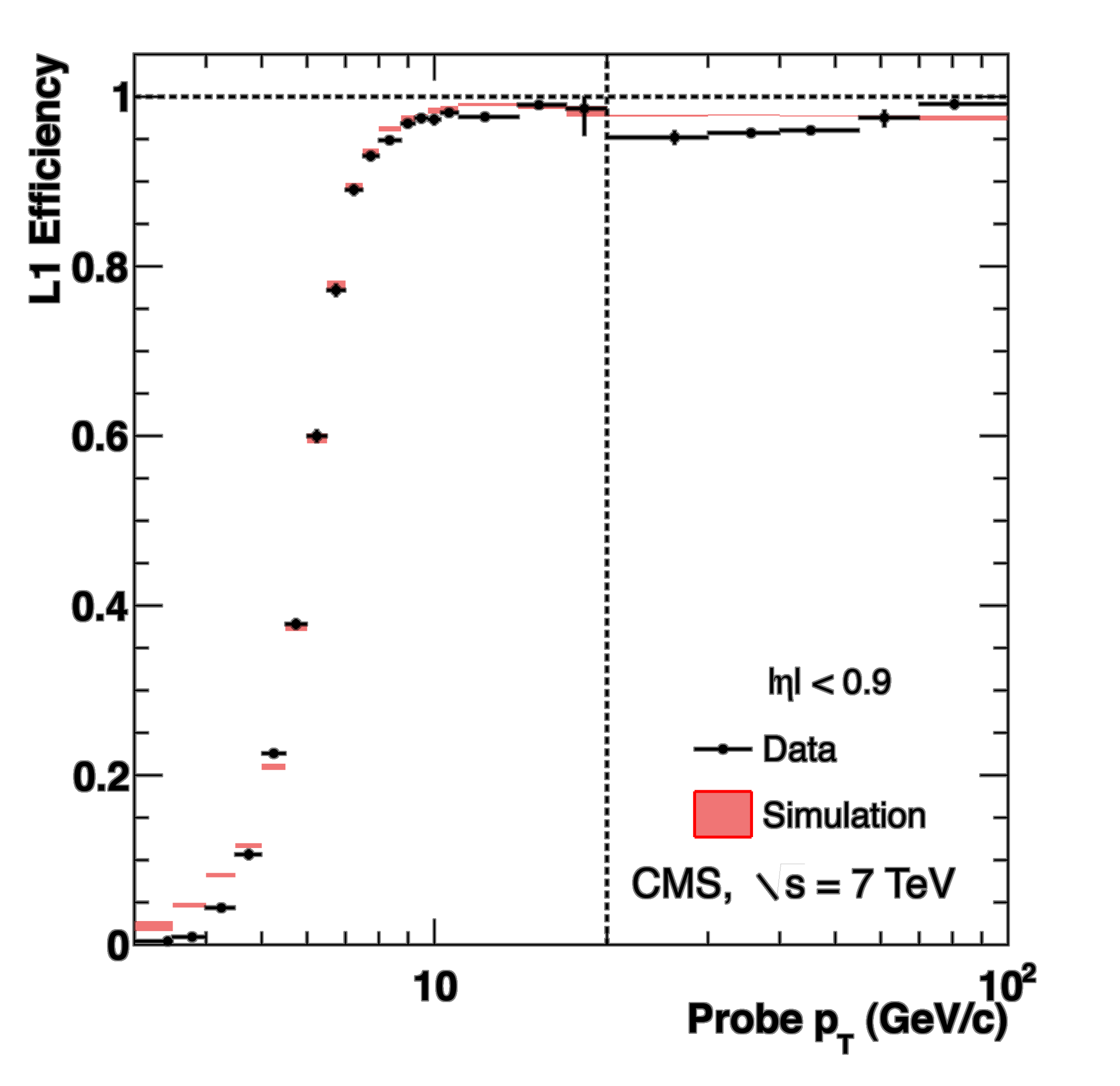}
    \includegraphics[width=0.32\textwidth]{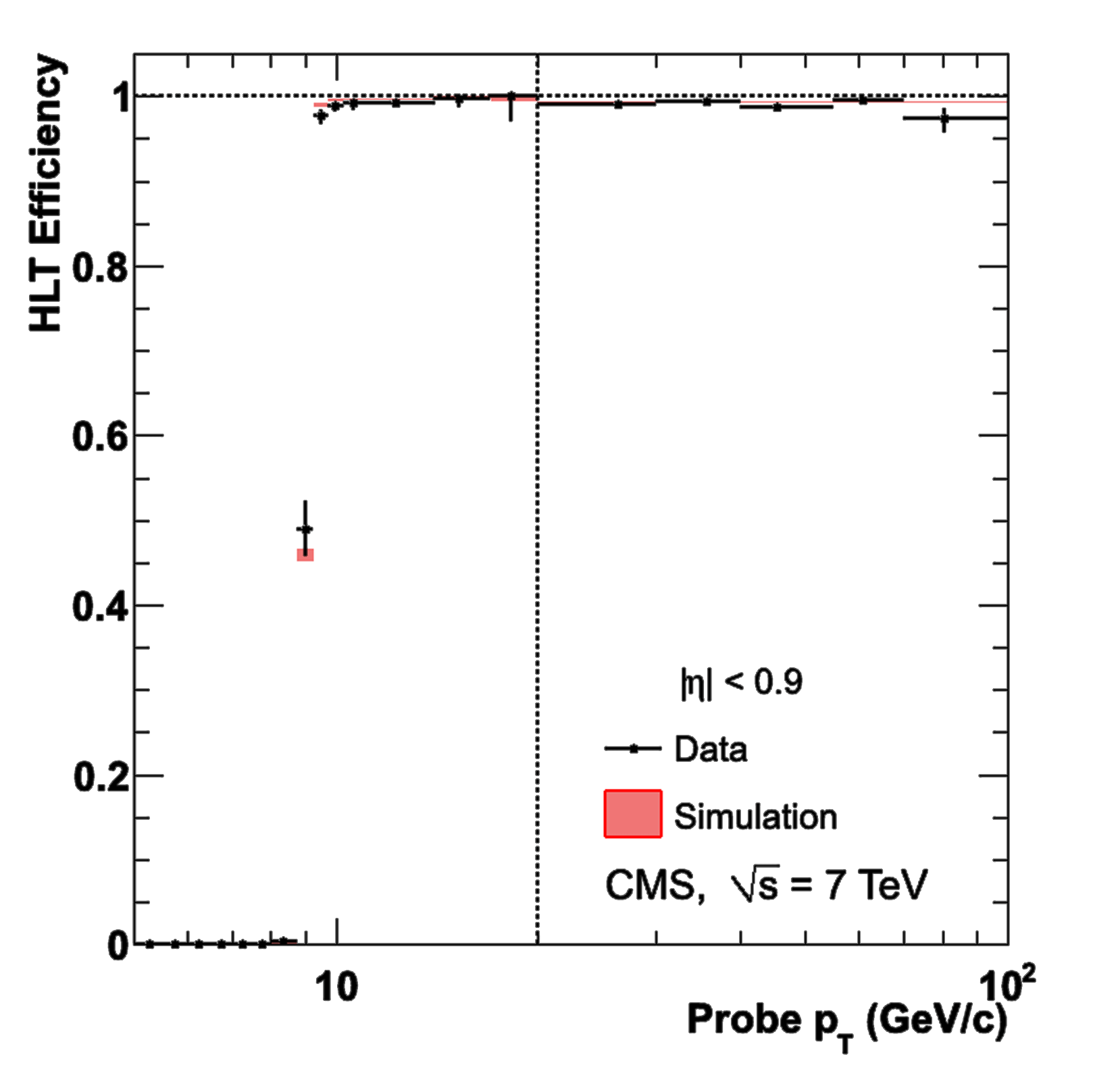}
    \includegraphics[width=0.32\textwidth]{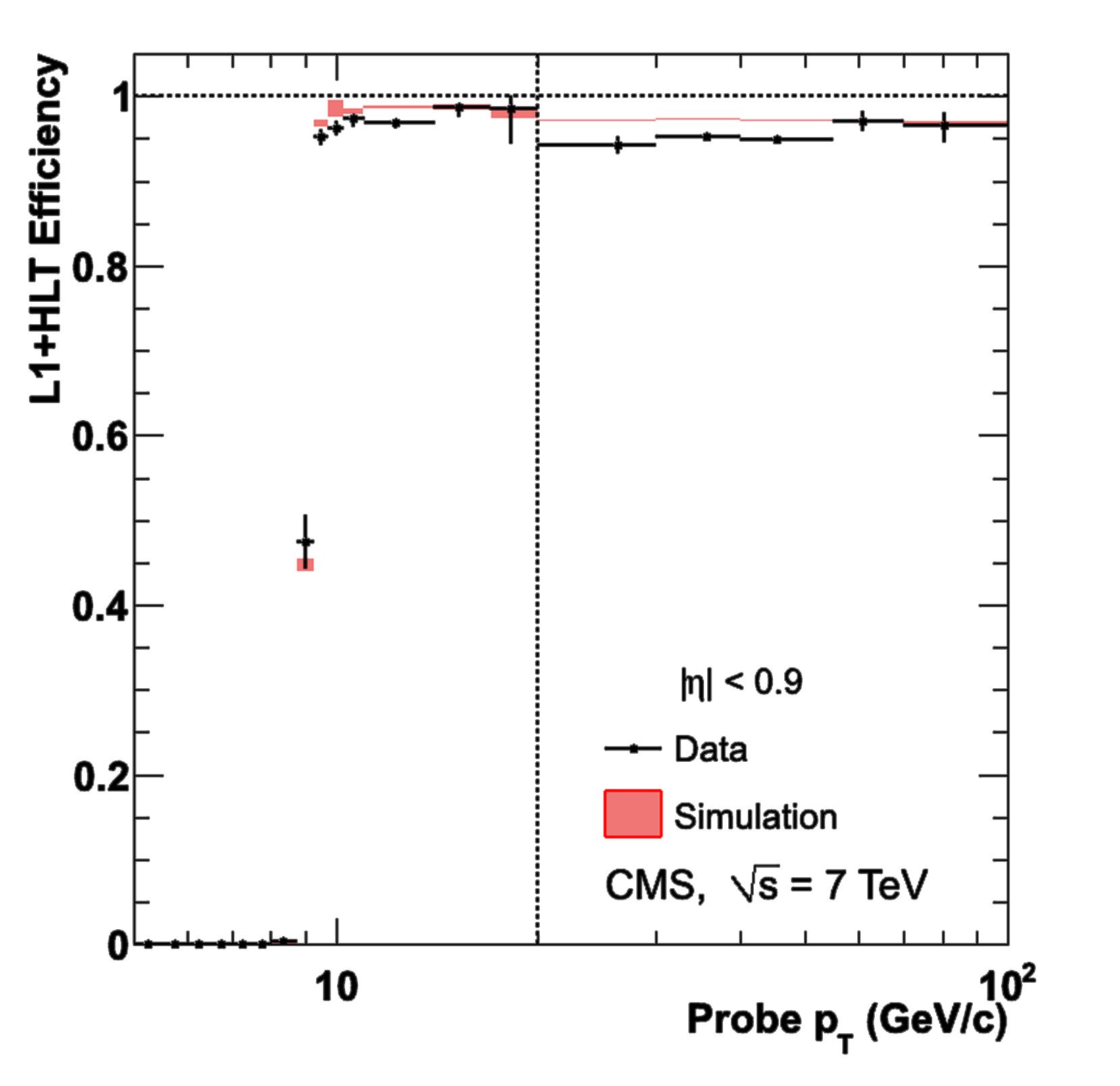}
    \includegraphics[width=0.32\textwidth]{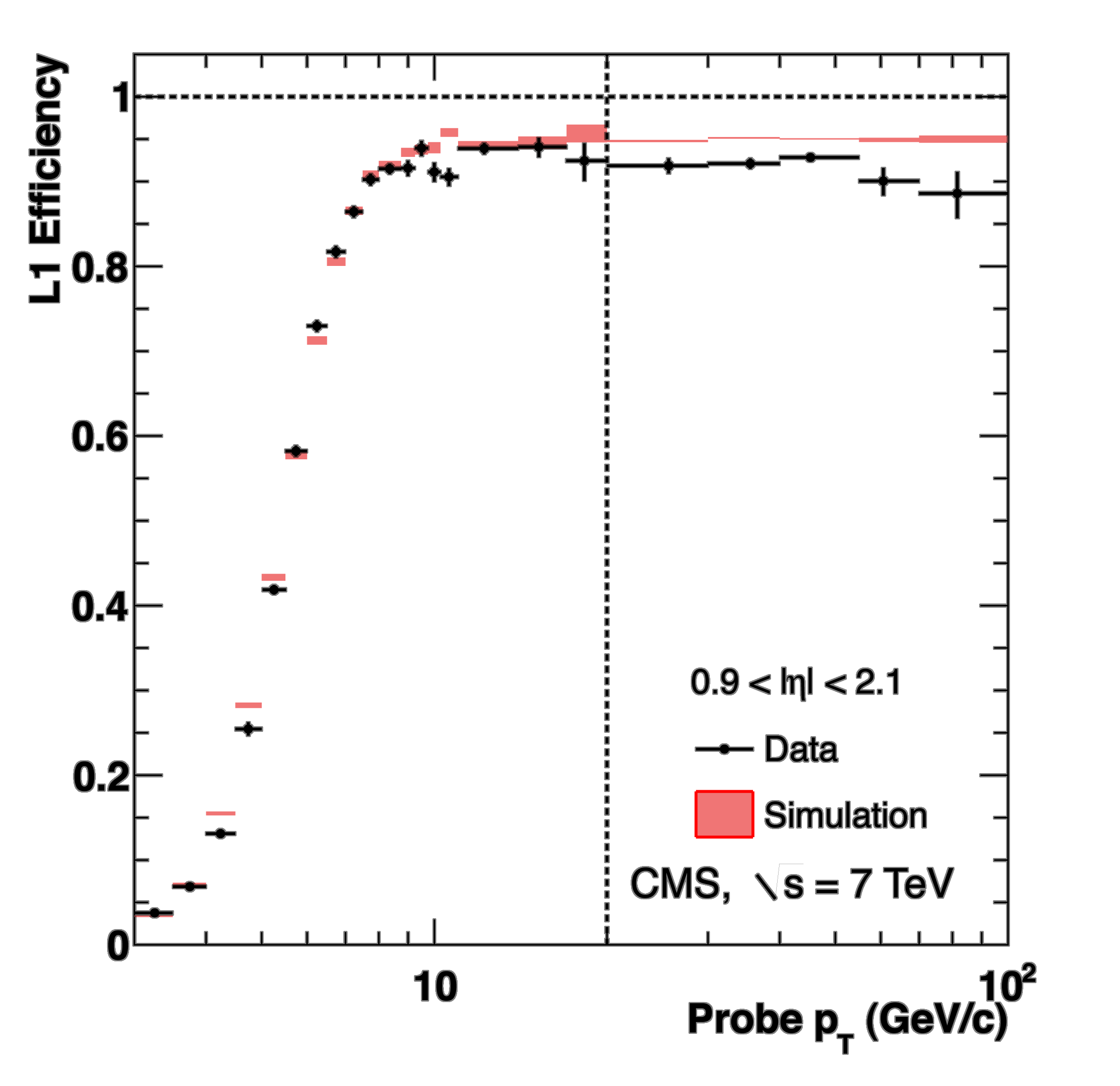}
    \includegraphics[width=0.32\textwidth]{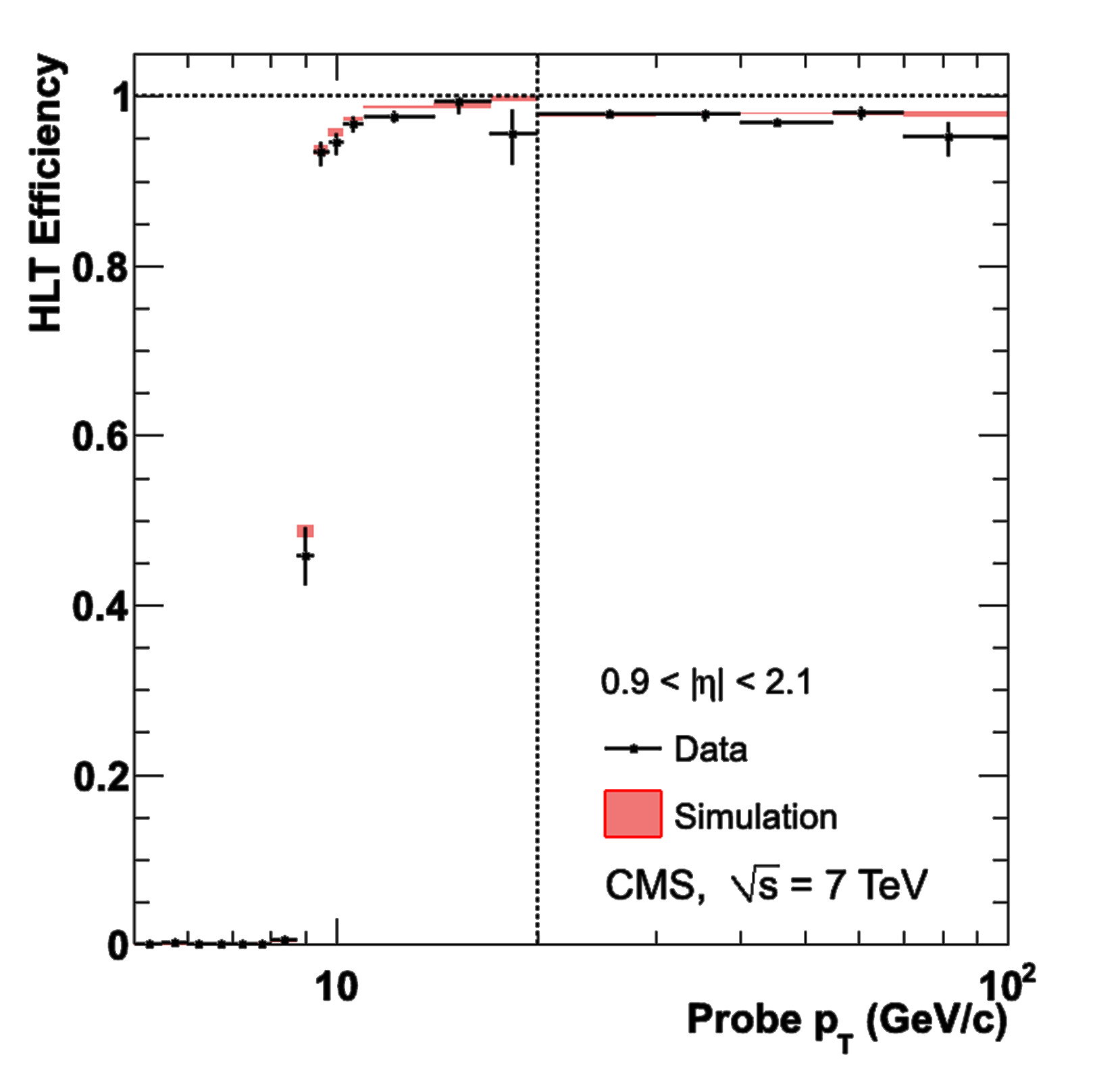}
    \includegraphics[width=0.32\textwidth]{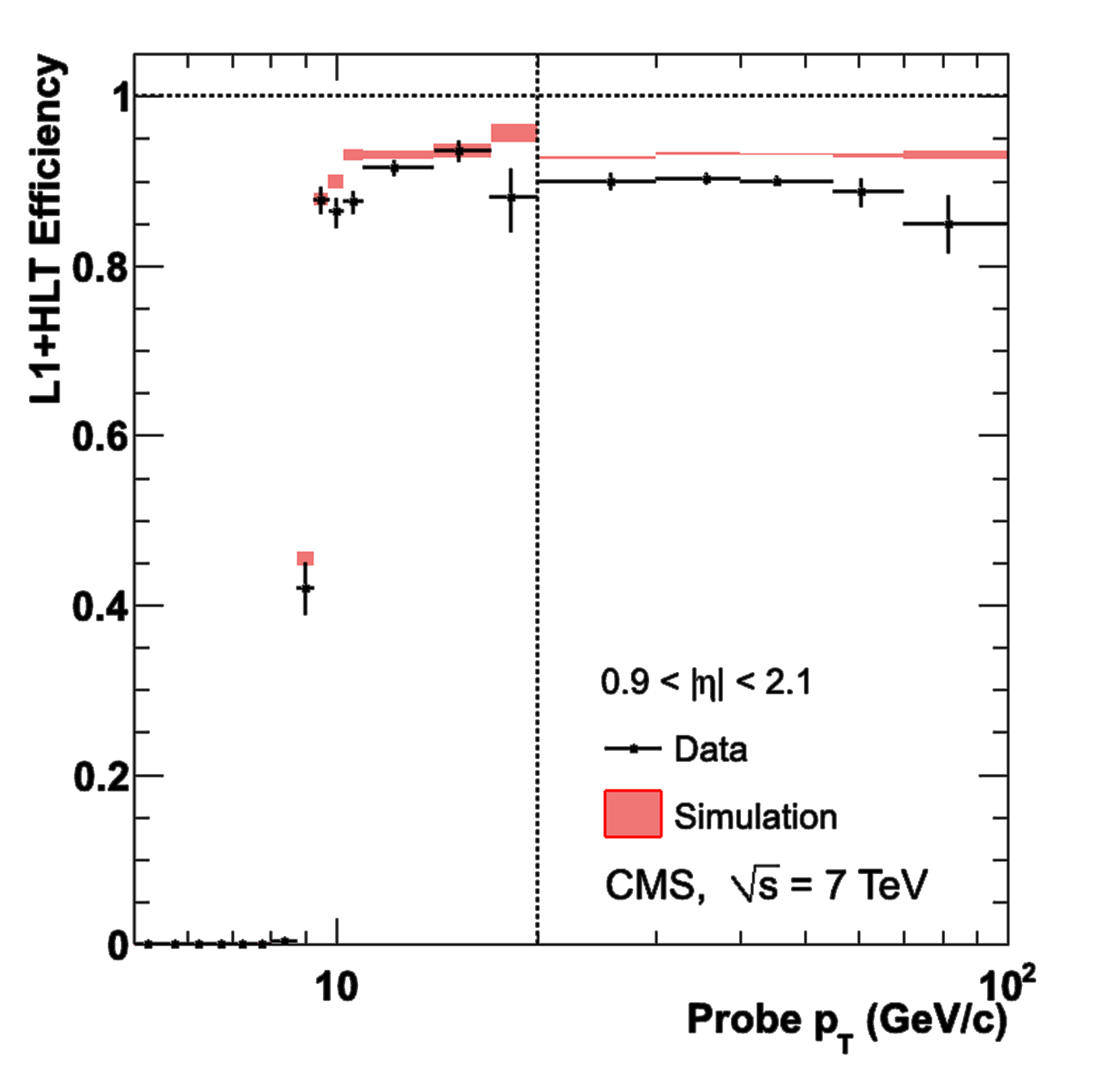}
    \caption{Single-muon trigger efficiencies for Tight Muons as a
      function of the Tight Muon $\pt$ in the barrel (top) and the
      overlap-endcap (bottom) regions.  The measurements are done with
      the tag-and-probe method, using $\mathrm{J}/\!\psi\to\mm$ events
      for $\pt$ below 20\GeVc and $\Zmm$ events above.  The
      efficiencies are shown for the following triggers: \Lone with
      $\pt>7 \GeVc$ threshold (left), HLT with $\pt$ threshold at 9\GeVc
      for $\pt$ below 20\GeVc and at 15\GeVc above
      (centre), and the combination of the above \Lone and HLT triggers
      (right).  The efficiencies in data (points with error bars) are
      compared with predictions from the simulation ($\pm 1\sigma$ bands);
      the uncertainties are statistical only.}
    \label{fig:JpsiZ}
  \end{center}
\end{figure}

\begin{figure}[thb]
  \begin{center}
    \includegraphics[width=0.32\textwidth]{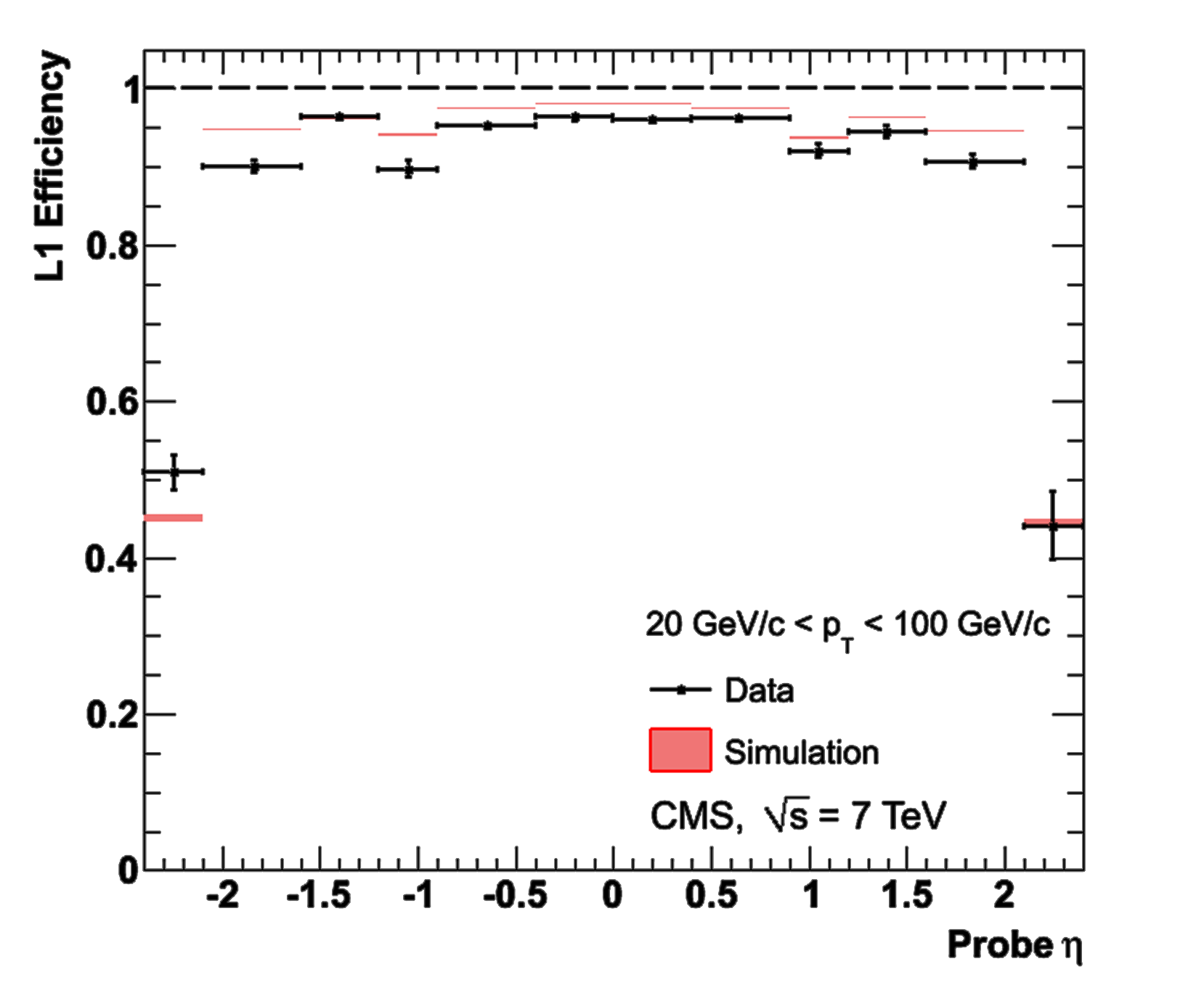}
    \includegraphics[width=0.32\textwidth]{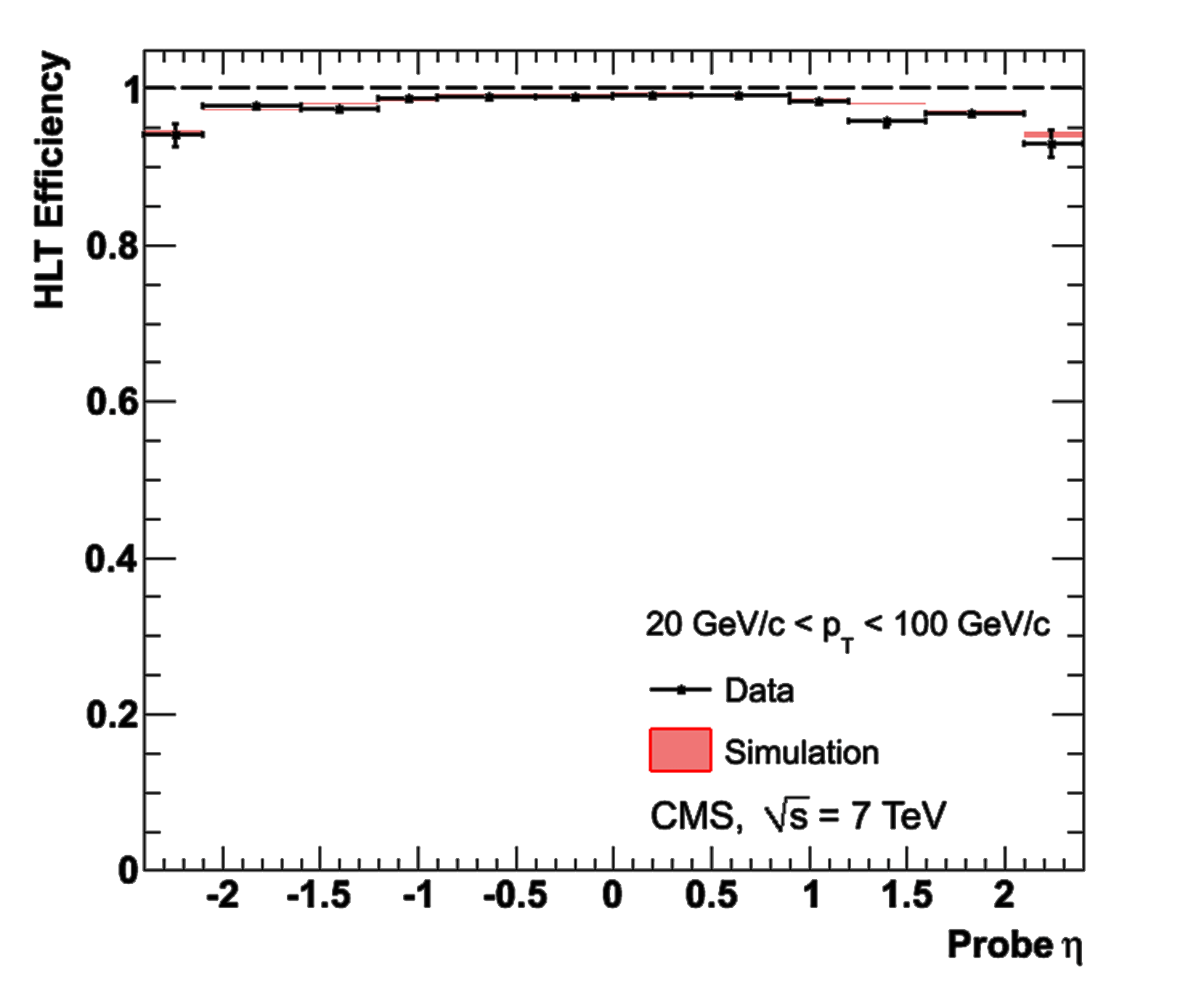}
    \includegraphics[width=0.32\textwidth]{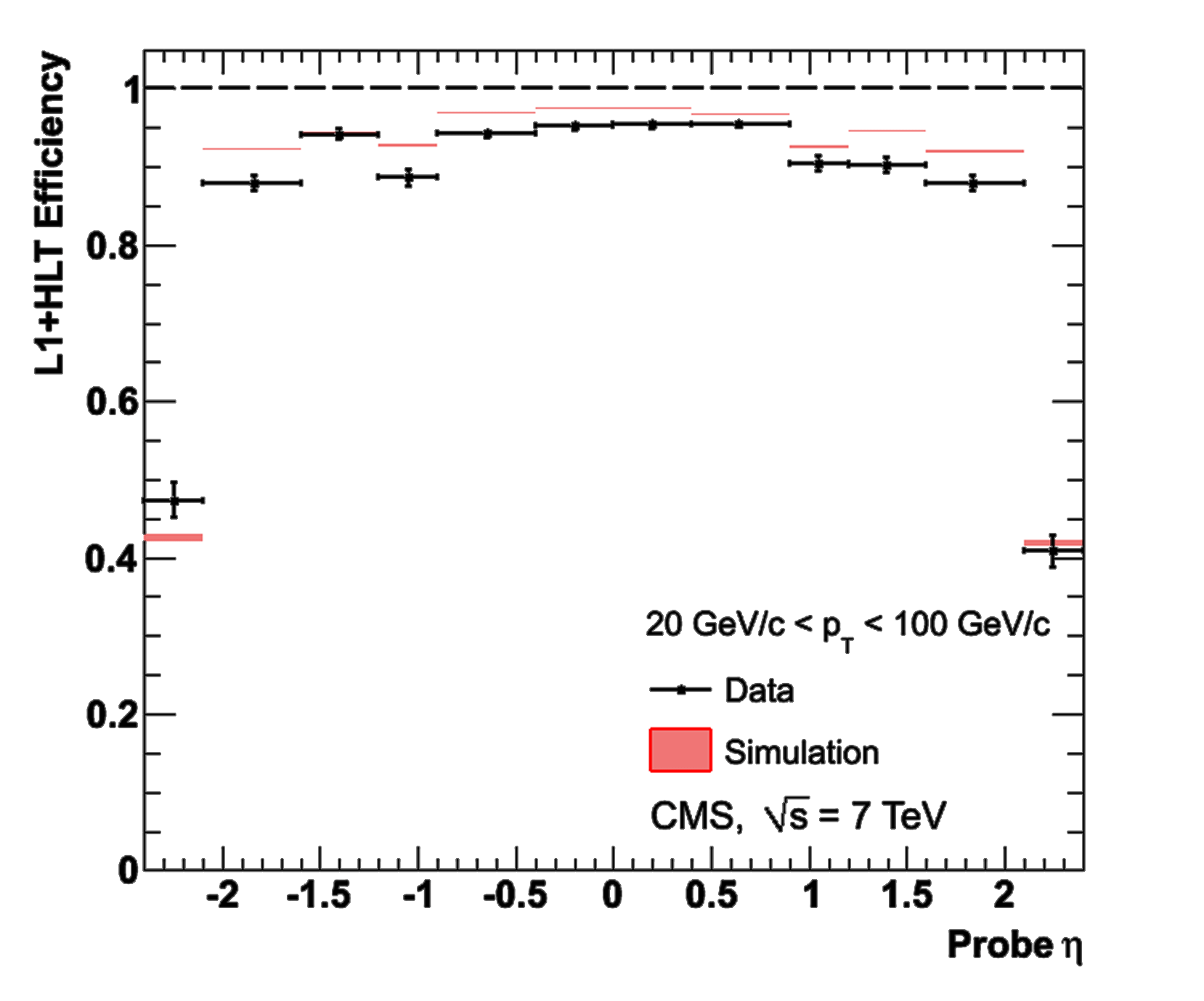}
    \caption{Single-muon trigger efficiencies for Tight Muons with $\pt >
      20\GeVc$ as a function of the Tight Muon $\eta$: the efficiency
      of the \Lone trigger with $\pt$ threshold at 7\GeVc (left), the
      efficiency of the HLT with $\pt$ threshold of 15\GeVc with
      respect to \Lone (middle), and the combined efficiency of \Lone and
      HLT (right).  The efficiencies obtained using $\Zmm$ events
      (points with error bars) are compared with predictions from the MC
      simulation ($\pm 1\sigma$ bands); the uncertainties are statistical
      only.}
    \label{fig:ZL1andHLTeta}
  \end{center}
\end{figure}

Using this combination of these two resonances, we can study efficiencies
of the \Lone trigger with $\pt$ threshold at 7\GeVc and of two high-level
triggers, with thresholds at 9 and 15\GeVc.  These efficiencies,
together with the combined efficiency of the \Lone and HLT, are shown
in Fig.~\ref{fig:JpsiZ} as a function of muon $\pt$.  The efficiencies
for muons with $\pt > 20 \GeVc$ evaluated using $\Zmm$ events
are shown as a function of muon $\eta$ in Fig.~\ref{fig:ZL1andHLTeta} and
are summarized in the first two columns of Table~\ref{tab:triggereffSummary}.

\begin{table}[thb!]
\begin{center}
\topcaption{Summary of \Lone, HLT, and overall trigger efficiencies for Tight
Muons at the efficiency plateau in different pseudorapidity regions.
The results obtained using muons from $\Z$ decays
and single isolated muons collected with jet triggers are averaged over the $\pt$ range of 20--100\GeVc; the
results from single muons contained in b-tagged jets are reported without an upper $\pt$ limit.
For each method, the first column shows
efficiencies measured from data; the second column shows the ratios
between the estimates in data and in simulation.
The quoted uncertainties are purely statistical.}
\begin{tabular}{|c|cc|cc||cc|} \hline
\multicolumn{1}{|c|}{Trigger Level}    & \multicolumn{2}{c|}{$\Zmm$}
                                       & \multicolumn{2}{c||} {Single isolated $\mu$}
                                       & \multicolumn{2}{c|}{Single $\mu$ in b jets} \\
\multicolumn{1}{|r|}{Region}           & Eff. [\%]    &  Data/MC &  Eff. [\%] &  Data/MC  &  Eff. [\%] &  Data/MC  \\ \hline \hline
\multicolumn{1}{|c|}{\Lone}
                                       &              &
                                       &              &
                                       &              &                 \\
\multicolumn{1}{|r|}{$|\eta|<2.1$}     & 94.1$\pm$0.2 & 0.976$\pm$0.002
                                       & 92.9$\pm$0.8 & 0.966$\pm$0.011
                                       & 94.1$\pm$0.3 & 0.975$\pm$0.003 \\
\multicolumn{1}{|r|}{$|\eta|<0.9$}     & 95.9$\pm$0.2 & 0.981$\pm$0.002
                                       & 94.7$\pm$1.0 & 0.971$\pm$0.013
                                       & 95.4$\pm$0.3 & 0.978$\pm$0.004 \\
\multicolumn{1}{|r|}{$0.9<|\eta|<2.1$} & 92.3$\pm$0.3 & 0.971$\pm$0.004
                                       & 91.1$\pm$1.2 & 0.961$\pm$0.017
                                       & 92.2$\pm$0.5 & 0.971$\pm$0.006 \\
                   \hline
\multicolumn{1}{|c|}{HLT}
                                       &              &
                                       &              &
                                       &              &                 \\
\multicolumn{1}{|r|}{$|\eta|<2.1$}     & 98.2$\pm$0.1 & 0.996$\pm$0.001
                                       & 97.8$\pm$0.5 & 0.993$\pm$0.007
                                       & 94.4$\pm$0.3 & 0.977$\pm$0.003 \\
\multicolumn{1}{|r|}{$|\eta|<0.9$}     & 99.0$\pm$0.1 & 0.996$\pm$0.001
                                       & 98.6$\pm$0.5 & 0.996$\pm$0.008
                                       & 96.9$\pm$0.3 & 0.988$\pm$0.003 \\
\multicolumn{1}{|r|}{$0.9<|\eta|<2.1$} & 97.4$\pm$0.2 & 0.995$\pm$0.002
                                       & 97.1$\pm$0.8 & 0.991$\pm$0.011
                                       & 90.5$\pm$0.6 & 0.959$\pm$0.007 \\
                   \hline
\multicolumn{1}{|c|}{Level-1+HLT}
                                       &              &
                                       &              &
                                       &              &                 \\
\multicolumn{1}{|r|}{$|\eta|<2.1$}     & 92.5$\pm$0.3 & 0.972$\pm$0.003
                                       & 90.8$\pm$0.9 & 0.959$\pm$0.013
                                       & 88.9$\pm$0.4 & 0.953$\pm$0.004 \\
\multicolumn{1}{|r|}{$|\eta|<0.9$}     & 95.0$\pm$0.3 & 0.978$\pm$0.003
                                       & 93.3$\pm$1.1 & 0.967$\pm$0.015
                                       & 92.5$\pm$0.4 & 0.967$\pm$0.005 \\
\multicolumn{1}{|r|}{$0.9<|\eta|<2.1$} & 89.9$\pm$0.4 & 0.966$\pm$0.004
                                       & 88.5$\pm$1.4 & 0.952$\pm$0.020
                                       & 83.5$\pm$0.8 & 0.931$\pm$0.008 \\
                   \hline
\end{tabular}
\label{tab:triggereffSummary}
\end{center}
\end{table}

Like the trigger efficiencies for Soft Muons, the efficiency
curves for Tight Muons show a rapid turn-on of efficiencies near the
applied threshold: for \Lone threshold at $\pt = 7\GeVc$ and HLT
threshold at $\pt = 9\GeVc$, the plateau is reached at $\pt \approx 10\GeVc$.
The turn-on region is well reproduced by the simulation.  The
combined Level-1 and HLT efficiency at the plateau is about 95\% in the
region of $|\eta| < 0.9$ and about 90\% in $0.9 < |\eta| < 2.1$.  Most
of the efficiency loss occurs at \Lone and, in particular, in the
overlap-endcap region for the reasons described in Section~\ref{sec:triggerSoft}.
In the very forward region, $2.1 < |\eta| < 2.4$, the single-muon trigger
efficiency is about 45\%.
Most of the losses are due to the
underestimation of $\pt$ by the \Lone $\pt$-assignment algorithm
described in Section~\ref{sec:triggerSoft}; this algorithm has been
modified during the 2010--11 winter technical stop of the LHC to yield
higher efficiency for muons with $\pt$ larger than 20\GeVc.
The difference between the measured and
predicted average plateau efficiencies (see
Table~\ref{tab:triggereffSummary}) is within 0.5\% for the HLT and
of the order of 2--3\% for the \Lone.

To estimate the effect of multiple interactions on the muon
trigger performance, the measurement of the efficiency at the plateau
is performed as a function of the number of reconstructed primary vertices in the event.
Just as for muon identification, no loss in efficiency has been observed for events containing
up to six reconstructed primary vertices.

\subsubsection{Systematic uncertainties}

The studies of systematic uncertainties in the data-to-simulation
ratios of trigger efficiencies closely followed those performed for
the muon reconstruction and identification efficiencies
(Section~\ref{sec:muonideff_syst}) and focused on lineshape modelling
and background subtraction.

The efficiencies obtained by applying the tag-and-probe method to the
simulated samples of mu\-ons are compared with the ``true'' efficiencies
computed by simple counting of the passing and failing probes in simulated
$\mathrm{J}/\!\psi\to\mm$ and $\Zmm$ events. The results are in good
agreement: the difference in the efficiencies is smaller than 0.1\%
for muons in both samples, well within the statistical uncertainties
of the measurements.

For $\mathrm{J}/\!\psi\to\mm$ events in data, the efficiencies are recomputed
using a simple Gaussian instead of a Crystal Ball function~\cite{crystalball} to model the resonance and with a quadratic
polynomial instead of an exponential to model the background.
Like the results reported in Section~\ref{sec:muonideff},
the differences in the efficiencies resulting from this variation in assumed signal shape are smaller than 0.1\%;
the changes in efficiencies due to a different background parametrization
are less than 0.2\%. The same
test has been made for $\Zmm$ events: using a quadratic
polynomial instead of an exponential to model the background results in the efficiency changes of 0.05\%. In all cases,
the difference between the two results gives an estimate
of possible systematic uncertainty due to the modelling of signal and backgrounds.

Finally, for the efficiencies calculated using $\mathrm{J}/\!\psi$ events,
the uncertainty due to residual
correlation effects between the two muons has been studied by
changing the separation criteria as described in
Section~\ref{sec:muonideff_syst}.
The effect
on the measured ratio of efficiencies in data and in simulation is about 0.3\% in the barrel
and less than 0.1\% in the endcaps. % this difference has been added to

\subsubsection{Efficiency of online muon isolation requirements}
To further reduce the trigger rate, isolation criteria can be applied at the
HLT. The isolation requirements are based on the calorimeter information for
the \Ltwo muons and on the pixel tracks for the \Lthree muons~\cite{HLTTDR}.

The efficiency of online isolation requirements was evaluated by
applying the tag-and-probe technique to muons from $\Z$ decays.  To
obtain the trigger efficiency for muons that are commonly used in
physics analyses, the probe is required to pass the Tight Muon
selection.  Furthermore, to determine the additional effect of
online isolation criteria, we require that the probe muon match the trigger
object that passed a non-isolated muon trigger of the same threshold.

\begin{figure}[hbtp]
  \begin{center}
    \includegraphics[width=0.32\textwidth]{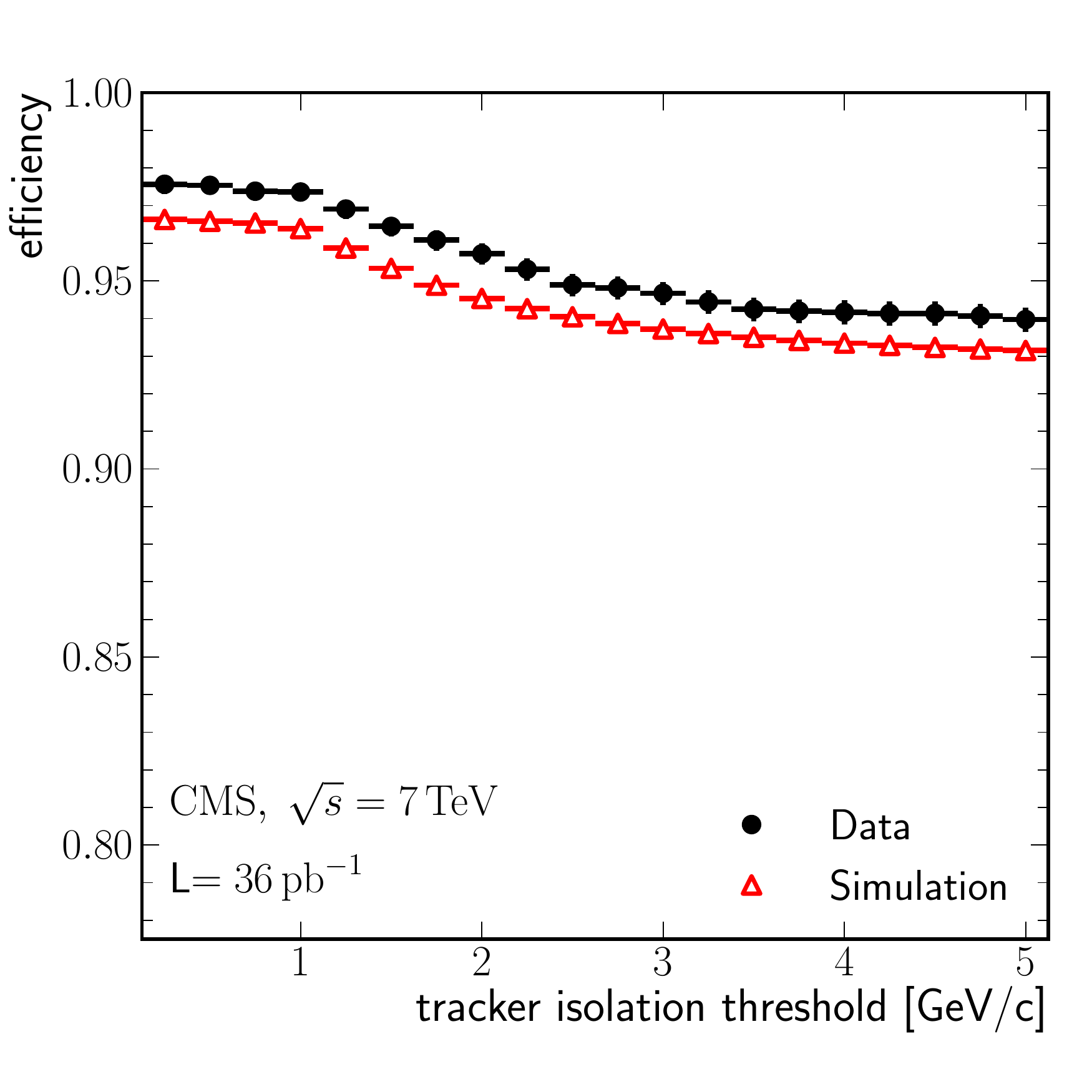}
    \includegraphics[width=0.32\textwidth]{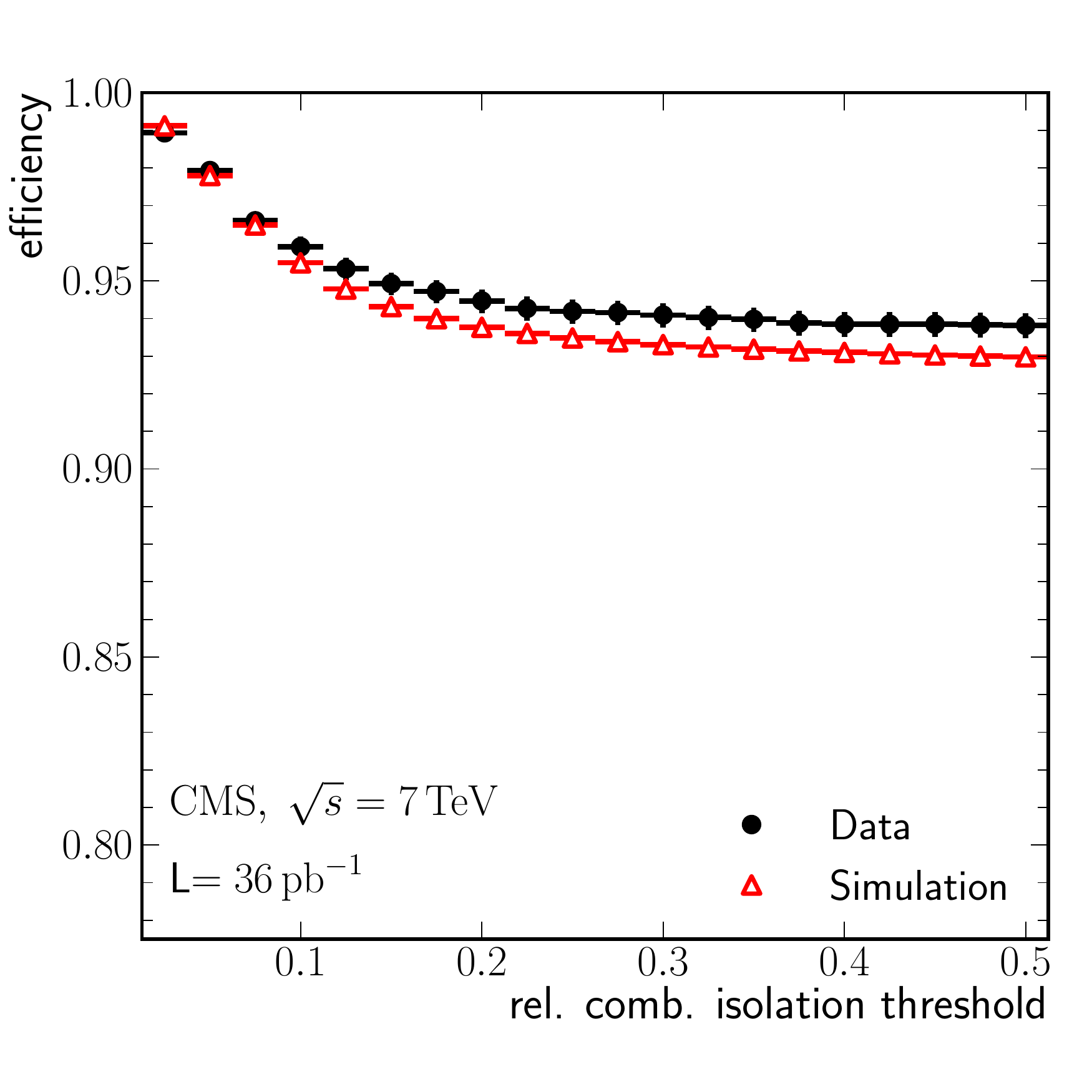}
    \includegraphics[width=0.32\textwidth]{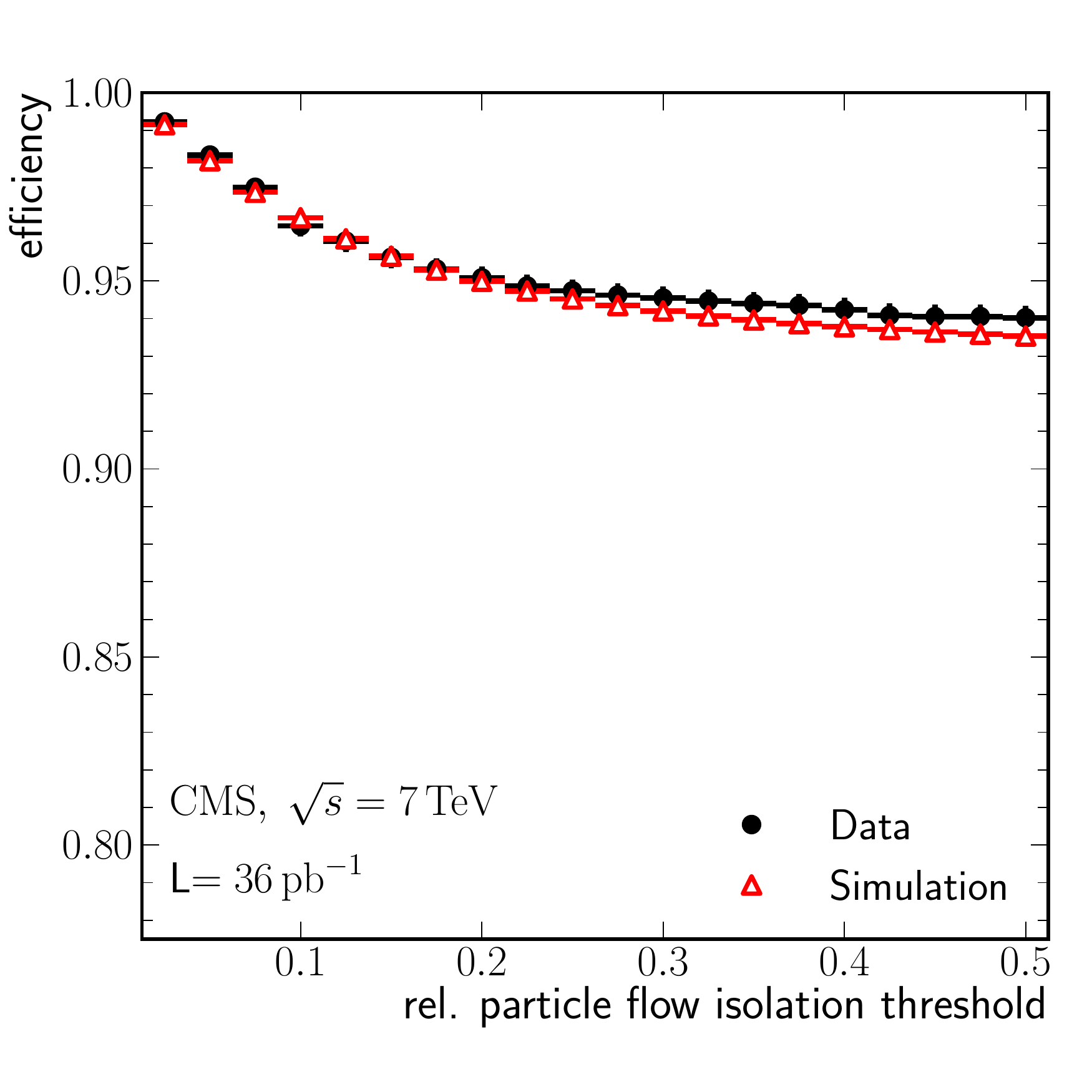}
    \caption{Efficiency of single-muon trigger including isolation
      requirements for Tight Muons matched with the trigger
      objects passing trigger without isolation.  Efficiencies
      measured in data (circles) are compared with the predictions of
      the MC simulation (triangles) as a function of the thresholds
      applied to the offline isolation variables: tracker absolute
      isolation (left), tracker-plus-calorimeters relative isolation
      (middle), and the particle-flow relative isolation (right).}
    \label{fig:isolated_muon_trigger_absolute1}
  \end{center}
\end{figure}

Figure~\ref{fig:isolated_muon_trigger_absolute1} shows the efficiency
of the online isolation selection as a function of the threshold
applied to each of the three offline isolation variables described in
Section~\ref{sec:isolation}: the tracker absolute isolation (the
numerator of $\IRelTrk$), the tracker-plus-calorimeters relative
isolation ($\IRelComb$), and the particle-flow relative isolation ($\IPF$).
The probes are Tight Muons with $|\eta|<2.1$ and $\pt > 20 \GeVc$,
matched with the muon trigger candidates that passed the \Lone trigger
with a $\pt$ threshold of 7\GeVc and HLT with a $\pt$ threshold of
9\GeVc.  We then ask whether the matched trigger
object passes the combination of the \Lone and HLT triggers with the
same $\pt$ thresholds, but also including the isolation criteria.

The efficiency as a function of the threshold on the tracker-based
isolation shows a drop around 1\GeVc. This is a
direct consequence of the thresholds implemented in the
pixel-track-based isolation at \Lthree.  For the typical values of the thresholds
applied offline (see Section~\ref{sec:isolation}), an additional
efficiency loss due to online isolation requirements is
of the order of 4--5\%.  The measured efficiency distributions are well
described by the simulation, with differences of the order of 1\%
or less over the wide range of threshold values.

\subsection{Muon trigger efficiencies from jet-triggered samples}
\label{sec:trigger_inclmu}

While the tag-and-probe technique using muons from J/$\psi$ or $\Z$ allows one to determine
trigger efficiencies for prompt signal muons with sufficiently small bias, 
the momentum range over which the efficiency can be probed is 
restricted to the momentum range covered by muons from the decays of these resonances.

As an alternative to the tag-and-probe method, it is useful to study the inclusive trigger efficiency for single muons 
reconstructed offline, in a sample collected
using a trigger not involving the muon and tracker systems.  In this section,
we report trigger efficiencies obtained from 2010 data samples recorded
with jet triggers, which use only energy measurements in the calorimeters.
We use data collected using two single-jet triggers, with thresholds
on uncorrected jet $\pt$ of 70 and 100\GeVc, as well as events
in which the sum of all uncorrected transverse energy $\HT$ from jets
is larger than 100\GeV.
To reduce background, the probes are required to be Tight Muons. They are matched to the HLT muon using 
the track direction at the vertex. Mismatches are reduced by requiring
that only one reconstructed muon be present in the event.

Backgrounds from pion and kaon decays, as well as remaining hadron
punch-through, lower the estimated efficiency.
From Table~\ref{tab:compositionHighPt}
we expect these backgrounds to be at the level of about 10\% for Tight Muons with $\pt > 20$\GeVc.
A higher rejection of these backgrounds can be achieved by applying two alternative and independent additional selections:
\begin{itemize}
\item Isolation, which selects 
 mainly muons from $\W$ decays and suppresses the heavy-flavour contributions that are typically non-isolated.
\item B tagging, which selects muons from semileptonic heavy-flavour decays.
 Jets are reconstructed by the particle-flow algorithm~\cite{PFT-09-001},
 and a b-tagging criterion~\cite{BTV-11-001} is imposed on the jet associated
 with the muon 
 by requiring a secondary vertex formed from at least two high-quality tracks.
\end{itemize} 

\begin{figure}[thb!]
\begin{center}
\includegraphics[width=0.48\textwidth]{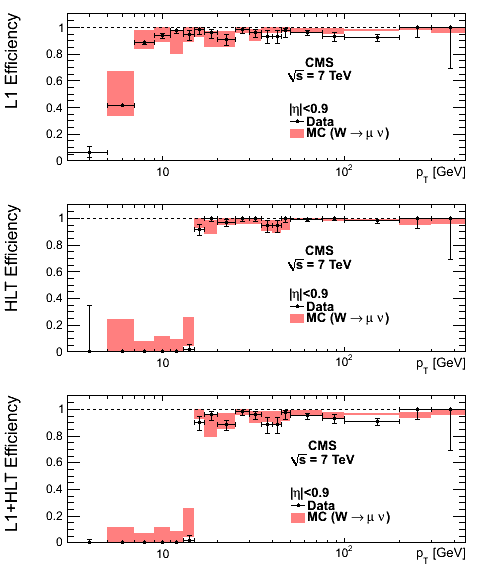}
\includegraphics[width=0.48\textwidth]{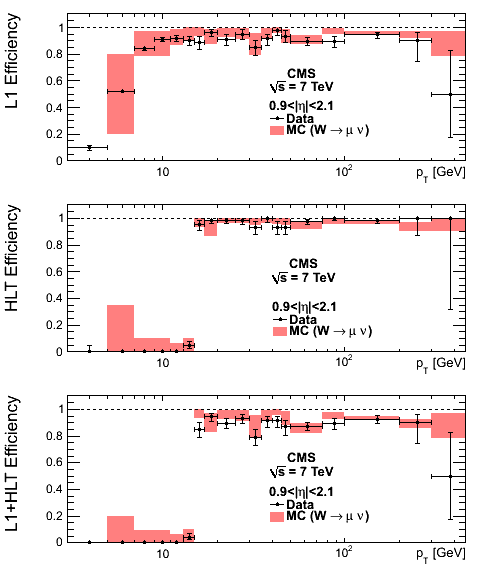}
\caption{Single-muon trigger efficiencies for Tight Muons passing the offline isolation selection
        as a function of the muon $\pt$, in the barrel ($|\eta|<0.9$, left) 
        and the overlap-endcap ($0.9<|\eta|<2.1$, right) regions. In both regions 
        the efficiencies for the following triggers are shown: the \Lone trigger with $\pt$ threshold at 7\GeVc (top), 
        the HLT with $\pt$ threshold at 15\GeVc with respect to the \Lone (middle),
	and the combination of above \Lone and HLT triggers (bottom).
        The efficiencies in data (points with error bars) are
        compared with predictions from the simulation ($\pm 1\sigma$ bands);
        the uncertainties are statistical only.}
\label{fig:TriggerEffsIsoMuons}
\end{center}
\end{figure}

Trigger efficiencies obtained using isolated muons are shown in Fig.~\ref{fig:TriggerEffsIsoMuons} as a function of the reconstructed muon $\pt$, both
for data and for simulated $\Wmn$ decays.
To select isolated muons, the tracker-plus-calorimeters relative isolation is required to be $\IRelComb < 0.15$.
The $\pt$ spectrum of muons in the sample thus obtained extends beyond 300\GeVc, thus probing a significantly wider 
momentum range than the tag-and-probe method. 
The efficiency beyond $\pt > 20$\GeVc is consistent with being flat.
The average plateau efficiencies are reported in the third and fourth columns
of Table~\ref{tab:triggereffSummary}. 
They can be directly compared to the results from the tag-and-probe
method at the $\Zmm$ resonance, which effectively uses isolated muons as well.
The results from the two methods are in good agreement.

\begin{figure}[thb!]
\begin{center}
\includegraphics[width=0.32\textwidth]{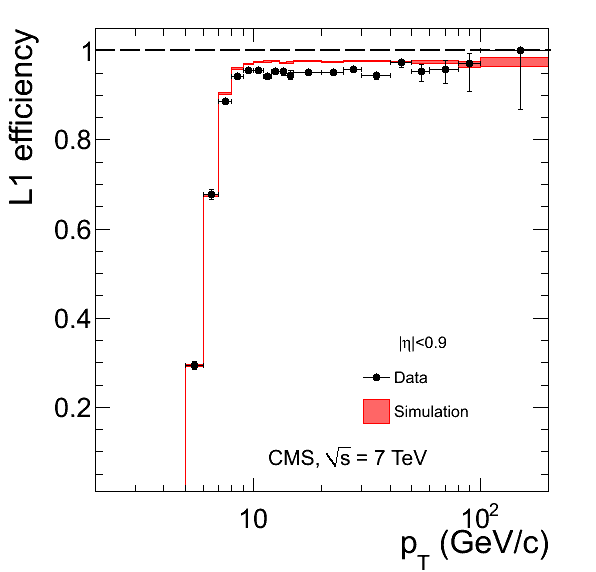}
\includegraphics[width=0.32\textwidth]{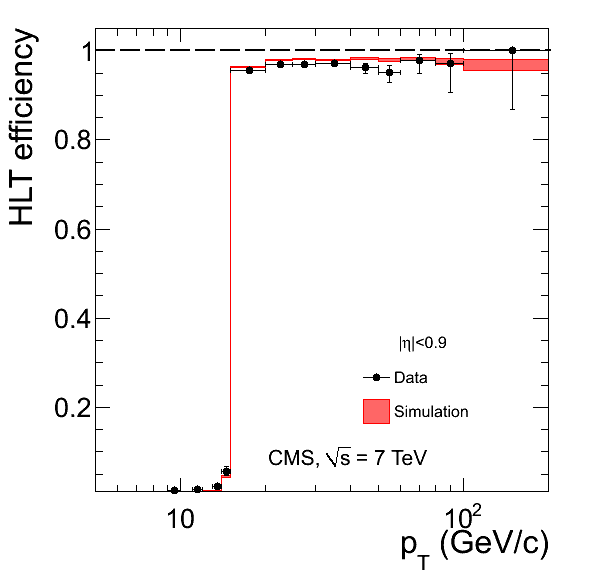}
\includegraphics[width=0.32\textwidth]{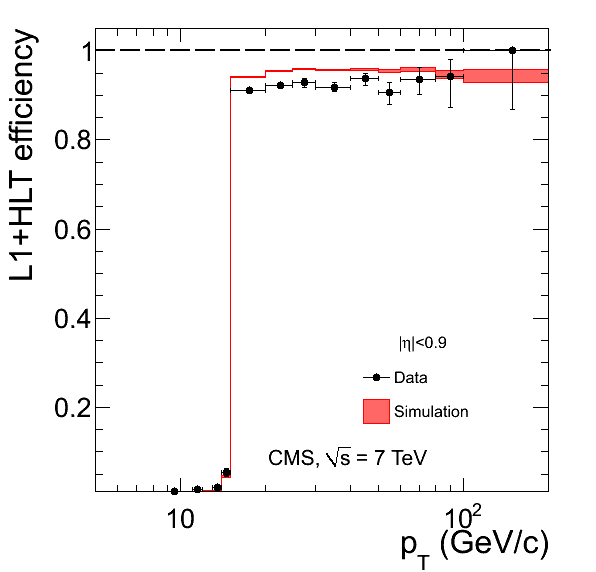}
\includegraphics[width=0.32\textwidth]{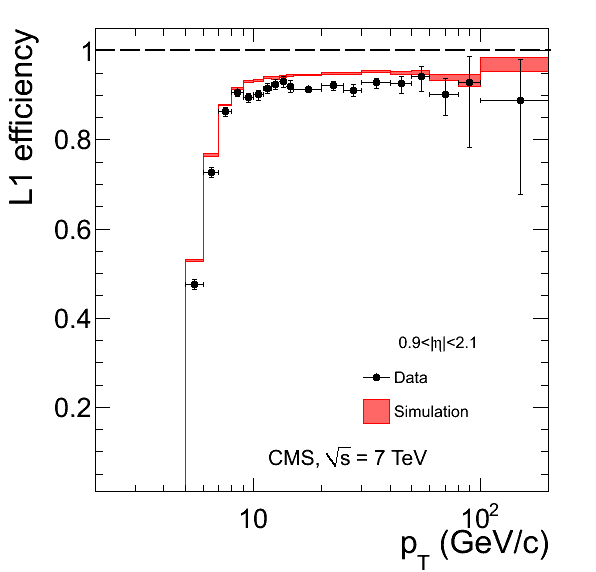}
\includegraphics[width=0.32\textwidth]{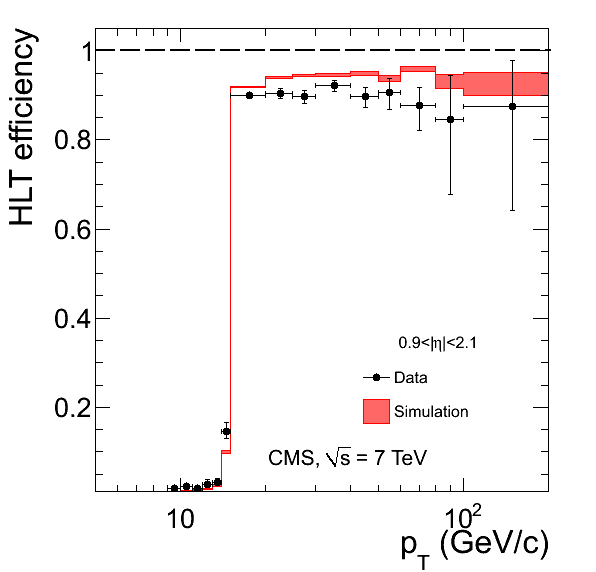}
\includegraphics[width=0.32\textwidth]{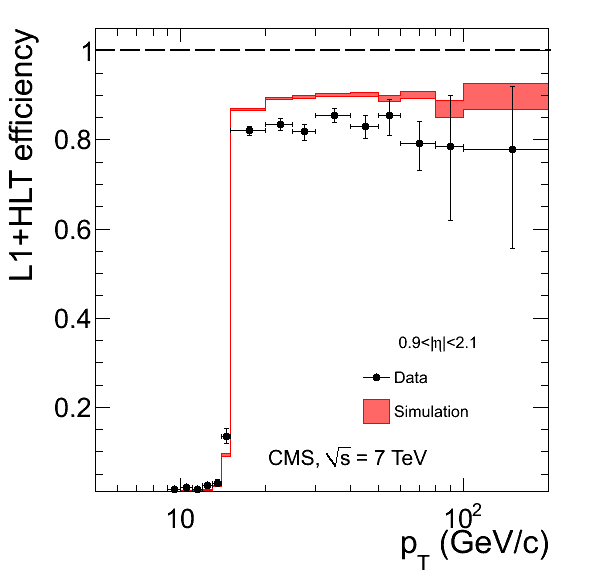}
\caption{Single-muon trigger efficiencies for Tight Muons contained in b-tagged jets
        as a function of the muon $\pt$, in the barrel (top) 
        and the overlap-endcap (bottom) regions.
        The efficiencies are shown for the following triggers: the \Lone trigger with $\pt$ threshold at 7\GeVc (left),
        the HLT with $\pt$ threshold at 15\GeVc with respect to the \Lone (middle),
	and the combination of above \Lone and HLT triggers (right).
        The efficiencies in data (points with error bars) are
        compared with predictions from the simulation ($\pm 1\sigma$ bands);
        the uncertainties are statistical only.}
\label{fig:TriggerEffsBtagJet}
\end{center}
\end{figure}

\begin{figure}[thb!]
\begin{center}
\includegraphics[width=0.32\textwidth]{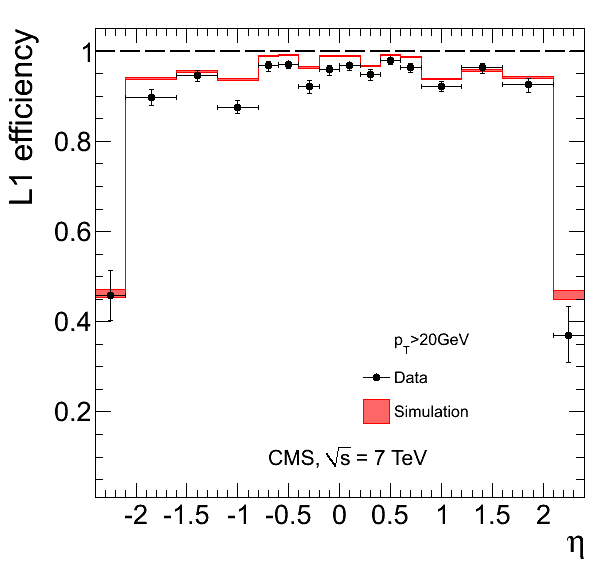}
\includegraphics[width=0.32\textwidth]{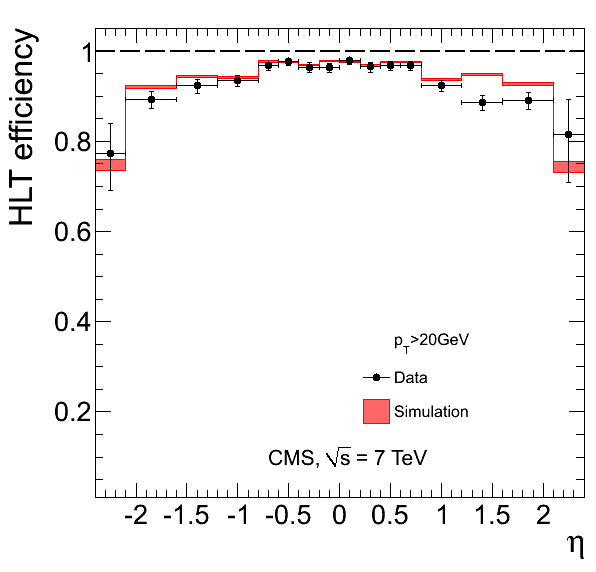}
\includegraphics[width=0.32\textwidth]{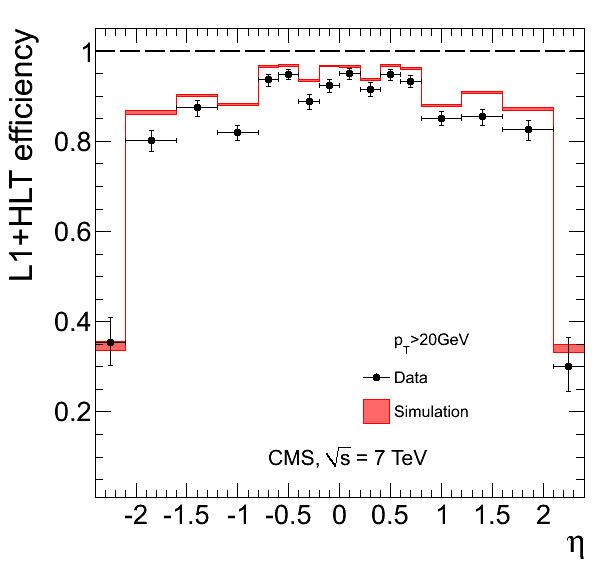}
\caption{Single-muon trigger efficiencies for Tight Muons contained in b-tagged jets
        as a function of the muon pseudorapidity, for muons with $\pt > 20$\GeVc:
        the efficiency of the \Lone trigger with $\pt$ threshold at 7\GeVc
        (left), the efficiency of HLT with $\pt$ threshold at 15\GeVc with
        respect to the \Lone (middle), and
	the combined efficiency of \Lone and HLT (right).
        The efficiencies in data (points with error bars) are
        compared with predictions from the simulation ($\pm 1\sigma$ bands);
        the uncertainties are statistical only.}
\label{fig:TriggerEffsBtagJetEta}
\end{center}
\end{figure}

The trigger efficiencies for Tight Muons in jets, obtained from the b-tagging approach,
are shown in Fig.~\ref{fig:TriggerEffsBtagJet} for data and for a muon-enriched sample of minimum-bias events generated using \PYTHIA.
The simulation includes the same jet-trigger requirements as used in the data.
The shape of the efficiency curves is well reproduced by the simulation.
In the plateau above 20\GeVc the data and simulation agree to about the 5\% level.
The plateau trigger efficiency as a function of the muon pseudorapidity is shown in Fig.~\ref{fig:TriggerEffsBtagJetEta}
and summarized in the last two columns of Table~\ref{tab:triggereffSummary}.

The systematic uncertainty due to possible residual background from hadron punch-through and muons from light flavours surviving the b-tagging selection was studied by varying the selection criteria.
The results do not change significantly if
\begin{itemize}
\item before applying the b-tagging condition, the offline muon requirements are further tightened
by imposing more stringent selection criteria on the $\chi^2$ of the tracker-only and the global 
track fits, and by increasing the minimum required number of pixel and strip tracker hits;
\item the b-tagging condition is tightened,
by requiring at least three instead of the default two high-quality tracks associated with the secondary vertex;
\item only those events are used in which the muon track is matched to the secondary vertex by the b-tagging algorithm.
\end{itemize}
Performing the same checks on the sample of simulated events leads to similar results.
In addition, we verified in simulation that the efficiency determined for muons from heavy-flavour decays
agrees with that determined by applying the b tagging to the full Monte Carlo sample including backgrounds.
This demonstrates that b tagging is very effective in suppressing the background. 
However, due to pre-selection requirements applied at the generator level, the simulated sample of muon-enriched minimum-bias events used in this study lacks 
punch-through and light-hadron decays occurring downstream of the ECAL, 
which results in a smaller impact of the b tagging on the muon selection in simulation than in data.
Therefore we consider the difference between the results obtained with and without the b tagging, which is 
smaller than 1\% both in the barrel and endcap, to be a conservative estimate of the systematic 
uncertainty due to background.

\begin{figure}[thb!]
\begin{center}
\includegraphics[width=0.45\textwidth]{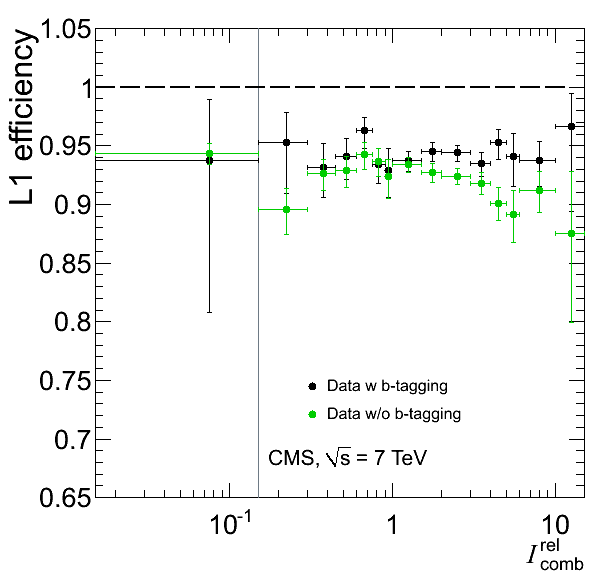}
\includegraphics[width=0.45\textwidth]{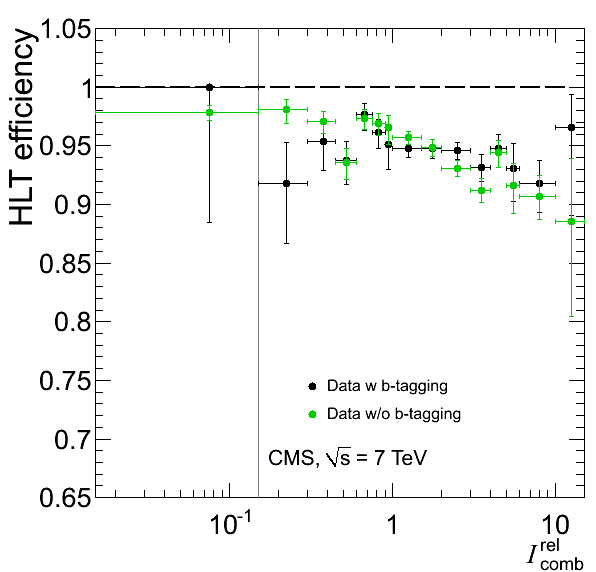}
\caption{Single-muon trigger efficiencies for Tight Muons with $\pt > 20 \GeVc$
and $|\eta| < 2.1$, with and without the further b-tagging
selection, as a function of the muon isolation variable $\IRelComb$: 
  the \Lone trigger (left), and the HLT with respect to the \Lone (right).}
\label{fig:effVsIso}
\end{center}
\end{figure}

The trigger efficiencies determined for muons in b jets are lower than those obtained for isolated muons
in similar ranges of muon transverse momentum. The dependence of trigger efficiency on isolation is demonstrated in Fig.~\ref{fig:effVsIso}, which shows
\Lone and HLT efficiencies for muons in a jet-triggered sample versus the tracker-plus-calorimeter relative isolation $\IRelComb$.
The thin vertical lines at 0.15 show the typical value defining the isolated muon selection. 
Without the b tagging, a slight dependence of the \Lone efficiency
on isolation can be seen, due to the residual background contamination;
the efficiency becomes flat once the b tagging is applied. 
The HLT efficiency, on the other hand, has a clear dependence on the muon isolation, even after b tagging. 
The downward trend has a steeper slope without the b-tagging condition due to the larger background. 
Similar conclusions are obtained by studying other isolation variables, 
\eg, the track multiplicity within an angular cone around the muon direction.
A worse performance of the single-muon trigger at HLT for muons within high-multiplicity jets is not unexpected.
The precision in the matching of the Level-2 muon candidate from the muon chambers with the tracker track from the inner tracker is limited by
time constraints dictated by the online operation.
Hence, with higher track multiplicity in the tracker matching region the probability of wrong association increases.
Nevertheless, the HLT efficiency remains greater than 90\% even in the most unfavourable environment of high multiplicities, 
corresponding to the highest values of $\IRelComb$ in Fig.~\ref{fig:effVsIso}.

The ratios of muon trigger efficiencies derived from jet-triggered
events in data and in simulation are shown in the last column of
Table~\ref{tab:triggereffSummary}.  Since muons in jet-triggered
events tend to be less isolated than those in muon-triggered events,
these corrections are not directly applicable to the muon-triggered
events and must themselves be corrected for the difference in
isolation in the two samples. This is done by reweighting the jet-triggered 
events to match the
distribution of a given isolation quantity in the muon-triggered
events, and then recalculating the corrections.
After the reweighting is done, the data-to-simulation efficiency ratios are closer to unity
than those before reweighting by 0.01 in the region of $|\eta|<0.9$ and by
0.03 in the region of $0.9<|\eta|<2.1$. These final ratios are in
agreement with those obtained with the other two methods, which are shown
in Table~\ref{tab:triggereffSummary}. Reweighting the events using either the $\IRelComb$ distribution or the distribution of the number of tracks in the vicinity of the muon gives statistically compatible results.

The possible effect of event pile-up on the trigger efficiencies was examined by performing the efficiency measurements as a function of 
the number of good reconstructed primary vertices in the event. No significant dependence, within the statistical uncertainty of about 2\%, 
was found in the available range of one to seven reconstructed primary vertices.

\subsection{Trigger rejection rates}

The main task of the trigger is to reduce the rate of events to be recorded, while keeping high
efficiency for the physics-signal events. The two previous sections focused
on the trigger efficiency; in this section
we show what fraction of minimum-bias events is rejected as a function of muon trigger $\pt$ threshold
at \Lone, \Ltwo, and \Lthree, and compare to the prediction from simulation of minimum-bias events.

\begin{figure}[hbtp]
  \begin{center}
     \includegraphics[width=0.60\textwidth]{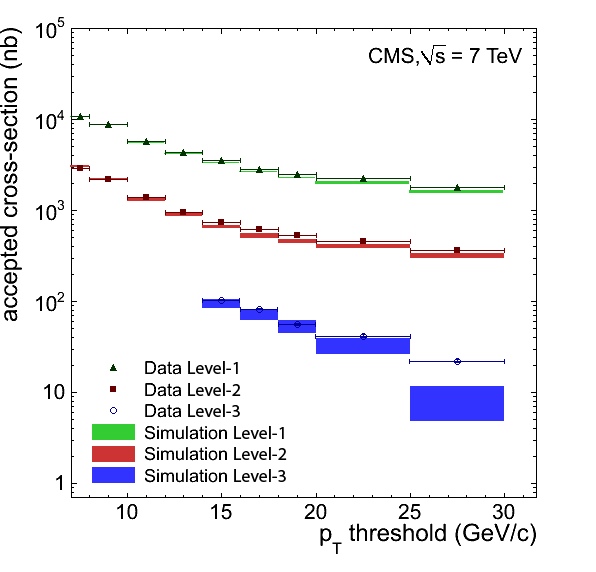}
     \caption{The accepted cross section of events as a function of muon trigger $\pt$ threshold for
              the actual \Lone, \Ltwo, and \Lthree muon trigger objects processed online in data, compared to the emulated
              \Lone and HLT trigger in simulation. Points corresponding to thresholds of prescaled \Lthree triggers are not plotted.
             }
    \label{fig:TriggerRejection}
  \end{center}
\end{figure}

The rates of accepted muon triggers were evaluated on a data sample
collected using the \Lone muon trigger with a $\pt$ threshold of 7\GeVc and
corresponding to an integrated luminosity of 26.5~$\nbinv$.  The resulting cross sections are
shown in Fig.~\ref{fig:TriggerRejection}. At the peak instantaneous
luminosity of about $2 \times 10^{32}$ cm$^{-2}$ Hz, achieved at the end of the LHC operation in 2010,
the rate of \Lone muon trigger with the $\pt$ threshold of 7\GeVc was about 2~kHz and the rate of the HLT
trigger with the $\pt$ threshold of 15\GeVc was slightly below 20~Hz.

To reproduce the trigger rate, the simulation must reproduce the
correct sample composition and the correct trigger efficiency, and also must correctly
describe the resolution for muons from different sources including the resolution
tails. The level of agreement between the \Lone and HLT trigger performance in data
and the Monte Carlo simulation is generally fairly good, with the simulation slightly underestimating the rates at higher thresholds.
These results demonstrate that the trigger simulation is a useful tool
to predict muon trigger rates and confirm that the performance of the CMS muon trigger system
is close to the design expectation.

\section{Conclusions}
\label{sec:conclusions}

The performance of muon reconstruction, identification, and triggering in CMS
has been studied extensively using 40~pb$^{-1}$ of data collected in pp collisions 
at $\sqrt{s}$ = 7\TeV at the LHC in 2010. These data were used to 
study several representative muon selections, which were chosen as benchmarks
covering a wide range of physics analysis needs.

The distributions of kinematic and identification variables for muons
reconstructed from inclusive data samples are generally in good
agreement with predictions from simulation over the momentum
range of $\pt \lesssim 200\GeVc$, 
including variables sensitive to large hit occupancies in the muon detectors.
Identification efficiencies, for muons with $\pt$ larger than a few $\GeVc$, are 
above 95\% for all selections studied. They are correctly reproduced 
by the Monte Carlo simulations.
Misidentification is lower than $1\%$ for the loosest selection and 
below $0.1\%$ for the tightest. In the specific case of several nearby muons, where the tightest selection becomes less efficient, an optimized 
selection was developed to preserve high efficiency with only a minimal
increase of misidentification.

Two complementary methods have been developed to evaluate muon momentum scale biases and resolution
in the range $20 < \pt < 100 \GeVc$, where the momentum measurement is provided by the tracker.
The average bias $\Delta(\pt)/\pt$ in the muon momentum scale was measured with a precision
of better than 0.2\% and was found to be consistent with zero.
The relative $\pt$ resolution is between 1.3\% to
2.0\% for muons in the barrel and better than $6\%$ in the endcaps, in good agreement with simulation.
The transverse momentum resolution of tracks reconstructed using information from only muon detectors 
is better than 10\% in the barrel region. At high momenta, the best measurement of muon $\pt$ is
obtained by selective use of information from the muon system in addition
to that from the inner tracker.
The $\pt$ resolution, evaluated in the barrel region using muons from cosmic rays, is better than 10\% up to 1\TeVc.

Algorithms to identify cosmic and beam-halo backgrounds among collision
events were developed and successfully used in physics analyses of 2010 data.
The performance of various muon isolation algorithms was shown to be reasonably
well modelled by the simulation.

The muon trigger efficiency for isolated muons is better than $90\%$ over the full $\eta$ range, and is typically substantially better.

In this document we have shown that the performance specifications set out
for the measurement of muons in CMS have largely been met.
The good performance and detailed understanding of the muon reconstruction, identification, and triggering provides the necessary 
confidence in all elements of the chain from muon detection to muon analysis,
which is essential
for searches for physics beyond the Standard Model
as well as accurate Standard Model measurements.

\section{Acknowledgments}
\label{sec:ack}
\hyphenation{Bundes-ministerium Forschungs-gemeinschaft Forschungs-zentren} We wish to congratulate our colleagues in the CERN accelerator departments for the excellent performance of the LHC machine. We thank the technical and administrative staff at CERN and other CMS institutes. This work was supported by the Austrian Federal Ministry of Science and Research; the Belgium Fonds de la Recherche Scientifique, and Fonds voor Wetenschappelijk Onderzoek; the Brazilian Funding Agencies (CNPq, CAPES, FAPERJ, and FAPESP); the Bulgarian Ministry of Education and Science; CERN; the Chinese Academy of Sciences, Ministry of Science and Technology, and National Natural Science Foundation of China; the Colombian Funding Agency (COLCIENCIAS); the Croatian Ministry of Science, Education and Sport; the Research Promotion Foundation, Cyprus; the Estonian Academy of Sciences and NICPB; the Academy of Finland, Finnish Ministry of Education and Culture, and Helsinki Institute of Physics; the Institut National de Physique Nucl\'eaire et de Physique des Particules~/~CNRS, and Commissariat \`a l'\'Energie Atomique et aux \'Energies Alternatives~/~CEA, France; the Bundesministerium f\"ur Bildung und Forschung, Deutsche Forschungsgemeinschaft, and Helmholtz-Gemeinschaft Deutscher Forschungszentren, Germany; the General Secretariat for Research and Technology, Greece; the National Scientific Research Foundation, and National Office for Research and Technology, Hungary; the Department of Atomic Energy and the Department of Science and Technology, India; the Institute for Studies in Theoretical Physics and Mathematics, Iran; the Science Foundation, Ireland; the Istituto Nazionale di Fisica Nucleare, Italy; the Korean Ministry of Education, Science and Technology and the World Class University program of NRF, Korea; the Lithuanian Academy of Sciences; the Mexican Funding Agencies (CINVESTAV, CONACYT, SEP, and UASLP-FAI); the Ministry of Science and Innovation, New Zealand; the Pakistan Atomic Energy Commission; the State Commission for Scientific Research, Poland; the Funda\c{c}\~ao para a Ci\^encia e a Tecnologia, Portugal; JINR (Armenia, Belarus, Georgia, Ukraine, Uzbekistan); the Ministry of Science and Technologies of the Russian Federation, and Russian Ministry of Atomic Energy; the Ministry of Science and Technological Development of Serbia; the Ministerio de Ciencia e Innovaci\'on, and Programa Consolider-Ingenio 2010, Spain; the Swiss Funding Agencies (ETH Board, ETH Zurich, PSI, SNF, UniZH, Canton Zurich, and SER); the National Science Council, Taipei; the Scientific and Technical Research Council of Turkey, and Turkish Atomic Energy Authority; the Science and Technology Facilities Council, UK; the US Department of Energy, and the US National Science Foundation.

Individuals have received support from the Marie-Curie programme and the European Research Council (European Union); the Leventis Foundation; the A. P. Sloan Foundation; the Alexander von Humboldt Foundation; the Associazione per lo Sviluppo Scientifico e Tecnologico del Piemonte (Italy); the Belgian Federal Science Policy Office; the Fonds pour la Formation \`a la Recherche dans l'Industrie et dans l'Agriculture (FRIA-Belgium); the Agentschap voor Innovatie door Wetenschap en Technologie (IWT-Belgium); and the Council of Science and Industrial Research, India. 

\bibliography{auto_generated}   % will be created by the tdr script.

\cleardoublepage \appendix\section{The CMS Collaboration \label{app:collab}}\begin{sloppypar}\hyphenpenalty=5000\widowpenalty=500\clubpenalty=5000\textbf{Yerevan Physics Institute,  Yerevan,  Armenia}\\*[0pt]
S.~Chatrchyan, V.~Khachatryan, A.M.~Sirunyan, A.~Tumasyan
\vskip\cmsinstskip
\textbf{Institut f\"{u}r Hochenergiephysik der OeAW,  Wien,  Austria}\\*[0pt]
W.~Adam, T.~Bergauer, M.~Dragicevic, J.~Er\"{o}, C.~Fabjan, M.~Friedl, R.~Fr\"{u}hwirth, V.M.~Ghete, J.~Hammer\cmsAuthorMark{1}, M.~Hoch, N.~H\"{o}rmann, J.~Hrubec, M.~Jeitler, W.~Kiesenhofer, M.~Krammer, D.~Liko, I.~Mikulec, M.~Pernicka$^{\textrm{\dag}}$, B.~Rahbaran, C.~Rohringer, H.~Rohringer, R.~Sch\"{o}fbeck, J.~Strauss, A.~Taurok, F.~Teischinger, P.~Wagner, W.~Waltenberger, G.~Walzel, E.~Widl, C.-E.~Wulz
% ...
\vskip\cmsinstskip
\dag:~Deceased\\
1:~~Also at CERN, European Organization for Nuclear Research, Geneva, Switzerland\\
% ...

\end{sloppypar}
\end{document}